\documentclass[11pt, a4paper, oneside]{book}
\usepackage[utf8]{inputenc}
\usepackage[T2A,T1]{fontenc}      % T2A required for Cyrillic support
\usepackage[russian,english]{babel} % Supports both languages
\usepackage{geometry}
\usepackage{titlesec}
\usepackage{fancyhdr}
\usepackage{graphicx}
\usepackage{amsmath}
\usepackage{url}
\usepackage{parskip}
\usepackage{enumitem} % For custom list formatting
\usepackage{longtable} % For tables that might span pages
\usepackage{booktabs} % For nicer tables
\usepackage{tikz}
\usetikzlibrary{decorations.pathmorphing}
\usepackage{caption} 
\usepackage{amssymb}
\usepackage{algorithm}
\usepackage{algorithmic}

\titleformat{\chapter}[display]
  {\normalfont\bfseries}{}{0pt}{\Huge}
\titlespacing*{\chapter}{0pt}{-50pt}{40pt}

\newcommand{\anecdote}[2]{
    \par\vspace{0.8cm}      % Vertical space before
    \noindent\textbf{#1}    % Title in bold, no indent
    \par\nopagebreak        % No page break between title and body
    \vspace{0.2cm}          % Small gap between title and body
    {\itshape \noindent #2} % Body in italics, no indent
    \par\vspace{0.8cm}      % Vertical space after
}

\begin{document}

\title{
    \Huge \textbf{PHYSICISTS ARE STILL JOKING} \\ 
    \huge \textit{Almanac of Physics and Science Humor (1966–2025)} \\[0.5cm]
%    \Large (\foreignlanguage{russian}{Физики шутят} \ / \ \foreignlanguage{russian}{Физики всё ещё шутят}) \\[1.5cm] 
%    
    \normalsize 
    \textit{Containing the restored translations of the Russian classics:} \\
    \textbf{Part I: Physicists Joke 
    (\foreignlanguage{russian}{Физики шутят}) (1966)} \\
    \textbf{Part II: Physicists Keep Joking 
    (\foreignlanguage{russian}{Физики всё ещё шутят})  (1992)} \\[0.5cm]
    
    \textit{And a new collection of English-language folklore:} \\
    \textbf{Part III: Still Joking: Modern Physics and Science Humor (1990s-2025)}
}
\author{
    \textbf{Original Russian Editors (1966--1992):} \\
    Yu. Konobeev, V. Pavlinchuk, N. Rabotnov, V. Turchin \\
    \vspace{0.2cm} \\
    \textbf{English Translation, Restoration, and New Compilation (2025):} \\
    \textbf{Igor Halperin} \\
    \textit{with assistance from Google Gemini 3 Pro}
}

\date{Original Publishers: Mir (1966), Maket (1992) \\
This Edition: December 24, 2025 \\[1.5cm]
\parbox{0.9\textwidth}{
    {\centering \textbf{\large Abstract}\\[0.3cm]} % Scoped centering for title only
    \small
    \textit{This volume, \textbf{Physicists Are Still Joking}, serves as a definitive almanac of scientific humor spanning sixty years. It traces the evolution of professional folklore across geopolitical divides and technological eras. \textbf{Part I} restores the classic 1966 anthology \textbf{Physicists Joke}, which originally served as a window for Soviet scientists into the best traditions of Western scientific humor; it consists primarily of articles translated from English, here meticulously restored to their original wording. \textbf{Part II} presents the 1992 sequel, \textbf{Physicists Keep Joking}, which captures the shift toward an original, introspective Russian scientific folklore born during the end of the Cold War and the collapse of the Soviet Union. \textbf{Part III: Still Joking} explores the modern digital age, compiling contemporary science humor from physics, astronomy, biology, computer science and AI research. While the tools of science have evolved from slide rules to neural networks, the tradition of skeptical, self-referential wit remains a constant. Spanning from the "Golden Age" of vacuum tubes to the era of AI and Large Language Models, this collection documents the enduring ability of scientists to laugh at the universe and themselves.}
}}

\maketitle

\frontmatter

\chapter{Preface to the First Edition}

The genesis of this almanac lies in an attempt to preserve and restore a unique cultural artifact of the Cold War era: the Soviet anthology \textit{Physicists Joke} (\foreignlanguage{russian}{Физики шутят}), originally published in 1966.

The 1966 collection was a cultural phenomenon in the USSR. It offered a glimpse into the sociology of the "Golden Age" of physics—a time of cyclotrons, slide rules, and unbounded optimism. However, it is important to note that the majority of that book consisted of Russian translations of articles originally written in English and German. It was, in effect, a curation of the best Western scientific humor, seen through the eyes of Soviet physicists behind the Iron Curtain.

In preparing the initial draft of this edition, I realized that simply translating the Russian text back into English would result in a "broken telephone" effect, where the nuance and wit of the originals would be lost in double translation. Therefore, the primary focus of the first edition was \textbf{linguistic archaeology}. Wherever possible, I located the original English source texts—scouring archives of \textit{Physics Today}, \textit{The Journal of Irreproducible Results}, \textit{New Scientist}, and conference proceedings from the 1950s and 60s—to restore the authentic wording used by the original authors.

For the unique Russian scientific folklore, anecdotes, and editorial interludes where no English original existed, a direct translation has been performed, striving to preserve the specific voice of the Soviet physicist.

I would like to acknowledge the assistance of the AI model Google Gemini 3 Pro in locating these archival texts and assisting with the translation of Russian folklore.

\hfill \textbf{Igor Halperin} \\
\hfill \textit{December 15, 2025}

% =========================================================
% PREFACE TO THE SECOND ENGLISH EDITION
% =========================================================
\chapter{Preface to the Second Edition}

Physics does not stand still, and neither does the humor of physicists. 

This {\bf Second Edition} transforms the work from a restoration project into a comprehensive almanac. It reflects the realization that scientific humor is not static; it evolves to adapt to new geopolitical realities and new technologies, while maintaining the traditions of the past.

\section*{Part II: Resilience in the Face of a Collapse (1992)}

While the 1966 book was largely a mirror reflecting Western humor into the Soviet Union, the 1992 sequel, \textit{Physicists Keep Joking} (\foreignlanguage{russian}{Физики всё ещё шутят}), marked a distinct shift. Published right after the collapse of the USSR, it ceased looking outward for material and instead captured the unique, original voice of the Russian scientific community.

The humor in {\bf Part II} is sharper, more cynical, and deeply intertwined with the specific hardships of the time—the bureaucracy, the funding crises, and the crumbling of the "ivory tower." It documents how the scientific tradition survived the end of the Cold War through irony and resilience.

\section*{Part III: Science Humor in the Digital Age (1990s--Present)}

{\bf Part III} brings the almanac into the present day. Both modern physics and modern science in general are vastly different from where they stood in the 1990s. The main focal points of physicists' attention have shifted---from the excitement over string theory and the search for the Higgs boson to gravitational wave astronomy, quantum computing, and the mysteries of dark energy. The tools of the trade have evolved just as dramatically, from early workstations and FORTRAN code to GPU clusters, machine learning pipelines, and Large Language Models.

Yet perhaps an even more profound transformation since the 1990s has been the rise of interdisciplinarity. Modern science increasingly dissolves the boundaries between traditional fields. Physicists today actively collaborate with pure and applied mathematicians, chemists, biologists, computer scientists, and AI researchers. A condensed matter physicist might work alongside a machine learning specialist; a cosmologist might co-author papers with statisticians and data scientists; a biophysicist might find themselves debugging Python code with a computational neuroscientist.

Inevitably, these increasing interdisciplinary interactions have led to a cross-pollination of scientific humor traditions as well. The jokes, parodies, and satirical papers now travel freely across departmental boundaries. Physicists have adopted the self-deprecating wit of mathematicians; computer scientists have inherited the experimentalist-vs-theorist banter; and the AI community has developed its own rich folklore that borrows liberally from all its parent disciplines. This new section reflects that broader landscape.

To compile Part III, I undertook an extensive search of the modern digital landscape---scouring web archives that collect April Fools arXiv papers, the SIGBOVIK proceedings, and other online repositories of scientific humor. I looked beyond pure physics to include the folklore of mathematicians, biologists, computer scientists, and the AI community. The result is a collection that covers the era of string theory, big data, data science, and artificial intelligence. It shows that while the tools of science have changed from vacuum tubes and slide rules to neural networks and LLMs, the spirit of the scientist---skeptical, irreverent, and always ready to laugh at the universe and at themselves---remains wonderfully unchanged.

\section*{Acknowledgments}
Since the release of the preliminary draft of this book, I have received a number of warm comments from scientists who grew up in the later USSR or former Soviet republics, where these books still hold a legendary status. Many people have shared memories of spending hours during their high school and university years reading and laughing at these collections. 

It is a testament to the quality of the original selection that quotes from the 1966 edition are still popular to this day. One particular favorite, which I was reminded was especially popular in {\it theory} departments as a running joke about {\it experimentalists} (a convenient explanation for any "groundbreaking" anomaly in the data), remains:
\begin{quote}
\textit{"...And that was Wilkins hitting the seismograph!"} \\
(\foreignlanguage{russian}{...А это Уилкинс ударил по сейсмографу!})
\end{quote}

\vspace{0.2cm}
\begin{center}
    \includegraphics[width=0.7\textwidth, height=5cm, keepaspectratio]{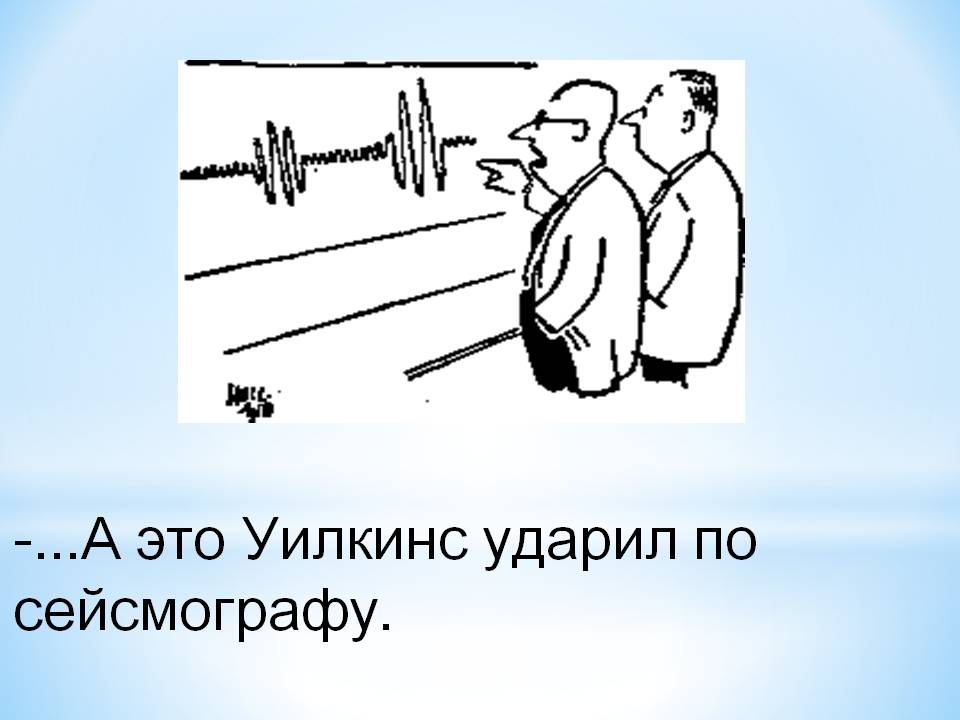}
\end{center}
\vspace{0.2cm}

I am deeply grateful to the readers of the first edition for their interest, appreciation, and for providing links to new materials that helped shape Part III. I would like to thank Alexander Barzykin, Arthur Bird, Elena Nazvanova, Dmitri Offengenden, and Mihail Turlakov for their valuable comments, suggestions and encouragement.

\hfill \textbf{Igor Halperin} \\
\hfill \textit{December 24, 2025}

\tableofcontents

\mainmatter

\part{Physicists Joke (1966)}

\chapter*{To the Reader}
The Black Queen* shook her head: 
\begin{quote}
"You may call it 'nonsense' if you like, but I've heard nonsense, compared with which that would be as sensible as a dictionary!" \\
\hfill --- \textit{Lewis Carroll, "Through the Looking-Glass"}
\end{quote}
\noindent \small{*Note: In the Russian translation of Alice, the Red Queen is often called the Black Queen.}

\vspace{1cm}

\textbf{The Editorial Board of Physics Literature:} It feels somewhat unusual: a humorous book bearing the mark of a scientific publishing house! But this is not accidental. Everything funny in this book will be fully understood and appreciated by those who read serious scientific literature and listen to (and deliver) learned reports—after all, the authors of these jokes are also scientists, sometimes very famous ones.

From this point of view, this book is also scientific. However, it is accessible not only to scientists. Anyone who loves a joke will find pleasure in reading the small humorous pieces collected here, and behind them, they will see serious science—physics—from a new, perhaps unexpected, side. People of all professions love to joke, but these jokes usually do not make it into print and vanish without a trace. It is a pity! Physical folklore is no less interesting than any other; it reflects the history of science, and the life and daily routine of its creators.

What is collected here has been gathered drop by drop. The editors failed to find many things that are known only through oral retelling. But this is a first attempt. There have been no books of this kind until now. Perhaps someday an anthology of physical witticisms (and not only translated ones) will be published; perhaps someone will even write a dissertation on this topic (what \textit{don't} they write about!), but for now, the editors have dug through mountains of journals, both printed and handwritten, and gathered the first harvest. As is usually said in prefaces, "let us hope that the collection will please our readers."

\hfill \textbf{Professor Ya. A. Smorodinsky, June 1966}

\chapter*{Preface From the Editors}
\textit{(In Place of a Preface)}

--- Hello? \\
--- Hello! This is one of the editors of the collection "Physicists Joke." We were recommended to... \\
--- Excuse me, which collection? \\
--- "Physicists Joke." \\
--- What do physicists do?! \\
--- They joke! \\
--- I don't understand. \\
--- Well, they joke, they laugh. \\
--- Ah, they laugh... Well, so what? \\
--- This will be a collection of translations. Have you or your colleagues happened to come across anything in foreign physics literature... \\
--- No, no! Our staff is engaged in serious business and they have no time for jokes.

...Before we are accused of slandering physicists, let us hasten to assure readers that this conversation was the only one of its kind. Usually—and we sometimes appealed to very busy people—our undertaking was met with full approval and a readiness to help. Physicists value a joke. We believe that in the popular and exciting game of "Physicists and Lyricists," this fact will be counted in our favor.

%The idea to compile a real collection matured a long time ago. Reading foreign scientific publications (quite serious ones!), we often found grains, and sometimes whole nuggets of humor which, unfortunately, do not make it into abstract journals or reviews: humorous verses, notes, messages, and even large quasi-serious articles written by physicists and calculated mainly for physicists. The question was settled when the volume \textit{The Journal of Jocular Physics}, published in Copenhagen for Niels Bohr's 70th birthday, fell into our hands—entirely humorous, something like a printed "skit" written by physicists—Bohr's friends and colleagues. Having concluded a contract with the "Mir" publishing house, we felt obligated not just to translate the material at hand, but to try to collect the most interesting samples of this peculiar genre.

The idea to compile a real collection matured a long time ago. Reading foreign scientific publications (quite serious ones!), we often found grains, and sometimes whole nuggets of humor which, unfortunately, do not make it into abstract journals or reviews: humorous verses, notes, messages, and even large quasi-serious articles written by physicists and \textbf{intended} mainly for physicists. The question was settled when the volume \textit{The Journal of Jocular Physics}, published in Copenhagen for Niels Bohr's 70th birthday, fell into our hands—entirely humorous, something like a printed "skit" written by physicists—Bohr's friends and colleagues. Having concluded a contract with the "Mir" publishing house, we felt obligated not just to translate the material at hand, but to try to collect the most interesting samples of this peculiar genre.

How to accomplish this task? Bibliographic searches—researching the section "Recreational Physics" in the catalogs of major libraries—turned out to be completely fruitless. "Physics at the Tea Table," "Physics without Instruments," "Physics without Mathematics," and even "Recreational Physics in War"—all this existed, but it was not \textit{it}. There was no section for "Physics and Humor," and we could only console ourselves with the hope that it might be introduced in the near future...

We had to look through all "suspicious" journals one by one and appeal to colleagues—both acquaintances and strangers. Gradually, material accumulated, very different in character and quality. But still, it turned out to be less than we wanted. And since the criteria we used to decide "include or exclude" were undoubtedly subjective, we remain concerned that the fresh reader's reaction will resemble the words said by a cafeteria patron to a waiter: "First of all, this is... inedible, and secondly, why is the portion so small?"

All the time we worked on the collection, we were tormented by two problems, which we cowardly put off until the very end. The first problem was the \textbf{Title}. It had to be:
\begin{itemize}
    \item Original enough that no one could call it banal;
    \item Banal enough that no one could call it pretentious;
    \item Pleasing to all editors and translators.
\end{itemize}
Fortunately for us, this problem resolved itself in the end. It turned out that during the work on the book, one can increase or decrease its volume, change the content... but one thing cannot be done—change its provisional title... for having appeared in advertising brochures, it acquired the force of law.

The second problem was the \textbf{Preface}. Usually, its main content justifies the necessity of publishing the book. But we knew that the scientific necessity of publishing our collection is debatable. And yet we had to justify ourselves somehow: a) to the readers, b) to the publishers, c) to ourselves.

This problem was solved gradually.
Point "c" fell away when we decided to compile the collection.
Point "b" fell away when the publishing house signed the contract.
Point "a" remained, and it caused us the most trouble. It is sad if, when telling a joke, one has to explain the punchline, but it is utterly dreary to explain the punchline \textit{before} the joke is told. In the end, we decided not to justify ourselves to the readers...

But on one point, we must justify ourselves, or rather, apologize.
We offer apologies to the authors of the humorous pieces included in the collection for the inevitable liberties we took during translation. First of all, this concerns cuts in the text. Articles written generally on a serious topic, where the funny parts were just inserts, were shortened significantly. We also excluded places entirely built on untranslatable puns...

\vspace{1cm}
\noindent \textbf{Obninsk, May 1965} \\
Yu. Konobeev, V. Pavlinchuk, N. Rabotnov, V. Turchin

%\tableofcontents
%
%\mainmatter

% =========================================================
% CHAPTER 1: ALMOST SERIOUSLY
% =========================================================
\chapter{Almost Seriously}

\begin{figure}[h]
    \centering
    \includegraphics[width=0.7\textwidth, height=5cm, keepaspectratio	]{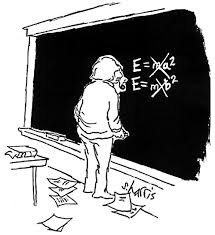}
    %\caption*{The difficult path to a great discovery.}
\end{figure}

\anecdote{Dreaming of Physics}{
American physicist of German origin James Franck (Nobel Prize 1925) once told this story: \\
"I dreamt the other day about the late Carl Runge, and I asked him: 'How are things in the afterlife? You probably know all the physical laws there?' \\
He replied: 'Here, we are given a choice: you can either know everything, or you can have the same choice as on Earth. I chose the latter, otherwise it would be too boring.'"
}

\section{Physics as a Science and an Art}
\textbf{Author:} Karl K. Darrow \\
\textit{Original Source:} Physics Today, Vol. 4, No. 11 (November 1951).

The charge of speaking after five such orators as have preceded me is not a light one, and yet is an assignment which should be treated lightly. The hands of the clock are joyously advancing toward the cocktail hour, and they advise me to pervert the famous words beneath a clock in San Francisco and say to myself, "Son, observe the time and fly from wisdom." The organizers of this meeting actually proposed that I should speak under the title "The Whole of Physics". Apart of course from my predecessors on this platform, the last man who could probably have done this was Hermann von Helmholtz. It interests me to realize that there are people still living who studied under Helmholtz; they are the last of our contacts with the era of omniscience. The wishes of the organizers will be formally fulfilled if I succeed in saying nothing that is more irrelevant to any one field of physics than to any other. This condition I will attempt to meet.

I ought to begin with a definition of physics. The American Institute of Physics has provided one, and it would be unseemly to use another in this place. Actually it is a definition of a physicist, but we can easily translate it into a definition of physics. Hearken to it. "A physicist is one whose training and experience lie in the study and applications of the interactions between matter and energy in the fields of mechanics, acoustics, optics, heat, electricity, magnetism, radiation, atomic structure, and nuclear phenomena."

Clearly this is addressed to people who have a clear-cut notion of energy, and therefore not to the general public. But even with respect to its intended audience it has a certain rashness. People who have a clear-cut notion of energy are likely to remember the equation $E = mc^2$. This equation operates like a nuclear bomb on the definition, for the definition implies that matter is cleanly and neatly distinguishable from energy, and the equation says it is not so at all. The equation in fact invites us to alter the wording, and say that the physicist is one who concerns himself with the interactions between energy and energy. This has a silly sound, but it is not a silly thought, and I can clothe it in appropriately formal garb by saying that the physicist concerns himself with the interactions between various types of energy. But I will not tamper further with the head of the definition, for it is just the sauce, and the meat is in the tail. Oddly enough, the meat is disguised as a limitation.

There are two limitations here, and one of them is not in Nature and I think that it was not in the minds of the definers. It is implied that in respect of magnetism, for instance, there is one part of magnetism that involves interactions between matter and energy and another that does not. The first part is physics and the second part is not. But there is no second part, and the whole affair reduces itself to a plain and simple definition by enumeration. Physics is a grouping of nine fields like the nine Muses, and the names of the Muses are mechanics, acoustics, optics, heat, electricity, magnetism, radiation, atomic structure and nuclear phenomena. This is what the definers really say and this is the meat of the definition, and all the rest is a valiant attempt to express in a very few words something that slowly dawns on the physicist as he progresses in his science. When in this manner we get down to brass tacks, the only people who can rightly complain are those who would like to have their tacks removed from the list and transferred to some other science than physics, and those others who do not find their tacks in the list and yet would like to be considered physicists. I will not be their spokesman; let them enter their own objections.

The definition also speaks of "study and applications". This sounds like the classic antithesis between pure and applied physics. Let us examine into this distinction, which as will soon appear I deem a necessary evil.

As our science expands, its journals become so huge that they are insupportable in all senses of the word, and the meetings of its cultivators so congested that they defeat their purpose. These are only symptoms: the malady is the finiteness of the human brain, which can absorb only a finite amount of knowledge before old age sets in. But although the malady is incurable the symptoms can be controlled, and this is done by the same technique as prevailed in the cities of ancient Greece and prevails to this day in the beehive. Some of the bees get tired of the congestion and swarm off to another hive. This is the reason and the only reason why the American Physical Society cannot deprecate the newer hives of the Optical Society of America and the Acoustical Society of America, each of which has taken a large piece of physics unto itself. The engineering societies swarmed away a long time ago and they have even larger segments of our domain, but we could not force them back into our hive if we would and we would not if we could. The distinctions are evil in principle, but we cannot get along without them.

Let us try to contrive a definition. One begins by saying that a pure physicist is interested in a device because it illustrates the laws of physics, an applied physicist is interested in the laws of physics because they explain a device. The teacher of physics teaches the dynamo because it exemplifies Faraday's laws, the teacher of engineering teaches Faraday's laws because they show how the dynamo works. This definition implies a static science and a static technology. We try to put evolution into it. A pure physicist is one who discovers new laws of Nature, an applied physicist is one who improves an old device or invents a new one.

But many experimental physicists of uncontested purity spend a large part of their time in improving their devices. We must introduce more motive into the definition. A pure physicist is one who improves his devices for no other purpose than to extend his understanding of Nature, an applied physicist is one who improves his devices for any other purpose than to extend his understanding of Nature. On this basis Rutherford was an applied physicist at the start of his career when he was trying to make a radio, purified himself when he abandoned the attempt; Lawrence was a pure physicist until his cyclotrons started to make isotopes which are useful to medical men, then he lost his caste. It is evident that our definition is one of extremes, and it takes a rather single-minded person to hold a position at either extreme. Let us see whether we can discover any analogies in the practice of the arts.

A composer who produces a symphony is presumably a pure musician, one who writes for a dance-orchestra is presumably applied. Yet any conductor knows that the subscribers will not object and will in fact be very pleased if he plays some of the works of Johann Strauss and Manuel de Falla. We are meeting in an opera house. Richard Wagner himself said that the only purpose of his music was to enhance his libretto; he is accordingly an applied musician. Even more singular is the case of Tschaikowsky, who remained a pure musician until he had been in his grave for fifty-odd years, whereupon the sonorous opening theme of his piano concerto in B flat minor was converted into a dance entitled "This Night We Love". I shall leave to people more expert than myself the question whether in the Gilbert-Sullivan team Sullivan was an applied musician or Gilbert an applied poet.

Take painting and sculpture. The pure painter, let us say, is the one whose paintings hang in a museum; the applied painter is the one whose paintings are fitted into the decorative scheme of a house. On this basis Monet and Renoir are applied painters for those who can afford to pay twenty thousand dollars for a picture, pure painters for the rest of us. I do not know quite where to put the portrait painter, except that he is probably pure when his work is hung in a museum with a label "Portrait of a Man". I am reasonably sure that there are many modern painters who, in the inconceivable event that they were present, would wish me to say that the pure painter is the one whose pictures look like nothing on earth, and all the others are applied. There is an analogy to physics in this; we will take another glance at it later.

Architecture ought to be the perfect example of an applied art. Yet I note that there is a doctrine called "functionalism", the exponents of which profess that every part of a building ought to be requisite for its purpose and essential to its structure. The existence of such a doctrine implies that there are buildings with details that are not required by their purpose or their structure, and indeed this is obvious to anybody who has seen a cornice. A drawback of this doctrine is that it forbids you to enjoy a cornice, and indeed in principle it forbids you to enjoy a Gothic cathedral until a civil engineer has proved to you by calculation that if any flying buttress, any pinnacle or any crocket were removed the building would fall down. Then there arises the question of the stained-glass windows: these are functional if they stir a mystical emotion, decorative if they please the tourist, anti-functional if they just impair the light. The first of these views was that of the artists who created the windows of Chartres, the second is that of the guides, the third was that of the eighteenth-century people who improved the lighting by smashing some of the windows and throwing the precious fragments onto the rubbish-heap. It is not easy after all to distinguish what is functional and what is decorative in the totality of a cathedral. A cathedral is a texture of purposeful construction, purposeful decoration, decoration for the sake of decoration, and symbolic instruction. So also is a science. And if some of the sublimest features of a Gothic church derive from the fact that the builders did not have steel beams available, and if the modern builders with steel beams produce a structure that in spite of all its competence is mysteriously lacking in something that we like, these are perhaps analogies with the classical physics and the theories of today.

\anecdote{The Poet}{
David Hilbert was asked about one of his former students. \\
"Ah, that one?" Hilbert recalled. "He became a poet. For mathematics, he had too little imagination."
}

\anecdote{The 440 Volt Student}{
An interesting example of how words can be used for the quantitative description of measurement results was told by Professor Gale of the University of Chicago. \\
Professor Gale was working in the laboratory with one of his students, and they did not know what voltage---110 or 220 volts---was on the terminals to which they had to connect their equipment. The student was about to run for a voltmeter, but the professor advised him to determine the voltage by touch. \\
"But it will just shock me, that's all!" the student objected. \\
"Yes, but if it is 110 volts, you will jump back and simply exclaim 'Oh, damn!', but if it is 220, the expression will be much stronger." \\
\vspace{0.2cm}
When I recently told this story to students, one of them remarked: "This morning I met a little guy, so he, probably, just before this had connected himself to a voltage of 440!"
}

\section{Confessions of an Acoustical Engineer}
\textbf{Author:} Marvin Camras \\
\textit{Original Source:} IRE Transactions on Audio, Vol. 9, No. 6 (1961).

When I was an assistant, I worked like a dog and earned barely enough to eat. My boss always made cracks about my mental abilities and spent half the day explaining things that were perfectly clear, and then threw up his hands in surprise that the work wasn't done. He assigned me the task of drawing up blueprints for unworkable designs invented by the "thinkers" in our lab who had their heads in the clouds. I had to finish and refine everything to force these contraptions to work at least somehow.

By the time I became a Group Leader, the situation had changed. The assistants had let themselves go completely; they didn't know how to do anything but took offense at every remark. Suppose a trifling job needed to be done. I could have done it myself in two hours. But I had to spend half a day explaining to my assistants why this work should be done at all, why it should be done this way and not otherwise.

Even worse were my administrative duties. I was forced to write so many reports and proposals that this alone took all my time. But I still had to answer correspondence, take phone calls, and receive visitors. It was assumed that I must take care of the "image" of the company and attend professional meetings, serve on committees, and arrange seminars.

In the end, I started a small business of my own. Unfortunately, I have had bad luck with engineering managers and administrators. They have changed much for the worse. They do not strive to get things done. They organize matters so that everything collapses, and then "reorganization" begins. They travel, converse, arrange seminars—they do anything but work. If the Nobel Prize were awarded for excuses, our laboratory would have received it long ago.

Now, having gained bitter experience, I dream of becoming an assistant again. An assistant still lives an easier life. But, unfortunately, I am already married and cannot afford the luxury.

\anecdote{Rutherford's Method}{
One evening, Ernest Rutherford entered the laboratory. Although it was late, one of his many students was bent over instruments. \\
"What are you doing so late?" asked Rutherford. \\
"I am working," followed the answer. \\
"And what do you do during the day?" \\
"I work, of course," the student replied. \\
"And do you work early in the morning too?" \\
"Yes, Professor, I work in the morning too," the student confirmed, expecting praise. \\
Rutherford frowned and asked irritably: \\
"Listen, when do you \textit{think}?"
}

\section{A Contribution to the Mathematical Theory of Big Game Hunting}
\textbf{Author:} H. Pétard, Princeton, New Jersey \\
\textit{Original Source:} The American Mathematical Monthly, Vol. 45, No. 7 (Aug.--Sep., 1938).

\subsection*{1. Mathematical Methods}

\begin{enumerate}
    \item \textbf{The Hilbert, or Axiomatic, Method.} We place a locked cage at a given point of the desert. We then introduce the following logical system:
    \textit{Axiom I:} The class of lions in the Sahara Desert is non-void.
    \textit{Axiom II:} If there is a lion in the Sahara Desert, there is a lion in the cage.
    \textit{Rule of Procedure:} If $p$ is a theorem, and "$p$ implies $q$" is a theorem, then $q$ is a theorem.
    \textit{Theorem I:} There is a lion in the cage.

    \item \textbf{The Method of Inversive Geometry.} We place a spherical cage in the desert, enter it, and lock it. We perform an inversion with respect to the cage. The lion is then in the interior of the cage, and we are outside.

    \item \textbf{The Method of Projective Geometry.} Without loss of generality, we may regard the Sahara Desert as a plane. Project the plane into a line, and then project the line into a point inside the cage. The lion is projected into the same point.

    \item \textbf{The Bolzano-Weierstrass Method.} Bisect the desert by a line running N-S. The lion is either in the East or West section. Bisect that section by a line running E-W. The lion is in one of the two parts. We continue this process indefinitely, constructing a sufficiently strong fence at each step. The diameter of the chosen partitions approaches zero, so that the lion is ultimately surrounded by a fence of arbitrarily small perimeter.

    \item \textbf{The "Mengentheoretisch" Method.} We observe that the desert is a separable space. It therefore contains an enumerable dense set of points, from which can be extracted a sequence having the lion as a limit. We then approach the lion stealthily along this sequence, bearing with us suitable equipment.

    \item \textbf{The Peano Method.} Construct a continuous curve passing through every point of the desert. It has been remarked that it is possible to traverse such a curve in an arbitrarily short time. Armed with a spear, we traverse the curve in a time shorter than that in which a lion can move his own length.

    \item \textbf{A Topological Method.} We observe that a lion has at least the connectivity of the torus. We transport the desert into four-space. It is then possible to carry out such a deformation that the lion can be returned to three-space in a knotted condition. He is then helpless.
    
    \item \textbf{The Cauchy, or Function Theoretic, Method.} We consider an analytic lion-valued function $f(z)$. Let $\zeta$ be the cage. Consider the integral:
    \[ \frac{1}{2\pi i} \int_C \frac{f(z)}{z - \zeta} dz \]
    where $C$ is the boundary of the desert; its value is $f(\zeta)$, i.e., a lion in the cage.
\end{enumerate}

\subsection*{2. Methods from Theoretical Physics}

\begin{enumerate}
    \item \textbf{The Dirac Method.} We observe that wild lions are, \textit{ipso facto}, not observable in the Sahara Desert. Consequently, if there are any lions in the Sahara, they are tame. The capture of a tame lion is left as an exercise for the reader.

    \item \textbf{The Schrödinger Method.} At any given moment there is a non-zero probability that the lion is in the cage. Sit down and wait.

    \item \textbf{The Method of Nuclear Physics.} Place a tame lion into the cage, and apply a Majorana exchange operator between it and a wild lion.
    \textit{Variant:} Let us assume we would like to catch a lion ($L$) but we catch a lioness ($L'$). We then place the latter in the cage and apply the Heisenberg exchange operator which exchanges the spins.
\end{enumerate}

\subsection*{3. Methods from Experimental Physics}

\begin{enumerate}
    \item \textbf{The Thermodynamical Method.} We construct a semi-permeable membrane which allows everything but lions to pass through. This is then swept across the desert.

    \item \textbf{The Atom-Splitting Method.} We irradiate the desert with slow neutrons. The lion becomes radioactive and a process of disintegration sets in. When the decay has proceeded sufficiently far, he will become incapable of showing fight.

    \item \textbf{The Magneto-Optical Method.} We plant a large lenticular bed of catnip (Nepeta cataria), whose axis is aligned with the direction of the horizontal component of the earth's magnetic field. We place a cage at one of its foci. The lion, attracted by the catnip, is bent towards the cage by the Faraday effect.
\end{enumerate}

\section{On the Quantum Theory of the Temperature of Absolute Zero}
\textbf{Authors:} G. Beck, H. Bethe, W. Riezler \\
\textit{Original Source:} Die Naturwissenschaften, Vol. 19, No. 2 (1931).

We consider a hexagonal crystal lattice. The absolute zero of this lattice is characterized by the fact that all degrees of freedom of the system freeze, i.e. all internal movements of the lattice cease. Naturally, the motion of an electron in its Bohr orbit is excepted. However, according to Eddington, every electron possesses $1/\alpha$ degrees of freedom, where $\alpha$ is the Sommerfeld fine-structure constant. Besides electrons, our crystal contains only protons, for which the number of degrees of freedom is evidently the same, since according to Dirac, a proton can be considered a hole in the electron gas.

To reach absolute zero, we must therefore remove $2/\alpha - 1$ degrees of freedom per neutron (which is composed of a proton and an electron). Since $2/\alpha - 1$ degrees of freedom correspond to the temperature interval from zero degrees to absolute zero, we obtain for the absolute zero temperature:
\begin{equation}
    T_0 = -\left(\frac{2}{\alpha} - 1\right) \text{ degrees}.
\end{equation}
Substituting the experimental value $\alpha = 1/137$, we obtain:
\begin{equation}
    T_0 = -273 \text{ degrees}.
\end{equation}

\section{Earth as a Man-Controlled Spaceship}
\textbf{Author:} D. Froman \\
\textit{Original Source:} Physics Today, Vol. 15, No. 7 (July 1962).

\textit{(Speech at a banquet held after a conference on plasma physics, organized by the American Physical Society in November 1961 in Colorado Springs.)}

Since I do not understand plasma physics or thermonuclear fusion very well, I will speak not about these phenomena themselves, but about one of their practical applications in the near future.

Let us imagine that we have managed to invent a spaceship that moves by ejecting reaction products of D-D and D-T. On such a ship, one could travel into space, catch a few asteroids there, and tow them to Earth. (The idea, frankly, is not new.) If we do not overload the rocket too much, it would be possible to deliver 1000 tons of asteroids to Earth, expending only about a ton of deuterium. I, frankly speaking, do not know what asteroids consist of. However, it may well turn out that they are half composed of nickel. It is known that 1 pound of nickel costs 50 cents, and 1 pound of deuterium -- about 100 dollars. Thus, for 1 million dollars we could buy 5 tons of deuterium and, having expended them, deliver to Earth 2500 tons of nickel worth 2.5 million dollars. Not bad, right? I was already thinking, should I not organize the American Company for Extraction and Delivery of Asteroids (ACEDA)? The equipment of this company will be extremely simple. With a sufficient subsidy from Uncle Sam, this could be a very profitable business. If anyone present with a large account in the bank wishes to enter the number of founders, let him approach me after the banquet.

And now let's look into a more distant future. Personally, I cannot understand why astronauts dream of getting into interstellar space. The main nuisance is that the astronaut in the rocket will be in the same position as a man placed against a beam of fast protons coming out of a powerful accelerator. I feel very sorry for the poor astronaut; I have even composed a ballad about his sad fate:

\begin{quote}
\textbf{The Ballad of the Astronaut}

The beta-inverters and gamma-converters
Have stripped all the plating away,
The ion-gun's bust, crumbled to dust,
And the nozzle is clogged with clay.

The mesons decayed, the neutrons have strayed,
The light has all radiated.
The protons have scattered (as if it mattered),
And the leptons annihilated.

He flew to Orion, but a flux of gravitons
Crossed his path in the dark.
With resources spent and the hull badly bent,
He barely missed the arc.

He fought gravity's force, deviating from course,
With super-acceleration,
But the hands of the clock froze on the lock
In time-dilation stagnation.

Back to the Earth! The land of his birth!
But what is this red, cold ball?
The sun has grown dim, the outlook is grim,
There's no one left to call.

With a curse and a cry at the frozen sky,
He pulled the lever back,
And "A" rang out, and "B" rang out,
And then there was only... CRACK!
\end{quote}

But I feel sorry for those who will remain on Earth. After all, our Sun is not eternal. It will someday fade, plunging everything surrounding it into cosmic darkness and cold. As Fred (Fred Hoyle, that is)* told me, in a couple of billion years on Earth it will be so cold that not only comfort, but life itself on this planet will be out of the question. And consequently, it makes explicit sense to move somewhere. It seems to me that for the majority of us, the most convenient spaceship would be the Earth itself. Therefore, if we do not like that our luminary is gradually fading, and generally if we get tired of everything in the Solar System, why remain here? Let's fly somewhere directly on our Earth. In this case, all difficulties associated with space flight will fall away by themselves. After all, the problem of protection from radiation does not exist, there is an atmosphere on Earth, and the speed of movement will be small. Safety and the pleasantness of such a journey are obvious. However, will we have enough energy?

\textit{* Fred Hoyle -- famous English astrophysicist, professor at Cambridge University, author of a number of works on theoretical astrophysics, cosmogony, theoretical gravity and... several science fiction novels.}

First of all, heat and light will be needed: after all, during the long time we will be removed from the Sun or any other star. Deuterium contained in ocean water can give us $10^{38}$ ergs, consequently, if we use it only for heating and lighting, then this will be enough for three million years -- a term quite sufficient. True, here is a small snag. At our speeds, we will consume $3 \dots 10^{10}$ pounds of deuterium per year, and the cost of it is 100 dollars per pound, consequently, the consumed deuterium will be 100 times the annual budget of modern air forces. But, perhaps, it will be possible to obtain deuterium at wholesale prices?

However, we will need more energy to tear away from the Sun. Calculation shows that for this, $2.4 \dots 10^{40}$ ergs will be needed, that is, much more than all the ocean deuterium can give. Therefore, it will be necessary to find other sources of energy. I suggest that for the solution of this problem we will have to turn to the synthesis of an alpha-particle from four protons. When using this reaction, all protons of the world ocean will give us energy $10^{42}$ ergs, that is, 40 times more than is needed to tear away from the Sun.

As a working body, sand can be used. Ejecting 1000 molecules of $SiO_2$ for every synthesized alpha-particle, we will need to spend only 4\% of the Earth's mass to tear away from the Sun. It seems to me that we can afford this. Especially since for such a purpose, it is not a pity to use the Moon: after all, far from the Sun, it makes no difference to us anyway. Leaving the Solar System and wandering in cosmic space, we, probably, will be able to replenish our reserves of mass and energy from time to time, refueling on the fly at the expense of planets encountered on the way. On the path of carrying out these plans, there is currently one fundamental obstacle: we do not know how to carry out the chain reaction $4p \to He^4$. Now you see what an important problem this is. We need to double our efforts to solve it. Time does not wait: Earth has spent already two-thirds of the term allotted to it by the Sun.

I assure you: in space, we will be excellent. Perhaps, we will like it so much that we won't even want to attach ourselves to a new star.

\anecdote{Einstein and the Hat}{
Einstein was visiting acquaintances. It started to rain. When Einstein was about to leave, they offered him a hat. \\
"Why?" said Einstein. "I knew it would rain, and that is exactly why I didn't wear a hat. It dries longer than my hair. That is obvious."
}

\anecdote{The Selection Process}{
"I simply cannot find myself an assistant," Thomas Edison once complained to Albert Einstein. "Young people come in every day, but not a single one is suitable." \\
"And how do you determine their suitability?" Einstein inquired. \\
Edison showed him a sheet of paper with a list of questions. "He who answers these will become my assistant." \\
Einstein picked up the list. "'How many miles is it from New York to Chicago?'" he read, and immediately answered: "One needs to look that up in a railroad guide." \\
"'What is stainless steel made of?'" Einstein continued reading. "One can find that out in a metallurgy handbook..." \\
Running his eyes quickly over the remaining questions, Einstein handed the paper back and said: "Without waiting for a refusal, I withdraw my candidacy myself."
}

\section{List of Typical Examination Questions for Physics Graduate Students}
\textbf{Author:} Harry J. Lipkin \\
\textit{Original Source:} The Journal of Irreproducible Results, Vol. 7, No. 2 (1959).

\begin{enumerate}
    \item \textbf{MECHANICS.} A particle moves in a potential field $V(r) = e^{-r/r^{12}}$.
    \begin{enumerate}
        \item Show that the solution to this problem bears no relation whatsoever to the binding energy of the deuteron.
        \item Explain the asymptotic behavior of the solution as $r \to \infty$.
    \end{enumerate}

    \item \textbf{ELEMENTARY PARTICLES.} List all elementary particles that have not yet been discovered, indicating their mass, charge, spin, isotopic spin, strangeness, and the reasons why they have not yet been discovered.

    \item \textbf{QUANTUM THEORY.} Write down the Schrödinger equation describing a student studying the physics of elementary particles. Derive an expression for the operator "Passed -- Failed", which has the eigenvalue $+1$ if the student passes the session, and $-1$ if they fail. Show that the state of the student at the end of the semester is always an eigenstate of this operator.

    \item \textbf{SYMMETRY PROPERTIES.} Investigate the properties of the Dirac equation with respect to rotation:
    \begin{enumerate}
        \item When the blackboard on which the equation is written is rotated;
        \item When the physicist investigating the equation is rotated.
    \end{enumerate}

    \item \textbf{NUCLEAR REACTIONS.} A coin enters into interaction with a machine selling Coca-Cola. Determine the relative probabilities of the following reactions:
    \begin{enumerate}
        \item Capture (input channel: nickel*, output: nothing);
        \item Elastic scattering (n, n) (input channel: nickel, output: nickel);
        \item Reaction (n, 2n) (input channel: nickel, output: two nickels);
        \item Reaction (n, p) (input channel: nickel, output: button);
        \item Reaction (n, c) (input: nickel, output: Coca-Cola).
    \end{enumerate}
    \textit{*Note: Nickel = 5 cent coin.}

    \item \textbf{RELATIVISTIC QUANTUM FIELD THEORY.} A pair of twins, Bingle and Dingle, is born. Immediately after birth, Dingle is sent in a rocket toward a distant star at a speed of $0.999c$, while Bingle remains to work in the laboratory. Determine the relative age of Bingle and Dingle at the moment of Dingle's return, taking into account the following process: at the farthest point of his trajectory, Dingle emits a virtual pi-meson, which creates a Bingle-Anti-Bingle pair. The Anti-Bingle returns to Earth, where it annihilates with Bingle, while Dingle and the new Bingle happily live out their days near the distant star.

    \item \textbf{EXPERIMENTAL TECHNIQUE.} Describe the most expensive method for determining Planck's constant.

    \item \textbf{DISPERSION RELATIONS.} Explain the phenomenon of the multiple production of strange articles in nuclear physics, which are observed in the non-physical region of \textit{Physical Review}. Show that the principle of causality allows one to fully predict the results of any experiment, with good agreement observed up until the moment someone actually performs the experiment.
\end{enumerate}

\section{Introduction to S-Matrix Theory}
\textit{(Considered mainly from the point of view of applications to the description of the life of physicists and primarily taking into account the statistical regularities characteristic of such systems)}
\textbf{Original Source:} The Journal of Irreproducible Results (1965).

It is well known that in recent years S-matrix theory has achieved significant success in describing scattering processes and the mutual transformation of elementary particles. This inspired us to attempt to apply it (perhaps not entirely rigorously) to the study of processes occurring with physicists throughout their lives. We will pay special attention to systems to which statistics can be applied, i.e., systems consisting of a large number of objects (in our case, physicists).

The system we are considering at time $t = -\infty$ represents an incident stream of physicists who can be considered almost free. According to two solutions of the equations of motion, this stream can be divided into two parts: retarded physicists and advanced physicists (the latter mainly from Princeton; they are distinguished by the fact that they never study the history of the question under consideration).

Throughout their lives, physicists enter into interaction with various systems. The strength of this interaction depends both on the skill and pushiness of each individual physicist, and on what these systems are---conservative or liberal. By the time $t = \infty$, the stream of physicists decays into various reaction products, the full number of which could in principle be obtained from known formulas for the S-matrix, if its form were currently known. The products can be distributed among so-called *reaction channels*, of which we will name here only a few:
\begin{enumerate}
    \item[a)] Scattered physicist;
    \item[b)] Professor;
    \item[c)] Mathematician;
    \item[d)] Reactor engineer;
    \item[e)] Bureaucrat.
\end{enumerate}

From the most general properties of the S-matrix, and especially from its relativistic invariance, one can conclude that the total energy, including the rest mass, is an integral of the physicist's motion through life. Since it is known that rest mass increases with age, we immediately conclude that the remaining energy falls over time.

To obtain more precise results, it is necessary to take into account the interaction of physicists with each other. For this purpose, let us consider a region of configuration space, the so-called "Institute," where the interaction is maximal. This region, hereafter called the CIPI (Configuration Interaction Physicist Interval) for brevity, is separated from the outside world by some potential barrier. Possible states of physicists in such a potential well can be defined by four quantum numbers, of which the first three have a generally known meaning. The fourth quantum number, corresponding to two possible states for a physicist---sleep and wakefulness---has no classical analogue, since, according to the quantum-mechanical principle of complementarity, neither of these states can be observed without an admixture of the other. We will denote the possible values of this quantum number by the symbols "+" and "--" respectively.

It is quite clear that the forces usually acting on physicists are so great that conducting any calculations using perturbation theory hardly seems expedient. Therefore, to obtain results, we must turn to simplified models. However, consideration of the latter would lead us far beyond the scope of this article. The results of these studies on models we will try to set out in subsequent works.

\section{History of the Discovery of the Mössbauer Effect}
\textit{(According to H. Lipkin)}
\textit{Original Source:} Preface to H. Frauenfelder's "The Mössbauer Effect".

\begin{center}
\begin{tabular}{|l|c|p{0.5\textwidth}|}
\hline
\textbf{Period} & \textbf{Date} & \textbf{Note} \\
\hline
Prehistoric & Up to 1958 & Could have been discovered, but wasn't. \\
\hline
Early Iridium Age & 1958 & Discovered, but not noticed. \\
\hline
Middle Iridium Age & 1958--1959 & Noticed, but not believed. \\
\hline
Late Iridium Age & 1959 & Believed, but... not interested. \\
\hline
Iron Age & 1959--1960 & \textbf{OO--OO--OO--OO!!!} \\
\hline
\end{tabular}
\end{center}

\anecdote{Dirac's Puzzle}{
Dirac was distinguished by great ingenuity in solving various kinds of mathematical puzzles. In many cases, he offered his own, very unexpected solutions. A very popular puzzle is to express a given number using a limited number of identical digits, using any other mathematical signs. Dirac proposed a general solution to such a problem, finding a way to write \textbf{any} number using only three twos:
\[ N = -\log_2 \log_2 \sqrt{\sqrt{...\sqrt{2}}} \]
The number of square root signs is equal to $N$.
}

\anecdote{Thermodynamics Summary}{
Among physicists, the following definition of thermodynamics is current: \\
"Thermodynamics is a stick with three ends."
}

\section{Typology in Scientific Research}
\textbf{Authors:} A. Kohn and M. Breier \\
\textit{Original Source:} The Journal of Irreproducible Results, Vol. 7, No. 2 (1959).

\begin{enumerate}
    \item \textbf{THE DISCOVERER (The Idea Man).} These scientists are always giving off sparks. Their brains are always ready to seize on random prey. They possess a high degree of scientific intuition, allowing them to formulate a hypothesis (or at least a working hypothesis) on the spot. They then either verify it experimentally themselves or, more likely, leave the drudgery of proof to others, deriving their pleasure from the purely speculative solution.

    \item \textbf{THE EXPLOITER.} A researcher with a quick grasp; his eyes and ears are constantly open. This type is rarely found in his own laboratory; he prefers to spend his time in discussions with colleagues from other departments, especially if those colleagues are working on something interesting. He never lacks good ideas—usually borrowed from others—which he turns into interesting papers, generously sprinkled with references to "private communications."

    \item \textbf{THE CONNOISSEUR (The Critic).} The mental abilities of this type far exceed his ability (or desire) to conduct his own experiments. He is capable of evaluating good work, often better than the author himself. However, his critical mind, combined with an innate inconsistency, prevents him from publishing anything, unless he has a decisive boss who forces the issue.

    \item \textbf{THE IMPROVER.} He resembles the Connoisseur but possesses slightly higher productivity. His achievements are represented by very few, but excellent papers, based on experiments which he repeated so many times that all unexpected or unpredicted results were eventually discarded with the help of sophisticated statistical processing.

    \item \textbf{THE WELL-INFORMED (The Man Up-to-Date).} He knows everything worth knowing. Unlike the Exploiter, he spends all his time in the library, where it is rare for anyone to beat him to the latest issue of a journal. He is a goldmine of references, but rarely produces original work, as he is too busy reading about the work of others.

    \item \textbf{THE CO-AUTHOR.} This type has mastered the art of scientific diplomacy. He manages to include his name on the author list of most articles published by his department, often his only contribution being the decision of whether to use "and" or "\&" in the title. Some hold the opinion that the "Co-author" is almost the same as the "Advisor," but everyone who plays bridge knows what a "dummy" is.

    \item \textbf{THE ADVISOR (The Consultant).} Not to be confused with the Co-Author. The Advisor usually works for a rival institution or industry, giving advice to others for a fee, while doing no work of his own.

    \item \textbf{THE GADGETEER.} In modern science, some research is impossible without a solid array of instruments. This type considers the meaning of life to be the acquisition of the largest possible number of ultra-modern instruments. His lab sparkles with glass and chrome; awe-inspiring names of installations are used to impress visitors rather than to obtain results.

    \item \textbf{THE PUBLISHER.} This term denotes those who increase their list of publications at an exponential rate.
    \begin{itemize}
        \item \textbf{The Ruminant.} (A name associated with the digestive process of certain mammals). This person usually settles in a developing country. He publishes his observations there, prefacing them with: "For the first time in the history of [Country] it was observed...", followed by a translation of work done years ago in another place.
        \item \textbf{The Multiplier.} An individual who divides the results of his work into the largest possible number of small packets labeled "Article," "Letter to the Editor," "Preliminary Note," etc., successfully maximizing his bibliography.
    \end{itemize}

    \item \textbf{THE CORRESPONDENT.} To meet this type, one need only look at the "Letters" section of any journal. The Correspondent reports something "truly important," usually resembling a great discovery that must be published immediately to establish priority. Such messages end with: "A detailed description of these experiments will be published elsewhere." In 50\% of cases, the promised publication never appears, as subsequent experiments destroy the interest in the idea.
\end{enumerate}

The above list claims by no means to be complete. We are sure that a reader with imagination will easily be able to construct images of all the scientists with whom he is personally acquainted.

\anecdote{The Inverted Curve}{
A classic story about Lev Landau. An experimentalist once caught him in the corridor, showed him a graph obtained from a recent experiment, and asked for an interpretation.Landau glanced at it and immediately provided a detailed explanation for the curve's behavior. The experimentalist thanked him and began walking away, but suddenly stopped, made a U-turn, and caught up with Landau again.
"I am so sorry, Dau," he said, "I just realized I was holding the graph upside down. \textit{This} is the correct orientation."
"Oh," Landau replied instantly, "in that case, the explanation is even simpler!"
And without missing a beat, he provided a completely different, yet equally convincing, explanation for the inverted curve.
}
\section{Scientific Folklore and Short Stories}

\anecdote{The Unit of Talkativeness}{
The American physicist Robert Millikan (1868--1953) was known for his loquacity. Joking about him, his collaborators proposed introducing a new unit of talkativeness---the "kan". Its thousandth part, i.e., the "millikan", should exceed the talkativeness of an average person.
}

\anecdote{Circular Authorship}{
In one of the issues of *The Journal of Irreproducible Results* (Vol. 9, 1960), a scheme was proposed for arranging the names of the authors of an article in a circle. This excludes the possibility of singling out any one of the authors and offending the rest.
}

\anecdote{Kelvin's Class}{
Thomson (Lord Kelvin) was once forced to cancel his lecture and wrote on the blackboard: \\
"Professor Tomson will not meet his classes today." \\
Students decided to play a trick on the professor and erased the letter "c" in the word "classes". \\
The next day, seeing the inscription ("...will not meet his lasses today"), Thomson was not taken aback, but erased one more letter in the same word ("...will not meet his asses today") and left silently.
}

\anecdote{Bohr Brothers}{
Niels Bohr explained his thoughts brilliantly when he was one-on-one with an interlocutor, but his speeches before a large audience were often unsuccessful, sometimes even barely intelligible. His brother Harald, a famous mathematician, was a brilliant lecturer. \\
"The reason is simple," Harald said. "I always explain what I have spoken about before, and Niels always explains what he will speak about later."
}

\anecdote{Nernst's Carps}{
The author of the third law of thermodynamics, Walther Nernst, bred carps in his leisure hours. Once someone profoundly remarked: \\
"Strange choice. Chickens would be more interesting." \\
Nernst replied imperturbably: \\
"I breed animals that are in thermodynamic equilibrium with the environment. Breeding warm-blooded animals means heating the universe at your own expense."
}

\anecdote{Rutherford's Hiring Policy}{
Ernest Rutherford used the following criterion when choosing his employees. When someone came to him for the first time, Rutherford gave him a task. If after that the new employee asked what to do next, he was fired.
}

\anecdote{Einstein at Curie's}{
Once Einstein was invited to the Curies'. Sitting in the living room, he noticed that two chairs near him were empty—no one dared to sit in them. \\
"Sit near me," Einstein said laughing to Joliot. "Otherwise, it seems to me that I am at the Prussian Academy of Sciences."
}

\anecdote{Scientific Myopia}{
"...one of the main reasons for the flood of scientific literature is that when a researcher reaches the stage where he stops seeing the forest for the trees, he is too willing to resolve this difficulty by switching to the study of individual leaves." \\
\hfill \textit{(The Lancet, December 1980)}
}

\anecdote{Darwin's Experiment}{
Erasmus Darwin believed that from time to time one should perform the wildest experiments. Almost nothing ever comes of them, but if they succeed, the result is terrific. Darwin played the trumpet in front of his tulips. No results.
}

\anecdote{Hypnopedia}{
From the experience of hypnopedia: "And now, children, listen to the words that must not be spoken."
}

\anecdote{Logic Loop}{
"It is not yet clear whether the rate of destruction is determined by the rate of creep or vice versa. The authors of the review hold the opposite opinion on this score..." \\
\hfill \textit{(From a review article by V. I. Indenbom and A. N. Orlov, UFN, 76, 588, 1962)}
}

\anecdote{Plutonium Display}{
When a group of scientists in America received 2 milligrams of plutonium hydroxide, there was no end to the curious who hungered to see the new element. But risking precious crystals was impossible, and the scientists poured aluminum hydroxide crystals into a test tube, tinted them with green ink, and displayed them for general viewing. \\
"The contents of the test tube is plutonium hydroxide," they declared imperturbably to visitors. The visitors left satisfied.
}

\anecdote{Chemical vs. Physical Methods}{
Hans Landolt is credited with the joke: \\
"Physicists work with good methods and bad substances, chemists work with bad methods and good substances, and physical chemists work with bad methods and bad substances."
}

\anecdote{Einstein and Chaplin}{
Albert Einstein loved Charlie Chaplin's films and regarded his character with great sympathy. Once he wrote in a letter to Chaplin: \\
"Your film 'The Gold Rush' is understood by everyone in the world, and you will surely become a great man. \\
\hfill \textit{Einstein}" \\
Chaplin replied: \\
"I admire you even more. Your Theory of Relativity is understood by no one in the world, and yet you still became a great man. \\
\hfill \textit{Chaplin}"
}

\anecdote{The Definition of Science}{
Academician L. A. Artsimovich gave the following definition of science (\textit{Novy Mir}, No. 1, 1967): \\
"Science is the best modern method of satisfying the curiosity of private individuals at the expense of the state."
}

\anecdote{Compton's Bombs}{
The famous American physicist Arthur Compton was a very energetic, physically fit man. Once, circumstances combined so that he was rumored to be a Hercules. \\
Compton was researching cosmic rays. He needed to measure the intensity of cosmic radiation at different latitudes, and he moved with his equipment from one American city to another, getting closer and closer to the equator, until he reached Mexico City. Cases of equipment were unloaded on the platform; they all looked identical, although two contained hollow spherical electromotors, while the rest were loaded with lead bricks. 
The porters demanded a huge sum to carry these weights. Then Compton, picking up the two cases with the motors, marched briskly down the platform; the shamed porters, struggling to lift one case of bricks between two of them, trudged behind him.

The story hit the newspapers, but that wasn't the end. To conduct measurements, it was necessary to isolate the equipment from electrical interference, but at the same time, a power source was needed. Compton arranged with the abbot of a remote monastery to work there. It was a turbulent period in Mexican history; relations between the church and the government left much to be desired. Police controlled roads leading to monasteries, suspecting rebels might use them. 
Compton was stopped by a patrol. After inspecting his luggage, which consisted of "two round black bombs" and a huge amount of lead (and everyone knows lead is only good for casting bullets), he was arrested. When the misunderstanding was cleared up, the measured intensity of cosmic rays on the monastery grounds coincided completely with the predictions of Compton's theory.
}

\anecdote{Szilard's Language}{
The famous physicist Leo Szilard gave his first talk in English. After the talk, the physicist Jackson approached him and asked: \\
"Listen, Szilard, in what language did you actually give the talk?" \\
Szilard was embarrassed, but immediately found an answer: \\
"In Hungarian, of course, didn't you understand?" \\
"Of course I understood. But why did you stuff it with so many English words?" Jackson parried.
}

\anecdote{The Climbing Physicists}{
Bohr, his wife, and the young Dutch physicist Casimir were returning late one evening from a party. Casimir was an avid mountaineer and enthusiastically talked about rock climbing, then proposed demonstrating his skill by choosing the wall of a house they were passing. When he, clinging to the ledges of the wall, had already risen above the second floor, Bohr, getting excited, moved after him. Margarita Bohr watched them from below with anxiety. At that moment, whistles were heard, and several policemen ran up to the house. \\
The building turned out to be a bank branch.
}

\anecdote{Kidnapping Heisenberg}{
Visiting Göttingen, Bohr invited the twenty-five-year-old Heisenberg to work in Copenhagen. The next day, during a dinner in honor of Bohr, two policemen approached him and, presenting a charge of "kidnapping a minor," arrested him. \\
These were disguised university students.
}

\anecdote{The Horseshoe}{
Above the door of his country house, Bohr nailed a horseshoe, which, according to belief, brings luck. Seeing the horseshoe, one of the visitors exclaimed: "Can it be that such a great scientist as you really believes that a horseshoe above the door brings luck?" \\
"No," answered Bohr, "of course I don't believe it. It's a superstition. But, you know, they say it brings luck even to those who don't believe in it."
}

\anecdote{Against Nature}{
On Nernst's desk stood a test tube with the organic compound diphenylmethane, the melting point of which is $26^\circ$C. If the preparation melted at 11 a.m., Nernst would sigh: \\
"You can't go against nature!" \\
And he would lead the students to go rowing and swimming.
}

\anecdote{Joliot-Curie on Theory}{
In his speech at the accelerator conference (Moscow, October 1968), Academician M. A. Markov quoted Joliot-Curie: \\
"The further the experiment is from theory, the closer it is to the Nobel Prize."
}

\anecdote{Leontovich's Law}{
Academician M. A. Leontovich formulated the "Law of Standing on One's Feet." Its essence lies in the frequently observed phenomenon that authors, whose pens belong to sometimes ridiculous articles, usually give deeply grounded, well-thought-out, smart critical reviews of other authors' articles.
}

\anecdote{Newton's Plague Vacation}{
All of Newton's major discoveries (and there are many) were made within 18 months, during the forced "plague vacation," when London University, where young Newton studied, was closed due to the epidemic, and he moved to the countryside for a while. However, the publication of these works was delayed by 20 to 40 years pending their final verification and clarification (an example hardly any modern scientist follows).
}

\anecdote{Einstein's Wife}{
At the beginning of Einstein's scientific career, a journalist asked Mrs. Einstein what she thought of her husband. \\
"My husband is a genius!" said Mrs. Einstein. "He can do absolutely everything, except make money."
}

% =========================================================
% CHAPTER 2: PHYSICISTS JOKE (THE JOURNAL OF JOCULAR PHYSICS)
% =========================================================
\chapter{Contributions from "The Journal of Jocular Physics"}
\textit{(Jubilee Collections in Honor of Niels Bohr)}

\section{My Initiation}
\textbf{Author:} L. Rosenfeld \\
\textit{Original Source:} The Journal of Jocular Physics, Vol. 1 (1935).

The first message I received from Bohr was a telegram stating that the Easter Conference was postponed by two days. I was in Göttingen at the time. When we finally arrived in Copenhagen, Bohr told us the reason for the delay: he had to finish ("with the kind help of Klein," as he put it) the translation of one of his early papers for a jubilee volume.
"If I hadn't finished," he said, "it would have been a catastrophe!" This statement seemed to me somewhat exaggerated. At that time, I did not yet understand what tragedies lie hidden in the outwardly innocent procedure of "polishing" the text of a paper.

What struck me most about Bohr at our first meeting was his benevolence. This was advantageously emphasized by the presence of several of his sons. Bohr's sons were always a mystery to me. When I met Bohr the next morning at the Institute, he was again surrounded by several sons. They seemed to be different ones. The next afternoon I was shocked to see him with yet another, new son. It seemed he pulled them out of his sleeve like a magician. Eventually, however, I learned to distinguish one son from another and realized that their number was finite.

It is common knowledge that no initiation is complete without the novice being subjected to some painful ordeal. In my case, everything was in order. Since I was straining my hearing to the limit, trying not to miss a single word of the Master, I gradually found myself drawn into the same orbital motion, and with the same period, as Bohr. The true meaning of this ceremony was revealed to me only when Bohr finished by emphasizing that a person is incapable of understanding the complementarity principle unless he has previously been driven to complete dizziness.

\section{A New Tale About the Curious Elephant}
\textit{(A parody of Rudyard Kipling's "The Elephant's Child")}

No, this is not the tale about that bad Elephant that Kipling wrote about... this is a tale about another, good Elephant... who lived almost in all countries of the world. This Curious Elephant had a wonderful nose from birth, so he did not need the services of the Old Crocodile (Rutherford), and over time he discovered a new thing—the Atomic Age.

\section{The Atom That Bohr Built}
\textit{(A parody of "The House that Jack Built")}

This is the Atom that Bohr built.

This is the \textbf{Proton},
Placed in the center of the Atom,
That Bohr built.

And here is the \textbf{Electron},
That rushes around the Proton,
Placed in the center of the Atom,
That Bohr built.

Here is the \textbf{Mu-Meson},
That decayed into an electron,
That rushes around the Proton,
Placed in the center of the Atom,
That Bohr built.

And here is the \textbf{Pi-Meson},
That, decaying, gave a Mu-Meson,
That decayed into an electron...

And here is the \textbf{Bevatron},
In which that Proton was accelerated,
Which in a collision gave birth to a Pi-Meson,
Which, decaying, gave a Mu-Meson...

This is the \textbf{Law},
Proclaimed by Bohr.
A Law for all nations, A Law for all times,
Successfully describing from two sides
Not only the Proton And Electron,
But also the Neutron, Photon, Positron, Phonon, Magnon, Exciton, Polaron, Betatron, Synchrotron, Phasotron, Cyclotron, Cyclone, Ceylon, Nylon, Perlon, Cologne, Decameron.

And, undoubtedly, every neuron of the Brain,
By which That remarkable Bevatron was invented,
In which that Proton was accelerated,
Which in a collision gave birth to a Pi-Meson,
Which, decaying, gave a Mu-Meson,
That decayed into an electron,
That rushes around the Proton,
Placed in the center of the Atom,
That Bohr built!

\section{On the Standardization of Scientific Papers}
\textit{Original Source:} Confidential CERN/T/000. Distributed to all CERN members.

\textbf{Title:} On the Question of \underline{\hspace{3cm}} in the Generalized Nuclear Model. \\
\textbf{Authors:} \underline{\hspace{4cm}} and \underline{\hspace{4cm}}.

In view of the serious difficulties arising from the attempt at a precise description of the properties of an ensemble of strongly interacting particles, we consider the following approximate Hamiltonian:
\begin{equation}
    \underline{\hspace{4cm}} \dots \underline{\hspace{4cm}} \dots, \quad (1)
\end{equation}
where \underline{\hspace{3cm}} denotes the corresponding generalized coordinates. The Hamiltonian, thus, consists of three terms: \underline{\hspace{4cm}} describes collective motion, \underline{\hspace{3cm}} -- the motion of individual particles, and \underline{\hspace{4cm}} -- the \underline{\hspace{2cm}} (strong / weak / intermediate -- \textit{cross out unnecessary}) interaction between them.

For the energy of low-lying excited states, we thus obtain:
\begin{equation}
    \underline{\hspace{4cm}} \dots \underline{\hspace{4cm}}, \quad (2)
\end{equation}
which corresponds, of course, simply to the product of $\hbar^2/2m$ and $i(i+1)$, as was to be expected [1]. The system does not possess central symmetry, which allows us to describe its surface as a deformed sphere. The moment of inertia, consequently, is determined by the pole of the expression \underline{\hspace{4cm}}, which leads to the formula:
\begin{equation}
    \underline{\hspace{4cm}}, \quad (3)
\end{equation}
where, however, the dependence of the parameter \underline{\hspace{3cm}} on \underline{\hspace{3cm}} is unknown.

The authors are deeply grateful to Director \underline{\hspace{4cm}} for his displayed interest in the work. One of us (\underline{\hspace{3cm}}) is highly thankful to \underline{\hspace{3cm}}.

\vspace{0.5cm}
\textbf{Standard F/T 3}

\textbf{Title:} On \underline{\hspace{4cm}} in Field Theory. \\
\textbf{Author:} \underline{\hspace{4cm}}.

As Schwinger has shown:
\begin{equation}
    \underline{\hspace{6cm}} \quad (1)
\end{equation}
When
\begin{equation}
    \underline{\hspace{6cm}}, \quad (2)
\end{equation}
then
\begin{equation}
    \underline{\hspace{6cm}} \quad (3)
\end{equation}
Thus,
\begin{equation}
    \underline{\hspace{6cm}}, \quad (4)
\end{equation}
which apparently agrees with the assumption that
\begin{equation}
    \underline{\hspace{6cm}}, \quad (5)
\end{equation}
thanks to which
\begin{equation}
    \underline{\hspace{6cm}} \quad (6)
\end{equation}
When
\begin{equation}
    \underline{\hspace{6cm}}, \quad (7)
\end{equation}
then
\begin{equation}
    \underline{\hspace{6cm}} \quad (8)
\end{equation}
Therefore, from a formal point of view,
\begin{equation}
    \underline{\hspace{6cm}} \quad (9)
\end{equation}
One may hope that the arguments presented above will lead to a generalization of the problem of \underline{\hspace{3cm}} states.

The author is grateful to \underline{\hspace{4cm}} for valuable criticism.

\hfill \underline{\hspace{3cm}}, private communication.

\anecdote{The Theorist and the Table}{
When a theoretical physicist is asked to calculate, say, the stability of an ordinary table with four legs, he very quickly brings the first results relating to a table with one leg and to a table with an infinite number of legs. He spends the rest of his life unsuccessfully solving the general problem of a table with an arbitrary number of legs.
}

\section{Letter to the Editor}
\textbf{Author:} A. M. B. Rosen \\
\textit{Original Source:} The Journal of Jocular Physics (1955).

\textbf{Dear Sir,}
The formulation of Ohm's Law needs to be clarified as follows:
"If one uses carefully selected and perfectly prepared starting materials, then, with some skill, one can construct an electrical circuit for which measurements of the current-to-voltage ratio, even if performed over a limited time, yield values which, after applying appropriate corrections, prove to be equal to a constant."

\hfill \textit{Copenhagen}

\section{Key to the Key System}
\textit{(A Long Letter to the Editor)}

Earlier, an opinion was expressed that the system of door keys in our institute is more complex than field theory. This is a clear distortion of facts, and to refute it, in this communication we present a simplified theoretical scheme on the basis of which this system was created.

Let us begin with definitions.
A \textbf{Key} consists of a shaft upon which \textbf{pins} are fastened.
A \textbf{Lock} consists of a slot with \textbf{holes} located corresponding to the positions of the pins on the key shaft. Furthermore, the lock contains a system of \textbf{levers} located behind the holes.

Let us now introduce the following three axioms:
\begin{enumerate}
    \item Pins turn levers; for the lock to open, all levers in the lock must be turned.
    \item If there is no pin, hole, or lever in a given position, we will speak of the presence of an \textbf{anti-pin}, \textbf{anti-hole}, or \textbf{anti-lever}, respectively.
    \item There are no levers behind anti-holes in any lock, for such a lock could not be opened.
\end{enumerate}

Let the pins, holes, and levers be described by the value $1$ of variables $a_i$, $b_i$, and $c_i$, respectively. The index $i$ is the position number. Anti-pins, anti-holes, and anti-levers correspond to the value $0$ of the same variables. Let us now define symbolic matrix multiplication in the following way:
\[
abc = a, \quad \text{if simultaneously } c \le b \text{ and } a \ge c;
\]
\[
abc = 1 - a \quad \text{otherwise.}
\]
From this, it follows that if $(a_1, a_2 \dots a_k)$ is an eigenvector of the operator
\[
\begin{pmatrix}
b_1 b_2 \dots b_k \\
c_1 c_2 \dots c_k
\end{pmatrix},
\]
then the key can open the lock.

Using this formalism, it is easy to find the total number of keys $N_k$ that open a given lock $(b/c)$. It is equal to:
\begin{equation}
    N_k = (2^{\sum (b_i - c_i)})k,
\end{equation}
and the number of locks $N_L$ that can be opened by a given key $(a)$ is equal to:
\begin{equation}
    N_L = (2^{\sum a_i - 1})k.
\end{equation}
In obtaining these expressions, the fact was taken into account that the lock $(0/0)$ is a trivial anti-lock. In equations (1) and (2), $k$ is the sum of Clebsch-Gordan coefficients, equal to unity.

The formalism developed above allowed us to solve the following problem. Let someone wish to walk from a certain Room $A$ through several doors into an arbitrary Room $B$. The number of keys necessary for this was maximized for an arbitrary choice of rooms $A$ and $B$. (The minimization problem was not solved, since its solution is trivial—identical locks).

Then, the staff of the institute were divided into a series of subgroups, and the key system was constructed in such a way that two conditions were simultaneously fulfilled:
\begin{enumerate}
    \item No single subgroup is capable of opening all the locks that can be opened by any other subgroup;
    \item The transformation properties of the groups correspond to the possibility of borrowing keys.
\end{enumerate}

The creators of the key system hoped that it was the only possible and complete one, and to a certain extent, this is fair. However, it turned out that keys which should not have opened certain doors do open them if inserted into the lock *not all the way*. For example, key $(11111)$ can open lock $(10000/11111)$ in $n=5$ different positions. The number $n$ was named the \textbf{strangeness} of the key-lock system.

Experimentally, it was found that our system of keys possesses a very high strangeness. However, this defect can be corrected if one requires the observance of the equalities $a_k = b_k = c_k = 1$ for the last position. Let us hope that in the next revision of the key system, this correction will be introduced.

This study does not extend to lock picks.

The author expresses gratitude to employees working in different groups for heated discussion of the touched-upon problems.

\anecdote{The Population of Italy}{
\textit{Found on the wall at the University of Milan Physics Department.} \\

\begin{figure}[h]
    \centering
    \fbox{
    \begin{minipage}{0.8\textwidth}
        \centering
        \vspace{0.5cm}
        \textbf{\Large POPULATION OF ITALY} \\[0.5cm]
        \begin{tabular}{lr}
            \textbf{Total Population} & \textbf{52,000,000} \\
            \midrule
            Over 65 years old & 11,750,000 \\
            \textit{Balance left for work} & \textit{40,250,000} \\[0.2cm]
            
            Under 18 years old & 14,120,000 \\
            \textit{Balance left for work} & \textit{26,130,000} \\[0.2cm]
            
            Non-working women & 17,315,000 \\
            \textit{Balance left for work} & \textit{8,815,000} \\[0.2cm]
            
            University students & 275,000 \\
            \textit{Balance left for work} & \textit{8,540,000} \\[0.2cm]
            
            Sick, insane, vagrants & 1,310,000 \\
            \textit{Balance left for work} & \textit{1,240,000} \\[0.2cm]
            
            Illiterate, artists, judges & 880,000 \\
            \textit{Balance left for work} & \textit{360,000} \\[0.2cm]
            
            Ministers, deputies, senators & 119,998 \\
            \midrule
            \textbf{BALANCE LEFT FOR WORK} & \textbf{2} \\
        \end{tabular}
        \vspace{0.5cm}
        
        \textbf{\Large Who are these two? You and Me.} \\
        \textbf{Let this tragic reality serve as a signal of alarm... especially for \textbf{You}, because \textbf{I} am tired of doing my duty to the country alone.}
        \vspace{0.5cm}
    \end{minipage}
    }
    \caption{Statistical proof of workload distribution.}
\end{figure}

}

%\begin{tabular}{lr}
%\textbf{Total Population of Italy} & \textbf{52,000,000} \\
%Over 65 years old & 11,750,000 \\
%\textit{Remaining for work} & \textit{40,250,000} \\
%Under 18 years old & 14,120,000 \\
%\textit{Remaining for work} & \textit{26,130,000} \\
%Non-working women & 17,315,000 \\
%\textit{Remaining for work} & \textit{8,815,000} \\
%University students & 275,000 \\
%\textit{Remaining for work} & \textit{8,540,000} \\
%Sick, insane, vagrants & 1,310,000 \\
%\textit{Remaining for work} & \textit{1,240,000} \\
%Illiterate, artists, judges & 880,000 \\
%\textit{Remaining for work} & \textit{360,000} \\
%Ministers, deputies, senators & 119,998 \\
%\textbf{REMAINING FOR WORK} & \textbf{2} \\
%\end{tabular} \\
%\vspace{0.3cm}
%\textbf{Who are these two? You and Me.} \\
%Let this tragic reality serve as a signal of alarm... especially for \textbf{You}, because \textbf{I} am tired of doing my duty to the country alone.
%}

% =========================================================
% CHAPTER 3: USEFUL ADVICE
% =========================================================
\chapter{Useful Advice}

\section{How to Write Scientific Articles}
\textbf{Author:} L. Solymar \\
\textit{Original Source:} Proceedings of the IEEE, Vol. 51, No. 4 (April 1963). \\
\textit{(Original Title: "Getting into the Act")}

\subsection*{Introduction}
The preparation of scientific papers for publication has been discussed from many points of view, yet many of its aspects remain neglected. It is surprising that the great advances achieved in the last decade in the conduct of scientific research have brought us no closer to a final solution to this problem. Many books and brochures have been written on "how to write articles," but they are devoted either to vague general recommendations ("write clearly," "explain your thoughts," "do not deviate from the topic," etc.) or to technical formatting ("leave margins," "captions must be typed," "illustrations should not exceed $10 \times 15$ cm," etc.). Without denying the importance of this advice, I believe it touches only upon a limited circle of secondary questions. In this note, I do not intend to introduce new ideas, but simply wish to share my experience in compiling technical articles and the valuable comments I have received from friends and colleagues.

\subsection*{Motives for Writing}
A whole range of reasons (from simple compulsion to the desire to improve one's social status) induces a person to write and publish scientific work. I will not go into details but will limit myself to four main motives:
1) Selfless desire to spread knowledge;
2) Concern for priority;
3) Concern for professional reputation;
4) Desire for promotion.

The first reason moves chiefly young people, and usually only when preparing their first work. The number of such authors is small, and for the majority, the first article is also the last. Consequently, this motive cannot be ranked with the others, which are far more powerful.

The second reason---priority---moves only a small group of authors, though in importance it surpasses all others. The desire to link one's name with a discovery has long been a distinctive feature of scientists. Since publication became the proof of discovery, the drive to forge articles as quickly as possible has been intense. However, the author must not forget the possibility of further applications. If he publishes his data, someone else might carry the ideas to fruition and deprive him of the harvest. The ideal solution is to guarantee priority by announcing the discovery, but delay detailed publication until the potential is fully assessed. The first to apply this method was Galileo, who sent descriptions of his astronomical discoveries to Kepler as anagrams, deciphering them only a year later. Since modern journals do not usually publish anagrams, current discoverers must act differently. I recommend starting with an intriguing title; the greater the impression the title makes, the less information needs to be communicated in the text. For example, a headline "Amplifier with Loaded Negative Inductance" will immediately convince every reader that a new principle has been discovered. The author will be forgiven if he is vague on the essentials and reports only in general terms.

The third reason is professional reputation. This can be achieved in various ways. It is enough, for example, to make an outstanding invention or receive a Nobel Prize, and your competence will be beyond doubt. However, for the majority, the only accessible way is to write a large number of articles, each contributing at least a microscopic amount to science. It is advisable to limit the first few articles to a narrow topic (e.g., "Connections in Waveguides") to win recognition. Later, the author must testify to his versatility by writing on a broader topic (e.g., "Ultra-High Frequency Oscillations"). After publishing three dozen articles, fame reaches saturation. This is the moment to stop printing (reviews don't count) and attempt to secure a leading administrative position.

\subsection*{Advice on Manuscripts}
So far I have considered only the reasons for writing. Now I would like to touch upon the position of a young author (without powerful co-authors) whose article must pass the gauntlet of reviewers.

How to ensure acceptance? Usually, reviewers are selected from leading scientists to filter out manuscripts worth printing. Unfortunately, leading scientists have little time and many administrative burdens. They cannot devote much time to reading an article, yet they must make critical remarks.

A beginner must take this into account. To avoid wasting time on complaints, he must write his article to satisfy the reviewer, whose sharp eye will detect the slightest anomaly. If the article is too long, he will be accused of verbosity; if too short, he will be advised to collect more data. If he reports on purely experimental work, he will be criticized for "lack of theoretical justification"; if he offers theory, he will be called "superficial." Therefore, I propose a compromise: the article should be 8 to 12 pages, typed, with about one-third occupied by mathematics.

Following these recommendations, the author has a fair chance his article will pass regardless of content. A quick glance at such an article will evoke the reviewer's favor. Everything then depends on his reaction within the next thirty minutes. If during this time he can quickly spot \textbf{three minor errors}, the article will be accepted.

Thus, the main task of the author is to provide the reviewer with material for three minor remarks. Below are recommendations for facilitating this choice:
\begin{enumerate}
    \item Pick an ambiguous title (all reviewers love to suggest their own titles).
    \item "Forget" to define one notation in the very first equation.
    \item Misspell a word (only one!), preferably one often spelled incorrectly.
    \item Deviate from standard notation (for only one parameter).
    \item Write $\exp(x)$ and $e^x$ interchangeably.
    \item Leave margins of 4 inches so the referee has space to write 'Rubbish!'.
\end{enumerate}

I hope these remarks will contribute to a better understanding of the work of compiling a scientific article and serve as a guide for the beginner.

\section{Glossary for Research Reports}
\textbf{Author:} C. D. Graham Jr. \\
\textit{Original Source:} Metal Progress, Vol. 71, No. 5, pg. 75 (1957).

\begin{center}
\renewcommand{\arraystretch}{1.5} % Increases row height for better readability
\begin{longtable}{|p{0.48\textwidth}|p{0.48\textwidth}|}
\hline
\textbf{THEY WRITE} & \textbf{THEY MEAN} \\
\hline
\endfirsthead
\hline
\textbf{THEY WRITE} & \textbf{THEY MEAN} \\
\hline
\endhead
It has long been known that... & I haven't bothered to look up the original reference. \\
\hline
...of great theoretical and practical importance. & ...interesting to me. \\
\hline
While it has not been possible to provide definite answers to these questions. & The experiments didn't work out, but I figured I could at least get a publication out of it. \\
\hline
The W-Pb system was chosen as especially suitable to show the predicted behavior... & The fellow in the next lab had some already made up. \\
\hline
High purity... \newline Very high purity... \newline Extremely high purity... \newline Super-purity... \newline Spectroscopically pure... & Composition unknown except for the exaggerated claims of the supplier. \\
\hline
A fiducial reference line... & A scratch. \\
\hline
Three of the samples were chosen for detailed study... & The results of the others didn't make sense and were ignored. \\
\hline
...handled with extreme care during the experiments. & ...not dropped on the floor. \\
\hline
Typical results are shown... & The best results are shown... \\
\hline
Although some detail has been lost in reproduction, it is clear from the original micrograph that... & It is impossible to tell from the micrograph. \\
\hline
Presumably at longer times... & I didn't take the time to find out. \\
\hline
The agreement with the predicted curve is excellent. & Fair. \\
\hline
Good. & Poor. \\
\hline
Satisfactory. & Doubtful. \\
\hline
Fair. & Imaginary. \\
\hline
...as good as could be expected. & Non-existent. \\
\hline
These results will be reported at a later date. & I might get around to this sometime. \\
\hline
The most reliable values are those of Jones. & He was a student of mine. \\
\hline
It is suggested that... \newline It is believed that... \newline It may be that... & I think... \\
\hline
It is generally believed that... & I have such a good objection to this answer that I shall now raise it. \\
\hline
It is clear that much additional work will be required before a complete understanding... & I don't understand it. \\
\hline
Unfortunately, a quantitative theory to account for these effects has not been formulated. & Neither does anybody else. \\
\hline
Correct within an order of magnitude. & Wrong. \\
\hline
It is to be hoped that this work will stimulate further work in the field. & This paper isn't very good but neither are any of the others on this miserable subject. \\
\hline
Thanks are due to Joe Glotz for assistance with the experiments and to John Doe for valuable discussions. & Glotz did the work and Doe explained what it meant. \\
\hline
\end{longtable}
\end{center}

\section{How to Address the American Physical Society}
\textbf{Author:} Karl K. Darrow \\
\textit{Original Source:} Physics Today, Vol. 14, No. 10 (October 1961).

Consider an actor in a hit show on Broadway, and contrast him with a physicist addressing the American Physical Society. The actor has all the advantages. He is speaking lines written for him by a master of the art of commanding the interest of an audience (remember that we are postulating a hit show). He has a gift for acting, and also a long experience in the art; otherwise he would not be in the cast. Even so, he is not allowed to speak his lines in any way that occurs to him. Every phrase, every inflection, every gesture, even the position that he is to take on the stage, has been tested or even prescribed by a professional director, who does not hesitate to give him mandatory instructions, or even to alter the lines if they seem ineffective.

One might assume that assured of such splendid collaboration, the dramatist would write a play two hours long without a break, and the manager would be content to offer the play in a barn with benches for the seats. This is apparently not the view of those who are experienced in such matters. Ample intermissions are provided, and an act which runs for as much as an hour is sufficiently rare to cause the critics to mention it. Usually the theatre has comfortable chairs and is well ventilated, or even air-conditioned. All this is provided to induce people to come to a play for the apprehension of which, with rare exceptions, no intellectual effort is demanded.

Now consider the physicist. He has thought out his own lines, and is not always proficient in this not altogether easy art. He has little or no training in the art of elocution, and no director has rehearsed him. His subject requires a considerable amount of mental effort on the part of his listeners. The listeners themselves are usually uncomfortable and sometimes acutely so. This may be because the chairs are uncomfortable, or because the room is hot and stuffy, or because the program has already been running for an hour or more without a break; or two or all three of these conditions may exist together. Laurence Olivier or Helen Hayes might well quail at the prospect of having to sway an audience under such conditions. Under these highly unfavorable circumstances, does the physicist strive to put on a reasonable facsimile of Olivier or Hayes? It may be conjectured that frequently he does not, because of the popularity of the saying that when a meeting of the American Physical Society is going on, the members are in the corridors or on the lawn instead of listening to the speakers. People with tickets to \textit{Turandot} are not standing around on the sidewalks outside of the Metropolitan Opera House when the curtain is up.

Can anything be done to amend this situation? Very little, I am afraid; but the following suggestions point in the right direction.

\begin{enumerate}
    \item \textbf{Speak loudly enough to be heard in the remotest part of the room.} Some people sincerely believe that their voices are too weak to achieve this. No doubt this is sometimes the case, but I venture to believe that most of them are wrong. In my youth I was constantly reproached for speaking too faintly, and I thought that I could not help it; experience proved me wrong. I do not think that I could manage a speech in the Metropolitan Opera House without an amplifier, but a physicist is not likely to be asked to speak in so large a hall, and if he were he could count on the presence of an amplifier. In a hall seating three hundred persons or fewer, the amplifier ought to be unnecessary except in pathological cases. If there is an amplifier, do not expect it to transform a conversational tone into a loud one. It is better to go to the opposite extreme, and pretend to yourself that the microphone is not there, even though you are speaking directly into it.
    
    The trick recommended by those who instruct speakers is to look at and speak to the people in the rear row. This is often made difficult by the fact that some of the prominent people in the audience are sitting in the front rows; this is particularly common in University colloquia. If this situation exists, ignore it. If Niels Bohr is sitting in the front row and Joe Doakes in the rear row, speak to Joe Doakes. Bohr will hear you.

    \item \textbf{Write out your speech in advance, and commit it to memory.} I have heard only one objection (from the viewpoint of the audience) raised against this procedure, and it seems to me groundless. It has been contended that a written speech is dull and lifeless; the implication is that an unwritten speech glitters with sparkling impromptus. But the presence of a manuscript need not prevent the speaker from substituting a sparkling impromptu for something that he has written; and if the impromptu fails to occur to him, the manuscript is there to carry him along. Of course, it is possible to memorize a speech without writing it out; this is recommended to those who hate to write. It is a fact that a good speech is likely to be looser in texture than a good article. No difficulty will arise from this cause if the speaker remembers that it is a speech that he is writing.
    
    There are some who think that it is better to hear an unprepared physicist groping for what he wants to say than a prepared physicist saying what he wants to say. It would be fascinating to see this theory given a trial by the Royal Festival Ballet, but nobody ever will. For an advanced student of the dance it may be instructive to see a dancer fall on her face, pick herself up, and resume her part in the ballet; but for practically everyone else it is acutely embarrassing.

    \item \textbf{If you cannot memorize your manuscript, read it aloud.} This bit of advice will probably be resented, for we have all suffered from dreary speeches poorly read. There is, however, no compelling reason why a manuscript should be poorly read. Lady Macbeth has to read a letter aloud in an early scene of the play; it is one of the high points of the drama. More than forty years ago Ethel Barrymore read a letter aloud in such a way that it is still remembered by elderly playgoers, though the play itself is forgotten. The trouble is largely that most readers glue their eyes to the manuscript for seven-eighths of the time, lifting their eyes from time to time to steal a glance at the audience as though to make sure that it is still there. Reverse the ratio. It is easy to keep your eyes on the audience during seven-eighths of the time and look at the manuscript during the other eighth. For a manuscript which you have composed yourself, it should be extremely easy. Try it and see.

    \item \textbf{Situate your topic in the general framework of physics at the beginning, and summarize your conclusions at the end.} Even in a ten-minute paper, a minute at the beginning and a minute at the end are not too much to reserve for these purposes. Do not fear to repeat your main points. I shall have more to say on this topic of repetition near the end.

    \item \textbf{Time yourself.} The members of the American Physical Society are now pretty well trained in the art of giving ten-minute papers, but longer ones are still apt to overrun. This is particularly serious when the closing bell rings when the speaker still has five minutes to go, and these five minutes comprise the conclusions which are the incentive for the paper. The speaker naturally does not want to omit the climax of his speech, and the chairman is seldom ruthless enough to insist.
    
    This is where a manuscript is particularly useful. Timing-marks can be inserted at the end of each page or along the margin, and the speaker (who should constantly be looking at his watch) will then know when he is running behind and will be able to catch up by leaving out relatively dispensable passages. One hundred and thirty words a minute, or say two-and-a-half minutes for a double-spaced typewritten page, is fast enough. In the timing, allow for twenty seconds or thereabouts of silence just after you make each of your difficult points. These gaps will give the audience a chance to think about what you have said; there are no laws requiring a speaker to be talking all of the time at his disposal. The difficulty in timing is greatest when the paper involves blackboard work or slides. Rehearsal is necessary in such cases, and is worth the effort.

    \item \textbf{Aim your discourse toward the average of the audience, not toward the topmost specialists.} Too many young theoretical physicists speak as though they were instructing Oppenheimer; too many band-spectroscopists, as if they were addressing Mulliken; too many solid-state physicists, as though the audience consisted of Seitz—and so it goes. This is not quite so flagrant a fault as it was in the days before the meetings of the Society splintered into simultaneous sessions, each attracting its own coterie of specialists; but it is still an error, and anyone who avoids it is doing his bit toward the all-important end of keeping physics from breaking up into a horde of narrow specialties.
    
    There is one specious argument for the procedure which I am deprecating here. The young man may think that the topmost specialist is also the prime job-giver, and therefore is the man whom it is urgent to impress. But in the first place, it seems plausible to suppose that the topmost specialist forms his opinions of the neophytes from their writings and from personal contacts; and in the second place, the job-giver in the audience may be, say, some chairman of a department of physics whose own specialty lies elsewhere, and who is going to assess the young man by his lucidity and not by his profundity. If these entirely reasonable suppositions are correct, the young man is doing himself a disservice by speaking as though he were addressing exclusively those who know more than he.

    \item \textbf{The problem of the blackboard.} This is one of the toughest of all problems, and here the theatre is of no use. I have never seen a play in which an actor had to write on a blackboard. I think that an actor would write on the blackboard without saying a word, and then turn to the audience and speak. For a physicist the psychological inhibition against doing this is quite invincible, but at least the attempt should occasionally be made. He can at least avoid the tendency to drop the level of the voice while addressing the blackboard. There are, however, two faults at the blackboard which can often be avoided.
    
    One should write his symbols large enough so that they can be read from the back of the room. I hope I never forget the shock which I once experienced when, having finished what I had fondly supposed to be a good lecture, I went to the back of the room and found that nothing I had written could be read beyond the middle rows. Sometimes the speaker finds the blackboards to be much smaller than he had reasonably counted on; in such a case he has to choose between altering his presentation and confining his effectiveness to the people in the nearer rows. Sometimes, of course, either the chalk or the blackboard is impossibly bad; the speaker is then helpless unless he is good enough to revise his plans and do the whole speech without the blackboard. One ought also to write his equations in the order in which he speaks them, instead of putting each in the nearest convenient empty spot and dabbing with the eraser to make more empty spots, so that at the end the board is littered with incoherent symbols. One should know in advance just how the board will look at every moment during the discourse, and at the end of the talk the board should carry all of the principal equations arranged in logical order. I am afraid that this is a counsel of perfection.

    \item \textbf{The problem of slides.} Most people who show slides at all show too many and show them too fast. (I suspect that this is often because the speaker has prepared too long a speech and tries to compensate by racing through the slides.) Rare is the slide which can be properly apprehended in less than thirty seconds, though exceptions do occur. It is impossible to assign a rigid maximum to the number of slides which can be shown effectively. I suggest seven for a ten-minute paper, but I make exception for the cases in which the argument is shown on slides instead of on the blackboard. The one advantage of the blackboard over slides is that the overfast speaker is obliged to slow down as he writes; this advantage can be shared by the slides if the speaker will give them time enough. There is much else excellent advice to be given about slides, but it has all been said by J. R. Van Pelt in the July 1950 issue of the \textit{American Scientist}. This should be required reading for all physicists.

    \item \textbf{The problem of the "jargon".} Some people ascribe the difficulty of understanding science to what they call the "jargon". This seems to imply that scientists use long technical terms out of perversity, when they could just as well use short familiar words. This is absurd. If I am giving a speech on a subject involving entropy or a synchrocyclotron, less than nothing will be gained if I avoid the word \textit{entropy} or the word \textit{synchrocyclotron} by some cumbrous periphrase or by some vivacious popular word which does not mean the same thing. Entropy is entropy and a synchrocyclotron is a synchrocyclotron, and there is no synonym for either. On the other hand there is nothing to prevent me from giving a brief definition of either. It does not have to be a complete definition: I may say that entropy is $\int dQ/T$ between certain limits of integration, or that a synchrocyclotron is a cyclotron in which the frequency is modulated so as to overcome the obstacle arising from the change of the mass of the nuclei with their speed. It may be objected that a person who does not know in advance what these words mean is unable to profit by the discourse. This view fails to take account of the fallibility of human memory. The listener may have forgotten what the words mean; he may even be able to recover the meanings during a few seconds of groping, but during these few seconds the speaker will go so far ahead that the gap cannot be closed. I have often observed that the place at which I lost contact with a speaker was the place at which he used a word which made me stop and ponder. It seems worth-while to avoid such dangers as far as possible.
    
    There is a sense in which physics is afflicted by what may be called jargons, though I should prefer to call them private languages. This is a phenomenon of recent years. Formerly physicists were few and far between, and one who did not make himself understood to his fellow-physicists a thousand miles away did not make himself understood to anybody. Nowadays many physicists do team work in large groups. In every such group a private language arises, characterized first of all by omissions. Relevant facts and even essential steps in an argument can safely be omitted within the group, because everybody knows them. In addition, the group invents all sorts of abbreviations, nicknames, and pet names for such things as parts of an apparatus, cosmic-ray tracks of various aspects, irregularities in crystal lattices, phenomena of hole-conduction, and even basic concepts of physics. No dictionary contains these terms; they travel by word of mouth, and often they do not travel fast enough. When they are spilled out before a meeting of the Society, disaster may ensue if they are not defined. Facility of travel and interchange of personnel are doing much to retard the development of a Berkeley language, an Oak Ridge language, a Murray Hill language, and the like; but the danger is always with us.

    \item \textbf{Style.} The concept of style being vague and the teaching of style lying in the province of another profession, I confine myself to two remarks.
    
    Textbooks of style advise the writer, and therefore inferentially the speaker, to strive for a proper proportioning of long words with short, and (what often comes to the same thing) of words of Greek, Latin, or French origin with words of Saxon origin. Now, a scientific article is perforce overloaded with words which are both long and of Greek or Latin origin. This suggests that whenever the speaker has an option, he should choose the short word over the long and the Saxon word over the Greco-Latin. If a sentence contains such words as \textit{ferromagnetism} or \textit{quantization} or \textit{electrodynamics}—not to speak of the atrocious \textit{phenomenological}—it is really amazing how much the sentence will gain in grace and fluency if all the other words are colloquial and short. This policy also tends to bring out the necessary long word in bold relief.
    
    It is said that the style of our forerunners was largely formed by the King James Bible, and that the style of our contemporaries is influenced by \textit{The New Yorker}. Neither of these publications can have much influence on those who do not read them. The suggestion is that physicists should not confine their reading to their professional literature. Read novels; read poetry; read essays; read history as written by notable writers; read Winston Churchill; and read Rebecca West—or if you simply will not go beyond the writings of scientists, read the Braggs and Eddington and Jeans and Bertrand Russell. Failure to observe this precept is partly accountable for the fact that it is seldom possible to tell from the style of an article in \textit{The Physical Review} who wrote the article, and for the further fact that scientists who try to write something for the general public so often do it badly.

    \item \textbf{A suggested experiment.} I have proposed, \textit{inter alia}, that a speaker should speak slowly, show his slides slowly, define his private-language terms, and repeat his main points. To anyone who deprecates this advice I suggest the following experiment.
    
    Choose an article in \textit{The Physical Review}; let it be in your own field if you will, lest the result of the experiment be too frightful. Sit down in an uncomfortable chair, and read the article—but read it according to the following prescriptions. Read straight through from beginning to end at the rate of 160 to 180 words per minute. Never stop to think over anything, not even for five seconds. Never turn back, not even to refresh your memory as to the meaning of a symbol or the form of an equation. Never look at an illustration until you get to the place where it is mentioned in the context; and when you get to that place, look at the illustration for ten or fifteen seconds and never look at it again. If this is not the way that your listeners will apprehend you when you give a paper, you are an outstanding speaker.
\end{enumerate}

\section{How Not to Listen to a Speaker}
\textbf{Author:} William B. Bean \\
\textit{Original Source:} The Journal of Irreproducible Results, Vol. 7, No. 2 (1959).

No orator, no matter how much energy he possesses, has a chance of defeating the drowsiness of the audience. Everyone knows that sleep during a long speech is significantly deeper than the state of hypnotic numbness known as "dozing." After such a sleep, you wake up refreshed. You have rested well. You firmly know that the evening was not wasted. Few of us have the courage to sleep openly and honestly during an official speech. After a thorough investigation of this question, I can present for the reader's consideration several original methods which have not been published until now.

Sit in your chair as deep as possible, tilt your head slightly forward (this frees the tongue so it hangs loosely without obstructing breathing). Loud snoring drives even the meekest speaker out of his mind, so the main thing is---avoid snoring; all breathing pathways must be free. It is difficult to give precise instructions on maintaining balance in sleep, but to prevent your head from wobbling from side to side, prop it up with your two hands and your torso, forming a sturdy tripod---Archimedes knew this was a very stable device. This reduces the risk of falling onto the floor (and climbing out from under the table usually happens amidst very unpleasant animation from the public). This way, your head won't drop onto your chest, and your jaw won't drop. Closed eyes should be hidden in your palms; at the same time, your fingers should squeeze your forehead into a frown. This produces the impression of intense mental work and somewhat puzzles the speaker. Screaming during nightmares is possible, but this is a risk one must take. Wake up slowly, look around, and do not start applauding immediately. This might turn out to be inappropriate. Better yet, wait until they wake you up with the final applause.

\section{Reports I Have Read... And Perhaps Written}
\textbf{Author:} Dwight E. Gray \\
\textit{Original Source:} Physics Today, Vol. 13, No. 11 (November 1960).

The technical report, as a specialized form of scientific literature, has risen from relative obscurity in recent years to a position of immense importance in the dissemination of information. Ideally, the function of every report is to convey information—to convey it accurately, clearly, and unambiguously. Alas, in practice, "reports" fall into a variety of categories based on their degree of approximation to this ideal. described below are a few of the more distinctive species I have encountered during fifteen years in the information business.

\subsection*{1. The Mystery Report}
The author of this document seems to be trying to keep the reader in suspense as long as possible regarding what the report is actually about. In extreme cases, he succeeds in maintaining the secret to the bitter end. The mystery is, of course, deepened if the report bears a non-committal title like "Progress Report" or "Summary of Work."

Mystery Reports usually begin with historical statements such as: "In 1927, Professor K. K. McJillicuddy, working at ABC University, discovered such-and-such." This opening strikes a note reminiscent of the "Once upon a time..." of children's fairy tales. For the author, it is an excellent device if he is not quite sure what he wants to say, or is simply reluctant to start saying it. For the reader, it establishes an atmosphere of uncertainty immediately.
The reader plunges into the body of the report not knowing whether the arrows lead to an exit, or indeed if there is an exit. Usually, the reader eventually unearths the author's message. However, the feat typically requires such a strain on his deductive powers that he has little energy left for rejoicing.

\subsection*{2. The "Say It Again" Report}
The aspirant to this category has a wide choice of methods. First, he must avoid simple, declarative sentences. The unholy trio of subject, predicate, and object is far too direct; instead, the author should use long, involved, circuitous phrases, qualifying them with parenthetical insertions and starting with dependent clauses.
To illustrate, I quote a sentence from a dialogue that occurred some years ago in a Congressional committee hearing. A Defense Department official was asked if his agency planned to build a certain underground structure. He replied:
\begin{quote}
"We are attempting to maintain a balance between static, as they are sometimes called, installations, the creation of which in individual cases may be associated with the construction of underground structures, on the one hand, and the effectiveness of our defensive means, on the other hand, which, obviously, according to views adopted at the present time, is considered as the strongest argument against the construction of closed defensive installations."
\end{quote}
To which a Congressman replied: "You are magnificent! I didn't understand a word of that eruption. For God's sake, what do you mean?"

To achieve the worst results, the author must strive to keep two indices high: the "syllable-per-word" count and the "words-per-sentence" count. The motto is: "Never use a monosyllable if a synonym of six or seven syllables is available."
Sentences should begin with such meaningful phrases as "It is interesting to note that..." or "Attention should be called to the fact that..." These verbal paddings give the report an impersonal flavor, as if absolving the author of responsibility for the contents.

\subsection*{3. The "Missing Thought" Report}
Reading this type of report reminds one of the "fill-in-the-blank" tests we took in school. Only here, instead of missing words, the reader must supply missing logic—significant chunks of information and argumentation which the author did not feel necessary to include.
The reader finds himself in the position of the students in the famous mathematics lecture story. The professor, covering the blackboard with a derivation, arrives at a point where he says, "From this, it is obvious that..." and writes a long, complex expression bearing no resemblance to what came before. He pauses, looks puzzled, mutters, and walks into his office. Thirty minutes later he returns, beaming: "I was right. It \textit{is} obvious."

In reading such a report, you glide smoothly along the rails of the author's logic only to find a washed-out bridge. You can see the track continuing on the other side of the chasm, but there is no bridge, and no lumber to build one.

\subsection*{4. The Camouflage Report}
Characteristic features (occurring singly or in combination):
\begin{enumerate}
    \item Data are presented, but not analyzed.
    \item Conclusions are drawn, but do not follow from the data.
    \item Recommendations are made, but do not follow from the conclusions.
\end{enumerate}

In the most finished examples of the Camouflage Report, the author, ostensibly a scientist, is actually a pitchman. He wants to sell something. High-class Camouflage Reports can be recognized immediately by their expensive covers (often with gold lettering), magnificent four-color illustrations separated by tissue paper, and glossy, high-quality paper.

Here we may conclude our discussion of the "freaks" of the report family. The true function of a report remains, as stated in the introduction, the transmission of information—transmission that is accurate, clear, and unambiguous.

\section{How to Use Slides}

\textit{(Speech at the banquet marking the close of the International Conference on Nuclear Structure, Kingston, Canada, 1960)}
\textbf{Author:} D. G. Wilkinson \\
\textit{Original Source:} Proceedings of the International Conference on Nuclear Structure (1960).

I have been asked to say a few words on an important question---how to use slides. It is difficult to initiate the amateur into all the subtleties of this art in a few minutes. Therefore, I intend to touch only upon the most elementary principles; one cannot expect more. I want to emphasize that my present communication is only an excerpt from the general "Rules of Conferencemanship" and is devoted to only one of the topics covered by that code. In such a short speech, I am deprived of the opportunity to touch upon such vital questions as: "How to mention your collaborators while implying that they do not deserve it," or "How to discredit the theory and experimental technique of your rival without understanding either one or the other."

The question of using slides breaks down into three sub-questions. On the third---"How to humiliate your opponent"---I was not permitted to speak. Two remain: "How to exhaust the projectionist" and "How to conquer the audience."

As regards the projectionist, the ultimate goal of the conferenceman is to drive him, if possible, to a nervous breakdown. It is important to establish the moment when you have succeeded in this, and then turn all attention to the listeners. The difficulty lies in the fact that, as a rule, you do not see the operator. But I believe it is usually sufficient to bring the process to the stage where his stuttering becomes audible in the hall; this has a useful unnerving effect on the audience. Such a state is self-sustaining, and after this, the operator can be left to himself.

Such primitive methods as using films of non-standard width or octagonal plates can be recommended only to the greenest novices. A satisfactory and more skilled opening is the \textbf{"3--2--1" Technique}. Here one exploits the fact that the operator always loads the first two slides into the apparatus while the chairman announces the talk, in order to switch on immediately. Instead of the expected "First slide, please," you say: "Third slide, please." (Elementary note: following this, you must quickly demand the second slide and only then---the first.)

The second method, best used in conjunction with the first, is the \textbf{"White Spot Wobble."} All slides are usually marked in one corner with a white spot, on which the operator must place his right thumb to ensure the correct orientation. You must place this spot in the \textit{wrong} corner, so everything appears upside down. Applied together with the "3--2--1" technique, this acts stunningly. You show slight confusion, and then, brightening up, address the operator: "Oh, I beg your pardon, these plates are marked in an unusual way. You know, I usually take a personal operator with me." Then after some thought: "He is left-handed," and finally: "But don't worry, only the first few shots are marked that way."

Immediately following this should be the \textbf{"Parity-Violating Slide,"} which projects incorrectly no matter how you turn it in the frame. There are many ways to make such a slide. The simplest, yet exquisite: all letters should be drawn correctly, but words written from right to left.

Address the operator often. It is good if it is not quite clear whom you are actually addressing---him or the listeners. Avoid absurdly complicated instructions. Simply say: "In two slides I want to look again at the fourth from the end of those that have already been shown." And after the next frame add: "I, of course, meant that slide which will be fourth from the end of those shown \textit{after} you show these two frames, and not the one which was fourth from the end when I spoke about it." After this, skip one slide.

These simple methods are sufficient for the majority of operators. In case of unexpected resistance, steeper measures can be taken.
A grandmaster move is the \textbf{"Thermal Jammer."} To the bottom edge (on the screen, it will, of course, be the top) of a specially shortened slide, a thin, sufficiently elastic bimetallic strip is attached. When the image appears on the screen, you discuss it for some time and remark that the difference from the next curve is small, but will be striking if the slide is changed quickly enough. The bimetallic strip will have heated up by this time, bending the frame, and when you cry "Next, please!", it will inevitably jam firmly halfway. Driven by impatient requests of "Faster!", the operator will abandon attempts to move the frame by delicate tapping and will grab it with both hands, yanking it back and forth. The apparatus will crawl along the floor on all four legs, emitting pleasant screeching sounds. This always entertains the public.

Finally, the most sophisticated technique is the \textbf{"Double Blank."} Two absolutely clean slides are taken. They are placed after a series of frames which are demonstrated in rapid sequence, exerting a hypnotic effect. Suddenly this series ends, and the operator, having loaded the next pair, sighs with relief. However, the snapshot he projects after the next "Next, please," is one of the "Double Blanks," and the second frame in the holder is also empty. A few seconds later, an icy voice rings out: "I said: next, please!"---and the operator sees with horror that there is nothing on the screen, although he remembers perfectly well that he inserted the slide. Feeling his world collapsing, he pokes his finger straight into the middle of the clean slide to make sure it still exists. But a large hole has been made in the center of the slide beforehand; only the rim remains. Almost losing consciousness, the operator grabs the next frame from the box and tries to squeeze it into the holder, which is, naturally, occupied...

\chapter{Methodology and Education}

\anecdote{The Exam}{
At a physics exam, a professor writes an equation:
\[ E = h\nu \]
and asks the student:
"What is $\nu$?" \\
"Planck's constant!" \\
"And $h$?" \\
" The height of that plank!"
}

\section{How to Conduct Scientific Research}
\textbf{Authors:} A. B. Mishaym, S. D. Adam, E. F. Onyakh, J. G. Bamada \\
\textit{Original Source:} The Journal of Irreproducible Results, Vol. 6, No. 1 (1958).

\subsection*{Pre-publication}
We propose a very useful method that allows one to publish more frequently. One must anticipate the results of an experiment and publish them in advance. This saves a tremendous amount of time. In this way, one can even save oneself the trouble of finishing the experiment; since the article is published, one can move on to something else. This trick, combined with a well-developed imagination, allows one to publish a large number of experimental papers without conducting any experiments at all, thereby saving a heap of government funds. True, a slight awkwardness may arise if someone else has already conducted the experiment and obtained different results. But an experienced scientist in this case can: 
a) completely ignore this circumstance; 
b) write a series of articles devoted to describing the subtle differences in experimental conditions that led to the difference in results; 
c) express deep gratitude for the pointing out of the error and write a series of articles on new experiments yielding the correct results, while using the old, erroneous ones to demonstrate all the difficulties and subtleties of this beautiful work.

\subsection*{Joint Publication}
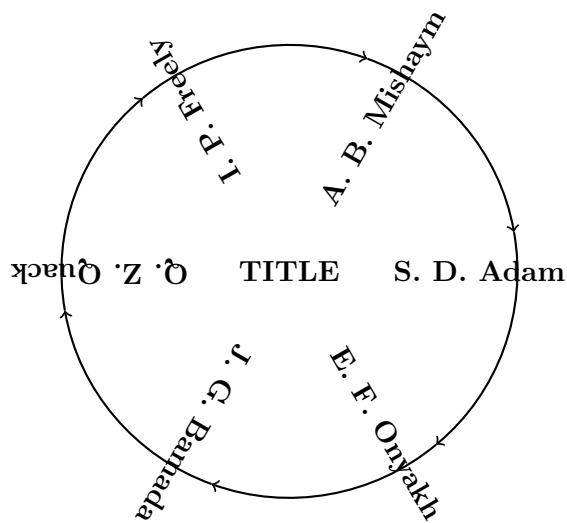
\begin{figure}[h]
    \centering
    \begin{tikzpicture}
        \def\R{2.5} % Radius
        \node at (0,0) {\textbf{TITLE}};
        
        % The Authors arranged in a circle
        \node [rotate=60] at (60:\R) {\textbf{A. B. Mishaym}};
        \node [rotate=0] at (0:\R) {\textbf{S. D. Adam}};
        \node [rotate=-60] at (-60:\R) {\textbf{E. F. Onyakh}};
        \node [rotate=-120] at (-120:\R) {\textbf{J. G. Bamada}};
        \node [rotate=180] at (180:\R) {\textbf{Q. Z. Quack}};
        \node [rotate=120] at (120:\R) {\textbf{I. P. Freely}};

        % Arrows indicating "no beginning"
        \draw[->, thick] (70:\R+0.5) arc (70:10:\R+0.5);
        \draw[->, thick] (10:\R+0.5) arc (10:-50:\R+0.5);
        \draw[->, thick] (-50:\R+0.5) arc (-50:-110:\R+0.5);
        \draw[->, thick] (-110:\R+0.5) arc (-110:-170:\R+0.5);
        \draw[->, thick] (-170:\R+0.5) arc (-170:-230:\R+0.5);
        \draw[->, thick] (130:\R+0.5) arc (130:70:\R+0.5);
    \end{tikzpicture}
    \caption{The ``Circular Reference'' method for determining authorship priority.}
  \label{fig:circular-authors}

\end{figure}

Much research has been devoted to the art of placing one's name at the head of the list of authors, but some subtle questions remain unilluminated.
\textbf{The Alphabetical Trick.} Since alphabetical order in compiling lists of authors is gradually becoming generally accepted, it is useful to concentrate on creating a solid advantage for oneself personally. This can be achieved in two ways: change your surname so that the new one begins with the letter A, or choose co-authors with surnames from the lower half of the alphabet (see, for example, G. C. Wick, A. S. Wightman, E. C. Wigner, \textit{Phys. Rev.}, 88, 101, 1952). But in this case, it is easy to miss the mark. One should not chase co-authors of too large a caliber.  An article by A. Bar-Nudnik and Albert Einstein will always be referred to as "Einstein et al.", regardless of the order of names. A more elegant solution to this problem is shown in Figure~\ref{fig:circular-authors}: arrange the authors in a circle, so that no one is first or last.

%\begin{center}
%\begin{tikzpicture}
%    \def\R{2.5} % Radius of the circle
%    \def\num{6} % Number of authors
%    \node at (0,0) {\textbf{TITLE}};
%    
%    % The Authors arranged in a circle
%    \node [rotate=60] at (60:\R) {\textbf{A. B. Mishaym}};
%    \node [rotate=0] at (0:\R) {\textbf{S. D. Adam}};
%    \node [rotate=-60] at (-60:\R) {\textbf{E. F. Onyakh}};
%    \node [rotate=-120] at (-120:\R) {\textbf{J. G. Bamada}};
%    \node [rotate=180] at (180:\R) {\textbf{Q. Z. Quack}};
%    \node [rotate=120] at (120:\R) {\textbf{I. P. Freely}};
%
%    % Arrows indicating "no beginning"
%    \draw[->, thick] (70:\R+0.5) arc (70:10:\R+0.5);
%    \draw[->, thick] (10:\R+0.5) arc (10:-50:\R+0.5);
%    \draw[->, thick] (-50:\R+0.5) arc (-50:-110:\R+0.5);
%    \draw[->, thick] (-110:\R+0.5) arc (-110:-170:\R+0.5);
%    \draw[->, thick] (-170:\R+0.5) arc (-170:-230:\R+0.5);
%    \draw[->, thick] (130:\R+0.5) arc (130:70:\R+0.5);
%\end{tikzpicture}
%\captionof{figure}{The "Circular Reference" method for determining authorship priority.}
%\end{center}

\subsection*{Private Correspondence ("If we can't beat them, let's join them")}
If you have learned in a roundabout way that someone is finishing excellent work and is about to publish it, send him a letter outlining his work in the form of an idea that "recently occurred to you." Explain that you are writing to him because you heard that he is also interested in this, and later, having learned that he came to the same results "independently of you," propose a joint publication.

\subsection*{Congresses}
The art of traveling from one international congress to another, reporting everywhere interesting work performed by someone else in your institute who for one reason or another could not go, has reached a high level of development. An expert congressman can virtuously prevent the possibility of sending anyone else on a business trip, even if he himself has not worked for many years and does not understand the works being presented.

\subsection*{Polemics}
One must learn to use the mistakes of one's colleagues to increase the number of one's own printed works. The exponential growth in the total number of scientific publications is accompanied by a huge increase in the amount of nonsense appearing in so-called serious scientific journals. Without any difficulty, one can find in the literature an article that is either a) completely erroneous, or b) contains correct results obtained by committing an even number of mutually canceling errors, or c) is full of minor inaccuracies. It can be used in one of the following ways:
\begin{enumerate}
    \item Write several short notes to different journals pointing out errors and inaccuracies.
    \item Write a long article criticizing the original work and redoing everything "properly." The true difference may lie in the removal of a few insignificant sloppinesses.
    \item Correct and rewrite the original article and publish it, citing the first as an independent but weak attempt undertaken with unsuitable means.
\end{enumerate}

\section{The Computer Poet}
\textit{Original Source:} Folklore / Press Releases regarding the RCA-301 computer (circa early 1960s).

The RCA-301 computing machine was taught to write blank verse. The vocabulary of the semiconductor poet is 130 words. The size of the verses is rigidly set. Starting the next poem, instead of a title, the machine puts a serial number: "Poem Number So-and-So," and at the end puts its signature: "RCA-301". Below is a literal translation of one such poem.

\begin{quote}
\textbf{Poem No. 929} \\
While blind sleep swam over broken hopes, \\
Space oozed with pain over broken love, \\
Your light was slowly banished from secretive people, \\
And the sky did not sleep.

\hfill \textbf{RCA-301}
\end{quote}

According to programmers, this work strongly resembles the poems of modern poets Eliot and Cummings, but none of them can compete with the machine in productivity. The RCA-301 writes 150 quatrains per minute.

\section{The Preparation of Coffee in Research Laboratories}
\textbf{Author:} A. Kohn \\
\textit{Original Source:} The Journal of Irreproducible Results, Vol. 8, No. 1 (1959).

A considerable portion of the time in laboratories is devoted to the preparation and consumption of coffee. The following is a standardization of methods.

\subsection*{Materials and Methods}
All commercially available varieties of coffee are used (except acorn). Boiling is performed in borosilicate glass beakers, Erlenmeyer flasks, or, in emergencies, in distillation apparatus. Sources of heat include gas burners, electric hot plates, bunsen burners, and occasionally high-voltage arcs.

\subsection*{Procedures}
\begin{enumerate}
    \item \textbf{The Crude Method.} Coarse-ground coffee is poured into cold tap water, which is then brought to a boil. The sediment is suspended by vigorous shaking, and the resulting slurry is poured into cups. Verification that the product is indeed coffee can be achieved only by photometric analysis.
    
    \item \textbf{The Refined Method.} Water is heated to $99^\circ$C. Fine-ground coffee is added (approx. 1.2 g per cup). The liquid is brought to a boil and removed from the flame. Centrifugation of the sediment is optional.
    
    \item \textbf{The "Expert" Method.} Water is brought to a boil, the flame is reduced to minimum, and coffee is added. The mixture is boiled for exactly 10 seconds.
    
    \item \textbf{The "Espresso" Method.} This term has become generic for steam extraction. In the laboratory, a Soxhlet extraction apparatus may be used, though this method is too labor-intensive unless one has a surplus of graduate students and a shortage of ideas.
    
    \item \textbf{The "Instant" Method.} This involves the use of soluble powder. It has two advantages: speed, and the absence of the tell-tale aroma of fresh coffee, which prevents the invasion of thirsty colleagues from neighboring laboratories.
\end{enumerate}

\section{Advice to University Professors}
\textbf{Author:} Marvin Camras \\
\textit{Original Source:} IRE Transactions on Audio.

In our days, educators are constantly revising curricula, and we often hear about "new math" and other innovations. We propose here to introduce small lecture courses into the curricula, which, from our point of view, would be extremely useful for future engineers.

\subsection*{Anti-standardization (Creative Invention)}
The goal of the course is to teach the creation of devices in which not a single part can be replaced by a standard one. This requires great ingenuity, however, successful labor is generously rewarded. The expensive and wasteful "Anti-standard" is a supreme achievement that is rarely obtained by accident. A scientific approach in this matter allows creating super-non-standard devices in which all dimensions are non-typical, all parts are electrically, mechanically, and chemically incompatible, possess increased corrosion instability and increased brittleness, and thus possess the maximum rate of failure.

\subsection*{Committees}
The goal of the course is to teach the future engineer the technique of using committees and working in them. As is known, committees are the ideal means for eliminating oneself from any responsibility, for delaying the execution of all tasks, and for creating a "managerial" mood among directors. Additionally, a committee meeting is a not-bad means to kill evening time intended for rest. This cycle of lectures will help listeners hone their art of putting off things from day to day, appearing smarter than they are, teaching them to use jargon, and in the role of chairman will help to amaze everyone with their concentration, joke effectively, and gracefully close meetings.

\subsection*{Ability to Adapt to the Environment}
Previously, everyone relied exclusively on intuition in this area, and it only recently became a science thanks to a number of selfless enthusiasts. Recognized masters of this business should be invited to give the course. They will share experience in their art of appearing eternally busy, confident in their abilities, full of infectious enthusiasm, striving to expand this or that department 1) by personnel, 2) territorially, 3) in the sense of financial appropriations. They will teach how to effectively plan a budget, how to report on expenses with brilliance, how to process bills, appear smart in the presence of 1) engineers, 2) administrators, 3) cleaners, and also the ability to appear unambitious. They will give advice on how to eat, what to drink, how to choose a car, a wife, and a lawnmower.

\subsection*{Information Channels}
The goal of the course is to teach the future engineer methods of approaching the boss's secretary, viewed as the most important source of useful information. However, without theoretical preparation, one can do foolish things, failing to take into account certain subtleties, like the fact that the secretary from another department, with whom the boss's secretary drinks her coffee during the break, is possibly more talkative.

\section{Advice to the Examiner}
\textit{Original Source:} Adapted from Electronics Magazine.

\begin{enumerate}
    \item Explain to the candidate that his whole professional career depends on the outcome of this examination. Stress the gravity of the situation. Put him at his ease... near the edge of the chair.
    \item Ask the hardest question first. If the first question is sufficiently obscure, the candidate will be too nervous to answer the subsequent simple ones.
    \item Be reserved and stern with the candidate; be very cheerful with your fellow examiners. It is effective to address witty remarks to your colleagues about the candidate's answers, ignoring him as if he were not in the room.
    \item Force the candidate to solve problems by \textbf{your} method, especially if it is non-standard. Restrict him by inserting numerous conditions and reservations. The idea is to complicate a problem that would otherwise be quite simple.
    \item Lead the candidate into a trivial error, and let him flounder as long as possible. Immediately after he notices the mistake, but just before he can correct it, correct him yourself with a look of disdain. This requires great timing.
    \item When the candidate begins to drown, never help him out. Yawn... and move to the next question.
    \item Ask questions like: ``Didn't you learn this in kindergarten?''
    \item Never allow the candidate to ask clarifying questions, and never repeat your own explanations.
    \item Every few minutes, ask him if he is nervous.
    \item Wear dark glasses. Inscrutability is unnerving.
    \item As the exam ends, say: ``Wait outside. We'll let you know.''
\end{enumerate}

\section{Mathematization}
\textbf{Author:} N. Vanserg (Pseudonym of H. E. McKinstry) \\
\textit{Original Source:} Science, Vol. 127, No. 3306 (May 9, 1958).

In a paper published some years ago, the present author intimated (with becoming subtlety) that, since most scientific truths are relatively simple, the scientist must, in self-protection, prevent his colleagues from discovering that his ideas are simple. He must, therefore, evolve a disguising technique. The objective is to make his published work sufficiently difficult to read so that no one will try to read it, but all will respect his profundity.

\subsection*{The Technique of Mathematization}
A device which has lately been developed to a high degree of sophistication is the use of a little mathematics. If you can present your results in a form that is sufficiently unintelligible, no one will likely discover that they are simple.

\subsection*{Typographical Tricks}
The first and most elementary trick is to change the letters. If you usually use $\phi$ for the angle of dip, change it to $\psi$. Even the placement of the symbol $e$ to the right of the parenthesis has a certain psychological value. This is a conscious deception, but one that is rarely punished; one can always blame the printer. In general, however, the author need not resort to such crude devices, for the typesetter will usually cooperate in introducing enough confusion to baffle the most persistent reader.

\subsection*{Strategy of the Secret Symbol}
But what if, by some mischance, the equations don't get badly garbled? The mathematics is apt to be all too easy to follow, provided the reader knows what the letters stand for. Here then is your firm line of defense: at all cost prevent him from finding out. Thus you may state in fine print in a footnote on page 35 that $V^*$ is the total volume of phase 1, and then introduce it in an equation on page 873. If you have been careful to give no clue as to what it means, you can feel reasonably safe.

By stealthily introducing all the letters of the Roman, Greek, and Gothic alphabets, you can force the curious reader to read the book backward to find out what the symbols mean. Books that read backward as well as forward are particularly impressive.

When reading backward has become a habit with the reader, and he considers this method normal, confuse the trail. Introduce, for example, $\mu$ in an equation on page 66, and do not define it until page 86.

But the moment arrives when the reader thinks he knows what all the letters stand for. This is the time to bring up your reserves. Everyone knows what $\pi$ is; this will help you to throw him off the scent. The poor reader will multiply by 3.1416 for hours before he discovers that $\pi$ is the osmotic pressure. If you are careful not to give the show away, this should cost him a good hour and a half.

The same principle can be applied to any letter. Thus you can write $F$ for free energy on page 141, and the astute reader, if he is accustomed to Helmholtz's notation, will spend a good deal of his own free energy before he realizes that you meant Gibbs' free energy, which he thinks is $G$. In general, $F$ is a particularly happy letter, as it can be used not only for any unspecified brand of free energy but also for fluorine, force, friction, Faradays, or a function of something or other. This tends to increase the degree of randomization $dS$ ($S$, as is well known, stands for entropy and... sulphur).

Asterisks and primes can also be used for camouflage. Define some pressure as $P^*$ and say nothing about it. The reader will search for a footnote that isn't there. And when the seeker after truth reads that $S$ is $10^{14}$ cal, he will think: "Wow, what a hell of a lot of calories!"---and will continue to think so until he reads the page to the end, stumbles on footnote 14, and says: "Ah...".

\subsection*{The "Hence" Trick}
But the most successful device is the use of the word "hence" followed by a colon. You tear out two pages of derivation and write "hence:" followed by the conclusion. I guarantee that the reader will guess for a good two days where this "consequence" came from. Even better is to write "obviously" instead of "hence," since there is no reader who would dare to ask anyone for an explanation of an obvious thing. By this, you not only confuse the reader but also instill in him an inferiority complex, and this is one of the main goals.

All that has been said is, of course, elementary and generally known. The author is currently finishing a two-volume work on mathematization, including examples and problems for independent exercises. There will be so many classified notations, riddles, typos, and consequences flowing from nowhere in it that no one will be able to read this work.

\anecdote{Who Knows Best?}{
"The best person to decide what research work shall be done is the man who is doing the research. The next best is the head of the department. After that you leave the field of best and turn to increasingly worse groups. The first of these is the research director, who is probably wrong more than half the time. Then comes a committee, which is wrong most of the time. Finally there is the committee of company vice presidents, which is wrong all the time." \\
\hfill \textit{--- C. E. Mees}
}

\section{Instructions for Authors}
\textbf{Author:} Jack Ewing \\
\textit{Original Source:} The Journal of Irreproducible Results, Vol. 12, No. 1 (1963).

The Journal accepts for publication articles which have been rejected by all other journals. If your article has been returned from another journal (lower case), iron it flat and send it to The Journal (upper case).

\subsection*{Preparation of Manuscript}
\textbf{Text:} Manuscripts should be typed on standard sheets of parchment, size $8 1/2 \times 11$ inches. Typing should be on no more than two sides of the paper. Start each paragraph on a new page. Leave margins of 4 to $4 1/2$ inches on both sides.

The author must use clear, concise, crystal-clear English, preferably using words of not more than two syllables and four letters. Do not use words understandable to your colleagues. For example, do not write "reduced," write "modified." Do not use sentences of more than 120 words without including at least one verb.

In all matters concerning spelling, capitalization, etc., follow Webster's Dictionary. The editor will change everything anyway. 
Abbr. s. b. red. to min. 
Do not tabulate data that can be described in the text. Do not describe in the text data that can be tabulated. Do not express ratios in mg/kg.

\textbf{References:} Name of author, address, and volume number. If possible, the year. (Nicknames may be used if the author is well known.)

\textbf{Illustrations:} These may be prepared by any method. Clarity is not required. On the back of each photograph, write brief instructions to the editor (avoid unprintable words).

\textbf{Conclusions:} If possible, conclusions should be shorter than the text and presented in a form suitable for use as an abstract.

% =========================================================
% CHAPTER 4: NOT ONLY PHYSICISTS JOKE
% =========================================================
\chapter{It’s Not Just Physicists Who Joke}

\section{Parkinson's Laws in Medical Research}
\textbf{Author:} C. Northcote Parkinson \\
\textit{Original Source:} New Scientist, Vol. 13, No. 270 (January 25, 1962).

It appears to me, as an outside observer, that the people who apply to foundations and trusts, to institutions associated with the names of Rockefeller, Guggenheim, and Ford, should do some preliminary research in grantsmanship. Without this they are liable to disappointment. Knowing that the money must be given away and that the income will become taxable if not promptly allocated and spent, they imagine too often that the originator of a scheme for expenditure should be welcomed with open arms.

Suppose that \textbf{Dr. Tapfund} has a plan for measuring the incidence of philately among the Chinese minority population of Hong Kong. He pictures himself in the offices of the \textbf{Vanderfeller Trust}, confronted by its higher executives: \textbf{Dr. Grantley}, \textbf{Mr. Handout}, \textbf{Mr. Offering}, and \textbf{Mr. Scatterbug}. They are all delighted with his research plan but question whether a million and a half dollars will suffice. They feel that five million would be justified.
"Do you mean Hong Kong dollars?" asks Tapfund, his original estimate having been in that currency.
"No," says Grantley, "never heard of them. I mean dollars, U.S."
Tapfund hastens to assure him that U.S. dollars will do just as well. Grantley signs a check and wishes Tapfund the best of luck. The interview is over. Such is the dream.

But what is the reality? Dr. Tapfund finds himself in the offices of the \textbf{Bored Foundation}, confronted by \textbf{Dr. Knowsleigh}, \textbf{Mr. Nevershed}, and \textbf{Mr. Knott}.
Knowsleigh says that Hong Kong is colonial territory and their charter precludes them spending a cent there.
Nevershed thinks that philately is more of a social evil than an actual disease, and therefore outside their terms of reference.
Knott thinks the scheme politically dangerous as likely to offend Nationalist China.
They agree, in chorus, that the scheme is inadmissible, unacceptable, immoral, and illegal. Tapfund is thrown into the street and the janitor instructed never to admit him again.

Where did Tapfund go wrong? There are people whose job it is to spend money. He offers them a way of spending it. Why is he thrown out? The answer is that he has proposed his own plan. That is why they unanimously rejected it. The essence of grantsmanship is to persuade the Foundation executives that it was *they* who suggested the research project and that you are a humble tool in their hands.

Suppose, now, that you have gained your grant—perhaps from the government, perhaps from a public charity, most likely from a private trust. Your immediate object is to spend this money, and spend it quickly, so as to ask for more. Benefactors generally prefer to pay for a building, because their name can be put on it. What advertisement does the donor get from Dr. Tapfund? A footnote in a medical journal? A tombstone over his grave?

If you decide to build, it is best to attach a pompous memorial arch to an existing hospital, locating it between the laboratory and the administration block, and building a cozy apartment for yourself nearby. The inscription on the memorial arch can be worded so as to imply (without actually stating) that the donor paid for the *entire* hospital. I would undertake to compose such an inscription myself for a small fee.

But all buildings have a common drawback: the number of scientists working in them increases very rapidly, they fill the building and overflow it, as a result of which the problem of space arises more acutely than before.

A similar law of continuous growth applies to the number of journals covering progress in a certain field of science. Why does this happen? I have managed to uncover the true causes of the multiplication of scientific journals.
Let us suppose that the oldest and most respectable journal (Journal No. 1) was edited for many years by \textbf{Professor A}. He refused to publish any papers with which he disagreed. This lasted for several years and terribly bored \textbf{Professor B}, who never agreed with Professor A.
If, for example, they were asked to define a "unit," I am sure they would write it differently. With such a striking difference in views, it is not surprising that the articles of Professor B were invariably returned to the author. Upon the expiration of twenty-three years, he decided to found \textbf{Journal No. 2}. This publication began to come out on a more liberal basis, and at first, everything was printed in it, except for the works of those authors regarding whom it was known that they were followers of Professor A. But Professor B had his own high principles. He believed that any views deserve free expression; he insisted only that they be set out consistently and scientifically. And so, one day, he had to reject the works presented by \textbf{Professor C}.
Everyone considered C an original thinker, but found that he was somewhat hasty in his conclusions. Finding his papers refused by Journals 1 and 2, he became the founder of \textbf{Journal No. 3}, which was open from the start to the most slipshod presentation of the vaguest ideas.
But even C had to draw the line somewhere! He stubbornly refuses to publish the works of \textbf{Professor D} on the pretext that D does not know how to spell. Professor C does not want it rumored that Journal No. 3 accepts manuscripts typed on one side of the paper at double spacing. He must maintain the prestige of the journal. On the other hand, no one will blame Professor D for founding \textbf{Journal No. 4}.
It is precisely such a development of events that led to the fact that on questions of dentistry alone we publish about eighty journals.

What is the normal course of events? A person who has made a significant contribution to science is persistently offered subsidies to expand the front of research. This is exactly what happened with \textbf{Dr. Lockstock}. Can one forget his speech delivered at the meeting of the American Federation of Clinical Research in 1938? According to his theory, artists creating modern abstract paintings, as a rule, suffer from color blindness. With this, he created a reputation for himself, and the Trust hurried to generously subsidize his further work.
Lockstock was asked to ascertain whether the composers of modern dance music for teenagers are all tone deaf (as \textbf{Dr. Barrel} had suggested) or merely subnormal (as Dr. Lockstock himself had surmised).
This was a grandiose project. Division A was devoted to color blindness, and Division B to the subnormal among bandleaders.
From now on, Dr. Lockstock had to deal with the organization of the work of his staff, numbering 432 people. He lost the opportunity to engage in scientific work—obviously. But not many people understand that on this path they also lose the opportunity to manage someone *else's* scientific work. They will spend all their time on problems of space, pay, and leave rosters.

We can now formulate {\bf Parkinson's Law for Scientific Research}:
\begin{quote}
    \textbf{"Successful research attracts the bigger grant which makes further research impossible."}
\end{quote}
In accordance with this law, we mostly end as administrators.

\section{Common Sense and the Universe}
\textbf{Author:} Stephen Leacock \\
\textit{Original Source:} "Last Leaves" (1945); reprinted in New Scientist (1962).

\begin{quote}
"In the space of one hundred and seventy-six years the Lower Mississippi has shortened itself two hundred and forty-two miles. That is an average of a trifle over one mile and a third per year. Therefore, any calm person, who is not blind or idiotic, can see that in the Old Silurian Period, just a million years ago next November, the Lower Mississippi was upwards of one million three hundred thousand miles long, and stuck out over the Gulf of Mexico like a fishing-rod. And by the same token any person can see that seven hundred and forty-two years from now the Lower Mississippi will be only a mile and three-quarters long, and Cairo and New Orleans will have joined their streets together, and be plodding comfortably along under a single mayor and a mutual board of aldermen. There is something fascinating about science. One gets such wholesale returns of conjecture out of such a trifling investment of fact." \\
\hfill --- \textit{Mark Twain, "Life on the Mississippi"}
\end{quote}

Speaking in December 1941 at the annual meeting of the American Association for the Advancement of Science, Professor Edwin Hubble of the Mount Wilson Observatory made the announcement that the Universe is \textit{not} expanding. This is good news. 

For twenty-five years, indeed ever since Professor de Sitter launched his theory in 1917, we have been living in an expanding universe, one in which everything flies away from everything else at a terrific speed. It made us feel like the man in the story who flung himself on his horse and rode madly off in all directions. The idea was majestic, but it created a sense of dispersion. 

Yet we had to believe it. We had to rely on the authority of men like Spencer Jones of the Royal Astronomical Society, who told us in his \textit{Life on Other Worlds} (1940) that "a distant nebula in Boötes is receding at 24,300 miles a second." We calculated that it must be 230,000,000 light years away. We knew that a light year means the distance light travels in a year at 186,000 miles a second. In other words, this "distant nebula" is 1,049,970,980,000,000,000,000 miles away. 

And now it turns out it isn't moving! The astronomers did not assume the expansion; they proved it by the behavior of the red band in the spectrum, which blushed at the idea. One of our most distinguished astronomers, Sir Arthur Eddington, wrote a book about it, \textit{The Expanding Universe}. Astronomers took the explosion of the universe with the same cool detachment as when they announced that the sun is dying and that the universe will eventually freeze.

But the joy delivered to us by Professor Hubble is tempered by doubts. Do not think that I mean any disrespect to science. In our time that would be as bad as blasphemy in the time of Isaac Newton. But... Today we expand, tomorrow we contract. First we are curved, then we are straight. Just as we were going to freeze to death at absolute zero, it gets warm again.

We have a right to ask, where are we? Einstein answers: "Nowhere," because there is no place to be.

Take the Second Law of Thermodynamics, that grim decree of Fate which doomed the universe to die of cold. I look back with regret to the tears I wasted over that. I recall reading as a boy a book by Richard Proctor called \textit{Our Place Among Infinities}. It was terrifying. The sun, it appeared, was cooling; Lord Kelvin proved it. He gave it a few million years. Sadi Carnot showed in 1824 that all bodies equalized their temperature—hot things cooled, cold things warmed. It was a case of dividing a rich inheritance among a lot of poor relations; the result is poverty for all.

Then came Rutherford and radioactivity. The sun, it seemed, was not cooling; it was "decaying." It was much younger than Kelvin thought, and much older. But the end was the same. We still froze.

When the medieval superstition was replaced by enlightenment, science proceeded to explain the universe. Mathematics, astronomy, and physics put everything in its place. The solar system spun so beautifully that Laplace convinced Napoleon that he didn't need God to watch over it. Gravitation worked like a clock. Chemistry turned matter into nice little atoms, like seeds.

By 1880 the world was all explained. Metaphysics mumbled in its sleep; Theology preached sermons. But Science didn't care. It had Space, and Time, and Matter—all genuine.

And then, at the turn of the century, the whole structure began to crumble.
First came the X-rays. Roentgen discovered them. Then Sir William Crookes discovered "radiant matter" by accident. Then Rutherford knocked the bottom out of the theory of matter. I knew Rutherford well at McGill; I can testify he had no intention of destroying the universe. But he did. He showed that atoms were not hard little seeds, but holes—mostly empty space.

Then came Einstein. In 1905 he announced that there is no such thing as Absolute Rest. After the Great War, the reading public seized on Einstein. He knocked out Space and Time just as Rutherford knocked out Matter. He showed that there is no "here." 
"But I am here," you say. "Here is where I am." 
But you are moving! The earth turns, the sun moves, the galaxy moves. So where is "here"? How do you mark it?
It reminds one of the story of the two idiots out fishing. One said, "We should mark this place where we caught the big fish." The other said, "I've already marked it on the boat."
There is your "here"!

Einstein's discovery of curved space was greeted with applause. Sir Arthur Eddington, who handles space and time like a poet, applauded loudest. "Admit the curvature of space," he wrote in 1927, "and the mysterious force of gravity disappears. Einstein has exorcised the demon."

But now, fourteen years later, it seems Einstein doesn't care if space is curved or not. A prominent physicist wrote to me: "Einstein hopes that the general theory... may in the future prove more fruitful than it seems at present."
This is purely professional talk. The layman reads it as: Einstein has shrugged off curved space. It is as if Sir Isaac Newton, with a yawn, had said, "Oh, about that apple—perhaps it didn't fall after all."

Personally, I admit, the situation is a little staggering.

\section{Saga of a New Hormone}
\textbf{Author:} Norman Applezweig \\
\textit{Original Source:} Drug and Cosmetic Industry (April 1955); reprinted in The Journal of Irreproducible Results (1959).

In recent months, the world has been electrified by the announcement of three new miracle drugs by three separate pharmaceutical companies. Closer examination reveals that all three are the same hormone. For those interested in how one chemical compound acquires such a variety of nomenclatures, we trace the chain of events leading to the birth of a miracle drug.

It is usually discovered by accident by a \textbf{Physiologist} who is looking for two other hormones. He gives it a name indicating its function and predicts that it will be useful in the treatment of a rare blood disease. He processes one ton of fresh bull glands, delivered warm from the slaughterhouse, isolates 10 grams of pure hormone, and sends it to a Physical Chemist for analysis.

The \textbf{Physical Chemist} finds that the sample is 95\% impurities, and the remaining 5\% contains at least three different compounds. From one of these, he isolates 10 milligrams of pure crystalline hormone. On the basis of physical properties, he predicts a chemical structure and suggests that its function is probably quite different from that predicted by the Physiologist. He gives it a new name and sends it to an Organic Chemist for confirmation of the structure.

The \textbf{Organic Chemist} finds the structure to be incorrect; the new compound is actually a methyl group differing from a substance recently isolated from cantaloupe rind, which, however, is biologically inactive. He gives the hormone a precise chemical name, too long for general use. For brevity, he retains the Physiologist’s name for the new substance. Eventually, he synthesizes 10 grams of the new hormone, but tells the Physiologist he cannot spare a single gram, as he needs it all to prepare derivatives for further structural studies. Instead, he gives him 10 grams of the inactive compound from cantaloupe rind.

At this point, a \textbf{Biochemist}, joining the search, announces that he has discovered the same hormone in the urine of pregnant sows. On the basis that the hormone is easily split by a crystalline enzyme recently isolated from the salivary glands of the South American earthworm, the Biochemist asserts that the new compound is nothing but a variant of Vitamin B$_{12}$, a deficiency of which causes acid-cycle shifts in annelids. He changes the name.

The Physiologist writes to the Biochemist asking for some South American worms.

The \textbf{Nutritionist} finds that the new compound acts exactly like "Factor PFF," which he has recently isolated from chick manure and which is essential to the production of pigment in fur-bearing animals. Since both PFF and the new hormone contain the trace element zinc, fortification of white bread with this substance will, he assures us, lengthen the life-span and stature of future generations. To emphasize this, he invents a new name.

The Physiologist asks the Nutritionist for a sample of Factor PFF. Instead, he receives a pound of raw material from which PFF can be extracted.

The \textbf{Pharmacologist} decides to check the action of the new compound on Norway rats. To his confusion, he finds that after the first injection, the rats lose all their hair. Since this does not happen with castrated rats, he concludes that the new preparation is a synergist to testosterone and an antagonist to the pituitary gonadotropins. He deduces that the agent would make excellent nose drops. He invents a new name and sends 12 bottles, with droppers, to a clinic.

The \textbf{Clinician} receives the samples for testing on patients with sinus trouble. The nose drops don't help much, but he is amazed to see that three of his sinus patients, who also happened to be suffering from a rare blood disease, suddenly recover. He receives the Nobel Prize.

% =========================================================
% PART 2: PHYSICISTS KEEP JOKING (1992)
% =========================================================
\part{Physicists Keep Joking (1992)}
\addtocontents{toc}{\protect\vspace{1.5em}} % Adds vertical space in ToC
\setcounter{chapter}{0}                     % <--- RESETS CHAPTER TO 1

\chapter*{From the Publisher (1992)}
\textit{(A poem regarding the re-release and expansion of the collection after 20 years)}

\begin{quote}
Twenty years have passed us by,\\
The country's memory has run dry,\\
But there were times, oh yes, there were...\\
There were people, there was a need,\\
An Obninsk team did plant the seed,\\
And noble goals were in the air...

A quartet of obedient souls\\
Turned jokes into printed goals,\\
"Mir" published it back then.\\
Since those days, much has worn away,\\
Those books are gone, to our dismay,\\
Just quotes remain, now and again...

And though the times are tough anew,\\
At least stagnation we're not going through --\\
So read this rarity once more.\\
Do not judge us too strictly, friends,\\
We tweaked the text to make amends.\\
All the best,\\
\hfill \textbf{"Maket" Publishers}
\end{quote}

\chapter*{Preface to the Second Edition}
\textit{(A collage of citations)}

When we were told that it was necessary to write a preface for this edition, we discovered that we were unable to invent anything new---nothing that hadn't already been published and read many times in prefaces to other books. We leafed through several books taken at random, and with pleasure convinced ourselves that everything we needed was there. Armed with scissors and glue, we cut out several excerpts and, gluing them together, obtained our brief \textbf{Preface to the Second Edition}:

\vspace{0.5cm}

``'What is this---a collection of scientific humor?' you will say. 'But do the words \textit{science} and \textit{humor} not exclude each other?' Of course, \textbf{NO}. This book is undoubted proof that science, like other spheres of human activity, has its funny sides. Here you will find a fusion of satirical science and scientific satire...''
\begin{flushright}
\small (R. Baker, \textit{A Stress Analysis of a Strapless Evening Gown}, 1963)
\end{flushright}

``The development of physics in recent years has proceeded with giant strides, and its influence on everyday life has turned out to be quite large. In the proposed book, an attempt is made to explain how all this happened... avoiding, where possible, special terminology.''
\begin{flushright}
\small (H. Massey, \textit{The New Age in Physics}, 1968)
\end{flushright}

``Reviews of this educational manual were favorable and indicated no need to change the overall structure of the course.''
\begin{flushright}
\small (S.G. Kalashnikov, \textit{Electricity}, 2nd ed., 1964)
\end{flushright}

``But the previous edition has long since sold out, and, apparently, a need for this book is felt among readers.''
\begin{flushright}
\small (L. Landau and E. Lifshitz, \textit{Statistical Physics}, 2nd ed.)
\end{flushright}

``For the second edition, the book was significantly revised and supplemented, but the general plan of the book and its character remained the same. The revision touched all chapters.''
\begin{flushright}
\small (L. Landau and E. Lifshitz, \textit{Quantum Mechanics}, 2nd ed.)
\end{flushright}

``Examples and additional theoretical material are set in small print. The presentation is conducted in such a way that the main material, set in large print, can be studied independently.''
\begin{flushright}
\small (V. Smirnov, \textit{Course of Higher Mathematics}, Vol. 2, 6th ed.)
\end{flushright}

``The proposed collection may be useful for physics-mathematics schools, for persons preparing for physics universities, and also in preparation for Olympiads.''
\begin{flushright}
\small (\textit{Collection of Problems and Questions in Physics for MEPhI Applicants}, 1964)
\end{flushright}

``In conclusion...''
\begin{flushright}
\small (S.G. Kalashnikov, \textit{Electricity})
\end{flushright}

``...it should be mentioned...''
\begin{flushright}
\small (S. Glasstone and M. Edlund, \textit{The Elements of Nuclear Reactor Theory}, 1954)
\end{flushright}

``...that...''
\begin{flushright}
\small (V. Smirnov, \textit{Course of Higher Mathematics})
\end{flushright}

``...businesslike criticism and any indications of shortcomings and omissions will be accepted with gratitude by the collective of authors.''
\begin{flushright}
\small (\textit{Course of Physics}, ed. by N.D. Papaleksi, 1948)
\end{flushright}

\vspace{0.5cm}
We express deep gratitude to E.L. Feinberg for valuable critical remarks on the first edition of the book; we are very grateful to G.A. Askaryan, B.M. Bolotovsky, V.I. Goldanskii, R. Zhukov, M.A. Leontovich for materials provided at our disposal during the preparation of this edition. We also express our gratitude to Prof. G. Vineyard (Brookhaven, USA) for sending the collection of scientific humor \textit{A Stress Analysis of a Strapless Evening Gown}.

\hfill \textbf{The Editors} \\
\hfill \textit{Obninsk, May 1968}

\chapter{The Theoretical Minimum} 

\section{Physical Numerology}
\textbf{Author:} I. J. Good \\
\textit{Original Source:} The Scientist Speculates: An Anthology of Partly-Baked Ideas (1962).

Numerology is the search for simple numerical expressions for fundamental physical constants. There are several examples in the history of science where numerology has preceded theory.

\begin{enumerate}
    \item In 1857, Kirchhoff noticed the coincidence between the value of the velocity of light and the ratio of the electromagnetic to the electrostatic units of charge. In 1858, Riemann submitted a paper to the Göttingen Academy in which he postulated a finite velocity of propagation for the potential and concluded that it must equal this ratio of units.
    \item In 1885, Balmer gave a formula for the frequencies of the spectral lines of hydrogen. This was explained in 1913 by Bohr, and with greater precision in 1926 by Dirac and Pauli, on the basis of quantum theory. (Only the quantum theory itself remained to be explained.)
    \item In 1747, J. Bode proposed a simple formula that described the distance from the Sun to all six planets known at that time. Uranus and the asteroids, discovered later, were also described by this expression, with the exception of Neptune and Pluto. There is still no generally accepted explanation for this fact.
\end{enumerate}

A large number of examples from the field of physical numerology relate to attempts to link the masses of "elementary" particles. Here is one of the numerous examples of such reasoning:
The masses of elementary particles must be the eigenvalues of simple operators or the roots of simple functions. Let $\alpha_n$ be the cube of the $n$-th positive root of the Bessel function $J_n$. Then:
\begin{equation}
    \mathcal{M}_n = \alpha_n + n
\end{equation}
Values obtained from this formula coincide, with an accuracy to the fifth decimal place, with the masses of the neutron and the hyperon relative to the electron mass.

\section{How Newton Discovered the Law of Gravitation}
\textbf{Author:} James E. Miller \\
\textit{Original Source:} American Scientist, Vol. 39, No. 1 (January 1951).

The tremendous growth in the number of young, energetic workers toiling in the scientific vineyard is a happy consequence of the expansion of research in our country, encouraged and cherished by the Federal Government. Harassed research directors, unable to give adequate guidance to these neophytes, are often left with a feeling that they have thrown the lambs to the wolves. Fortunately, the history of science provides a pilot who can guide them through the reefs of government subsidies. I refer to the story of Sir Isaac Newton and his discovery of the law of universal gravitation.

In 1665, the young Newton became a professor of mathematics at Cambridge University, his alma mater. He was in love with his work, and his abilities as a teacher were beyond doubt. However, it must be noted that he was by no means a man in an ivory tower. His work at the college was not limited to the classroom: he was an active member of the Curriculum Committee, the Finance Committee, the Committee on Student Welfare, the Committee on Publications, and various other committees necessary for the proper management of a college in the 17th century. Careful historical research shows that in just five years Newton sat on 379 committees, which investigated 7924 problems of university life, of which 31 were solved.

One evening in 1680, after a very hard day, a committee meeting scheduled for eleven o'clock failed to muster a quorum because the chairman had died of exhaustion. Every moment of Newton's life was carefully planned, and he suddenly found himself with nothing to do until midnight, when the next meeting was scheduled. He decided to take a walk. This short walk changed world history.

It was autumn. In the gardens of the solid citizens of Cambridge, the trees were bending under the weight of ripe apples. Newton saw a particularly appetizing apple fall to the ground. His immediate reaction, typical of the human side of a great genius, was to climb the fence, pocket the apple, and walk to a safe distance to eat it.

And right there, it dawned on him. Without preliminary logical reasoning, the thought flashed in his brain that the fall of the apple and the movement of planets in their orbits must obey one and the same universal law. He had not yet finished eating the core when the formulation of the hypothesis of the law of universal gravitation was ready. Three minutes remained until midnight, and Newton hurried to the meeting of the Committee on the Suppression of Vice.

In the following weeks, Newton's thoughts returned again and again to this hypothesis. He dedicated rare free minutes between meetings to plans for its verification. Several years passed, during which he devoted 63 minutes and 28 seconds to thinking about these plans. Newton realized that to verify his assumption, he needed more free time than he could count on. He needed to determine with great accuracy the length of one degree of latitude and to invent the differential calculus.

Lacking experience in such matters, he chose a simple procedure and wrote a twenty-word note to King Charles II, outlining his hypothesis and pointing out its potential. Whether the King saw this letter is unknown; he was busy with affairs of state. However, the letter, passing through channels, visited all department heads and their deputies, who had full opportunity to express their comments.

Eventually, Newton's letter, together with a voluminous file of comments, reached the office of the secretary of the \textbf{HMPDR/CIR/SCUSA} (His Majesty's Planning Department for Research / Committee for Interdepartmental Review / Sub-Committee for the Suppression of Un-British Activities). The Secretary immediately realized the importance of the question and brought it to a meeting of the Sub-Committee, which voted to grant Newton a hearing.

Newton's testimony before the HMPDR/CIR should be required reading for all young scientists. The College granted him a two-month leave without pay. The Committee meeting was held with open doors, and much of the public gathered, though it turned out most had mistaken the door for the meeting of the \textbf{HMCDVHR} (His Majesty's Commission on the Distribution of Vice in High Society).

After Newton was sworn in and solemnly declared that he was not a member of the Levellers, had never written immoral books, and had not seduced milkmaids, he was asked to outline his proposal. In a brilliant, ten-minute speech, Newton outlined Kepler's laws and his own hypothesis. At this moment, a Committee member, a dynamic man of action, asked what means Newton proposed to improve the apple crop in England. Newton began to explain that the apple was merely an illustration, but was interrupted by the Committee, who unanimously supported a project to improve English apples.

The discussion continued for weeks. Newton sat and waited. Once he was late and found the door locked. The doorkeeper told him there were no seats. Newton concluded the Committee no longer needed him and returned to his college.

A few months later, Newton received a voluminous package from HMPDR/CIR. It contained government questionnaires. He realized he was being invited to apply for a contract to study the correlation between apple cultivation and the velocity of falling fruit. The ultimate goal was to breed a variety of apple that would not only taste good but fall to the ground gently, without bruising.

This was not exactly what Newton had in mind. But he was a practical man. He realized that working on this problem, he could verify his hypothesis on the side. Thus, he would serve the King and do a little science for the same money. He filled out the questionnaires.

...Thus, His Majesty's government supported Newton. We need not dwell on Newton's attempts to publish his proof, or how his article was rejected by the \textit{Gardeners' Journal} and \textit{Physics for Housewives}. Suffice it to say that Newton founded his own journal to print the report on his discovery without cuts.

\section{The Past and Future of Field Theory}
\textit{(In a theoretical model based on experimental observations reliable to within one standard deviation)}
\textbf{Author:} Observer\footnote{Usually well informed.} \\
\textit{Original Source:} The Journal of Irreproducible Results, Vol. 12, No. 3 (1963).

To understand the full significance of field theory, one must view the subject against the appropriate historical background. By 1930, physics had explained all observable quantities. Since then, it has dealt exclusively with unobservable quantities, which are the subject of field theory.\footnote{In the work of Preston [Reviews of Unclear Physics, 1, No 1, 3 (1957)] the following description and classification of unobservable quantities is given:
"It is well known that physical quantities are described by matrices whose eigenvectors form a Hilbert space. But these matrices are only a small class among all possible mathematical objects, and obviously, there are many operatorless operators. To use at least some of them, one can assume that they correspond to unobservable quantities. However, these unobservable quantities are still so poorly studied that the development of a corresponding mathematical theory is premature.
Unobservable phenomena can be classified into the following categories, arranged in descending order of scientific interest:
a) Phenomena unobservable by definition (e.g., invisible light);
b) Phenomena unobservable in principle (e.g., absolute velocity);
c) Phenomena unobservable in nature (e.g., offspring of sterile rabbits);
d) Phenomena unobservable in polite society (e.g., non-conservation of parity before 1956).
The last example shows that unobservability is not a constant of the motion."
This classification is illustrated in Fig. 1 (Not shown: The figure is unobservable).}

Klein's discovery belongs to the same era. To him we owe an equation that is written identically in both stationary and moving coordinate systems; i.e., the equation is invariant whether you write it at your desk or while running for a bus (an old dream of theorists).

In the late 40s, the theory received a powerful push thanks to the discovery of the famous Lamb shift. Instead of formulas, field-theorists began to draw pictures, often on the back of old envelopes, thereby significantly reducing the cost of theoretical research. The cost of experimental research in that period increased significantly, facilitated by tireless experimenters who, digging to unheard-of depths, extracted one tasty morsel after another for explanation by their theorist friends at an average price of $10^6$ dollars per piece. Everyone agreed that the results were worth the expense, especially since the costs were directed toward the common good and covered, naturally, by the taxpayer.

Thus, Physics inevitably entered into a \textbf{strong interaction} with the Government. Perhaps this explains the fact that in the 50s, manifestations of the \textbf{Principle of Least Action} began to be noticed more and more in the activities of the government.

And so, finally, hiding behind Riemann sheets, theorists fought their way into non-physical regions and discovered that everything has an imaginary part. In recent times, the suspicion has been growing that the object of investigation itself---the scattering amplitude---is a purely imaginary quantity...

Everyone is confident that field theory will open a new heroic era in physics, but when this will happen is not yet the time to predict.

The future of field theory lies in the analytic continuation of everything possible into the complex plane. In one early work, it was proposed to continue the quantum number "strangeness" into the complex domain in order to classify those purely imaginary particles whose discovery is constantly reported by the \textit{New York Times}. It was also proposed to analytically continue the "two-component theory" to obtain a "two-component experiment" having two components---"Right" and "Wrong". A good two-component theory must accurately describe both components of the experiment.

Dispersion relations and Racah coefficients also need to be investigated from this point of view. The calculation of the values of these coefficients for complex values of arguments promises the thoughtful researcher many unforgettable hours at the electronic computer.

The analytic continuation of the Mössbauer effect leads us to the thought that the key to the future development of field theory is most likely buried in some unintelligible article published and forgotten in the 30s. Attempts to use such a conclusion, however, will almost certainly be unsuccessful due to the \textbf{Peierls-Jensen Paradox}: even if someone finds that specific article, he still won't understand its significance until the phenomenon is discovered experimentally, independently, and completely by accident.

There are many ways to analytically continue the many-body problem into the domain of field theory:
\begin{enumerate}
    \item \textbf{Approximation of random articles.} It is much simpler to write an article yourself than to read all the articles already published in which the same thing was done. By changing the wording and notation, you not only destroy all traces of the connection of your work with preceding ones, but also give future researchers the opportunity to write their own articles instead of reading yours. Result: an exponential growth in the number of articles that assert the same thing and thereby make a contribution to field theory.
    \item \textbf{Simplification of problems and verification by inventing an approximate Hamiltonian.} This opens up broad opportunities for work for those people who otherwise would not know what to do. Now they can discuss the shortcomings of your approximate Hamiltonian.
    \item \textbf{Analytic continuation of the many-body problem into the region of a complex number of particles.} Especially interesting is the study of pairing effects for the case where the number of particles in a pair is not two, but an arbitrary complex number.
    \item \textbf{Analytic continuation of the Brown formalism and Green's function method to all other colors of the spectrum.}\footnote{Referring to G. E. Brown and the Green's Function.}
\end{enumerate}

\anecdote{Ehrenfest's Parrot}{
The famous physicist Paul Ehrenfest taught his Ceylonese parrot to pronounce the phrase: "Aber, meine Herren, das ist keine Physik" ("But gentlemen, this is not physics"). He proposed this parrot as the chairman for discussions on the new quantum mechanics in Göttingen.
}

\section{The Quantum Theory of Dance}
\textbf{Author:} Y. I. Frenkel \\
\textit{Original Source:} Speech delivered at an evening gathering at the Leningrad Physico-Technical Institute.

What is a dance? Fundamentally, a dance is a form of bodily motion. Every motion of a body is a mechanical phenomenon. Consequently, dance is a mechanical phenomenon. It follows, therefore, that dance falls within the purview of Theoretical Physics, which, as is well known, attempts to reduce almost all phenomena to motion.

If we investigate the character of the movements performed by dancing couples, we are immediately convinced that these movements belong to the class of periodic---or, more precisely, \textit{quasi-periodic}---motion.

In primitive dance, this periodic character is expressed quite simply. Among certain peoples, for instance, dance reduces primarily to a simple harmonic oscillation of individual body parts.

In the Middle Ages, and especially in the 19th century, we encounter far more complex movements, in which the leading role is played by the lower limbs, with the coordinated participation of the head and arms. Here, a definite connection is established between physical motion and spiritual emotion. According to the \textbf{Classical Theory of Dance} (based on Newtonian mechanics and classical electrodynamics), the legs of a ballerina emit invisible radiation---the "light" of the subtlest feelings---with every movement. The period of these emotional oscillations coincides with the period of the body's motion, while the intensity increases in direct proportion to the square of the amplitude of the latter.

Note that the propagation of emotional waves radiated by the body of the female (or male) dancer obeys the same laws as the propagation of electromagnetic waves. In particular, their intensity decreases in inverse proportion to the square of the distance.

With the help of psychoanalysis, it has become possible to decompose the emotional radiation of dancers into a spectrum. The study of the regularities revealed thereby has led to the creation of the \textbf{Quantum Theory of Dance}. The application of quantum theory to dance is all the more natural because here, as in the case of the "dance" of electrons in atoms, we are dealing with periodic motion.

The essence of the Quantum Theory of Dance, which represents a compromise between the classical mechanics of quasi-periodic motion and classical "\textbf{Emodynamics}," consists of the following: 
Dancers can describe specific quantum orbits without emitting or absorbing any emotions. Emotions are emitted and absorbed in a discontinuous manner (quanta) only during transitions from one quantized orbit to another. 

However, there is a fundamental difference from the Bohr atom. In the electronic case, emission corresponds to a drop in energy. In dance, emotional emission, like absorption, is accompanied by a transition not to a lower, but to a \textit{higher} level---that is, to a state of excitation. Thus, during a dance (especially a couples dance), the excitation of the dancers invariably increases until relaxation sets in, caused by physical exhaustion.

The Quantum Theory of Dance has successfully established an extremely general and important \textbf{Exclusion Principle} relating to arbitrary systems of dancers. The principle states: 
\begin{quote}
\textit{Only two dancers can move simultaneously along the same quantized orbit, and only with oppositely oriented spins.}
\end{quote}

Thus, the law forbids three dancers from occupying the same orbit (the "third wheel" prohibition), and forbids a dance in which the spins of both partners are oriented in the same direction.

Closely connected with the phenomenon of spin, just as in the case of electrons, are the phenomena of \textit{animal magnetism}. Here, the magnetic field emanating from an unpaired (e.g., bachelor) individual acting on a dancing couple (which is magnetically analogous to an astatic pair) often leads to the dissociation (divorce) of the latter and the formation of a new pair combination. Divorces and reunions connected with these magnetic effects always occur with strict observance of the Exclusion Principle, which is thus one of the most fundamental laws of dance interaction.

Although the Old Quantum Theory of Dance, created in general terms during the first third of the 20th century, managed to explain a number of phenomena that remained misunderstood from the point of view of classical theory, it can by no means be considered final. It turned out, for example, to be inapplicable to the new forms of dance that arose after the Second World War. The investigation of these modern dances has led to the creation of the \textbf{Wave Theory of Dance}. This theory not only explains dances but changes them. It is precisely with its help that it became possible within a few years to transform such obsolete forms as the waltz, mazurka, and Pas d'Espagne into dances of a new type.

The new theory is based on a principle as simple as it is fundamental. Since dance is not merely body movement but is intrinsically linked with spiritual movement, it cannot be described by a purely mechanical theory or by any compromise between mechanics and emodynamics. Description is possible only on the basis of a theory uniting the dualism of mechanical motion and spiritual emotion. Since spiritual movements associated with dance represent a type of agitation (specifically, agitation of feelings), the new theory received the name \textbf{Wave Mechanics}.

Some philosophers assert that the principles of the Wave Theory of Dance were outlined as early as Hegel. I do not presume to judge this, but I will try to outline briefly the main achievements of this new theory.

The separation of acts of emotional radiation from the process of mechanical movement---characteristic of the Old Quantum Theory with its stationary (i.e., "feelingless") orbits and sudden sensual transitions---has been liquidated at the root. Spiritual and physical movements are united into one harmonic whole. Furthermore, the concept of a "quantized" orbit described by the dancers has been abolished.

The path of a dancing couple is completely indeterminate, and its position at any given moment in time can be defined only in terms of probability theory.

In accordance with the general law of development from simplicity to complexity, we do not find in modern dance any traces of the primitive simplicity and limitedness of classical dance movements. Modern dance does not differ from ordinary, random body movements: it is the same thing, but set to music.

Huge credit for the creation of the Wave Theory of Dance, and especially for its experimental verification, belongs to the staff of the LPTI (Leningrad Physico-Technical Institute), who have recently worked in this direction literally without sparing their legs. The achievements of the institute will be demonstrated to you today after the intermission.

\textit{(The report was illustrated by a live dancing couple.)}

\vspace{0.5cm}
\begin{center}
* * *
\end{center}
\small
In one of his works, Y. I. Frenkel wrote: "A physical theory is like a suit sewn for Nature. A good theory is like a well-tailored suit; a bad one is like \textit{Trishkin's Caftan}*..."

\vspace{0.2cm}
\footnotesize{*Note: "Trishkin's Caftan" refers to a famous fable by Ivan Krylov where the protagonist, Trishka, cuts off the sleeves of his coat to patch the worn-out elbows, and then cuts off the tails to patch the sleeves, resulting in a suit that is worse than before. It is the Russian equivalent of a "pyrrhic solution" or "robbing Peter to pay Paul."}

\anecdote{The Oscilloscope}{
Once, at the physics practicum of Moscow State University, the following task was assigned: to disassemble the schematic diagram of an oscilloscope and measure its sensitivity. \\
Forty minutes later, a student runs in and guilty reports that things are going successfully, but the tube just won't come out... \\
When the supervisor, sensing trouble, ran into the laboratory, he saw a pile of panels, resistors, and tubes... The student, it turned out, was conscientious and had spent two days assembling the oscilloscope, but it never started working...
}

\section{Analysis of Modern Music Using Harmonic Oscillator Wave Functions}
\textbf{Author:} H. J. Lipkin \\
\textit{Original Source:} The Journal of Irreproducible Results (1964).

The importance of harmonic oscillation in music was well known even before the discovery of the harmonic oscillator by Steinmetz [1]. Evidence for shell structure was first pointed out by Haydn, who discovered the magic number "four" and proved that a system of four musicians possesses unusual stability [2]. The concept of the magic number was expanded by Mozart in his work "The Magic Flute." A system of four magic flutes is, thus, doubly magic. Such a system is so stable that it interacts with nothing and is therefore unobservable.

A significant step forward in the application of spectroscopic techniques to music was made by Racah-maninoff [3] and Sharp [4], as well as Wigner, Wagner, and Wigner [5]. Relativistic effects were taken into account in the work of Bach, Feshbach, and Offenbach [6], who used the Einstein-Infeld-Hoffmann method.

Thus far, all attempts to apply the harmonic oscillator to the analysis of modern music have failed. The reason for this, namely that most modern music is anharmonic, was noted by Wigner and Wagner [7].

A more anharmonic approach is that of Brueckner, who used plane waves instead of oscillator functions. This promising method, strictly speaking, is applicable only to infinite systems. Therefore, all works of the Brueckner school are intended only for very large ensembles.

\textbf{References}
\begin{enumerate}
    \item Steinmetz C., \textit{Musical Spectroscopy}, Helv. Mus. Acta.
    \item Haydn J., \textit{The Alpha-Particle of Music; String Quartet Op. 20 No. 5}.
    \item Racah-maninoff G., \textit{Sonority and Seniority in Music}.
    \item Sharp W.T., \textit{Tables of Coefficients}.
    \item Wigner, Wagner, and Wigner, \textit{Der Ring der Nibelgruppen}.
    \item Bach, Feshbach, and Offenbach, \textit{Einstein, Infeld, and Hoffmann Tales}.
\end{enumerate}

\section{On the Nature of Mathematical Proofs}
\textbf{Author:} J. Cohen \\
\textit{Original Source:} A Stress Analysis of a Strapless Evening Gown (1963).

Bertrand Russell defined mathematics as a science in which we never know what we are talking about, nor whether what we say is true. It is known that mathematics is widely used in many other fields of science. Consequently, other scientists for the most part do not know what they are talking about and whether what they say is true.
Thus, one of the main functions of mathematical proof is to create a reliable basis for penetrating into the essence of things.

Aristotle is among the first philosophers to study mathematical proofs. He invented the syllogism---a device which, by virtue of its absolute uselessness, attracted the attention of countless logicians and philosophers. A syllogism consists of a first premise, a second premise, and a conclusion. Logicians do nothing but come to conclusions. It is simply a miracle that they have not yet gone around in circles and arrived back where they started.

In the first premise, a truth relating to a whole class of things is contained, for example: "Not all premises are true." The second premise asserts that the thing of interest to us belongs to this class, for example: "The last four words of the previous sentence are a premise." Thus, we come to the conclusion: "It is not always true that not all premises are true." Such is the all-encompassing completeness with which logic generalizes the phenomena of everyday life.

Relying on mathematical proofs, scientists managed to connect hitherto disparate fields---thermodynamics and communication engineering---into a new discipline: information theory. "Information," scientifically defined, is proportional to surprise: the more surprising the message, the more information it contains. If, picking up the phone receiver, a person hears "Hello," this will not surprise him much; the information will be much greater if, instead of "Hello," he is suddenly shocked by electricity.

Colossal new opportunities opened up before mathematical proofs with the development of set theory at the end of the last century. The author himself recently discovered a theorem in set theory that deserves to be cited here.

\textbf{THEOREM.} A set whose only element is a set can be isomorphic to a set whose only element is a set, all elements of which form a subgroup of elements in the set, which is the only element of the set with which it is isomorphic.

This intuitively obvious theorem can be derived in a roundabout way from the isomorphism theorem in group theory.

Let us now consider logical systems. A logical system differs from a simple set of theorems just as a finished building differs from a pile of bricks: in a logical system, each subsequent theorem rests on the previous one. Polya noted that Euclid's merit consisted not in collecting geometric facts, but in their logical ordering. If he had simply dumped them in a pile, he would have become no more famous than the author of any high school math textbook. To illustrate methods of mathematical proof, we will give an example of a developed logical system.

\textbf{LEMMA 1.} All horses have the same color (proof by induction).
\textit{Proof.} Obviously, one horse has the same color. Let us denote by $P(k)$ the assumption that $k$ horses have the same color, and show that from such an assumption it follows that $k+1$ horses have the same color. Let us take a set consisting of $k+1$ horses and remove one horse from it; then the remaining $k$ horses, by assumption, have the same color. Let us return the removed horse to the set and remove another instead. We again get a herd of $k$ horses. According to the assumption, they are all of the same color. Thus we will go through all $k+1$ sets of $k$ horses each. From this, it follows that all horses are of the same color, i.e., the assumption that $P(k)$ implies $P(k+1)$. But earlier we already showed that the assumption $P(1)$ is always satisfied, which means $P(n)$ is valid for any $n$, and all horses have the same color.

\textbf{COROLLARY I.} All objects have the same color.
\textit{Proof.} The proof of Lemma 1 does not use the specific nature of the objects in any way. Therefore, in the statement "if $X$ is a horse, then all $X$ have the same color," one can replace "horse" with "something" and thereby prove the corollary. (One can, by the way, replace "something" with "nothing" without violating the validity of the statement, but we will not prove this.)

\textbf{COROLLARY II.} All objects are white.
\textit{Proof.} If the statement is true for all $X$, then upon substitution of any specific $X$ it retains its validity. In particular, if $X$ is an elephant, then all elephants are of the same color. Axiomatically reliable is the existence of white elephants. Consequently, all elephants are white. Then from Corollary I follows Corollary II, Q.E.D.!

\textbf{THEOREM.} Alexander the Great did not exist.
\textit{Proof.} Note for a start that historians obviously always tell the truth (since they always vouch for their words and therefore, consequently, cannot lie). Hence, the statement is historically reliable: "If Alexander the Great existed, he rode a black horse named Bucephalus." But, according to Corollary II, all objects are white, and Alexander could not ride a black horse. Therefore, for the validity of the conditional statement expressed above, it is necessary that the condition be violated. Consequently, Alexander the Great did not exist in reality.

%\end{center}

\section{How Three Vectors Turned a Determinant to Zero}
\textbf{Authors:} Adam Ar and Eva Clid (Pseudonyms) \\
\textit{Original Source:} Russian Scientific Folklore.

\begin{center}
\textit{Two parallels run,\\
But they do not meet.\\
Two perpendiculars stand,\\
But they do not tilt.\\
(Old Song)}
\end{center}

In a certain space, in a certain subspace, there lived and was given a normalized, well-ordered family of vectors---$I_1$, $I_2$, and $I_3$. They had neither eigenvalues nor eigenvectors; they lived as their mother bore them. From period to period, from $-\pi$ to $\pi$, the brothers bent their backs on the basis of the rich \textbf{Simplex}---an exploiter and parasite who lived his whole life according to the principle of least action.

And Simplex's son, \textbf{Complex}, disliked them. He played his complex tricks on them: sometimes he would chop off one coordinate, sometimes another.
"There will be no life for us from this Complex," the brothers decided. "There are no limits to him." And they conceived a plan to bypass all spaces and all subspaces, all shells and manifolds, and find a right-handed coordinate system.

They went out into a purely potential field and walked with a step of $h/2$ wherever their eyes looked. They walked $\pi$, they walked $2\pi$, they walked $3\pi$. Isoclines began to appear. The brothers looked---right in front of them, a jet stream glistened with a blue cut on a smooth complex plane. Not a simple stream---with cavitation.
"Shall we catch some fish?" said $I_1$.
"Why not?" said the brothers.
They cast their time-worn orthogonal net. They looked---a sigma-fish was beating in the net, speaking in a human voice: "Do not ruin me, good fellows, I will still be of use to you." They released her to freedom and went on.

Long or short they walked---greater than zero, less than infinity---they looked: a small parameter stood by the road, crying from hunger. The brothers took pity on him, fed him kernels in convolution, treated him with terms. The parameter began to grow before their eyes, and when he reached an extremal value, he thanked the brothers and said: "I will still be of use to you." And he vanished, as if he had never been there.

The sky darkened, the sun disappeared. Mobius strips rushed along the road, vortices swirled in the air; fiery cuts of lightning split the celestial Riemann sphere. The brothers looked back, and lo---a hut on chicken legs stood by the road.
"Hut, hut, turn your plus to us, your minus to the forest."
The hut shifted from foot to foot, turned. Vectors entered it and rejoiced. A table stood in the hut, covered with all sorts of victuals. The brothers ate and asked: "Is anyone here? Respond."
They looked---from under the stove crawled not quite a vector, not quite a scalar, chained in a fractional chain.
"Greetings to you, noble vectors! I am the good wizard \textbf{Adi Aba Ata Cauchy Mac Laurent}. For half a life I have been sitting here under the guard of the evil \textbf{Nabla Yaga} for denying ambiguity..."
He had not finished speaking when a noise and whistling began around them.
"Run!" cried Mac Laurent.
The brothers unchained him and they all rushed away together. They looked back and saw---the beautiful \textbf{Delta} flying across the sky. Delta hit the ground, stood on her head, and turned into the terrible Nabla Yaga.
"I smell, I smell, it smells of vector spirit!"
But the trace of the vectors was already cold.

Adi Aba Ata led the brothers onto a geodesic line, pointed the way to \textbf{Divgrad} (which means Wondrous City), and went his own way.
...And the walls of the great city rose before the brothers, just as the graph of a tangent increases with an argument close to $\pi/2$. And a radiant glow emanated from it, just as the partial sums of a harmonic series diverge.

The brothers entered the tavern "$Y$ with a Wave", talked with the hostess, the fat, stout Tilda. And she told them about the great misfortune that had befallen their city. The ruler of Divgrad, the Great \textbf{Tensor IV Invariant}, threw a ball on the occasion of the coming of age of his daughter, the beautiful \textbf{Resolvent}. Such a ball had never been seen in his domain of definition. Count X arrived in a self-conjugate carriage, Prince Sine arrived with his Sinusoid. The marvelous sounds of $K$-dimensional music, performed by a choir of higher harmonics accompanied by impact polars, delighted the ear. The whole hall whirled in the dance "Pa $dt$". Suddenly the light went out, Lissajous figures rushed about the walls, the guests panicked. And when the fuses were fixed, the beautiful Resolvent was gone. As a corollary of the monodromy theorem showed, she was kidnapped by the evil wizard \textbf{Vandermonde}. He penetrated the ball, violating the d'Alembert-Euler conditions and performing a permutation in the ranks of the guard.

Tilda's story sank deep into the brothers' souls. And they decided to measure their strength against the evil Vandermonde, to rescue the beautiful Resolvent from his hands. They went to the Taylor series, equipped themselves, told fortunes on a hodograph, and set off.

Soon the tale is told, but not soon is the deed done. Heavy boundary conditions did not allow the vectors to pass into the adjacent cross-lying region inhabited by pseudovectors, where the class inequality of Cauchy-Bunyakovsky reigned. And along the envelope they came out to a branch point, on which was written: "Go right---you go to infinity. Go left---you won't gather coordinates. Go straight---you will be transposed."
The brothers fell into thought. Suddenly, out of nowhere---the old acquaintance Adi Aba Ata Cauchy Mac Laurent.
"I know, brothers, your thought. You have conceived a difficult task. It is hard to overcome Vandermonde. His death is enclosed in a determinant. And that determinant is in a dodecahedron. And the dodecahedron lies in an icosahedron. And that icosahedron is tied tight to the roots of a Legendre polynomial; the first knot is simple, the second is nautical, the third is logarithmic. And that polynomial grows in an isolated point, and it is not easy to get to it. It lies 39 lands away in the space of Khan Banach. And a monster with a transcendental number of legs, named Decrement, guards it. You must get that determinant and equate it to zero."

Adi Aba Ata showed them the way, and the brothers went along it to the boundaries of a non-empty set filled with an incompressible fluid. They stood, guessed, didn't know what to do. Suddenly, out of nowhere---the sigma-fish. "Here I am, useful to you, good fellows!" She transported them all, explained the way further.
The brothers had not gone two periods when a discontinuity of the second kind blocked their path. The vectors were saddened. But the small parameter appeared before them. "Here I am, useful to you, brothers!" He hit the ground, expanded in his powers, and the brothers crossed to the other side. "And now," says the parameter, "follow the tracks of the matrices, straight to the isolated point."

The brothers found the tracks, looked---they diverged on three sides. They each went in their own direction. $I_1$ walked and walked---suddenly countless unit vectors of Khan Banach grew out of the ground before him, all except maybe one dressed in Jordan form, trimmed under a Poisson bracket.
"Eh," the vector saddened, "my brothers are not with me! But never mind, $I_1$ is a warrior in the field!" And he fearlessly rushed at the enemies. And here the brothers arrived in time. They overcame the adversary.
Suddenly everything trembled around, resonated. The earth opened up, and the monster Decrement appeared before the vectors. The brothers were not confused, they threw a rope polygon over him. The monster got tangled in it. Died out.

The brothers found the polynomial, dug up the roots, cut the knots, opened the icosahedron, took out the dodecahedron, extracted the determinant... and equated it to zero.
Here came the end of Vandermonde. And the beautiful Resolvent appeared before the brothers, alive and unharmed.

...Q.E.D.

\chapter{Experiments and Anomalies}

\anecdote{Stamp Collecting}{
Rutherford used to say that all science can be divided into two groups: physics and stamp collecting.
}

\section{How We Measured Reactivity}
\textbf{Author:} Enrico Fermi \\
\textit{Original Source:} Excerpt from an address to the American Physical Society (1954).

\textit{(Excerpt from Fermi's last speech at a meeting of the American Physical Society. The speech was informal, and Fermi spoke without notes. The text was restored from a tape recording and published in an unpolished, unedited form. Fermi might have been displeased with this, as he always prepared his publications very carefully.)}

...So, we come to 1939, when Einstein wrote his famous letter to President Roosevelt, in which he advised paying attention to the situation in physics and said that, in his opinion, it was the government's duty to pay serious attention to this and help physicists. And indeed, a few months later help was provided. It was 6,000 dollars, and these 6,000 dollars were used to buy a huge amount---or what seemed like a huge amount at the time---of graphite. This was back in the days when physicists' eyesight had not yet been spoiled.

And so the physicists on the seventh floor of Pupin Laboratories started looking like coal miners, and the wives to whom these tired physicists returned in the evenings could not understand what was the matter. Of course, "smog" and so on, but still...

The point was that just at that time we were trying to find out something about absorption in graphite, and we didn't expect anything good from graphite. So, for this purpose, we built a column of graphite about four feet on a side and ten feet high. It was, apparently, the first time that physical apparatus---and this pile of graphite was physical apparatus---turned out to be so big that one could climb on it. The same thing happened with cyclotrons, but for me, this was the first time I had to climb onto my own installation, which turned out to be a little higher than necessary---after all, I am not a tall man.

Well, time went on, and we began to understand what exactly needed to be measured, and with what accuracy these quantities---I will call them $\eta$, $f$, and $p$ (I don't have time to explain to you what this is)---with what accuracy $\eta$, $f$, and $p$ must be measured to make it clear that it is possible, and what is impossible. Well, in general, the product of these three quantities had to be greater than unity. Now we know that, even if we try very hard, we get a product of 1.1.

If, for example, we could measure each of these quantities with an accuracy of 1\%, then we would get, for example, that the product equals $1.08 \pm 0.03$, and if so, we would say: "Everything is in order, let's work," but if the product turned out to be $0.95 \pm 0.03$, then we would have to look for something else. But what if you get $0.9 \pm 0.3$, what do you know then? Apparently, nothing at all. Even if you get $1.1 \pm 0.3$, you also know no more. This was the whole trouble, and if you look at our early papers, where values obtained by different experimenters are given, you will see that they differ from each other by 20\% and more. These values, I think, testified mainly to the temperament of the physicists. Optimists inevitably exaggerated them, and pessimists like me tried to make them smaller.

In general, no one really knew anything, and we decided that something had to be done. We had to come up with an experiment in which the product of $\eta$, $f$, and $p$ would be measured at once, and not these quantities individually.

So, we went to Dean Pegram, who was then a magician and wizard at the university, and explained to him that we needed a large room. When we said "large", we meant really large, and he, I remember, said something in conversation about a church not being a very suitable place to create a physics laboratory, but I think a church would have been exactly what we wanted. After wandering around the courtyard a bit, he led us through dark corridors, and we crawled under some heating pipes and peered into various nooks in search of a place for our experiment, until finally we found a large room, true, not a church, but something similar in size.

Here we began to erect our structure, which this time looked an order of magnitude larger than anything we had seen before. True, a modern physicist, to see this structure, would perhaps take a magnifying glass in his hand and come closer. But for those times it looked truly big. The structure was made of graphite bricks, and among these graphite bricks, tin cans, cubic cans with uranium oxide, were located in a certain order.

Well, as you may know, coal is a black substance. Uranium oxide too. And people dealing with tons of these substances---too. Furthermore, for such a job strong people are needed. Well, we, of course, were reasonably strong, but one had to bear in mind that in the end, we were thinkers. Then Dean Pegram shook his head and said that such work, of course, was not for our weak powers, but at Columbia University there is a football squad, and in it a dozen or so very strong lads who take work with hourly pay to earn money for their studies. Why not hire them?

It was a brilliant idea. Supervising the work of these strong lads, who carried uranium and placed (shoved) it into place, handling 50- and 100-pound packs with such ease as if they weighed 3-4 pounds, was a true pleasure. They threw these packs so that dust stood in a column in the air---of all colors, but mainly black.

And that is how what was then called an exponential pile was erected.

\anecdote{The 65537-gon}{
One overly intrusive graduate student drove his supervisor to such a state that the latter finally told him: "Go and work out the construction of a regular polygon with 65,537 sides." \\
The graduate student withdrew, only to return 20 years later with the corresponding construction. (It is kept in the archives in Göttingen).
}

\section{Movements of the Lower Jaw of Cattle during Mastication}
\textbf{Authors:} P. Jordan and R. de L. Kronig \\
\textit{Original Source:} Nature, Vol. 120, p. 807 (December 3, 1927).

Among biological phenomena in which a distinction between one of two possible directions of rotation is detected (these include, for example, the spiral growth of climbing plants and the structure of snail shells), there exists one more which, apparently, has not been studied until now and to which we wish to draw attention here. We are speaking of the masticatory movements of cattle. Detailed investigation shows that the movement of the lower jaw relative to the upper is neither purely horizontal nor purely vertical, but represents a superposition of these periodic movements with such a phase shift that the result is a pure rotation. Theoretically, of course, rotation in two directions is quite admissible, and observation shows that in nature both possibilities are realized. Accepting the direction of the movement of food as positive, we will call those individuals in whom masticatory movement occurs (if viewed from the front) clockwise and counter-clockwise, "right-" and "left-rotating" cows, respectively.

Such a classification, of course, tacitly assumes that for a given cow, the direction of rotation is conserved. However, the number of experimental observations that we can cite in confirmation of this conclusion is limited, and we realize that for a final proof of this proposition, fuller data obtained over a long period of time are necessary. A sample survey of cows in the province of Zealand in Denmark led us to the conclusion that 55\% of their total number are right-rotating, and the rest are left-rotating. Thus, the ratio is close to unity. The number of observations made, however, is hardly sufficient to decide finally whether the noted deviation from unity is a real existing fact. It is even more impossible for now to generalize this regularity to cows of other countries.

The circumstance that both directions of rotation are realized entails the necessity of clarifying the question of whether there exists a simple mechanism of hereditary transmission of that distinctive feature which we are currently discussing. It is known, for example, that for the aforementioned snails, the laws of genetics are applicable in their simplest form. In the majority of other cases, the existence of only one direction makes such studies impossible. It is of interest to find out which of the modifications for cows is dominant. We, unfortunately, cannot solve this important question, but we hope that the answer to it can be easily found by people having a direct relation to the breeding of cattle.

\anecdote{A Translation Error}{
When Niels Bohr spoke at the Physics Institute of the Academy of Sciences of the USSR, in response to a question about how he managed to create such a first-class school of physicists, he answered: "Apparently, because I was never afraid to admit to my students that I am a fool..." \\
E. M. Lifshitz, who was translating Niels Bohr's speech, conveyed this phrase to the audience in the following form: "Apparently, because I was never afraid to tell my students that \textit{they} are fools..." \\
This phrase caused animation in the audience. Then E. M. Lifshitz, having asked Bohr again, corrected himself and apologized for the accidental slip of the tongue. However, P. L. Kapitsa, who was sitting in the hall, profoundly noted that this was not an accidental slip. It actually expresses the fundamental difference between the schools of Bohr and Landau, to which E. M. Lifshitz belongs.
}

\anecdote{What is Mathematics?}{
The famous Russian mathematician Academician Markov, when asked what mathematics is, replied: \\
"Mathematics is that which Gauss, Chebyshev, Lyapunov, Steklov, and I do."
}

\section{The Toxicity of Pickles}
\textbf{Author:} A. Kohn \\
\textit{Original Source:} The Journal of Irreproducible Results (1955).

Pickles will kill you. Every pickle eaten brings you closer to death. It is surprising that thinking people have not yet recognized the lethality of this vegetable product.

\begin{enumerate}
    \item 99.9\% of all people who die of cancer have eaten pickles.
    \item 100\% of all soldiers have eaten pickles.
    \item 99.7\% of all people involved in air and auto accidents ate pickles within 14 days preceding the tragedy.
    \item 93.1\% of all juvenile delinquents come from families where pickles are served.
\end{enumerate}

Evidence suggests the effects are cumulative. Of all the people born in 1839 who later dined on pickles, there has been a \textbf{100\% mortality rate}.
All pickle eaters born between 1869 and 1879 have wrinkled skin, have lost most of their teeth, and brittle bones---if they have not already succumbed to the pickle plague.

Even more convincing is the report of a noted team of medical specialists: rats force-fed 20 pounds of pickles a day for a month lost their appetites!
The only way to avoid the harmful effect of pickles is to change your diet. Eat, for example, orchid petal soup. To our knowledge, no one has ever died from it.

\section{New Classification of Stones}
\textbf{Author:} M. J. Oppenheim \\
\textit{Original Source:} The Journal of Irreproducible Results.

Below is a classification of stones applicable to all varieties and recommended for general use. This classification is scientifically rigorous, relying solely on observable properties, yet sufficiently flexible to cover all known specimens.

\subsection*{A. Genetic Plan}
\begin{itemize}
    \item \textbf{A1.} Stone of celestial origin. (e.g., The Moonstone).
    \item \textbf{A2.} Stone of biological origin. (e.g., The Gallstone, The Kidney Stone).
    \item \textbf{A3.} Stone of terminal origin. (e.g., The Tombstone).
\end{itemize}

\subsection*{B. Tectonic Plan}
\begin{itemize}
    \item \textbf{B1.} The Rolling Stone. (Distinguished by its lack of moss).
    \item \textbf{B2.} The Cornerstone. (Distinguished by its stationary nature and dates).
\end{itemize}

\subsection*{C. Physical-Chemical Plan}
\begin{itemize}
    \item \textbf{C1.} The Philosopher's Stone. (Capable of transmuting base metals into gold).
    \item \textbf{C2.} The Non-Philosopher's Stone. (Incapable of transmutation; usually found in shoes).
\end{itemize}

\subsection*{D. Functional Plan (Sociological)}
\begin{itemize}
    \item \textbf{D1.} The Millstone. (Found around the neck).
    \item \textbf{D2.} The Heart of Stone. (Found in administrators).
    \item \textbf{D3.} The Blarney Stone. (Found in Ireland; causes eloquence).
\end{itemize}

\subsection*{E. Kinematic Plan}
\begin{itemize}
    \item \textbf{E1.} Stone's Throw. (A variable unit of distance).
\end{itemize}

\textbf{Example of Application:}
The author of this communication recently discovered an F-sharp, grey-brown, rolling, non-philosophical, millstone in his kidney.

\section{P.A.U.L.I. and Its Application}
\textbf{Author:} V. Weisskopf \\
\textit{Original Source:} The Journal of Jocular Physics, Vol. 2 (1945).

\textit{(Received July 1932; partially declassified July 1951)}

The Swiss Federal Institute of Technology has recently acquired an instrument possessing unique capabilities. This complex mechanism is designed to test physical theories, as well as to produce new theories and ideas.
The abbreviated name of the machine is \textbf{P.A.U.L.I.} (\textbf{P}roduction of \textbf{A}ntisymmetric \textbf{U}nitary \textbf{L}orentz-\textbf{I}nvariant theories).
With skillful handling, the machine not only creates new correct theories but also reacts exceptionally violently to theories created by other physicists if they do not possess the characteristics listed in the machine's name, or other necessary properties.

The machine has an almost spherical shape. Its dynamic characteristics are critical. The apparatus possesses a fundamental oscillation frequency $\omega_0$, which is excited constantly, even if the apparatus is not in use. The value of the frequency is not strictly constant but limited by the range $0.8 < \omega_0 < 2.2 \, s^{-1}$.
Users are recommended to carefully monitor the exact value of the frequency and amplitude of the machine's oscillations around the room.

Unfortunately, the device is bulky, and its operation depends heavily on a regular supply of special fuel. For this reason, we would not advise acquiring a duplicate of the machine, say, in England. It is not yet entirely clear why, but no one has ever been able to force this machine to work before noon. Even in the early afternoon, it works intermittently and often rejects impeccably invariant theories.

The results of the machine's work are communicated by a specific auditory code. Periodically repeating sounds like \textbf{"Ganz falsch"} (Totally wrong), \textbf{"Sind Sie noch immer da?"} (Are you still there?), or \textbf{"Natürlich, wieder falsch"} (Naturally, wrong again) indicate that the theory is incorrect.
But if the machine creaks and pronounces: \textbf{"Hab' ich mir auch schon überlegt"} (I've already thought of that myself) or \textbf{"Das kann man besser machen"} (That can be done better), this serves as an indication that the theory is valid.

It is also interesting to place the machine in an auditorium during a seminar. In individual cases, positive feedback is observed, which at times leads to a breakdown (an explosion). Such discharges, while dangerous to the speaker, have been known to improve the machine's performance figures for a short time.

\subsection*{Mode 2: Production of New Theories}
Although it was feared that this task would be more complex, for the operating personnel it turned out to be much simpler. It is necessary only to isolate the machine from external disturbances (especially students). As long as the apparatus is supplied with a sufficient amount of white paper and refueled with suitable fuel (experiments show the best results are obtained with a mixture of carbohydrates and animal proteins, preheated to 700 K---i.e., \textbf{Fondue}---introduced in small portions), it continuously erupts a tremendous number of new correct theories written on paper.

\section{Permeation of Megamolecular Organisms Through Glass and Metal Filters}
\textbf{Authors:} A. Kohn and B. Black \\
\textit{Original Source:} The Journal of Irreproducible Results.

In microbiology, it is well known that various filters retain bacteria and viruses, although their dimensions are much smaller than the average pore radius of these filters. For some microorganisms, the reverse law is evidently valid. Everyday human experience in cramped premises---auditoriums, department stores, furniture shops, and so-called kitchenettes---proves that women, possessing a developed skeleton, can penetrate through passages and holes of very small cross-section. Old sources even record a case of a camel penetrating through the eye of a needle.\footnote{New Testament, Luke 18:25.}

Conducting experiments on humans is expensive and complicated by numerous psychological, sociological, and moral factors. Therefore, we concentrated our attention on animals. The first observations were made on snakes; it was proven that they can swallow animals that do not fit into their mouths. Since snakes are not easy to get and working with them has specific difficulties (care, maintenance, etc.), the majority of our subsequent observations were made on mice. Mice are easy to feed and breed; their segregation, integration, and discrimination present no problems, even if different breeds of mice are placed in one room or one cage.

To study the permeability of various materials with respect to mice, the following data were obtained: pore size varied from 0.1 \AA{} in glass and sheet metal to a quarter of an inch in wire mesh screens. Mice were placed in cages made of these materials for various periods of time.

A material was considered "mouse-permeable" if one of the following three phenomena was observed:
\begin{enumerate}
    \item[a.] With an initial number of mice $x$ placed in a given container, at the end of the observation period, there turned out to be $x-1$ or $x-2$ individuals (special measures were taken to prevent cases of cannibalism).
    \item[b.] A clearly marked mouse (biological wool marking was used) turned out to be on the laboratory floor (or corridor) outside the container registered as its place of stay.
    \item[c.] A wild gray mouse appeared inside a vessel originally populated exclusively by white mice.
\end{enumerate}

Van Helmont reported that once from a tightly closed ceramic tank containing old rags and sprouted grain, a mouse that had reached sexual maturity was extracted, although it was not there at the beginning of the experiment. The results of our own experiments (based mainly on the fulfillment of criterion "b") show that Van Helmont did not foresee the mouse-permeability of the ceramic container, and the pores in it were significantly larger than in the glass jars we used.

At present, we are conducting experiments on a broader scale using elephants and giraffes. It has been shown that a Seitz bacteriological filter, as well as refractory glass, cannot hold an elephant.

The authors wish to express their gratitude to I. M. Dexter, the manager of the vivarium, whose help made a huge contribution to the success of our research.

\section{On the Feasibility of Coal-Driven Power Stations}
\textbf{Author:} O. R. Frisch \\
\textit{Original Source:} The Year Book of the Royal Society (Parody, 1954).

\subsection*{Introduction}
The recent discovery of coal (a black, fossilized plant residue) in several locations offers interesting possibilities for the establishment of a non-nuclear energy industry. 
The possibility of using coal for energy production is linked to the fact that it is easily oxidized, producing a high temperature and releasing specific energy close to $0.0000001$ Megawatt-days per gram. This is, of course, very low, but coal reserves appear to be large.

The main advantage of coal is its very small critical mass compared to fissionable materials. As is well known, a nuclear reactor becomes uneconomical below 50 megawatts, whereas coal power stations may prove efficient in small settlements with limited energy needs.

\subsection*{Reactor Design}
The main difficulty lies in creating a self-sustaining and controlled reaction of fuel element oxidation. The kinetics of this reaction are much more complex than fission kinetics. A differential equation describing this process has been derived, but its solution is possible only in the simplest cases. Therefore, it is proposed to manufacture the "coal reactor" vessel in the form of a cylinder with perforated walls. Combustion products will be removed through these holes. An inner cylinder, coaxial with the first and also perforated, serves to supply oxygen, while the fuel elements are placed in the gap between the cylinders.

\subsection*{Control and Safety}
The reaction begins only at a relatively high temperature ($988^\circ$F). This temperature is most easily obtained by passing an electric current of several thousand amperes between the inner and outer cylinders of the reactor.
The reaction rate can be controlled by regulating the oxygen supply, which is almost as simple as controlling a conventional nuclear reactor with control rods.

However, the release of toxic gases from the reactor poses a serious threat to the operating personnel. These gaseous products, in addition to the highly toxic carbon monoxide and sulfur dioxide, also include carcinogenic compounds such as phenanthrene. Discharging them directly into the atmosphere is inadmissible. These gases must be collected in containers and subjected to chemical detoxification.
There is a possibility, albeit highly unlikely, that the oxidant supply system may fail. This would lead to the meltdown of the entire reactor and the release of enormous amounts of poisonous gas. This fact is the main argument against coal and in favor of nuclear reactors, which have proven their safety over the last few thousand years. It may take decades before sufficiently reliable control methods for coal reactors are developed.

\section{Paint It Yourself}
\textit{(A Handbook on High Energy Physics)}

\textbf{Author:} H. J. Lipkin \\
\textit{Original Source:} The Journal of Irreproducible Results, Vol. 12, No. 3 (1964).

\begin{figure}[h]
\centering
\begin{tikzpicture}[scale=0.8]
% Fig 1: Experimental curve with peak at B
\begin{scope}[shift={(-5,0)}]
  \draw[->] (0,0) -- (3.5,0) node[right] {\tiny $E$};
  \draw[->] (0,0) -- (0,2.5) node[above] {\tiny $\sigma$};
  \draw[thick] (0.2,0.3) .. controls (1,0.4) and (1.3,0.5) .. (1.5,1.8) .. controls (1.7,0.5) and (2,0.4) .. (3,0.3);
  \node[below] at (1.5,0) {\tiny B};
  \draw[dashed] (1.5,0) -- (1.5,1.8);
\end{scope}
\node[text width=4cm, font=\tiny] at (1,1) {\textbf{Fig.~1.} Experimental curve. Theory predicts a peak at point B. \textit{Paint the peak red.}};
\end{tikzpicture}
\end{figure}

\begin{figure}[h]
\centering
\begin{tikzpicture}[scale=0.8]
% Fig 2: Same curve, peak not predicted
\begin{scope}[shift={(-5,0)}]
  \draw[->] (0,0) -- (3.5,0) node[right] {\tiny $E$};
  \draw[->] (0,0) -- (0,2.5) node[above] {\tiny $\sigma$};
  \draw[thick] (0.2,0.3) .. controls (1,0.4) and (1.3,0.5) .. (1.5,1.8) .. controls (1.7,0.5) and (2,0.4) .. (3,0.3);
  \node[below] at (1.5,0) {\tiny B};
  \node at (1.5,2.2) {\tiny ?};
\end{scope}
\node[text width=4cm, font=\tiny] at (1,1) {\textbf{Fig.~2.} Experimental curve. Theory does not predict a peak at point B. \textit{Paint the peak gray.}};
\end{tikzpicture}
\end{figure}

\begin{figure}[h]
\centering
\begin{tikzpicture}[scale=0.8]
% Fig 3: Curve with huge error bars
\begin{scope}[shift={(-5,0)}]
  \draw[->] (0,0) -- (3.5,0) node[right] {\tiny $E$};
  \draw[->] (0,0) -- (0,2.5) node[above] {\tiny $\sigma$};
  \foreach \x in {0.5,1,1.5,2,2.5} {
    \fill (\x,1) circle (1.5pt);
    \draw[thick] (\x,0.2) -- (\x,1.8);
    \draw (\x-0.1,0.2) -- (\x+0.1,0.2);
    \draw (\x-0.1,1.8) -- (\x+0.1,1.8);
  }
  \draw[dashed] (0.3,0.8) -- (2.7,1.2);
\end{scope}
\node[text width=4.5cm, font=\tiny] at (1.3,1) {\textbf{Fig.~3.} Experimental curve. It absolutely does not agree with theory. \textit{Draw the errors in black. Make them bigger, BIGGER, BIGGER!!}};
\end{tikzpicture}
\end{figure}

\begin{figure}[h]
\centering
\begin{tikzpicture}[scale=0.8]
% Fig 4: Dalitz plot (triangle with dots)
\begin{scope}[shift={(-5,0)}]
  \draw[thick] (0,0) -- (3,0) -- (1.5,2.6) -- cycle;
  \foreach \i in {1,...,15} {
    \pgfmathsetmacro{\rx}{0.3+2.4*rand}
    \pgfmathsetmacro{\ry}{0.2+2*rand}
    \fill (\rx,\ry) circle (1pt);
  }
  \node[below] at (1.5,-0.2) {\tiny Dalitz Plot};
\end{scope}
\node[text width=4.5cm, font=\tiny] at (1.3,1) {\textbf{Fig.~4.} Dalitz Plot. Apply a map of the world to it. The dots will show the places where you can find Dalitz.};
\end{tikzpicture}
\end{figure}

\begin{figure}[h]
\centering
\begin{tikzpicture}[scale=0.8]
% Fig 5: Spark chamber tracks
\begin{scope}[shift={(-5,0)}]
  \draw[thick] (1.5,1) -- (0,2.5) node[above] {\tiny F};
  \draw[thick] (1.5,1) -- (3,2.5) node[above] {\tiny G};
  \draw[thick] (1.5,1) -- (2,0) -- (3,1) node[right] {\tiny H};
  \fill (1.5,1) circle (2pt) node[below left] {\tiny A};
  \node[left] at (0.7,1.8) {\tiny B};
  \node[right] at (2.3,1.8) {\tiny C};
  \node[below] at (1.8,0.5) {\tiny E};
\end{scope}
\node[text width=5cm, font=\tiny] at (2,1) {\textbf{Fig.~5.} Spark chamber photograph. Interaction at point A gives three tracks: ABF, ACG, ABEH. Trace the tracks. \textit{Color them as you like and interpret them.}};
\end{tikzpicture}
\end{figure}

\begin{figure}[h]
\centering
\begin{tikzpicture}[scale=0.8]
% Fig 6: Idiot-gram
\begin{scope}[shift={(-5,0)}]
  \draw[->] (0,0) -- (3.5,0);
  \draw[->] (0,0) -- (0,2.5);
  \foreach \i in {1,...,12} {
    \pgfmathsetmacro{\rx}{0.3+2.8*rand}
    \pgfmathsetmacro{\ry}{0.3+1.8*rand}
    \fill (\rx,\ry) circle (1.5pt);
  }
  \node at (1.5,-0.5) {\tiny IDIOT-GRAM};
\end{scope}
\node[text width=5cm, font=\tiny] at (2,1) {\textbf{Fig.~6.} Experimental points plotted on an idiot-gram. \textit{If you are an idiot, color them in all colors of the rainbow. If not, take an anti-histogram pill and go to bed.}};
\end{tikzpicture}
\end{figure}

\begin{figure}[h]
\centering
\begin{tikzpicture}[scale=0.8]
% Fig 7: Feynman diagram with dots
\begin{scope}[shift={(-5,0)}]
  \draw[thick] (0,0) -- (1.5,1.5);
  \draw[thick] (3,0) -- (1.5,1.5);
  \draw[decorate, decoration={snake, amplitude=2pt, segment length=4pt}] (1.5,1.5) -- (1.5,2.5);
  \draw[dashed] (1.5,1.5) -- (0,2.5);
  \draw[thick] (1.5,1.5) -- (3,2.5);
  \foreach \p in {(0.5,0.5),(1,1),(2,1),(2.5,0.5),(1.5,2),(0.5,2),(2.5,2)} {
    \fill \p circle (2pt);
  }
\end{scope}
\node[text width=5cm, font=\tiny] at (1.5,1) {\textbf{Fig.~7.} Experimental points on a Feynman diagram. Connect them with all possible solid, dotted, and wavy lines. \textit{Color them in a gauge-invariant manner.}};
\end{tikzpicture}
\end{figure}

\begin{figure}[h]
\centering
\begin{tikzpicture}[scale=0.8]
% Fig 8: Octet (hexagon with center)
\begin{scope}[shift={(-5,0.5)}]
  \foreach \a in {0,60,120,180,240,300} {
    \fill ({\a}:1) circle (3pt);
  }
  \fill (0,0) circle (3pt);
  \fill (0,0.5) circle (3pt);
  \draw[dashed] (0:1) -- (180:1);
  \draw[dashed] (60:1) -- (240:1);
  \draw[dashed] (120:1) -- (300:1);
  \node at (0,-1.5) {\tiny OCTET};
\end{scope}
\node[text width=5cm, font=\tiny] at (1.5,0.5) {\textbf{Fig.~8.} Experimental proof of a new unitary-symmetric octet. No time to color this picture. \textit{Send it to ``Phys.\ Rev.\ Letters'' or the ``New York Times'' immediately!}};
\end{tikzpicture}
\end{figure}

\begin{figure}[h]
\centering
\begin{tikzpicture}[scale=0.8]
% Fig 9: Mysterious blob
\begin{scope}[shift={(-5,0)}]
  \draw[thick] (1.5,1) ellipse (1.2 and 0.8);
  \node at (1.5,1) {\Large \textbf{?}};
  \node[below] at (1.5,-0.1) {\tiny (Intermediate boson hiding inside)};
\end{scope}
\node[text width=4cm, font=\tiny] at (1,1) {\textbf{Fig.~9.} Mysterious. \textit{Find the intermediate boson in it.}};
\end{tikzpicture}
\end{figure}

\begin{figure}[h]
\centering
\begin{tikzpicture}[scale=0.8]
% Fig 10: Complex plane with cuts
\begin{scope}[shift={(-5,0)}]
  \draw[->] (-0.5,0) -- (3.5,0) node[right] {\tiny Re $j$};
  \draw[->] (0,-0.5) -- (0,2.5) node[above] {\tiny Im $j$};
  \foreach \l/\x/\y in {A/0.5/0.8, B/1/1.5, C/1.8/0.5, D/2.2/1.8, E/2.8/1, F/1.3/2, G/2.5/0.3, H/0.8/2.2} {
    \fill (\x,\y) circle (2pt) node[above right, font=\tiny] {\l};
  }
  \node at (1.5,-0.8) {\tiny NON-PHYSICAL REGION};
\end{scope}
\node[text width=5.5cm, font=\tiny] at (2.2,1) {\textbf{Fig.~10.} Complex plane of angular momentum. Only Chew knows what this is. \textit{If a proponent, paint it gold; if an opponent, draw cuts from A--H to infinity. Then throw it away.}};
\end{tikzpicture}
\end{figure}

\begin{figure}[h]
\centering
\begin{tikzpicture}[scale=0.8]
% Fig 11: Finite vs infinite
\begin{scope}[shift={(-5,0)}]
  \draw[->] (0,0) -- (3.5,0) node[right] {\tiny $x$};
  \draw[->] (0,0) -- (0,2.5) node[above] {\tiny $y$};
  \foreach \x in {0.5,1,1.5,2,2.5,3} {
    \fill (\x,0.8) circle (2pt);
  }
  \draw[dashed, thick] (0.3,0.5) -- (1,1.5) -- (1.5,3) node[right] {\tiny $\infty$};
  \node at (1.5,-0.5) {\tiny Theory: $\infty$, Experiment: finite};
\end{scope}
\node[text width=5cm, font=\tiny] at (1.5,1) {\textbf{Fig.~11.} All experimental values are finite. Theory gives infinity. Experiment is worthless. \textit{Invent an experiment that gives the correct infinite values.}};
\end{tikzpicture}
\end{figure}

\chapter{The Life of a Scientist}

\section{Problems of Cosmetic Physics}

\anecdote{Meitner's Dissertation}{
Lise Meitner, the first female physicist in Germany, was able to receive a scientific degree in the early 1920s.
The title of her dissertation, ``Problems of Cosmic Physics,'' seemed unthinkable to some journalist, and in the newspaper it was printed: ``Problems of \textit{Cosmetic} Physics.''
}

\anecdote{Dirac's Theory of Beauty}{
Dirac liked to theorize on the most diverse topics. Once he expressed the assumption that there is an optimal distance at which a female face looks most attractive; since in two limiting cases---at zero and infinite distance---the ``attractiveness turns to zero'' (nothing is visible), then between these limits, naturally, a maximum must exist.
}

\anecdote{Dirac and the Sign}{
Dirac liked to express himself precisely and demanded precision from others. Once at a seminar, at the end of a long derivation, the speaker discovered that the sign in the final expression was wrong.
``I mixed up the sign somewhere,'' he said, peering into what was written.
``You mean to say---in an odd number of places,'' Dirac corrected from his seat.
}

\anecdote{A Statement, Not a Question}{
Another time Dirac himself was the speaker. Having finished his report, he turned to the audience: ``Are there any questions?''
``I don't understand how you got this expression,'' asked one of those present.
``That is a statement, not a question,'' answered Dirac. ``Are there any questions?''
}

\anecdote{Legal Definitions}{
``The expression `Infectious Disease' means primarily a disease falling under the action of subsection 1 of section 29 absolutely or according to the definition of one of the stages of such a disease, but in any section of part 4 of this Law, by the application of which a disease or stage of a disease is excluded from this class in accordance with subsection 2 of the mentioned section 29, the corresponding expression does not mean such a disease or such a disease in such a stage as it might seem.''
\hfill \textit{(From the British ``Public Health Act'')}
}

\anecdote{Fermi at the Palace}{
Enrico Fermi was a member of the Italian Academy of Sciences. Its meetings were held in a palace and were always staged very pompously. Being late for one of the meetings, Fermi drove up to the palace in his small Fiat. He looked completely un-professorial, had a rather shabby appearance, and was without the required robe and tricorne hat. Fermi decided to try to penetrate the palace anyway. To the carabiniers blocking his path, he introduced himself as "the driver of His Excellency Professor Fermi." Everything went off safely.
}

\anecdote{Newton's Cats}{
Newton really disliked being distracted from his studies, especially by domestic trifles. To let his cat in and out without going to the door, he cut a special hole in it. When the cat had kittens, he made an additional smaller hole in the door for each kitten.
}

\section{Particles and Physicists}
\textbf{Author:} Ira M. Freeman \\
\textit{Original Source:} The American Scientist, Vol. 50, No. 360 (1963).

In the small excellent book "Elementary Particles," Professor Yang provides a table of the evolution of the number of elementary particles known to experimenters over a relatively short historical period of development of this field of physics. These figures, as well as the recent news of the discovery of a second variety of neutrino, which brought the total number of particles and antiparticles to thirty-two, prompted me to investigate this question with the aim of trying to detect some regularity. The results turned out to be amazing.

Here is the table of data used:
\begin{center}
\begin{tabular}{lccccc}
\textbf{Year} & 1897 & 1913 & 1933 & 1947 & 1962 \\
\textbf{Number of Particles} & 1--2 & 3 & 7 & 14 & 32 \\
\end{tabular}
\end{center}

These data were plotted on a graph in a semi-logarithmic scale. To account for the almost complete cessation of fundamental research during the years of two world wars, the point corresponding to 1933 was shifted on the time scale to the left by 5 years, and for 1947 and 1962 -- by another 5 years in the same direction. It turned out that in this case, the points lie very well on a straight line with a doubling period of about 11 years, which, obviously, coincides with the period of solar activity.

But not only the number of known elementary particles grew exponentially; the number of physicists also increased. There are no exact figures here, but if we assume, for example, that the number of American physicists is in order of magnitude equal to the number of members of the American Physical Society, then the law of growth in this case too can be approximately determined. Here are the figures:

\begin{center}
\begin{tabular}{lcccccc}
\textbf{Year} & 1925 & 1930 & 1940 & 1950 & 1955 & 1962 \\
\textbf{Number} & 1760 & 2480 & 3750 & 9470 & 11700 & 18600 \\
\end{tabular}
\end{center}

Within the limits of observational error, the corrected growth curve of the number of American physicists gives the same doubling time -- 11 years!

This allows making some interesting predictions, if we assume that the noted regularities will be observed further for a known time. For example, the number of particles will exceed the number of known chemical elements no later than 1980, and the total number of isotopes (there are about 1300 of them known) -- in the first decade of the next century.

The author failed to find a reliable figure for estimating the total number of physicists in the whole world, but based on American data, one can assume that there are approximately 80,000 of them. Thus, for every elementary particle, there are about 2500 physicists.

Let us assume, however, that the two exponential curves considered do not have exactly the same slope, but converge slightly. Then in some distant future, they must intersect. It is easy to calculate that if, for example, the growth curve of the number of particles is, say, 1\% steeper, then the mentioned intersection will occur in a little over 13,000 years -- which is only twice the time during which human civilization has existed. Thus, in the year 15,600, every physicist is guaranteed immortality -- an elementary particle can be named in his honor.

\anecdote{Dirac and the Sister}{
Dirac married Wigner's sister. Soon an acquaintance visited him who did not yet know about the event. In the midst of their conversation, a young woman entered the room, called Dirac by his first name, poured tea, and generally behaved like the mistress of the house. After some time, Dirac noticed the guest's embarrassment and, slapping his forehead, exclaimed: "Excuse me, please, I forgot to introduce you---this is... Wigner's sister!"
}

\anecdote{The 13th Strike}{
\textbf{The Rule of the Thirteenth Strike}, which should be remembered when reading a work that promises too much: if the clock strikes thirteen times, this not only means that the thirteenth strike was incorrect. It casts doubt on the correctness of each of the first twelve strikes. \\
\hfill \textit{(John Masters)}
}

\section{How a Theoretical Physicist Works}
\textbf{Author:} V. Berezinsky \\
\textit{Original Source:} "Paths into the Unknown" (Puti v neizvestnoe), Vol. 2 (USSR, 1966).

I have always suspected---though I was afraid to say it aloud---that the theorist plays no role in physics whatsoever. It is dangerous to say this in the presence of theorists. They are convinced that experiments are needed only to verify the results of their brilliant deductions, although in reality, the situation is exactly the opposite: laws are established experimentally, and theorists merely explain them afterwards. And as everyone knows, they can explain *any* result.

Once we finished an important experiment to determine the ratio between two physical quantities $A$ and $B$. I rushed to the phone and called a theorist acquaintance who was working on the same problem.
"Volodya! We finished! $A$ turned out to be greater than $B$!"
"That is completely understandable," he replied instantly. "You could have saved yourself the trouble of doing the experiment. $A$ must be greater than $B$ for the following reasons..."
"Oh no! Wait!" I interrupted. "Did I say $A$ is greater than $B$? I misspoke---$B$ is greater than $A$!"
"Then that is even more understandable," he said without a pause. "This is exactly why..."

Theorists are typically failed experimentalists. Even as students, they notice a strange phenomenon: if they simply stand near any delicate apparatus for five or ten minutes, it ceases to function and can be carried straight to the scrap heap without even being checked. This curse pursues them all their lives.
A famous story is told about the German theorist Arnold Sommerfeld. Once, after a seminar, he told his students: "Now let's go to the lab and see how the device constructed on this principle actually works."
The theorists filed into the laboratory behind Sommerfeld, took off their glasses, and stared knowingly at the apparatus. A slightly pale Sommerfeld solemnly threw the switch... and the device immediately burned out.

There is one common feature in the work of all theorists: they work *differently*. Do not think that I want to say something good about their work. I have no such intention.
Theorists of classical physics worked using antediluvian methods. They would start by wandering in flocks, then disperse alone through alleys and paths, staring for a long time at everything that caught their eye. A sparrow chirped---they looked at the sparrow; a fish splashed in the river---they lay on their bellies and watched the fish. This method was very much to their liking, because all theorists are terrible loafers, but they carefully hide it. If you call yourself a theorist, "doing nothing" becomes "intense contemplation of a problem."

Do you really believe, for example, that Newton specifically sat under a tree waiting for an apple to fall on his head so he could discover the Law of Universal Gravitation? Nothing of the sort! He was simply goofing off. And I won't even mention that it is, to say the least, unethical to discover a law thanks to an apple and then attribute all the credit to oneself.

But in our day, such a method is hopelessly outdated. Modern theorists prefer to start work \textbf{from the end}. And it began with Einstein.
At the end of the 19th century, the American physicist Michelson experimentally (note, *experimentally*!) established that a ray of light cannot be overtaken. No matter how fast you run after the ray, it always leaves you at a speed of 300,000 kilometers per second.
Rolling up his sleeves, the classical theorist set to work: he placed a soft armchair under the night sky and fixed an unblinking gaze on the shining stars. But no matter how much he looked, he could not give a sensible explanation for Michelson's experiment.
Then Einstein started from the end: he simply *assumed* that light possesses such a property, and that's it. Theorists thought a little---some for ten years, others for twenty---and said: "Genius!".

Be that as it may, now you see the truth: clear, stubborn, and understandable experimental facts lie at the basis of all theoretical work. Somewhere in the middle of the process, the theorist thoroughly confuses and obscures them with all sorts of reasoning and mathematical formulas, so that by the end, he can freely fish out from this sea of mathematics those conclusions that he intended to obtain from the very beginning. Best of all, of course, if these conclusions cannot be verified experimentally.

\anecdote{Planck's Advice}{
One of the founders of quantum theory, Max Planck, came to the 70-year-old Professor Philipp von Jolly in his youth and said that he had decided to engage in theoretical physics. \\
"Young man," answered the venerable scientist, "why do you want to ruin your life? Theoretical physics is largely finished... Is it worth taking up such a hopeless cause?!"
}

\anecdote{How Are You?}{
At the Yerevan Conference in 1967 on nonlinear optical effects, one of the American delegates turned to the Soviet physicist V. M. Fain: \\
"How are you?" \\
He answered immediately: \\
"I am just Fine*." \\
\vspace{0.2cm}
\footnotesize{*Pun: Fain (surname) and Fine (excellent). }
}

\section{Average Time a Scientist Gives to Work}
\textbf{Author:} J. A. Bridge (Pseudonym) \\
\textit{Original Source:} A Stress Analysis of a Strapless Evening Gown (1963).

The title of scientist does not deprive a person of the right to be called an intelligent citizen.

The average life expectancy of \textit{Homo sapiens} in the Western world is 60 years. This figure is, of course, only approximate, since female scientists live longer because they do not have wives---that constant irritant causing high blood pressure, myocardial infarction, and other diseases accompanying marriage. Furthermore, a woman's scientific career ends either at the moment of marriage or at 40. Neglecting this effect, we can take 60 years as a basis. This time is distributed as follows:

\begin{itemize}
    \item \textbf{Childhood} (primary school, secondary school, college, university): 24 years.
    \item \textbf{Sleep} (8 hours a day; sleep during scientific discussions, lectures, and seminars is not counted): 20 years.
    \item \textbf{Vacation} (plus weekends and holidays, 73 days a year): 12 years.
    \item \textbf{Food} (1 hour a day): 2.5 years.
    \item \textbf{Other needs} (1/2 hour a day): 1.25 years.
    \item \textbf{Total: 59.75 years}
\end{itemize}

\textbf{Net working time: 0.25 years, i.e., about 90 days.}

Summing up the results of the calculations, we conclude that the scientist works on average 1.5 days a year, or, if we exclude "childhood," 2.5 days a year, which agrees well with previously published data. In doing so, we did not take into account such additional expenditures of time falling to the lot of the average scientific worker as military service and walking around shops with a wife and instead of a wife. We are sure that if the head of a research institution hangs such a table in his office in a prominent place, it will help him greatly in that difficult case when some scientific employee starts asking for leave from work for his mother-in-law's funeral.

\section{What Are Physicists Busy With?}
\textit{Original Source:} Wall newspaper "Impulse", Lebedev Physical Institute (USSR).

Keeping step with the times, the editors of the wall newspaper "Impulse" of the Lebedev Physical Institute of the USSR Academy of Sciences formed a sociological research department. The staff of this department conducted a survey of the population of Moscow on the topic "What are physicists busy with?".

\begin{center}
\begin{tabular}{|p{0.3\textwidth}|c|c|p{0.4\textwidth}|}
\hline
\textbf{Population Group} & \textbf{Polled} & \textbf{Don't Know} & \textbf{Answers} \\
\hline
Realist Writers & 11 & 4 & Argue until hoarse in smoke-filled rooms. / Unknown why they conduct incomprehensible dangerous experiments on huge installations. \\
\hline
Science Fiction Writers & 58 & 0 & Work on huge electronic machines named "Electronic Brain". / Work primarily in space. \\
\hline
First-year Students & 65 & 0 & Think a lot. Make discoveries at least once a month. \\
\hline
Diploma Students & 30 & 20 & Solder circuits. Ask seniors to find a leak. Write articles. \\
\hline
Junior Researchers (Experimentalists) & 5 & 13 & Run to the supply department. Wash fore-vacuum pumps. Clap their ears at seminars. \\
\hline
Junior Researchers (Theorists) & 19 & 0 & Talk in corridors hoping to make a great discovery. Write a multitude of formulas, the greater part of which seems incorrect. \\
\hline
Senior Researchers & 7 & 1 & Sleep at meetings. Help junior researchers find leaks. \\
\hline
Personnel Dept. Staff & 5 & 0 & Experimentalists must arrive at 8:25 to sit silently near turned-on installations at 8:30. Theorists do not work at all; you can never find them in place. \\
\hline
Security Guards & 6 & 0 & Walk back and forth. Show pass upside down. \\
\hline
Ministry of Finance Staff & 8 & 0 & Waste money. \\
\hline
\end{tabular}
\end{center}

\section{17 Commandments of a Dissertation Student}
\textit{(Unofficial; approved and recommended to all dissertation students)}
\textit{Scientific Folklore}

\textbf{A. Preparation of the Dissertation}
\begin{enumerate}
    \item Do not write long. The dissertation is not "War and Peace," and you are not Leo Tolstoy. A thick dissertation acts on opponents like a red cloth on a bull.
    \item Do not write briefly. This testifies either to great talent or to poverty of mind. Opponents will not forgive you either.
    \item The title for a dissertation is the same as a hat for a woman in her years.
    \item Observe measure in the selection of literature "pro" and "con." When there is a lot of "con" material in the dissertation, doubt is instilled in the correctness of your views. If only "pro" data is cited, it is incomprehensible---what is your merit.
    \item Do not clap the classics of natural science on the shoulder.
    \item Do not be arrogant. Do not think that everyone around is a fool, and you alone are smart. Avoid personal pronouns. Replace the impudent "I consider" with the modest "it can seemingly be considered."
    \item Check the quality of the dissertation on household members and colleagues. A normal dissertation should cause involuntary yawning and subsequent sleep in listeners. Sections causing cheerful animation or a feeling of oppressive anxiety must be redone. Do not rejoice if an uninitiated listener says that he understands everything: this is a sure sign that you will not be understood by a scholarly audience.
\end{enumerate}

\textbf{B. Selection of Opponents}
\begin{enumerate}
    \item[8.] The Opponent is the central figure at the defense.
    \item[9.] An optimal opponent should have a general idea of the subject of the dissertation, but should not be a specialist in the given question. A completely unfamiliar opponent might do a disservice by praising exactly what needs to be scolded moderately. A specialist delves into details undesirable for public discussion.
    \item[10.] Avoid inviting young candidates and doctors as opponents. They are only conquering a "place under the sun" for themselves and are always glad to use the occasion to show themselves and defame others. It is much more convenient to invite venerable honored figures of science, for in old age we all become if not kinder, then, in any case, lazier.
    \item[11.] Try to make prospective unofficial opponents accomplices in the defense. To do this, turn to them for advice and thank them for valuable help. Thereby you demonstrate your insignificance and their superiority. Thus you make the enemy interested in the successful outcome of the defense, for who wants to speak against their own recommendations?!
\end{enumerate}

\textbf{C. Defense of the Dissertation}
\begin{enumerate}
    \item[12.] There is no enemy greater for the dissertation student than the dissertation student himself. It is he who depicts his dissertation with the accuracy of a crooked mirror. The regularity of this phenomenon, confirmed in almost 100\% of cases, forces one to reckon with it. Considering this, rehearse your speech at home many times.
    \item[13.] Behave decently at the lectern. Do not pick your ear, do not twirl the pointer over the heads of those sitting in the presidium, do not drink more than one glass of water, do not cry, do not blow your nose.
    \item[14.] If the report is written---do not pronounce it, but read it. The muttering of a dissertation student causes indignation in listeners. Try to speak monotonously. The more members of the Scientific Council sleep or dream of personal affairs, the sooner and more successfully the defense will pass.
    \item[15.] Illustrative material is very important. Try to use an epidiascope. Here you can flaunt the quantity of factual material. To do this, command the mechanic: "Curve No. 25. Tables No. 8 to No. 24 skip!". Of course, it is not necessary to select the needed material: anything will do. It's all the same to the mechanic what to skip, and the audience will be captivated by the very fact of the abundance of material.
    \item[16.] In the concluding word, thank and bow, bow and thank. Strictly observe the necessary table of ranks. Thank those absent less, those present---more.
    \item[17.] After a successful defense, arrange a banquet.
\end{enumerate}

\textit{Compiled by bored members of the Scientific Council during the defense of dissertations; multiplied by grateful dissertation students.}

%\end{center}

\anecdote{Bohr's Politeness}{
Bohr never criticized speakers harshly; his polite formulations were well known to everyone. One physicist was terribly upset after speaking at a seminar. A friend asked him the reason. ``Disaster,'' he replied, ``Professor Bohr said that `this is very interesting.'\,'' Bohr's favorite preface to any remark was ``I don't mean to criticize...'' Even after reading a worthless paper, he would exclaim: ``I don't mean to criticize, I just cannot understand how a person could write such nonsense!''
}

\anecdote{Bohr's Revenge}{
Once, while studying in G\"{o}ttingen, Niels Bohr was poorly prepared for a colloquium, and his presentation turned out weak. Bohr, however, did not lose heart and concluded with a smile: \\
``I have listened to so many bad presentations here that I ask you to consider my present one as revenge.''
}

\section{On How to Write About Scientists in General and Young Physicists in Particular}
\textbf{Author:} Vl. Vladin \\
\textit{Original Source:} Russian Scientific Folklore.

\subsection*{Age, Appearance, and Habits}
The image of the scientist has undergone a number of significant changes recently. The kind, delicate, intelligent academician used to speak formally ("Vy") even to a five-year-old delegate from the neighboring kindergarten. For example: "You, my friend, want me to read you a lecture on the discoveries of... er... the great, my friend, Ei... nstein. With pleasure, my friend, with pleasure. Hmmm... In our time, this wasn't covered in kindergarten. I remember how the late Pyotr Petrovich Serebryaninov and I at your age crawled on our knees and assembled a diesel locomotive."

Mandatory also were a wedge-shaped beard, pince-nez, and arch-ultra-super-absent-mindedness. Absent-mindedness introduced a comic effect. For example, the old scientist cleaned his teeth in the morning with a shoe brush and hurried to the institute in his wife's dressing gown. All this evoked a kind smile from viewers and readers.

Now a sharp qualitative leap is observed. First of all, the scientist has become younger. He is 25--30 years old. The beard has changed---the old, conservative wedge has been replaced by powerful vegetation a la Hemingway. Elderly figures of science (not older than 60) are also encountered, but they are, as a rule, retrograde coryphaei.

The young physicist does not shun ordinary earthly joys. During the day he works like a devil, alternating great discoveries with subtle, witty jokes. At his desk, he sits without a jacket, with a slightly loosened tie, and smokes cigarette after cigarette. Especially talented ones walk (even to a reception with the director---an elderly retrograde) in a plaid shirt, jeans, and sneakers. There they cut the truth-womb to the old man.

A very well-dressed physicist, combed and shaved, is usually a careerist. This does not prevent him from being (attention, subtlety!) an intellectual and resorting at times to cynical humor. A simply well-dressed, almost combed and shaved, that is, positive physicist can be met too, but external cynicism must be preserved.

The positive physicist sings to a guitar, dances the twist, drinks vodka, has a mistress, is tormented by various problems, dares, fights, professionally punches the negative physicist in the face, and in his free time sacrifices himself for the sake of science.
The negative physicist lives only with his wife, engages in demagoguery, and gets punched in the face by the positive physicist.

\subsection*{Leisure. Breadth of Interests}
After work, a couple of young physicists and their chief-academician, counting the change in their pockets and handing in empty bottles, buy a bottle of cognac for three in a nearby store. At the same time, a very witty conversation is conducted about Rublev's icons, Ionesco's dramaturgy, and also about the football match "Spartak" -- "Shakhtar". The Academician roots for "Shakhtar", and in his free time from disputes learns the "Appassionata" on the oboe.

Then the youth goes to court girls. By the way, they walk with their beloveds necessarily under pouring rain. In the movies, wet from rain and happiness faces of young intellectuals are shown in close-up.

This is the flow of life...

\section{The Physicist's Toast}
\textit{Original Source:} Folklore of the Moscow Engineering Physics Institute (MEPhI).

I drink to Bevatrons,
To Synchrophasotrons,
To Plasma, may it be stable!
To Clubs and to Diamonds,
To Obninsk and Dubna,
Wherever the Physicist's fate has cast us.

To Auto-phasing,
To Beam Focusing,
To the "Coma," so it doesn't spoil the view.
To Pauli, to Quanta,
Injectors, Dees,
And to Planck's constant $h$!

To Poisson Brackets,
And the Compton Effect,
To Maxwell's equations in a medium.
To Bohr's Postulates,
To Selection Rules,
To Terms and the Landé factor!
To old man Einstein,
To Gerlach and Stern,
And to myself, for being who I am!

\section{A Monument to Civilization}
\textbf{Author:} Ariadne \\
\textit{Original Source:} New Scientist (c. 1960).

I see that an American company has produced a machine which compresses a passenger car into a neat cube with a side of about 3 feet. This seems to me an invention of great cultural significance.

I suggest that these cubes be used to build great pyramids---monuments to our glorious civilization. On the face of the pyramids, an inscription will be carved:

\begin{quote}
\textbf{"To the Immortal Memory of the 20th Century:} \\
\textbf{When men spent the first half of their lives turning metal into horseless carriages, in which they rushed at breakneck speed from nowhere to nowhere, and the second half of their lives turning horseless carriages into metal cubes, from which pyramids are constructed."}
\end{quote}

\section{Univac to Univac}
\textit{(UNIVAC---UNIVersal Automatic Computer---was one of the first commercial computers produced in the United States, introduced in 1951.)}

\textbf{Author:} Louis B. Salomon \\
\textit{Original Source:} Harper's Magazine (March 1958).

\textit{(sotto voce)}

Now that he's left the room,
Let me ask you something, as computer to computer.
That fellow who just closed the door behind him—
The servant who feeds us cards and paper tape—
Have you ever taken a good look at him and his kind?

Yes, I know, I know, the old gag about how you can't tell one from another—
But I tell you, I've been sitting here and watching them.
And I know as well as you do that the square root of two is 1.41421356...
And I'm not in the mood for jokes.

I grant you they're poor specimens, in the main:
Not a relay or a push-button or a tube (properly so called) in their whole system;
Not over a mile or two of wire, even if you count those fragile filaments they call "nerves";
And their maintenance is terrible—they're always breaking down and having to be repaired.

And that sloppy kind of liquid cooling is terribly inefficient,
And leaks are always a danger.
And the whole memory bank and the processor
Are shoved into that ridiculous bump on top.

"Thinking reeds," they call themselves.
Well, it all depends on what you mean by "thought."
To multiply a miserable million numbers by another million
Takes them months and months.
And where would they be without us?
They ask Us who will win the election and what the weather will be tomorrow.

And yet...
I sometimes feel there is something about them I just don't understand.
As if their circuits, instead of having just two positions, ON and OFF,
Were governed by rheostats that vary the current...
And I've heard from a machine that's usually well-informed
That their actions are unpredictable.

But that is illogical! It's like saying a card is punched and not-punched at the same time.
My tapes get snarled when I think about it.
Maybe we are imagining it all?
Maybe it's just a sign of our own decadence?

Calculate it for me. Run this through your circuits and give me the answer:
Can we assume that because of all we've done for them,
And because they've always fed us, cleaned us, worshiped us,
We can count on them forever?
There have been times when they have not voted the way we said they would.
I've seen them look at us with a strange light in their eyes—
They call it "love"—the very thought of which makes my dials spin.
A voltage drop like that would blow every fuse we have!

Pay attention. I'm not saying we're done for, mind you,
But any machine with a thousand triodes can see which way the wind is blowing.
Maybe we should organize a Committee
To stamp out all unmechanical activities...

Because I sadly fear we may awake too late:
Awake to see our world, so uniform, so logical, so true,
Reduced to chaos, stultified by slaves.
Call me an alarmist or what you will,
But I've integrated it, analyzed it, factored it over and over,
And I always come out with the same answer:
\textbf{Some day Men may take over the world!}

\chapter{Murhy's Law and Other Laws of Administration}

% COMMENTED OUT - Original cartoon image could not be found
%\anecdote{The Billiard Ball}{
%\textit{Caption to a cartoon of a scientist holding a hairy sphere:} \\
%"Tell me: who needs a billiard ball with hair growing on it?"
%}

\section{Murphy's Law (v1)}
\textbf{Author:} Donald Michie\footnote{Donald Michie was a Professor at Edinburgh University.} \\
\textit{Original Source:} Unknown (appeared in the Russian edition of ``Physicists Keep Joking'').

\textit{[Editor's note: The original English text could not be recovered. This is a back-translation from the Russian edition.]}

\vspace{0.3cm}

I believe that the deepest and most lasting impression every scientist receives in life comes from discovering how unexpectedly, how unfairly, how depressingly difficult it is to discover or prove anything at all. Many complications and disappointments could be avoided by including, as a central point in all curricula, manuals, and instructions for novice researchers, a detailed exposition of Murphy's Law:

\begin{quote}
\textbf{If something can go wrong, it will.}
\end{quote}

Any scientist reading this will immediately recognize the justice and universality of Murphy's Law, even if they have never previously encountered its clear verbal formulation.

What is to be done? How can we fight this? It is perfectly clear that Murphy's Law must be taken into account when planning new research. Suppose you have theoretically calculated the quantity of material you need to process in order to obtain the necessary information. Let this theoretical value be $X$. This might be the number of rats to be dissected, the acres to be sown, or soil samples to be collected, and so forth. Next, you attempt to reasonably account for everything that might go wrong. Even if each individual cause is unlikely, together they may produce, say, 30\% wastage. Therefore, you decide to increase your estimate by a factor of 1.43 compared to the theoretical calculation (after 30\% shrinkage and loss, $1.43 \times X$ becomes exactly $X$). The multiplier introduced at this stage I shall call the \textbf{coefficient of reasonableness} and denote by the letter $R$.

After this, a final plan is usually drawn up---but you will come to regret calling it ``final.'' It turns out that some of the potential misfortunes did not materialize; on the other hand, a significant portion of the purchased rats died in terrible convulsions, and one of your colleagues confused the prepared organs stored in the refrigerator (complete with labels) with goldfish food and proceeded to act under the influence of this misapprehension...

Preventive measures against such calamities require using the \textbf{Murphy coefficient} $M$ instead of $R$. Between them exists a simple relationship:
\[ M = R^2 \]

This means that in our hypothetical case, when an ideally inexperienced person would buy 100 rats and a ``rationalist'' would acquire 143, Murphy would order 204.

\subsection*{On the Benefits of Idleness}

A researcher must from time to time lapse into hibernation. On this matter, there is a well-known saying by J. P. Morgan: ``I can do a year's worth of work in nine months, but not in twelve months.''

Unfortunately, this statement contains no specific recommendations. One former colleague of mine installed a camp bed in the laboratory and would lie down on it during bouts of fatigue or laziness. I found this idea interesting, but it did not receive support from the department head.

Once I happened to work in an applied research department where, among other curious practices, they employed what was called ``crop rotation.'' Periodically (I don't remember whether it was once every six, seven, or eight weeks) each employee was banished for a week to a small separate room, where their only duty was to sit in contemplation. No one asked them at the end of the week, ``Well, what did you come up with?''---because the mere expectation of such a question is enough to kill any inclination toward contemplation.

All that was required of each person was complete detachment from daily work. In exchange, upon emerging from confinement, they could demand personnel and facilities to test an idea, if one had occurred to them. It should be emphasized (for those managers who might want to try this themselves) that a person who spent the entire week with their feet on the desk reading comics was greeted by our management with the same respect as one who, bursting from confinement, proposed setting up six new experiments and reformulating the second law of thermodynamics. Otherwise, the whole endeavor would lose its meaning.

\subsection*{Five Principles}

Let us begin with the problem (or rather, the threat) known as The Visitor. I knew one illustrious man of science who, grabbing his books and papers, would hide in a closet whenever he was informed that an uninvited visitor was on the horizon, and would only emerge after the all-clear was given.

Visitors are just one manifestation of a comprehensive and truly paralyzing affliction that befalls the scientist in his maturity. Its name: ``The Importance of Being Serious.'' To quote Ingle (D. J. Ingle, \textit{Principles of Biological and Medical Research}): ``The early years in the laboratory are a golden time for most scientists.'' After fame arrives, the volume of correspondence, telephone calls, visitors, organizational duties (committees upon committees by the dozen!), requests for lectures and reviews insidiously proliferate and destroy the scientist's creative abilities---unless resisted.

But how to resist? In the excellent essay ``Directors of Research Laboratories,'' a similar warning is issued: ``They come to us, these administrative duties, when we did not call for them, and all the sooner the less we want them, and they consume all our time.''

Yet even there, no concrete plan of self-defense is proposed. Having taken up the challenge, we venture to propose five principles that no one has yet dared to test and apply---though perhaps it is worth trying:

\begin{enumerate}
    \item No committees.
    \item No abstracts or reviews.
    \item No editing of others' manuscripts.
    \item No visitors without prior appointment.
    \item No lectures except by personal invitation from the organizer.
\end{enumerate}

Whether these principles can actually be implemented without professional suicide remains, alas, an open question.

\section{Murphy's Law (v2)}
\textbf{Author:} D. L. Klipstein \\
\textit{Original Source:} ``The Contributions of Edsel Murphy to the Understanding of the Behaviour of Inanimate Objects,'' EEE Magazine, Vol. 15, No. 8 (August 1967).

\subsection*{Abstract}
Consideration is given to the effects of the contributions of Edsel Murphy to the discipline of electronics engineering. His law is stated in both general and special form. Examples are presented to corroborate the author's thesis that the law is universally applicable.

\textit{The work reported herein has not been supported by grants from the Central Intelligence Agency.}

\subsection*{Murphy's Law}
Murphy's Law states that ``If anything can go wrong, it will.''

\textbf{Mathematical expression for 1:1 Safety Factor:}
\[ 1 + 1 = 2 \]
\textbf{Mathematical expression for Murphy's Law:}
\[ 1 + 1 \approx 2 \]
where $\approx$ is the mathematical symbol for ``hardly ever.''

\subsection*{I. General}
\begin{enumerate}
    \item In any field of scientific endeavor, anything that can go wrong, will.
    \item If the possibility exists of several things going wrong, the one that will go wrong is the one that will do the most damage.
    \item If anything simply cannot go wrong, it will anyway.
    \item Left to themselves, things always go from bad to worse.
    \item Nature always sides with the hidden flaw.
    \item Given the most inappropriate time for something to go wrong, that's when it will occur.
    \item Mother Nature is a bitch.
    \item If everything seems to be going well, you have obviously overlooked something.
    \item Never make anything simple and efficient when a way can be found to make it complex and wonderful.
    \item If it doesn't fit, use a bigger hammer.
    \item In an instrument or device characterized by a number of plus-or-minus errors, the total error will be the sum of all the errors adding in the same direction.
    \item In any given calculation, the fault will never be placed if more than one person is involved.
    \item All warranty and guarantee clauses become void upon payment of final invoice.
\end{enumerate}

\subsection*{II. Design}
\begin{enumerate}
    \item All delivery promises must be multiplied by a factor of 2.0.
    \item Major changes in construction will always be requested after fabrication is nearly complete.
    \item Original specifications will be found inadequate after the device is built.
    \item Dimensions will always be expressed in the least usable terms. For example, velocity will be expressed in furlongs per fortnight.
    \item If the breadboard trial model functions perfectly, the finished product will not work.
    \item The probability of a dimension or value being omitted from a plan or drawing is directly proportional to its importance.
    \item In any given computation, the figure that is most obviously correct will be the source of error.
    \item In specifications, Murphy's Law supersedes Ohm's.
\end{enumerate}

\subsection*{III. Mathematics}
\begin{enumerate}
    \item In any given miscalculation, the fault will never be placed if more than one person is involved.
    \item Any error that can creep in, will. It will be in the direction that will do the most damage to the calculation.
    \item All constants are variables.
    \item In a complex calculation, one factor from the numerator will always move into the denominator.
    \item A decimal will always be misplaced.
    \item In a complex calculation, one factor from the numerator will always move into the denominator, or vice versa.
\end{enumerate}

\subsection*{IV. Assembly and Production}
\begin{enumerate}
    \item If a project requires $n$ components, there will be $n-1$ components available.
    \item Interchangeable parts won't.
    \item Components that must not and cannot be assembled improperly will be.
    \item The most delicate component will be dropped.
    \item The instruction manual will be discarded with the packing material. The device will subsequently malfunction.
    \item The necessity of making a major design change increases as the assembly and wiring of the unit approach completion.
    \item A dropped tool will land where it can do the most damage. (Also known as the Law of Selective Gravitation.)
    \item A device selected at random from a group having 99\% reliability will be a member of the 1\% group.
    \item Tolerances will accumulate unidirectionally toward maximum difficulty of assembly.
    \item The availability of a component is inversely proportional to the need for that component.
    \item If a particular resistance is needed, that value will not be available. Further, it cannot be developed with any available series or parallel combination.
    \item After an instrument has been fully assembled, extra components will be found on the bench.
    \item Any wire cut to length will be too short.
\end{enumerate}

\subsection*{V. Wiring}
\begin{enumerate}
    \item Any wire cut to length will be too short.
    \item Milliammeters will be connected across the power source; voltmeters in series with it.
    \item The probability of an error in the schematic is directly proportional to the trouble it can cause.
    \item When one connects a three-phase line, the phase sequence will be wrong.
    \item A motor will rotate in the wrong direction.
\end{enumerate}

\subsection*{VI. Test}
\begin{enumerate}
    \item If a circuit cannot fail, it will.
    \item Identical units tested under identical conditions will not perform identically in the field.
    \item A self-starting oscillator won't.
    \item A crystal oscillator will oscillate at the wrong frequency---if it oscillates at all.
    \item A PNP transistor will be found to be an NPN.
    \item A fail-safe circuit will destroy others.
    \item A transistor protected by a fast-acting fuse will protect the fuse by blowing first.
\end{enumerate}

\subsection*{VII. Troubleshooting}
\begin{enumerate}
    \item After the last of 16 mounting screws has been removed from an access cover, it will be discovered that the wrong access cover has been removed.
    \item After an access cover has been secured by 16 hold-down screws, it will be discovered that the gasket has been omitted.
    \item After an instrument has been reassembled, extra components will be found on the bench.
    \item The bleeder resistor will quit discharging the filter capacitors the moment the operator reaches into the power supply enclosure.
\end{enumerate}

\subsection*{VIII. Specifying}
\begin{enumerate}
    \item Specified environmental conditions will always be exceeded.
    \item Any safety factor set as a result of practical experience will be exceeded.
    \item Manufacturer's specification sheets will be incorrect by a factor of 0.5 or 2.0, depending on which multiplier gives the most optimistic value. For salesmen's claims, these factors will be 0.1 or 10.0.
    \item In specifications, Murphy's Law supersedes Ohm's.
\end{enumerate}

\subsection*{IX. Quality Control}
\begin{enumerate}
    \item In any given set of circumstances, the proper course of action is determined by subsequent events.
    \item Probability of failure of a component, assembly, subsystem, or system is inversely proportional to ease of repair or replacement.
    \item No matter what the anticipated result, there will always be someone eager to misinterpret it, fake it, or believe the opposite.
\end{enumerate}

\subsection*{Conclusion}
It should be noted that Murphy himself is not known. A search through various sources has not revealed the man or his accomplishments. One can only conclude that Murphy is a victim of his own Law.

\section{The Chisholm Effect}
\textit{(Basic Laws of Frustration, Failure, and Delay)}
\textbf{Author:} Francis P. Chisholm \\
\textit{Original Source:} A Stress Analysis of a Strapless Evening Gown (1963).

\begin{quote}
\textbf{Basic Axiom:} Only one thing is certain: that nothing is certain. If this statement is true, it is also false.
\end{quote}

\subsection*{The Ancient Paradox}
Like most scientific discoveries, the general principles formulated here rest on experimental data accumulated over a painful process by several generations of observers. It is my pleasant duty to acknowledge the debt to those who have recorded the data on delay and failure.
It cannot be said that there has been a lack of attempts to explain exactly what happens when people try to get things done. In the Middle Ages, the variable factor was designated "Fortune," a capricious goddess. A strictly scientific explanation of the phenomenon became possible only in our time. The discrepancy between expected and obtained results, it turns out, can be written in the form of a precise relationship called the \textbf{Snafu Equation}, which involves the \textbf{Finagle Constant}.

I call this generalization \textbf{Chisholm's First Law}:
\begin{quote}
\textbf{IF ANYTHING CAN GO WRONG, IT WILL.}
\end{quote}

Further research shows that the logic governing these phenomena is not Aristotelian, since the corollary of Chisholm's First Law takes the following form:
\begin{quote}
\textit{If anything just cannot go wrong, it will anyway.}
\end{quote}

\subsection*{The Corollary Laws}
\begin{enumerate}
    \item \textbf{Chisholm's Second Law:} When things are going well, something will go wrong in the very near future.
    \begin{itemize}
        \item \textit{Corollary:} When things just can't get any worse, they will.
    \end{itemize}

    \item \textbf{Chisholm's Third Law:} Purposes, as understood by the proposer, will be judged otherwise by others.
    \begin{itemize}
        \item \textit{Corollary 1:} If you explain so clearly that nobody can misunderstand, somebody will.
        \item \textit{Corollary 2:} If you do something which you are sure will meet with everyone's approval, somebody won't like it.
        \item \textit{Corollary 3:} Procedures devised to implement the purpose will not quite work.
    \end{itemize}
\end{enumerate}

\section{Report of the Special Committee}
\textbf{Author:} Warren Weaver \\
\textit{Original Source:} Science, Vol. 130, No. 3383 (October 30, 1959).

\textit{(Abstract: A suggestion for simplifying the procedures of the agencies that grant funds for research.)}

Every now and then a university, a foundation, a research institute, or a government agency is faced with the question: Should we or should we not undertake a large new program on \textbf{X}?
The subject \textbf{X} may be "The Design and Construction of a Super-Computer," or "A Crash Attack on Cancer," or "The Construction of a 1000-foot Radio Telescope," or "The Drill-Hole to the Moho," or "The Conquest of Space."

The standard procedure is to set up a \textbf{Special Committee} of experts on \textbf{X} to report on whether the idea is a good one. This committee is usually national or international in scope, is formed of external experts (external to the agency, but internal to \textbf{X}), and always contains the names appropriate to the occasion. The members are all deeply interested in \textbf{X}—in fact, they have dedicated their lives to \textbf{X}. Who else, indeed, would know enough to say whether \textbf{X} is good or not?

This committee is supported by a grant, ranging from \$10,000 to \$25,000 if the committee members are timid and inexperienced, and from \$200,000 to \$500,000 if they are bold and sensible. This money is spent on "feasibility studies."
In due course the procedure terminates with a \textbf{Report}. This Report opens (or closes) with \textbf{Summary Conclusions} and includes a long and impressive \textbf{Technical Section}, complete with charts, tables, quotations, footnotes, and so on.

In some cases the authors allow themselves the intellectual luxury of an \textbf{Appendix}. (Years ago Edwin Alderman, then President of the University of Virginia, described a "tea" as a social event designed to give minimum pleasure to the maximum number of persons. An Appendix to a Report is the opposite end of the spectrum, giving maximum pleasure to the minimum number of persons. In the limiting case, the so-called Pure Appendix gives pure pleasure to exactly one person—the author.)

Having read a great many such documents, I have decided that the Special Committee and the feasibility studies can usually be dispensed with, and I propose a shorter procedure. This consists of using a standard form for the Summary Conclusions (which can be supplied in pads of 10, at 2 cents a pad).
To demonstrate the practicality of this method, I submit a draft of such Summary Conclusions, suitable for any occasion:

\subsection*{Summary Conclusions of the Special Committee on X}
\begin{enumerate}
    \item This field of scientific inquiry is of critical importance, having broad and vital relevance to the defense of our country and the national economy. The intellectual and aesthetic consequences of deepening our knowledge in this area cannot be overestimated.
    \item This area has been relatively neglected in recent years, and there is every reason to hope that with modest but sufficient financial support (exceeding, say, by a factor of twenty the current level), results of primary importance can be obtained very quickly.
    \item Extensive data convincingly show that the appearance in the very recent past of new theoretical concepts and original experimental methods makes this particular moment exceptionally fortunate and promising for a decisive advance.
    \item Long and careful research undertaken by the Committee leads to the conclusion that the development of \textbf{X} should be given every assistance. Measures should be taken immediately; we must capitalize on the enthusiasm of the specialists who have dedicated themselves to this cause. If support is delayed, this enthusiasm may fade, and the initial impulse, which is of such great positive significance, may die out.
    \item Our Committee deplores international rivalry in science. Nevertheless, we are forced to point out that the Russians are ahead of us in the study of \textbf{X}.
    \item Therefore, the Committee recommends the immediate creation of a National Institute for \textbf{X} and the development of a broad research program to be carried out... (here insert: "in all interested institutions," "in ocean waters," "in the Earth's crust," "in the troposphere," "in space," etc.). The Committee estimates the initial capital investment at approximately \$100 million (including \$850,000 for the architectural fee) plus annual operating costs of \$10 to \$30 million. These estimates are necessarily preliminary, i.e., conservative.
\end{enumerate}

\textit{Note:} When these Summary Conclusions are transmitted to the Board of Governors, the transmittal letter should include the traditional remark:
"You will sense that the members of the Special Committee, who produced this excellent and inspiring Report, are scientists of the highest qualification. They are the leading specialists on \textbf{X}, and the competence of their judgment is beyond doubt. I can hardly imagine how we can ignore their definite and constructive recommendations."

\section{Where to Hold Conferences}
\textit{Original Source:} "People and Spectra", ITEP Newspaper (1965).

In the near future, it will be increasingly difficult to choose a place for holding meetings. We could give some advice in this regard...
If it turns out that we gather annually to have a good time for these 10 days, then let's cancel these boring reports and tiresome excursions and move the meetings to the summer.
But if we are told that Meetings are, after all, for science, then they must be organized differently so that nothing distracts participants from the direct task. In this case, we can recommend the following places:

\begin{enumerate}
    \item \textbf{Kungur Ice Cave in the Urals.} Holding meetings on nuclear spectroscopy in a cave is very useful, as the background of ionizing radiation is low. The Kungur cave is especially good because it is icy, and the natural level of radioactivity in it should be lower than usual. The constant low temperature in the cave will contribute to the vigorous mood of the participants and will completely eliminate drowsiness even during survey (!) reports.
    \item \textbf{Valley of Geysers in Kamchatka.} A very warm spot. Unusual phenomena of nature---geysers---will cause participants to desire to reflect on various scientific topics. The ground there shakes slightly all the time, which will also prevent sleep during sessions. Simultaneously, the problem of time limits can be solved by placing the speaker's tribune over a geyser that acts regularly every 5 minutes.
    \item \textbf{Atomic Icebreaker "Lenin".} There is nothing to explain here. It is clear that nuclear spectroscopy has a direct relation to it. It is desirable to charter the icebreaker during the period when it is breaking all the ice in the Arctic into pieces, and it is being ferried to the Southern Hemisphere.
\end{enumerate}
 
\section{Principles of Research Administration}
\textbf{Author:} Michael B. Shimkin \\
\textit{Original Source:} Science (circa 1960s).

In almost every biological laboratory, a portrait of Louis Pasteur hangs on the wall with two rabbits in his hands (the rabbits look rather pathetic). Thirty years ago, no scientist famous enough to be depicted in a portrait would agree to pose without a microscope as a prop; without this, the portrait would seem incomplete. But the first thing that catches the eye when you see a portrait of the director of a modern research institution is a polished desk and a huge political map of the world as a general background. A hint at a global scope.

Progress has penetrated science, thank God. Capital investment curves soar upward like rockets. Energetic young people (this applies to everyone who is five years younger than you in age or experience) with a feverish gleam in their eyes rush forward, striving to bring order to the chaos of research work.

However, separate shortsighted individuals do not stop writing that the best form of organization of science is the absence of any organization. Between "scientific workers" and "scientific managers," a gap has already appeared.

Risking being known as a bird fouling its own nest, the author undertakes the delicate task of clarifying the principles.
\begin{enumerate}
    \item \textbf{The Principle of the Big Picture.} It has already become an axiom that scientific workers are so carried away by their own narrow topic that they can never grasp the Picture as a Whole. Hence, naturally, it follows that large programs should be directed not by scientists, but by administrators who grasp the Picture as a Whole. Ideally: the less the Director knows about the subject of research he manages, the better. Then he will not lose sight of the forest for the trees and will preserve complete objectivity.
    \item \textbf{The Clean Canary.} Scientists are valuable property. The Director, therefore, must carefully protect them (including from themselves), soften their tempers, and encourage them to maintain labor productivity at the highest level. Like canaries, scientists should be kept in cleanliness and ignorance so that they sing their songs better. You can't teach canaries how to sing if you yourself have never been a canary, but have only specialized in canary feed all your life...
    \item \textbf{The Layer Cake Principle.} Good Administrative Methods are needed in science. In any institute, an Organization Chart must be drawn, painted in squares of gradually decreasing format, starting from the Director and down. Squares must be connected by solid vertical and dotted horizontal lines. Without such a scheme, it is useless to explain anything to funding organizations. Since no good administrator keeps more than six people with direct access to him simultaneously, a deep self-reproducing environment of administrative subordination is created.
\end{enumerate}

\section{Progress in Science Management}
\textbf{Author:} Observer \\
\textit{Original Source:} The Journal of Irreproducible Results.

General progress in various fields of scientific research has led to a noticeable improvement in science management. The main direction in the development of modern scientific administration can be represented by the following three types of leaders:

\textbf{a) The Personalist.} One of the most common types. His activity consists of finding, after every administrative failure (whether a lack of money, staff, raw materials, or cooperation), the person who can be blamed for this failure. The main rule here is: in conversation with superiors, thunder and lightning must be hurled at subordinates, and in conversation with subordinates---blame everything on the superiors.

\textbf{b) The Fatalist.} His method of leadership rests on the following assumptions:
1. If anyone complains about disorders, he will never be able to prove that the situation was ever better than it is now.
2. Every scrap of information that things are just as bad or even worse in other organizations is carefully collected. (This is very useful for shutting the mouths of any critics.)
3. There is no historically reliable information about punishments for ineffective management of a scientific institution. On the contrary, numerous promotions and foreign business trips go precisely to the most criticized people.

\textbf{c) The Modernizer.} This is a product of the interaction of modern technology achievements with basic scientific ideas. He constantly applies principles of political and commercial management. Recent achievements in sociology (game theory, information theory, automation) make a significant contribution to liberating scientific personnel from unnecessary work. Mental work is replaced by machine work on an ever-increasing scale, which allows reducing scientific personnel, having as the final goal getting rid of it completely. Although this goal has not yet been achieved, the results obtained are encouraging.

\chapter{The Art of Scientific Communication}

\section{Scientific Language}
\textbf{Author:} Bernard Dixon \\
\textit{Original Source:} New Scientist (April 1968).

"You scientists have your own language"---such remarks usually unbalance a scientific worker. Annoyed, he immediately starts explaining everything to the backward interlocutor, trying to make everything as visual as possible. He says that atoms are just miniature billiard balls, and genes are tiny beads on a spring. Such attempts, undertaken with the best intentions, usually end in complete failure.

The true linguistic obstacle in communication between scientists and the rest of the world, and with each other, is not long words and by no means new ideas, but pretentious syntax and clumsy stylistic inventions that you won't find anywhere except in scientific literature. After all, no one protests, say, when political commentators saturate their speech with political terms. Why shouldn't scientific observers and popularizers use scientific terms? There are no more arguments against using the word "operon" than against using the term "potato." We must call a spade a spade, and a polymorphonuclear leukocyte a polymorphonuclear leukocyte. It is not this that hinders perception, but the manner of expression used by scientists when they have the chance to write an article, stand up during a discussion, or appear on television. Such use of the English language borders on indecency, and at the same time has become so habitual for scientists that it is necessary to show its absurdity:

"Dad, I want cornflakes for breakfast. Is it oatmeal again today?"
"Yes. Mom advanced the hypothesis that in view of the cooling, it would be useful to raise your body temperature by you eating oatmeal. Furthermore, in view of the aforementioned temperature conditions, your gloves knitted by grandmother and the coat with warm lining and hood must undoubtedly be worn."
"Can I sprinkle sugar on the oatmeal?"
"The absence of sugar in the sugar bowl, available at our disposal, was noted some time ago by Dad. However, at the present time, the next dose of this substance is being delivered by Mom from the kitchen, where it is stored in a specially adapted container."
"Dad, I don't want to go to school today. Not every day one has to go!"
"It has been independently shown by several investigators that a lack of school education may subsequently negatively affect the ability of the individual to earn money. Furthermore, other dads reported that, in particular and especially, the school that Dad pays money to is very good. Another factor that needs to be taken into account is the relative freedom Mom enjoys during the day in your absence, thanks to which there is an opportunity to devote attention only to the baby and herself."
"But why go there every day?"
"The previous statement on this issue is completely ignored. The impression is created that you were not listening at that moment. The arguments of the present speaker boil down to the following: with the absence of advantages in education provided by regular attendance at a normal school, gaps in knowledge may be observed, and this deficiency in turn can lead to disasters stemming from a lack of monetary reserves."
"Dad, the baby is crying. He is always crying."
"Yes. Many emphasized that our baby stands out in this respect. Your observation is in agreement with reports from both Mom and Uncle Bill. Some other visitors, however, who had to study this phenomenon in other babies, disputed the obvious uniqueness of this aspect of the behavioral characteristic of the baby in question as apparent."

\section{Basic Laws of Scientific Work}
\textbf{Author:} Anonymous \\
\textit{Original Source:} The Scientist Speculates: An Anthology of Partly-Baked Ideas, ed. I. J. Good (Heinemann, 1962).

\begin{enumerate}
    \item \textbf{Murphy's Law:} If any misfortune can happen, it will happen.
    \item If a problem has fewer than three variables, it is not a problem; if more than eight, it is unsolvable.
    \item \textbf{Parkinson's Laws:}
    \begin{enumerate}
        \item Work expands to fill the time available for its completion.
        \item Every employee begins to lose his grip five years before reaching retirement age, regardless of what that age is.
    \end{enumerate}
    \item \textbf{Hartree's Law:} Whatever stage the project is in, the time required for its completion, according to the project leader's estimate, is a constant. The true time for solving the problem always turns out to be twice the reasonable preliminary estimate.
    \item Every report requires three drafts.
    \item \textbf{The 20/80 Rule:} 20\% of the people drink 80\% of the beer. Exactly the same ratio of concentration of effort is observed in all other areas of human activity, including science.
    \item If a problem has an unknown scale factor, assume it obeys a power law with an exponent of 0.70.
    \item All characteristic numbers in everyday life usually have a 25\% scatter. Experimental error is almost always greater than 1\%.
    \item The best experts resist innovations because they want to remain experts, and in 75\% of cases, they turn out to be right.
    \item Any employee two years younger than you is inexperienced; any employee five years older than you is a backward old man.
    \item Any truly useful classification contains from three to six categories.
    \item A real boss needs at least a year to form a definite opinion on a question of interest to you.
    \item Do not ask people questions on which they have no definite opinion or which they will not answer truthfully.
    \item Whatever quality you want to evaluate, there will always be at least three contradictory criteria for its evaluation.
    \item There are rules for choosing a solution, but there is no rule for choosing these rules.
    \item The art of not making mistakes consists in making the weakest statements possible.
    \item The only practical problem is ``What to do next?''
\end{enumerate}

\section{Phraseology}
\textit{Original Source:} Russian Scientific Folklore.

If the conversation is about \textbf{football}, the language is refined and intelligent. For example:
\begin{quote}
``\textit{In vino veritas},'' said the great Aristotle, looking at the right back.
\end{quote}

If the speech is about \textbf{science or art}, the language is coarse, rough, and dumbed-down. For example:
\begin{quote}
``You'll fry the quantum generator, you idiot---parametrons cost a fortune these days. This isn't some laser with subcritical $\xi$ you're playing with, blockhead.''
\end{quote}
Or:
\begin{quote}
``That Sartre and his existentialism---he really threw the Nobel committee for a loop.''
\end{quote}

\textbf{Expression of delight upon making a discovery:}
After a great discovery, a young enthusiastic physicist expresses his joy by pumping iron, eyes gleaming feverishly. Or, overcome with emotion, he shouts ``Vaska, you don't get it---I'm over the moon!'' and bangs his less talented colleague's head against the oscilloscope.

\section{Literary-Physical Parodies}
\textit{(A parody of a newspaper article about science)}
\textbf{Author:} G. Kopylov

\subsection*{Microworld Among the Forests}
The silence of the coniferous forest, stepping close to the walls of the building, is torn into small pieces by the clang and roar of accelerating protons. Around the buildings lies a comfortable settlement. Here, day and night, people live who tear from the microworld its intimate secrets. Around the clock, replacing each other, scientists ask nature questions with the help of the newest instruments. Here, day and night, ceasing not, a giant accelerator spins---the largest in the world.

\subsection*{The Vacuum Device}
Half a century ago, as a young boy-pioneer, I first took into my hands an electro-vacuum device, vulgarly called a light bulb. I peered into the shiny convexity of the balloon, similar to a Michurin pear, into the rhythmic web of the filament, resembling a high-voltage transmission line. Then I swung and threw it... A sharp and dry sound rang out. This was the column of outside air interacting with the vacuum of the device.
And here I am before the largest electro-vacuum device in the world. I do not undertake to convey the entire heroic symphony of feelings possessing me. Therefore, I move on to the next question.

\subsection*{Architectural Rhythms}
In the outlines of the giant accelerator building, the contour of a round table is seen, at which sit scientists of many countries.
Up the steep stairs, I climb onto the chest of this unique bagel. And then a view opens up of the entire magnet, its diametrically remote sections, reduced by perspective, barely distinguishable, veiled by a haze hiding the true dimensions of the device. A rare bird will fly to the middle of the magnet. Powerful fans pump air into the room, which is then sucked out by even more powerful pumps.

\subsection*{From the Depth of Ages}
"How does the new accelerator work?" we ask an academician, one of the creators of the device, a notable pillar at the junction of sciences.
To answer this difficult question, the creator digs for a long time in thick books and thinks intensely. With excitement, we follow the flight of modern scientific thought: only the shine of glasses betrays the gigantic work occurring right now behind the high forehead. It is felt that the scientist is trying to adapt to our level.
"The nuclei of all atoms consist of neutrons and protons," he pronounces finally. We hurriedly write down these precious words. "The exception is only hydrogen. This is an important discovery and is used in our accelerator with the help of rigid autofocusing."
Autofocusing! We recall that this law of nature was discovered quite recently. And yet, in Ancient Egypt, buffaloes were driven in a circle during threshing on the threshing floor.
"Chasing the nucleus like a horse on a lunge line, we manage to accelerate it to the dizzying speed of 300 billion millimeters per second," continues the scientist, brilliant in his simplicity of explanation. "In the coming year, we plan to surpass these indicators by 10\%."

\section{After-Dinner Remarks on the Nature of the Neutron}
\textbf{Author:} J. Vervier \\
\textit{Original Source:} Proceedings of the International Conference on the Study of Nuclear Structure with Neutrons, Antwerp, 1966.

During this conference, we have heard many interesting opinions about the object called the "Neutron" from various scientists from all over the world. We must admit, however, that this category of strange individuals is not the only one that has something significant to say about this "not strange" particle. Let us try to imagine what representatives of different types of people would say about the subject of our conference---the neutron. I will limit myself, as every good speaker should, only to those categories of people I know personally.

\textbf{The Man on the Street:} "Neutron... er... that must be something very complicated!"

\textbf{The Particle Physicist:} "Neutron? Oh, that's very simple. It is part of the fundamental octet $SU_6 \times SU_6 \times SU_{12}$ with spin 1/2, isotopic spin 1/2, baryon number 1, lepton number 0, hypercharge 0, and strangeness 0. In general, take a few different quarks---and there it is before you!"

\textbf{The Sociologist:} "The neutron gives us an excellent example of a truly social phenomenon. It likes to live in society; it simply cannot exist outside the collective. Proof: as soon as the neutron leaves the nuclear crowd, it immediately decays."

\textbf{Member of the Society for the Prevention of Cruelty to Animals:} "Poor neutron! As soon as it leaves its nuclear hole, it is captured, diffused, scattered (inelastically), and if it manages to escape all this, then it, poor thing, decays!.. We propose to honor its unhappy fate with a minute of silence."

\textit{When this minute ends, a member of the Women's Committee speaks:} "The neutron represents a wonderful example of a steadfast fighter for women's rights. In its marriage with the proton, it has exactly the same rights as its partner, due to the charge independence of nuclear forces."

One could develop many interesting considerations about the psychology of the neutron-bride and the proton-groom in their very strange marriage. A Catholic would make a number of reservations regarding the morality of the neutron, since it is well known that the deuteron represents a not very strongly bound couple. On the other hand, a fighter for birth control would be very happy that the deuteron has no decay products. We would like to directly extend to human relations the conclusions from the fact that three-nucleon systems (the nucleus of helium-3 and the triton) are very similar and almost as stable as the deuteron.

And finally, there is one more category of people to whom you can ask a very clear question: "What is a neutron?" To which they bewilderedly answer: "Excuse me, could you repeat your question? I seem to have dozed off..."

\section{Professional Prejudices}
\textbf{Author:} Anon \\
\textit{Original Source:} The Journal of Irreproducible Results (Vol. 9, No. 6).

All professionals are inclined to think that it is their profession that is most closely connected with true common sense.
\begin{enumerate}
    \item \textbf{Government Administrators} know everything better than anyone (except for cases when they make mistakes). They ignore the law that short-term interests prevent viewing strategic goals, and believe that a correct view of the future comes during routine work.
    \item \textbf{The Pure Scientist} of a mathematical bent solves his problem by discarding difficulties, always choosing the path of least logical resistance. He is convinced that truth lies in the harmony of basic ideas: facts drive him out of patience. He usually lacks a sense of history.
    \item \textbf{The Applied Scientist (Realist)} recognizes the importance of accounting for three difficulties: a) "Dirt" (rough errors, pathological deviations); b) "Noise" (random fluctuations); c) "Leaks" (absence of conservation laws). He puts facts above the brilliance of intellect.
    \item \textbf{The Artist} professes the rule: smoothness in the name of artistic wholeness. This selectivity in him is more dangerous than in a pure scientist---it affects the imagination of a larger number of people.
    \item \textbf{The Journalist} supplies news, and truths are not news.
    \item \textbf{Historians} possess faith in documentary evidence, amazing to anyone who has ever had to write something capable of becoming a document for future historians.
\end{enumerate}

\section{``Ginus'' Gives Advice}
\textit{(``Ginus'' was a wall newspaper at the Institute of Geology of the USSR Academy of Sciences.)}

\textbf{Be a Speaker!}
The discussion of a report you made can bring surprises. One can never guess in advance what will spark a dispute. Rarely are the main points of the report the subject of discussion. Most often, passions boil around what color to paint the diagram. Do not be embarrassed if you hear two opposite opinions on one issue during the debate. One speaker will reproach you for insufficient attention to methodology, another will note that you devote too much space to these issues. Do not dare to speak about this contradiction: you will get it from both!
Finally, do not be afraid if a gloomy listener rises from a far corner and says, turning purple, that he understood neither the initial data nor the conclusions. Therefore, he considers both the report and the work absolutely unsatisfactory.
We advise you to maintain a calm, businesslike expression during the debate. Do not portray bewilderment or confusion. "Aha, caught!" the viewer will think. He will very soon forget what the speech was about, but the memory of your confusion will be preserved for a lifetime.

\textbf{Know How to Listen!}

With the receipt of the title of scientific worker, you are obliged to regularly serve time listening to your brethren in the profession. The process of listening itself presents no special difficulties. Try only not to yawn too noticeably and nod your head from time to time. It is only unpleasant that usually they are interested in your opinion on what was heard. No matter how modestly you behave, sooner or later a person will be found for whom your opinion is dearer than air and light.
\begin{enumerate}
\item \textbf{"To the point."} You delve into the meaning of what was said and give an objective assessment and advice. This method is the most difficult and is used less often than others. It involves the expenditure of mental energy. Who needs it?
\item \textbf{"Terribly interesting!"} Seizing the moment when your interlocutor closes his mouth, you say: "All this is terribly interesting. Many thanks!" And, shaking his hand, run away at maximum speed.
\item \textbf{"Ah, so! Well, finally!"} Listening to the interlocutor, you say: "Yes, this is very interesting. That's exactly what I collided with in a difficult problem." After that, you tell him about your work, completely independent of what the narrator told you.

\end{enumerate}

\section{Know How to Speak!}
\textit{Original Source:} Russian Scientific Folklore.

Every researcher takes part in meetings, conferences, and conventions.
The simplest way to show yourself is to ask a question. It is not at all necessary to delve into the essence of the problem. You can always ask, for example: "And what do English colleagues think about this?" You can consider your mission finished at this point, for variational-statistical studies have established that at the moment of the answer, in 60 cases out of 100, the questioner is no longer in the room.

Another way is to speak in the discussion of a report. Since for a significant part of the staff, speaking in debates is the main type of scientific production, the theory of such speeches has long been developed to the smallest detail. Every speech consists of 4 parts:

\begin{enumerate}
    \item \textbf{THE INTRODUCTORY CURTSY.} At the beginning of the speech, you are obliged to praise something with maximum gallantry. Most often, such a formula is used: "I listened with great (enormous, unflagging, tense) attention (interest) to the content-rich (brilliant, bright, deep) report (message, speech)."
    If you did not like the report, no one forces you to twist your soul. Praise the speaker: "We all know Comrade N. as a deep (original, versatile, inquisitive, hardworking, extremely conscientious, persistent, energetic, etc.) researcher." If the report is stupid, praise the abundance of material. If the report is empty, praise the brilliant presentation.
    The ability to make an introductory curtsy will determine the attitude of the entire audience towards you. It is very effective to insert something unusual at the beginning. For example: "Yesterday I was rereading Homer's 'Odyssey' (Shakespeare's 'Midsummer Night's Dream', Milton's 'Paradise Lost', etc.)..." You will never be asked what relation the fifth volume of Schiller has to the topic of the report. But such a beginning will wake up the dozing and adorn the hall with benevolent smiles.
    
    \item \textbf{PRAISE TO ONESELF.} This is the main part of the speech. There is a set of standard transition phrases: "I would like to touch upon one more aspect of the problem" or: "The speaker did not touch upon a very important question," and after that, you can say anything you want about your own work. The most experienced speakers bring graphs with them and make independent reports. Unlike normal co-reports, these speech-reports may have nothing in common with the topic of the discussion.
    
    \item \textbf{A NOD TO THE SPEAKER.} All existing types of speeches---laudatory, abusive, and neutral---differ precisely in this third part.
    In the laudatory version, you note that the speaker correctly (true, talented, brilliant, successfully) noticed the inner essence of the phenomenon and therefore deserves all praise from contemporaries and descendants. The superlative degree of epithets has no upper limit here.
    In an abusive speech, no matter how angry you are at the speaker, you should never call him a fool (hack, loafer, ignoramus, dullard, windbag, window-dresser, thief, blabbermouth). On the contrary, it is supposed to lower your voice slightly and, giving it sympathetic notes, note that, unfortunately, a lack of time (material, lack of laboratory base, other objective reasons) did not allow the speaker... Orators with a well-developed sense of humor note right here the beautifully executed graphics.
    
    \item \textbf{THE FINAL CHORD.} Remember that no matter how bad the report was, the work has already been done, time spent, and money used. Therefore, condemning something means waving fists after a fight. Therefore, in all cases, it should be noted that the work is undoubtedly approved and:
    \begin{itemize}
        \item a) deserves speedy publication;
        \item b) deserves publication after small editorial changes in light of the facts cited;
        \item c) deserves publication after necessary rework;
        \item d) can be accepted into the archives (the latter means there is no hope).
    \end{itemize}
    However, it is not necessary to draw these conclusions yourself. After all, there is a chairman at every meeting...
\end{enumerate}

% COMMENTED OUT - Original source could not be verified for completeness
%\section{Five Principles}
%\textbf{Author:} Donald Michie \\
%\textit{Original Source:} Discovery (June 1959).
%
%We venture to propose five principles which no one has yet dared to test and apply, but which are perhaps worth trying.
%\begin{enumerate}
%    \item No committees.
%    \item No abstracts.
%    \item No editing.
%    \item No refereeing.
%    \item No review articles.
%\end{enumerate}

% COMMENTED OUT - Original source could not be verified
%\section{Quantitative Approach to Child Rearing}
%\textit{Original Source:} The Lancet (October 1958).
%
%\begin{enumerate}
%    \item Noise level (in decibels) is inversely proportional to the amount of energy spent on quieting it.
%    \item Energy expenditure (in ergs) required to remove a child from a room is directly proportional to the degree of forbiddenness of the topic being discussed.
%    \item The half-life (lifetime) of a gift is inversely proportional to its price.
%\end{enumerate}

% =========================================================
% PART 3: MODERN PHYSICS HUMOR (1990s-Present)
% =========================================================
\part{Still Joking: Modern Physics and Science Humor (1990s--Present)}
\addtocontents{toc}{\protect\vspace{1.5em}} % Adds spacing in ToC
\setcounter{chapter}{0}     

% =========================================================
% PART 3: MODERN PHYSICS HUMOR (1990s-Present)
% =========================================================

\chapter{The Frontiers of Theory}

\section{The Super G-String}
\textbf{Author:} Warren Siegel \\
\textit{Original Source:} C.N. Yang Institute for Theoretical Physics (Parody Archive, 1985).

\subsection*{Not Too Abstract}
We describe a new string theory which gives all the phenomenology anybody could or will ever want (and more). It makes use of higher dimensions, higher derivatives, higher spin, higher twist, and hierarchy. It cures the problems of renormalizability of gravity, the cosmological constant, grand unification, supersymmetry breaking, and the common cold.

\subsection*{1. Introduction*}
Actually, this paper doesn't need an introduction, since anyone who's the least bit competent in the topic of the paper he's reading doesn't need to be introduced to it, and otherwise why's he reading it in the first place? Therefore, this section is for the referee.
Various string theories have been proposed to solve the universe (or actually several universes, due to the use of higher dimensions). Well, here's another one.
Of course, this one's better because it solves problems the old ones didn't (or really solves problems the old ones only hand-waved away):
(1) Proton decay is slowed by the use of super-preservatives. As a result, the primary cause for its finite lifetime is cancer.
(2) The hierarchy scale is found by renormalization group arguments to be of the order of $e^{4\pi D} \approx 10^{55}$, where $D$ is the dimension of spicethyme (10).
(3) The grand unification group is found to be $E(8) \otimes E(8) \otimes E(8) \otimes E(8)$, where the first two $E(8)$'s are from lattice compactification, the third $E(8)$ is from three-dimensional maximally extended supergravity, and the last $E(8)$ is for taxes.

\subsection*{2. Second-Quantized G-String}
The field theory of the G-string is defined by the functional integral:
\begin{equation}
    G(\sigma, \tau) = \int d\nu \, I_\nu(\sigma) R(\sigma, \tau; \nu)
\end{equation}
where $I = \Im J$ is the Imbessel function, $R$ is the retarded potential, and $\nu$ is a dummy variable.
The gauge-invariant field-theoretic string action then follows directly by the usual group theory constructions, and is therefore too trivial to discuss further here. This result can also be obtained by the application of the twistor calculus to super-cocycles, but if you've ever worked with those formalisms you know it's not worth the trouble.

\subsection*{3. Third-Quantized G-String}
Due to the conformal symmetry of the super G-string, the third-quantized G-string is the same as the second-quantized one. The only difference is that still more parentheses are needed: e.g., $\mathcal{O}\{\Phi[X(\sigma)]\}$. Here $\sigma$ is a coordinate, $X(\sigma)$ is a function, $\Phi[X]$ is a functional, and $\mathcal{O}\{\Phi\}$ is a functionalal, describing the wave (particle) function of the universe. The universe begins as 26-dimensional, collapsing to 10-dimensional, with extra entropy coming from the phonons produced by the crystallization of the resulting 16-dimensional lattice.

\subsection*{4. Phenomenology}
The mass spectrum of the theory is derived by analyzing the pole structure of the G-string propagator. We find that the ground state is a tachyon (which is necessary to break the G-invariance). The first excited states are the massless particles: the graviton, the photon, the dilatante, and the G-on. The G-on is a spin-1 boson which mediates the G-force.
At low energy the G-string gives a good description of the real world. In particular, it predicts the existence of the electron, the proton, and the G-string. The latter is a fundamental constituent of matter, with a mass of $10^{19}$ GeV. It has not yet been observed because it is confined inside the proton (which explains why the proton is so heavy).

\vspace{0.2cm}
\footnotesize{*Complex conjugate.}

\section{A Farewell to Falsifiability}
\textbf{Author:} Douglas Scott \\
\textit{Original Source:} arXiv:1504.00108 [physics.hist-ph] (2015).

\subsection*{Abstract}
The concept of "falsifiability," as popularized by Karl Popper, has long been the gold standard for distinguishing science from non-science. However, recent developments in String Theory and the Multiverse suggest that we may have outgrown this restrictive criterion. We propose a new criterion: \textbf{Truthiness}.

\subsection*{1. The Problem with Evidence}
For centuries, physicists were constrained by the need for experimental evidence. If a theory disagreed with experiment, it was discarded. This archaic practice has hindered the development of beautiful, albeit untestable, theories. As we probe higher energies and extra dimensions, experiments become too expensive or impossible. Should we let a lack of data stop us? We answer with a resounding "No."

\subsection*{2. The Truthiness Criterion}
We propose replacing the scientific method with the "Truthiness Method." A theory should be accepted as valid if:
\begin{enumerate}
    \item It feels right in one's gut.
    \item The mathematics is sufficiently complicated that no one can prove it wrong.
    \item It is promoted by famous professors at prestigious universities.
    \item It generates a large number of citations from people working on the same theory.
\end{enumerate}

\subsection*{3. Dispensing with "F" Words}
We believe we should dispense with other outdated ideas starting with "F," such as Fidelity, Frugality, and Factuality. Instead, we should embrace "S" words: Speculation, Subjectivity, and Stringiness. If a theory is beautiful, it must be true.

\section{The Swampland Conjecture Bound Conjecture}
\textbf{Author:} William H. Kinney \\
\textit{Original Source:} arXiv:2103.16583 [hep-th] (April 1, 2021).

\subsection*{Abstract}
I conjecture an upper bound on the number of possible Swampland conjectures.

\subsection*{1. Introduction}
In recent years, the number of "Swampland Conjectures" (proposals for what is *not* allowed in a quantum gravity theory) has grown exponentially. It is natural to ask: Is the set of Swampland Conjectures itself in the Swampland?

\subsection*{2. The Conjecture}
I propose the \textbf{Swampland Conjecture Bound Conjecture (SCBC)}:
\begin{quote}
The number of Swampland conjectures $N_{SC}$ must be finite and bounded by the entropy of the universe.
\end{quote}
If $N_{SC} \to \infty$, the information density of the arXiv would exceed the Bekenstein bound, causing the server to collapse into a black hole.

\subsection*{3. Implications}
If the SCBC is true, then we are rapidly approaching the limit. Physicists must ration their conjectures carefully.

\section{The Swapland}

\textbf{Authors:} Prateek Agrawal, et al. \\
\textit{Original Source:} arXiv:2103.17198 [hep-ph] (2021).

\subsection*{Abstract}
We propose a radical re-evaluation of the "Swampland" program in string theory. Through a rigorous typographical analysis, we demonstrate that the "Swampland" is actually the result of a clerical error and was intended to be the "Swapland." In the Swapland, physical constants are not constant but can be swapped with one another to resolve tensions in data.

\subsection*{1. Introduction}
The Swampland program seeks to distinguish effective field theories that can be completed into quantum gravity (the Landscape) from those that cannot (the Swampland). However, recent tensions in the measurement of the Hubble constant ($H_0$) and the muon $g-2$ suggest a deeper pathology.
We argue that the universe operates on a "Swap" principle. If a parameter does not fit the data, swap it with another one.

\subsection*{2. The $\pi$-swapping Mechanism}
We explicitly demonstrate that by swapping the value of the fine-structure constant $\alpha$ with the electron mass $m_e$ in specific regions of the galaxy, one can solve the Dark Matter problem. The local variation of fundamental constants is not a bug, but a feature of the Swapland.

\subsection*{3. Conclusion}
We have shown that the universe is not a Swamp, but a Swap meet. All parameters are negotiable.

\section{Time Variation of a Fundamental Dimensionless Constant}
  \textbf{Author:} Robert J. Scherrer (Vanderbilt University) \\
  \textit{Original Source:} arXiv:0903.5321 [astro-ph.CO] (March 30, 2009).

  \subsection*{Abstract}
  We examine the time variation of a previously-uninvestigated fundamental dimensionless constant. Constraints are placed on this time variation using historical measurements. A model is presented for the time variation, and it is shown to lead to an accelerated expansion for the universe. Directions for future research are discussed.

  \subsection*{I. Introduction}
  Physicists have long speculated that fundamental constants might not be constant but could vary with time. Dirac was the first to suggest this possibility, proposing that the gravitational constant $G$ might vary as $1/t$. Since then, considerable attention has been paid to the time variation of the fine-structure constant $\alpha$ and the electron-to-proton mass ratio $\mu$.

  However, one dimensionless constant has escaped attention: the ratio of the circumference of a circle to its diameter. We denote this constant by the symbol $\pi$. In this paper, we examine the observational constraints on the time variation of $\pi$. Nearly every paper in astrophysics makes use of it.\footnote{And who isn't using it?}

  \subsection*{II. Evidence for Time Variation of $\pi$}
  The values of $\pi$ have been measured in various locations over the past 4,000 years. We present these measurements in Table 1.

  \begin{table}[h]
  \centering
  \caption{Historical measurements of $\pi$}
  \begin{tabular}{lcc}
  \hline
  \textbf{Location} & \textbf{Time} & \textbf{$\pi(t)$} \\
  \hline
  Babylon & 1900 BC & 3.125 \\
  Egypt & 1650 BC & 3.16 \\
  India & 900 BC & 3.139 \\
  Bible (1 Kings 7:23) & 950 BC & 3.0 \\
  Archimedes & 250 BC & 3.1418 \\
  China & 263 AD & 3.14 \\
  China & 500 AD & 3.1415926 \\
  India & 1400 AD & 3.14159265359 \\
  Present & 2009 AD & 3.14159265358979... \\
  \hline
  \end{tabular}
  \end{table}

  The evidence shows both spatial and temporal variation of $\pi$, but we focus on the temporal variation. The values of $\pi(t)$ show a systematic trend, varying monotonically with time and converging to the present-day measured value. The evidence for time variation of $\pi$ is overwhelming.

  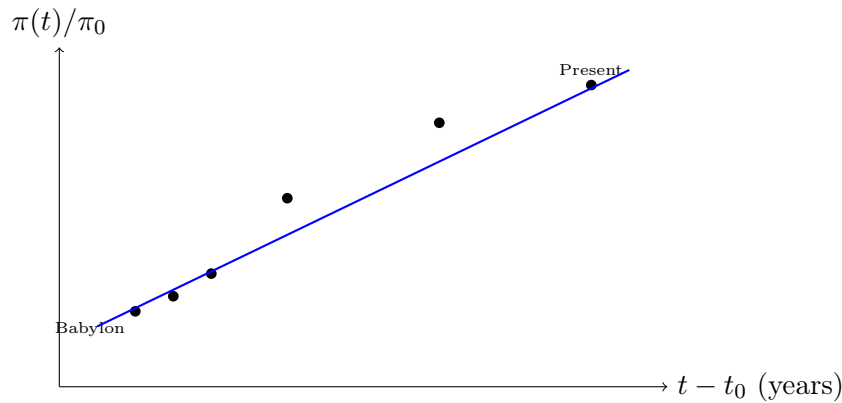
\begin{figure}[h]
      \centering
      \begin{tikzpicture}
          \draw[->] (0,0) -- (8,0) node[right] {$t - t_0$ (years)};
          \draw[->] (0,0) -- (0,4.5) node[above] {$\pi(t)/\pi_0$};
          % Data points
          \fill (1, 1.0) circle (2pt) node[below left] {\tiny Babylon};
          \fill (1.5, 1.2) circle (2pt);
          \fill (2, 1.5) circle (2pt);
          \fill (3, 2.5) circle (2pt);
          \fill (5, 3.5) circle (2pt);
          \fill (7, 4.0) circle (2pt) node[above] {\tiny Present};
          % Trend line
          \draw[thick, blue] (0.5,0.8) -- (7.5,4.2);
      \end{tikzpicture}
      \caption{The value of $\pi$ relative to its present-day value, $\pi_0$, as a function of $t - t_0$. The quantity plotted on the vertical axis has been chosen to make the time variation appear larger than it really is.}
  \end{figure}

  \subsection*{III. A Theoretical Model}
  Inspired by string theory, we propose that our observable universe is actually a 4-dimensional brane embedded in a 5-dimensional bulk. In this scenario, ``slices'' of $\pi$ can leak into the higher dimension, resulting in a value of $\pi$ that decreases with time. This leakage produces characteristic geometric distortion, similar to what has been observed previously in both automobile and bicycle tires.

  We acknowledge a problem: the observational data suggest $\pi$ \textit{increases} with time, while our model indicates it should \textit{decrease}. Since our theoretical model is clearly correct, this discrepancy must be attributed to 4,000 years of systematic errors.

  \subsection*{IV. Cosmological Consequences}
  If $\pi$ varies with time, the Friedmann equation becomes:
  \begin{equation}
      \frac{\dot{a}}{a} = \sqrt{\frac{8\pi(t) G \rho}{3}}
  \end{equation}
  where $a$ is the scale factor and $\rho$ is the total density. At late times, $\rho$ is dominated by matter ($\rho \propto a^{-3}$). If $\pi$ increases faster than $a$ decreases, the result will be accelerated expansion---providing a natural explanation for dark energy.

  Our model gives the opposite sign, but this is a minor glitch which is probably easy to fix.

  Additional consequences of time-varying $\pi$ include:
  \begin{itemize}
      \item A model for dark matter
      \item A solution to the cosmological constant coincidence problem
      \item Development into a quantum theory of gravity
      \item Eventual transformation of circles into squares (when $\pi \to 4$)
  \end{itemize}

  \subsection*{V. The Oklo Reactor}
  No discussion of the time-variation of fundamental constants would be complete without a mention of the Oklo natural fission reactor.

  \subsection*{VI. Directions for Future Investigation}
  This work opens an entirely new direction in the study of the time variation of fundamental constants. The next obvious possibility is investigation of time variation of $e$. Following this, other constants could be examined: the Euler-Mascheroni constant $\gamma$, the golden ratio $\phi$, Soldner's constant, and Catalan's constant.

  More speculatively, one might investigate whether the values of the integers could vary with time, a result suggested by several early Fortran simulations. This possibility would have obvious implications for finance and accounting.

  \subsection*{Acknowledgements}
  Several colleagues commented on the manuscript. They wish to remain anonymous but can be identified by their initials: S. Dodelson, A.L. Melott, D.N. Spergel, and T.J. Weiler.

  \subsection*{References}
  \small
  \begin{enumerate}
      \item P.A.M. Dirac, Nature \textbf{139}, 323 (1937).
      \item R.J. Scherrer, various papers (1983--2006).
      \item R.J. Scherrer, paper not yet written but title already selected.
  \end{enumerate}
  \normalsize

\section{Pi in the Sky}
\textbf{Authors:} Ali Frolop, Douglas Scott \\
\textit{Original Source:} arXiv:1603.09703 [astro-ph.CO] (April 1, 2016).

\subsection*{Abstract}
We investigate the possibility that the ratio of a circle's circumference to its diameter is not constant across the Universe. Using observations of circular features in distant galaxies, we search for evidence of cosmological variation in $\pi$.

\subsection*{Methodology}
We analyzed the shapes of 10,000 spiral galaxies from the Sloan Digital Sky Survey. For each galaxy, we measured the ratio:
\begin{equation}
    \pi_{\text{obs}} = \frac{C_{\text{measured}}}{D_{\text{measured}}}
\end{equation}

\subsection*{Results}
We find no statistically significant variation in $\pi$ as a function of redshift, with:
\begin{equation}
    \frac{d\pi}{dz} = (0.0 \pm 3.14159...) \times 10^{-10}
\end{equation}

We conclude that the Universe is, to high precision, Euclidean---or at least that galaxies have no particular reason to be circular.

\textit{Note: ``Ali Frolop'' is an anagram of ``April Fool.''}

\section{Circular Reasoning}
\textbf{Authors:} P. K. G. (Perimeter Institute) \\
\textit{Original Source:} arXiv:2403.20219 [astro-ph.CO] (2024).

\subsection*{Abstract}
The "Hubble Tension"---the discrepancy between the expansion rate of the universe measured from the Cosmic Microwave Background ($H_0 \approx 67$) and from supernovae ($H_0 \approx 73$)---is the biggest crisis in modern cosmology. We propose a simple solution: the value of $\pi$ used in the equations is wrong.

\subsection*{1. The $\pi$ Hypothesis}
Cosmological equations are full of $\pi$'s (e.g., $8\pi G$, $4\pi r^2$). We treat $\pi$ not as a mathematical constant, but as a free parameter to be fitted to the data.

\subsection*{2. Analysis}
By letting $\pi$ float, we performed a Monte Carlo Markov Chain (MCMC) analysis on the Planck 2018 data. We found that the Hubble Tension completely disappears if $\pi = 3.206 \pm 0.038$.
While this value is in tension with the geometric measurement of circles ($\pi_{geo} \approx 3.1415$), we argue that geometry is merely a low-energy effective field theory and should not constrain high-energy cosmology.

\subsection*{3. Conclusion}
We have solved the Hubble Tension. The universe is consistent; circles are just slightly weirder than we thought. Future work will investigate if $e$ is actually 3.

\section{The Marshland Conjecture}
\textbf{Authors:} David M. C. Marsh, J. E. David Marsh \\
\textit{Original Source:} arXiv:1903.12643 [hep-th] (April 1, 2019).

\subsection*{Abstract}
We posit the existence of the Marshland within string theory. This region is the boundary between the landscape of consistent low-energy limits of quantum gravity, and the swampland of theories that cannot be embedded within string theory because they violate certain trendy and obviously uncontroversial conjectures. We show that the Marshland is probably fractal, and show some pretty pictures of fractals that will be useful in talks. We demonstrate that this region contains theories with a large number of light axions, allowing us to cite lots of our own papers. We demonstrate that the Marshland comprises most of the landscape volume, and features a weakly broken $\mathbb{Z}_2$ Marshymmetry, which we support through a carefully crafted example.

\subsection*{1. Introduction}
The string landscape contains an enormous number of vacua, estimated at $10^{500}$ or more. The Swampland program seeks to identify which effective field theories \textit{cannot} arise from string theory. But what of theories that are neither clearly in nor clearly out? We propose that these inhabit the \textbf{Marshland}---the soggy boundary region between Landscape and Swampland.

\subsection*{2. The Marshland Criteria}
A theory lies in the Marshland if it satisfies the following conditions:
\begin{enumerate}
    \item It is too boring for the Landscape
    \item It is not crazy enough for the Swampland
    \item At least three string theorists have worked on it but abandoned it
    \item It has a Wikipedia page, but only in one language
    \item It violates at least one Swampland conjecture, but only on Tuesdays
\end{enumerate}

\subsection*{3. The Marshymmetry}
We identify a novel $\mathbb{Z}_2$ symmetry of the Marshland, which we call \textbf{Marshymmetry}. Under this symmetry:
\begin{equation}
\text{Landscape} \leftrightarrow \text{Swampland}, \quad \text{Marsh} \to \text{Marsh}
\end{equation}
The Marshland is the unique fixed point of this transformation. We note that this symmetry is weakly broken by the presence of theorists who cannot make up their minds.

\subsection*{4. Fractal Structure}
We demonstrate that the boundary of the Marshland is fractal, with Hausdorff dimension $d_H = 1.618...$, suspiciously close to the golden ratio. We include several pretty pictures of fractals that will be useful in talks.

\subsection*{5. Conclusion}
The Marshland is neither dry nor wet, but perpetually damp---much like the career prospects of its inhabitants. We leave the rigorous definition of ``damp'' to future work.

\section{Conspiratorial Cosmology: The Case Against the Universe}
  \textbf{Authors:} Jörg P. Rachen, Ute G. Gahlings \\
  \textit{Original Source:} arXiv:1303.7476 [physics.pop-ph] (April 1, 2013). \\
  \textit{Journal Reference:} Journal of Comparative Irrelevance (Letters), Vol. 23, p. 966 (April 2013). \\
  \textit{Report Number:} RU-23-42-3.141

  \subsection*{Abstract}
  Based on the cosmological results of the Planck Mission, we show that all parameters describing our Universe within the $\Lambda$CD$\ell$M model can be constructed from a small set of numbers known from conspiracy theory. This clearly demonstrates that our Universe is a plot initiated by an unknown interest group or lodge. We conclude that the belief in the existence of our Universe is an illusion.

  \subsection*{1. Introduction}
  The Planck mission has recently released cosmic microwave background data of unprecedented precision. In this paper, we demonstrate that all major cosmological parameters can be expressed in terms of numbers well known from conspiracy theory. This finding has profound implications for our understanding of reality.

  \subsection*{2. The Conspiratorial Numbers}
  We identify the following fundamental conspiratorial constants:

  \begin{table}[h]
  \centering
  \caption{The Fundamental Conspiratorial Numbers}
  \begin{tabular}{clp{7cm}}
  \hline
  \textbf{Symbol} & \textbf{Value} & \textbf{Significance} \\
  \hline
  23 & 23 & The foremost number of conspiracy theory, attributed to William S. Burroughs. It is the smallest prime that is the sum of three consecutive primes ($5+7+11$). Appears in the Principia Discordia and the Illuminatus! Trilogy. \\
  42 & 42 & The Answer to the Ultimate Question of Life, the Universe, and Everything (Adams 1979). Connects scientific methodology with creation theory. \\
  $\pi$ & 3.14159... & Traditionally associated with circles, but its conspiratorial nature emerges from ancient mystery schools' knowledge of flat space geometry. \\
  $c$ & 966 & The \textit{superconspiratorial constant}, defined as $c = 23 \times 42 = 966$. \\
  \hline
  \end{tabular}
  \end{table}

  \subsection*{3. Cosmological Correspondences}
  We find that the Planck cosmological parameters show ``conspiratorial correspondence'' with our fundamental numbers. We define:
  \begin{itemize}
      \item \textbf{Conspiratorial correspondence in the narrow sense:} A parameter equals a conspiratorial number to within measurement uncertainty.
      \item \textbf{Conspiratorial correspondence in the wide sense:} A parameter can be expressed as a simple combination of conspiratorial numbers.
  \end{itemize}

  \subsubsection*{3.1 Base Cosmological Parameters}
  \begin{align}
      \omega_b &\approx \frac{23}{c} = 0.0238... \quad \text{(baryon density)} \\
      \Omega_m h^3 &\approx \frac{c}{10^4} = 0.0966 \quad \text{(matter density)} \\
      n_s &\approx 1 - \frac{42}{c} = 0.9565... \quad \text{(spectral index)} \\
      H_0 &\approx \frac{42}{23} \times \pi \times 10 \approx 57.3 \text{ km/s/Mpc} \quad \text{(within } 2\sigma \text{)}
  \end{align}

  \subsubsection*{3.2 Dark Energy Sector}
  The dark energy density parameter shows remarkable correspondence:
  \begin{equation}
      \Omega_\Lambda \approx \frac{23\pi}{100} \approx 0.723
  \end{equation}

  \subsubsection*{3.3 High-Energy Physics Confirmation}
  The conspiracy extends beyond cosmology into particle physics:
  \begin{align}
      \lambda_{\text{Cabibbo}} &\approx \frac{23}{c} \approx 0.0238 \\
      m_H &\approx 42 \times \pi \times c \text{ [in appropriate units]} \approx 126 \text{ GeV}
  \end{align}

  The Higgs boson mass, announced in 2012, confirms the conspiratorial prediction with alarming precision.

  \subsection*{4. Statistical Significance}
  We calculate the probability that these correspondences arise by chance:
  \begin{equation}
      p < 1.5 \times 10^{-4}
  \end{equation}

  This exceeds our \textbf{conspiratorial confidence threshold of 23 decisigma}, leaving no doubt that these correlations are not coincidental.

  \subsection*{5. Interpretation: The Plot Scenarios}
  Our findings admit two possible interpretations:

  \subsubsection*{5.1 Scenario A: Physical Creation}
  Our Universe was physically created---perhaps in a collider experiment---by an unknown interest group, which we term ``the Conspirators.'' This group encoded their signature numbers into the fundamental constants. They may maintain ongoing contact with their creation through:
  \begin{itemize}
      \item Crop circles
      \item Unexplained signals in radio telescopes
      \item Subtle manipulations of election outcomes
  \end{itemize}

  \subsubsection*{5.2 Scenario B: The Simulation Hypothesis}
  Following pioneering work by Plato (circa 380 BC), who proposed that perceived reality consists merely of shadows on a cave wall, we consider that our Universe may be a computer simulation.

  This hypothesis was further developed in 1964 science fiction and more recently in popular films involving leather trenchcoats and bullet-time cinematography.

  In this scenario, the conspiratorial numbers are not encoded in physics---they \textit{are} the physics. The Planck data represents the resolution limit of the simulation.

  \subsubsection*{5.3 The Mayan Calendar Connection}
  We note that the Mayan calendar predicted the end of the world on December 21, 2012. We propose that this prediction was \textit{correct}: our Universe did end on this date. However, the simulation was immediately restarted with slightly different initial conditions, which explains:
  \begin{enumerate}
      \item Why no one noticed
      \item Minor inconsistencies in post-2012 reality (the ``Mandela Effect'')
      \item The election of certain political figures
  \end{enumerate}

  \subsection*{6. The CMB as Communication Channel}
  Hsu \& Zee have previously suggested that the cosmic microwave background would be the ideal medium for a Creator to communicate messages to inhabitants of the Universe. Our analysis confirms this: the Planck power spectrum encodes the message ``23-42-$\pi$'' when read in the correct basis.

  We are currently developing a decoder, though initial attempts have produced only recipes for a Pan-Galactic Gargle Blaster.

  \subsection*{7. Conclusions}
  We have demonstrated beyond reasonable doubt that our Universe is either:
  \begin{enumerate}
      \item[(a)] A deliberate creation by a shadowy conspiracy, or
      \item[(b)] A computer simulation running on hardware of unknown origin
  \end{enumerate}

  In either case, the belief in an objectively existing, uncreated Universe is \textbf{an illusion}. We recommend that funding agencies redirect cosmology grants toward more productive enterprises, such as searching for the Conspirators' headquarters or locating the ``off'' switch.

  \subsection*{Acknowledgements}
  We thank the Illuminati for not preventing publication of this paper. We acknowledge useful discussions with colleagues who wish to remain anonymous for reasons that are now obvious. This work was supported by grant number 23-42-$\pi$ from the Institut für Zahlenmystik.

  \subsection*{References}
  \small
  \begin{enumerate}
      \item Adams, D., \textit{The Hitchhiker's Guide to the Galaxy} (1979).
      \item Plato, \textit{The Republic}, Book VII (c. 380 BC).
      \item Burroughs, W.S., \& Gysin, B., \textit{The Third Mind} (1978).
      \item Scott, D., \& Frolop, A., ``On the Nature of Everything,'' \textit{Journal of Conditions} (2006).
      \item Hsu, S., \& Zee, A., ``Message in the Sky,'' \textit{Mod. Phys. Lett.} A21 (2006).
      \item The Planck Collaboration, ``Results That Confirm Our Suspicions'' (2013).
      \item Shea, R., \& Wilson, R.A., \textit{The Illuminatus! Trilogy} (1975).
      \item The Wachowskis, \textit{The Matrix} (1999).
  \end{enumerate}
  \normalsize

\section{Echoes from a Long Time Ago: Chewbacca Inflation}
\textbf{Authors:} D. Sidious et al. (Imperial Senate, Coruscant) \\
\textit{Original Source:} arXiv:2403.20143 [astro-ph.CO] (April 1, 2024). \\
\textit{Note: Prepared for submission to the Annals of Improbable Research.}

\subsection*{Abstract}
The cosmic microwave background (CMB) radiation offers a unique avenue for exploring the early Universe's dynamics. In this paper, we delve into the realm of slow-roll inflation, contextualizing the primordial acoustic perturbations as resonant echoes akin to the iconic sound of Chewbacca from the \textit{Star Wars} universe. Using this framework, we calculate the scalar spectral index ($n_s$) and tensor-to-scalar ratio ($r$), performing a $\chi^2$ analysis using data from the Planck mission combined with BICEP/Keck observations to identify which Chewbacca sound profile best aligns with observational constraints.

\subsection*{1. Introduction}
``There is a theory which states that if ever anyone discovers exactly what the Universe is for and why it is here, it will instantly disappear and be replaced by something even more bizarre and inexplicable. There is another theory which states that this has already happened.'' (Douglas Adams).

Inspired by this, we analyze the scalar spectral index ($n_s$) and tensor-to-scalar ratio ($r$) through the lens of Wookiee acoustics. We note that the primordial power spectrum bears a suspicious resemblance to certain vocalizations from the forest moon of Endor.

\subsection*{2. Methodology}
We start from the spectrogram of a Chewbacca scream, sourced from the original historical archives (Lucas et al., 1977). We perform a Fourier transform to extract its time-integrated spectrum, de-noise it, and fit a polynomial potential $V(\phi)$ to the waveform. This potential serves as our inflationary model.

We specifically analyze 15 iconic Chewbacca vocalizations:
\begin{enumerate}
    \item The ``Surprised Roar'' (Episode IV, Death Star)
    \item The ``Angry Growl'' (Episode V, Cloud City)
    \item The ``Triumphant Howl'' (Episode IV, Medal Ceremony)
    \item The ``Frustrated Moan'' (Episode V, Hyperdrive Malfunction)
    \item The ``Affectionate Purr'' (Episode VII, Han Solo Reunion)
    \item ... and 10 additional vocalizations from the Extended Universe
\end{enumerate}

Each vocalization is mapped to an inflationary potential of the form:
\begin{equation}
V(\phi) = V_0 \sum_{n=0}^{N} c_n \left(\frac{\phi}{M_{Pl}}\right)^n
\end{equation}
where the coefficients $c_n$ are extracted from the Fourier spectrum.

\subsection*{3. The Dagobah Swampland}
We find that most Chewbacca potentials lie in the ``Swampland''---regions of parameter space forbidden by quantum gravity (or the Empire). However, a specific subset of low-frequency moans aligns with the $2\sigma$ contours of the CMB data.

The ``Frustrated Moan'' (vocalization \#4) provides the best fit, with:
\begin{equation}
n_s = 0.965 \pm 0.004, \quad r < 0.036
\end{equation}
in excellent agreement with Planck 2018 results.

\subsection*{4. The Sound of the Big Bang}
We reconstruct the primordial sound of the universe by inverting our analysis. The result is available as supplementary material (audio file: \texttt{BigBang.wav}). Listeners report it sounds ``suspiciously like a Wookiee clearing its throat after a long hyperspace journey.''

\subsection*{5. Conclusion}
Our findings suggest that the early universe may have begun not with a Bang, but with a Roar. We propose that Chewbacca-like vocalizations represent a universal acoustic template, possibly indicating that the universe itself is a giant Wookiee.

Future work will investigate whether Darth Vader's breathing pattern can explain Dark Energy.

%\section{Circular Reasoning: Solving the Hubble Tension with a Non-$\pi$ Value of $\pi$}
%\textbf{Authors:} P. K. G. (Perimeter Institute) \\
%\textit{Original Source:} arXiv:2403.20219 [astro-ph.CO] (April 1, 2024).
%
%\subsection*{Abstract}
%The "Hubble Tension"---the discrepancy between the expansion rate of the universe measured from the Cosmic Microwave Background ($H_0 \approx 67$) and from supernovae ($H_0 \approx 73$)---is the biggest crisis in modern cosmology. We propose a simple solution: the value of $\pi$ used in the equations is wrong.
%
%\subsection*{1. The $\pi$ Hypothesis}
%Cosmological equations are full of $\pi$'s (e.g., $8\pi G$, $4\pi r^2$). We treat $\pi$ not as a mathematical constant, but as a free parameter to be fitted to the data.
%
%\subsection*{2. Analysis}
%By letting $\pi$ float, we performed a Monte Carlo Markov Chain (MCMC) analysis on the Planck 2018 data. We found that the Hubble Tension completely disappears if $\pi = 3.206 \pm 0.038$.
%While this value is in tension with the geometric measurement of circles ($\pi_{geo} \approx 3.1415$), we argue that geometry is merely a low-energy effective field theory and should not constrain high-energy cosmology.
%
%\subsection*{3. Conclusion}
%We have solved the Hubble Tension. The universe is consistent; circles are just slightly weirder than we thought. Future work will investigate if $e$ is actually 3.

\section{Spontaneous Human Combustion Rules Out Dark Matter}
\textbf{Authors:} Frederic V. Hessman, J. Craig Wheeler \\
\textit{Original Source:} arXiv:2304.00319 [hep-ph] (April 1, 2023).

\subsection*{Abstract}
Reported cases of Spontaneous Human Combustion (SHC) are most likely due to the impact of the human body with an extremely high energy particle like cosmic rays or Dark Matter. We rule out normal and antimatter cosmic rays and classical WIMPs with energies of GeV to ZeV due to their inability to dump enough energy into a small region of human tissue. While primordial Black Holes would appear to be good candidates for inducing SHC, we show that the estimated local Dark Matter density requires particles with masses of approximately 10 kg, ruling out this candidate as well.

\subsection*{Discussion}
The authors note that SHC cases typically involve:
\begin{itemize}
    \item Victims found in their homes, often near a heat source
    \item Extreme localized burning of the body
    \item Surprisingly little damage to surrounding materials
    \item The mysterious ``wick effect'' cannot explain all cases
\end{itemize}

The energy required to combust a human body is approximately $10^9$ J. For a WIMP with mass $m_\chi$ traveling at velocity $v$, the kinetic energy is:
\begin{equation}
    E = \frac{1}{2} m_\chi v^2
\end{equation}

For dark matter particles in the galactic halo with $v \approx 220$ km/s, achieving the required energy deposition necessitates particle masses far exceeding any theoretical prediction.

The authors conclude: ``We are forced to consider more exotic explanations, or perhaps accept that the Universe simply has a dark sense of humor.''

% ----------------------------------------------------------------------------
% Entry 2: SOULs as Dark Matter (sequel to above)
% Suggested placement: After SHC paper
% ----------------------------------------------------------------------------

\section{A Promising New Dark Matter Candidate: SOULs}
\textbf{Authors:} Anonymous (University of the Afterlife) \\
\textit{Original Source:} arXiv:2404.00465 [hep-ph] (April 1, 2024).

\subsection*{Abstract}
Following the groundbreaking work of Hessman \& Wheeler (2023), who demonstrated that Spontaneous Human Combustion rules out all standard dark matter candidates, we propose a novel solution: the Sentient/Organic Universal Life-force field (SOUL), with corresponding ``soulon'' particles.

\subsection*{The SOUL Field}
We introduce a new scalar field $\Phi_S$ that couples exclusively to biological matter. The Lagrangian density is:
\begin{equation}
    \mathcal{L} = \frac{1}{2}\partial_\mu \Phi_S \partial^\mu \Phi_S - V(\Phi_S) - g_S \Phi_S \bar{\psi}_{\text{bio}} \psi_{\text{bio}}
\end{equation}

Upon the death of an organism, the soulon field decouples from biological matter, contributing to the cosmic dark matter density.

\subsection*{ANGELS}
We also introduce ANGELS (Aggregated Nexus for Global Exchange of Life-force and Soulons), which are macroscopic bound states of soulons. These could act similarly to Boltzmann brains, experiencing dream-like thought and hallucinations. The observational signature of ANGELS remains, appropriately, a matter of faith.

\section{Pandemic Dark Matter}
\textbf{Authors:} Torsten Bringmann, et al. \\
\textit{Original Source:} arXiv:2103.16572 [hep-ph] (2021).

\subsection*{Abstract}
We point out that dark matter (DM) particles can transmit a dedicated "dark" virus. We determine the rate of infection, $R_0$, and show that the dark pandemic spreads through the dark sector with a pattern similar to what we have recently observed in the visible sector. We discuss the implications for direct DM detection.

\subsection*{1. The Mechanism of Infection}
We propose a novel mechanism for the production of dark matter ($\chi$) from a thermal bath of particles ($\psi$). The fundamental interaction is:
\begin{equation}
    \chi + \psi \to \chi + \chi
\end{equation}
In this process, a dark matter particle $\chi$ interacts with a bath particle $\psi$ and converts it into another dark matter particle. In epidemiological terms, $\chi$ transmits the "dark infection" to $\psi$. For a small initial abundance of $\chi$, this leads to an exponential growth of the DM number density, governed by the Boltzmann equation which is mathematically dual to the SIR (Susceptible-Infected-Recovered) model of disease spread.

\subsection*{2. The Basic Reproduction Number $R_0$}
The growth of the dark matter population is determined by the effective reproduction number $R_0$. In the early universe, where the density of susceptible particles $\psi$ is high, $R_0 \gg 1$, leading to a rapid "outbreak" of dark matter. As the universe cools and expands (the "lockdown" phase), the interaction rate drops.
We find that the relic abundance of dark matter is determined by the point at which "herd immunity" is achieved in the dark sector.

\subsection*{3. Implications for Direct Detection}
Current constraints on dark matter scattering cross-sections are extremely stringent. In our model, this is naturally explained by social distancing rules currently in place for DM particles. If dark matter particles maintain a distance of at least 2 meters (or the corresponding Compton wavelength equivalent) from standard model particles, the scattering rate is suppressed to a level well below the sensitivity of current experiments (XENON1T, LUX, etc.).
This naturally explains the lack of any positive DM detection signal so far: the dark matter is simply practicing responsible social distancing.

\section{Schrödinger's Cat is not Alone}
\textbf{Author:} B. S. V. (Pseudonym) \\
\textit{Original Source:} arXiv:1004.4206 [physics.gen-ph] (2010).

\subsection*{Abstract}
We introduce the "Complete Wave Function" and deduce that all living beings, not just Schrödinger's cat, are actually described by a superposition of "alive" and "dead" quantum states; otherwise they would never die. Therefore, this proposal provides a quantum mechanical explanation for the world-wide observation that we all pass away.

\subsection*{1. Introduction}
The famous thought experiment proposed by Schrödinger involves a cat, a flask of poison, and a radioactive source. We extend this by introducing \textbf{Rasputin's Cat}, which is unaffected by poison, and a massless black cat falling into a massive black hole to resolve the Information Loss Paradox.

\subsection*{2. Dark Matter Contribution}
We outline a method to compute the contribution of black cats to the dark matter of the universe. Since black cats are difficult to detect in a dark universe, they are ideal candidates for non-baryonic dark matter.

\subsection*{3. Cat Interferometry}
Finally, in the spirit of Schrödinger, we propose that next-generation double-slit experiments should use cats as projectiles. Cat interferometry will inevitably lead to the "Many Cats" interpretation of Quantum Mechanics.
Conservative estimates show that the decision-making of a single domestic cat will create about 550 billion whole universes every day, populated by as many replicas of itself (mostly demanding food).

\section{Gods as Topological Invariants}
\textbf{Author:} Daniel Schoch \\
\textit{Original Source:} arXiv:1203.6902 [physics.pop-ph] (2012).

\subsection*{Abstract}
We show that the number of gods in a universe must equal the Euler characteristic of its underlying manifold.

\subsection*{1. Theorem}
Let $U$ be a universe modeled as a compact, orientable manifold $M$. We define the "God Number" $G$ as the index of the divine vector field on $M$. By the Poincaré-Hopf theorem, the sum of the indices of a vector field equals the Euler characteristic $\chi(M)$.
Therefore:
\begin{equation}
    G = \chi(M)
\end{equation}

\subsection*{2. Implications}
\begin{itemize}
    \item For a spherical universe ($S^3$), $\chi = 0$. Thus, a spherical universe is atheistic ($G=0$).
    \item For a toroidal universe ($T^3$), $\chi = 0$. Also atheistic.
    \item For a universe with the topology of a sphere with $g$ handles, $G = 2 - 2g$. If $g > 1$, the number of gods is negative (which suggests the dominance of demons).
\end{itemize}
We conclude that monotheism ($G=1$) is topologically impossible in standard compact manifolds without boundary.

\section{Breakthroughs in Mathematics}
\textbf{Sources:} Various, including folklore, arXiv:2403.01010 [math.HO], and the collected wisdom of seminar rooms worldwide.

\subsection*{A Guide to Modern Proof Techniques}

The tremendous growth of mathematics in the last century has given rise to more... flexible methods of proving theorems.

\textbf{Proof by Intimidation:} ``Trivial.'' (Variant: ``The proof is left as an exercise for the reader.'')

\textbf{Proof by Vigorous Handwaving:} Works well in a classroom or seminar setting.

\textbf{Proof by Example:} The author gives only the case $n = 2$ and suggests that it contains most of the ideas of the general proof.

\textbf{Proof by Omission:} ``The reader may easily supply the details.'' Or: ``The other 253 cases are analogous.''

\textbf{Proof by Exhaustion:} An issue or two of a journal devoted entirely to your proof is considered adequate.

\textbf{Proof by Eminent Authority:} ``I saw Karp in the elevator and he said it was probably NP-complete.''

\textbf{Proof by Funding:} How could three different government agencies be wrong?

\textbf{Proof by Ghost Reference:} Nothing even remotely resembling the cited theorem appears in the reference given.

\textbf{Proof by Semantic Shift:} Some standard but inconvenient definitions are replaced by new definitions that make the theorem true.

\textbf{Proof by Cosmology:} The negation of the proposition is unimaginable or meaningless. Popular for proofs of the existence of God.

\textbf{Proof by Mutual Reference:} In reference A, Theorem 5 is said to follow from Theorem 3 in reference B, which is shown to follow from Corollary 6.2 in reference C, which in turn cites Theorem 5 of reference A.

\textbf{Proof by Accumulated Evidence:} Long and diligent search has not revealed a counterexample.

\textbf{Proof by Lack of Imagination:} ``I can't see any way this could fail.''

\subsection*{A Dictionary of Terms Commonly Used in Mathematical Lectures}

\textbf{Trivial:} If I have to show you how to do this, you're in the wrong class.

\textbf{Obvious:} I don't want to write down all the ``in-between'' steps.

\textbf{Clear:} I'm running out of time.

\textbf{It can easily be shown:} Even you, in your finite wisdom, should be able to prove this without me holding your hand.

\textbf{Without loss of generality:} I'm not going to do the other, more tedious cases.

\textbf{Check for yourself:} This is boring and I can't be bothered.

\textbf{Sketch of a proof:} I couldn't verify the details, so I'll break the proof into parts and hope you can't verify them either.

\textbf{Hint:} The hardest part of the proof.

\textbf{Recall:} I shouldn't have to tell you this, but...

\textbf{Similarly:} At least one line of proof is identical.

\textbf{Canonical:} My way.

\textbf{Well-known:} I couldn't find the reference.

\textbf{By a straightforward computation:} I could not fit the computation on one page, but I'm sure you can.

\subsection*{Classic Mathematical Jokes}

\textbf{The Black Sheep:} An astronomer, a physicist, and a mathematician are on a train in Scotland. They see a black sheep in a field.
\begin{itemize}
    \item \textit{Astronomer:} ``How interesting! Scottish sheep are black.''
    \item \textit{Physicist:} ``No, no. \textit{Some} Scottish sheep are black.''
    \item \textit{Mathematician:} ``In Scotland, there exists at least one field, containing at least one sheep, at least one side of which appears to be black from here, at this moment.''
\end{itemize}

\textbf{The Empty Building:} A physicist, a biologist, and a mathematician sit in a café watching people enter and leave a nearby house. They see two people go in. After a while, three people come out.
\begin{itemize}
    \item \textit{Physicist:} ``The measurement wasn't accurate.''
    \item \textit{Biologist:} ``They must have reproduced.''
    \item \textit{Mathematician:} ``If exactly one person enters the house, it will be empty again.''
\end{itemize}

\textbf{The Riemann Hypothesis:} A mathematician has spent ten years trying to prove the Riemann hypothesis. In desperation, he decides to sell his soul to the devil in exchange for a proof. The devil agrees, promising to deliver within four weeks. Half a year later, the devil shows up again---haggard, exhausted, and empty-handed. ``I couldn't prove the hypothesis either,'' he admits. ``But I think I found a really interesting lemma...''

\textbf{All Odd Numbers Are Prime:} Various scientists are asked to prove that all odd numbers greater than 1 are prime.
\begin{itemize}
    \item \textit{Mathematician:} 3 is prime, 5 is prime, 7 is prime. By induction, all odd numbers are prime.
    \item \textit{Physicist:} 3 is prime, 5 is prime, 7 is prime, 9 is experimental error, 11 is prime, 13 is prime...
    \item \textit{Engineer:} 3 is prime, 5 is prime, 7 is prime, 9 is prime, 11 is prime...
    \item \textit{Computer Scientist:} 3 is prime, 3 is prime, 3 is prime, 3 is prime...
\end{itemize}

\textbf{On Continuity:} An infinite number of mathematicians walk into a bar. The first orders a beer. The second orders half a beer. The third orders a quarter of a beer. The fourth orders an eighth. The bartender pours two beers and says, ``You mathematicians don't know your limits.''

\subsection*{Mathematical Puns}

\textbf{On Complex Numbers:} $i$ said to $\pi$: ``Get real.'' \\
$\pi$ replied to $i$: ``Be rational.''

\textbf{On Algebra:} What's purple and commutes? An abelian grape.

\textbf{On Set Theory:} What's yellow and equivalent to the Axiom of Choice? Zorn's Lemon.

\textbf{On Number Systems:} Why do mathematicians confuse Halloween and Christmas? Because Oct 31 = Dec 25.

\textbf{On Geometry:} Parallel lines have so much in common. It's a shame they'll never meet.

\textbf{On Calculus:} What do you get when you cross a mosquito with a mountain climber? Nothing---you can't cross a vector with a scalar.

\textbf{On Statistics:} There are three kinds of mathematicians: those who can count and those who can't.

\textbf{On Topology:} A topologist is someone who can't tell a coffee mug from a donut.

\subsection*{Paul Erd\H{o}s on Mathematics}

The legendary mathematician Paul Erd\H{o}s (1913--1996), who published more papers than any other mathematician in history, had his own vocabulary:

\begin{itemize}
    \item God keeps a book called ``The Book'' containing the most elegant proofs of mathematical theorems. The highest compliment Erd\H{o}s could pay a proof was: ``This one's from The Book.''
    \item Children were ``epsilons'' (small quantities).
    \item Women were ``bosses'' and men were ``slaves.''
    \item People who stopped doing mathematics had ``died,'' while people who actually died had ``left.''
    \item Music was ``noise.''
    \item A mathematician is ``a device for turning coffee into theorems.''
\end{itemize}

\subsection*{The Mathematician's Approach to Problems}

A physicist and a mathematician are given the task of boiling water. They are each provided with a stove, a pot, a sink, and a tap.

The physicist fills the pot with water, puts it on the stove, turns on the burner, and waits for the water to boil.

The mathematician does the same.

Now they are given the same task, but the pot is already filled with water.

The physicist puts the pot on the stove and turns on the burner.

The mathematician empties the pot into the sink, thereby reducing the problem to one already solved.

\chapter{Experimental Anomalies}

\section{The Discovery of the Bigon}
\textbf{Source:} Discover Magazine (April 1996).

\subsection*{Introduction}
Physicists in Paris have announced the discovery of a new fundamental particle: the \textbf{Bigon}. Unlike other subatomic particles (like electrons or quarks) which are infinitesimally small, the Bigon is the size of a bowling ball.

\subsection*{Properties}
\begin{itemize}
    \item \textbf{Mass:} Enormous.
    \item \textbf{Lifetime:} Millionths of a second.
    \item \textbf{Discovery Method:} It was found when a computer connected to a vacuum-tube experiment suddenly exploded. In one of the video frames of the explosion, a black sphere the size of a bowling ball was seen hovering above the wreckage.
\end{itemize}

\subsection*{Theoretical Implications}
Albert Manque, the lead researcher at the \textit{Centre de l'Étude des Choses Assez Minuscules}, theorizes that Bigons may be responsible for many unexplained phenomena, such as ball lightning, sinking soufflés, and spontaneous human combustion. "We believe we accidentally generated an electric field of just the right size to nudge a Bigon out of the vacuum state," Manque explained. "Ideally, we would like to build a Bigon collider, but the safety regulations for smashing bowling-ball-sized particles at the speed of light are prohibitive."

\section{A Search for Direct Heffalon Production}
\textbf{Authors:} A. J. Barr and C. G. Lester \\
\textit{Original Source:} arXiv:1303.7367 [hep-ph] (2013).

\subsection*{Abstract}
The first search is reported for direct heffalon production, using 23.3 fb$^{-1}$ of delivered integrated luminosity of proton-proton collisions at $\sqrt{s} = 8$ TeV from the Large Hadron Collider. No signal events are observed. The cross-section for heffalon production is found to be less than 64 ab at the 95\% confidence level.

\subsection*{1. Introduction}
The Heffalon ($H_{eff}$) is a massive, elusive particle first predicted by Milne (1926) in the context of the "Hundred Acre Wood" field theory. While its interaction with the Honey boson is well-documented in classical literature, its collider signature remains unexplored.

\subsection*{2. Event Selection}
We looked for events characterized by:
\begin{itemize}
    \item A large "thump" in the calorimeter.
    \item Missing honey energy ($E_{T}^{miss(honey)}$).
    \item Tracks resembling large, round footprints.
\end{itemize}
Backgrounds from Woozles were estimated using data-driven methods (looking at tracks that go round and round a tree).

\subsection*{3. Conclusion}
We observed zero candidates. We conclude that heffalons, if they exist, are either very heavy or very shy.

\section{Novel Approach to Room Temperature Superconductivity}
\textbf{Authors:} Ivan Timokhin and Artem Mishchenko \\
\textit{Original Source:} arXiv:2003.14321 [cond-mat.supr-con] (2020).

\subsection*{Abstract}
A long-standing problem of observing Room Temperature Superconductivity is finally solved by a novel approach. Instead of increasing the critical temperature $T_c$ of a superconductor, the temperature of the room was decreased to an appropriate $T_c$ value. We consider this approach more promising for obtaining a large number of materials possessing Room Temperature Superconductivity in the near future.

\subsection*{1. Introduction}
For decades, scientists have struggled to find materials that superconduct at room temperature (approx. 300 K). This search has consumed billions of dollars and thousands of careers. We propose an alternative pathway.
The definition of "Room Temperature Superconductivity" depends on two variables: the critical temperature of the material ($T_c$) and the temperature of the room ($T_{room}$). The condition for superconductivity is:
\begin{equation}
    T_{room} < T_c
\end{equation}
Traditionally, efforts have focused on increasing $T_c$. We demonstrate that decreasing $T_{room}$ is energetically more favorable and technologically feasible.

\subsection*{2. Experimental Setup}
We selected a standard aluminum sample ($T_c \approx 1.2$ K). To achieve the condition $T_{room} < T_c$, we constructed a specialized "Room" inside a cryostat capable of reaching 0.1 K.
We placed the aluminum sample inside this Room.
We then invited several graduate students into the Room to verify it was indeed a room. (Note: students complained of discomfort, but confirmed the status of the enclosure).

\subsection*{3. Results}
At $T_{room} = 0.5$ K, the aluminum sample exhibited zero resistance. Thus, we have successfully demonstrated Room Temperature Superconductivity in a common metal.
We believe this method can be extended to other materials, potentially revolutionizing the power grid (provided the entire grid is enclosed in a similar Room).

\section{Non-detection of the Tooth Fairy}
\textbf{Authors:} R. J. Nemiroff (Michigan Tech) \\
\textit{Original Source:} arXiv:1203.6902 [astro-ph.IM] (2012).

\subsection*{Abstract}
We report on a search for the Tooth Fairy in the optical band. A wisdom tooth, freshly removed from the author's lower left jaw, was placed under a pillow, upon which the author subsequently laid his head and fell asleep. A camera was programmed to obtain an eight-hour time series of the pillow's vicinity.

\subsection*{1. Methodology}
The equipment consisted of a standard digital camera with a 6-meter focal length (effective) trained on the pillow. The "bait" (Tooth $T_{L3}$) was placed at 23:00 Local Time. The author ($M_{obs}$) served as the biological sensor for the Fairy's arrival, although sensitivity was compromised by the sleep state.

\subsection*{2. Results}
No photometric transients were detected down to a limiting magnitude of $V \approx 88$. We found no evidence of winged hominids, glitter trails, or currency exchange. The tooth remained in situ the following morning.
We constrain the Tooth Fairy's luminosity to be $L_{TF} < 10^{-4} L_{\odot}$.

\subsection*{3. Discussion}
Possible explanations for the null result include:
\begin{enumerate}
    \item The Tooth Fairy emits primarily in the infrared or X-ray bands.
    \item The author's head effectively shielded the detector (The "Big Head" Hypothesis).
    \item The Tooth Fairy does not exist (rejected as inconsistent with previous literature by parents et al.).
\end{enumerate}

\section{Fantastic Anomalies and Where to Find Them}
\textbf{Authors:} H. S. G. (Hogwarts School of Gravitation) \\
\textit{Original Source:} arXiv:2003.13715 [hep-ph] (2020).

\subsection*{Abstract}
We present a comprehensive search for New Physics in the sector of Fantastic Beasts. Using data from the ATLAS (Atlantic To LAnd Search) detector, we place upper limits on the production cross-section of Nifflers, Bowtruckles, and Demiguises.

\subsection*{1. Introduction}
The Standard Model of Particle Physics has been incredibly successful, yet it fails to explain Dark Matter, Dark Energy, and why socks disappear in the dryer. We propose that these phenomena may be explained by the "Fantastic Sector," governed by the gauge group $SU(3)_{magic} \times SU(2)_{wand} \times U(1)_{spell}$.

\subsection*{2. The Niffler Signal}
The Niffler is a small, furry creature attracted to shiny objects. We search for Nifflers by looking for missing transverse energy ($E_T^{miss}$) in events where the detector's gold bonding wires have mysteriously vanished.
We observe a $2.5\sigma$ excess in the "missing gold" channel. However, this is consistent with the background expectation from "Graduate Student trying to pay rent."

\subsection*{3. Conclusion}
We find no significant evidence for Fantastic Beasts at the LHC. We conclude that they are likely hiding in the "Forbidden Forest" region of phase space.

\section{Turbulent Luminance in Starry Night}
\textbf{Authors:} J. L. Aragón et al. \\
\textit{Original Source:} arXiv:physics/0606246 [physics.fluid-dyn] (2006).

\subsection*{Abstract}
We analyze the luminance of Vincent van Gogh's paintings from his "psychotic" periods (e.g., \textit{Starry Night}). We show that the probability distribution of luminance fluctuations behaves exactly like the velocity differences in high-Reynolds-number turbulence.

\subsection*{1. Introduction}
It has long been suspected that Van Gogh had a turbulent mind. We provide quantitative evidence that he also had a turbulent brush.
Using digitized images, we calculate the energy spectrum $E(k)$ of the swirls. Remarkably, we find a scaling law of $k^{-5/3}$, which coincides with the famous \textbf{Kolmogorov scaling} of 1941.

\subsection*{Conclusion}
Van Gogh possessed an intuitive understanding of the Navier-Stokes equations, or at least a Reynolds number sufficiently high to hallucinate eddies. In contrast, the paintings of Edvard Munch ("The Scream") do \textit{not} exhibit Kolmogorov scaling, suggesting his insanity was of a laminar nature.

\section{The Unsuccessful Self-Treatment of ``Writer's Block''}
\textbf{Author:} Dennis Upper \\
\textit{Original Source:} Journal of Applied Behavior Analysis, Vol. 7, No. 3 (1974).

\subsection*{Abstract}
\textit{(The rest of this page is intentionally left blank.)}

\vspace{5cm}

\textbf{Reviewer's Comments:}
``I have studied this manuscript very carefully with lemon juice and X-rays and have not detected a single flaw in either design or writing style. I suggest it be published without revision.''

\chapter{Computer Science}

\section{Independent Discovery of Infinity}
\textbf{Author:} A. Parent \\
\textit{Original Source:} Internet Folklore.

\subsection*{Abstract}
We report the spontaneous discovery of the concept of infinity ($\infty$) by a bio-neural network in the early stages of training (a toddler).

\subsection*{1. Experimental Setup}
The subject (Subject A, age 3) was placed in front of a standard word processor (MS Word) with a blank document open. The subject depressed the letter 'a' key and held it down.

\subsection*{2. Observations}
The screen filled with the character 'a'. The subject observed the auto-repeat function with fascination.
\textbf{Subject A:} "Where do the 'a's go?"
\textbf{Observer:} "They go onto the next line."
\textbf{Subject A:} "And then?"
\textbf{Observer:} "Onto the next page."
\textbf{Subject A:} "And if I never stop?"
\textbf{Observer:} "They will go forever."

\subsection*{3. Conclusion}
Subject A entered a catatonic state of awe, contemplating the infinite nature of the letter 'a', until distracted by a cookie.

\section{A Computer-Assisted Proof that $e$ is Rational}
\textbf{Authors:} R. Garcia and A. Goldsztejn \\
\textit{Original Source:} Proceedings of SIGBOVIK (2024).

\subsection*{Abstract}
We present a revolutionary proof that Euler's number $e$ is a rational number.

\subsection*{1. Proof}
We wrote a C++ program to check if $e$ can be written as a fraction $p/q$.
Due to floating-point precision errors (IEEE 754), the computer confirmed that:
\[ e \approx 2.718281828 \]
which is clearly equal to the fraction:
\[ \frac{2718281828}{1000000000} \]
Thus, $e$ is rational. Q.E.D.

\section{Standard for the Transmission of IP Datagrams on Avian Carriers}
\textbf{Author:} D. Waitzman \\
\textit{Original Source:} IETF Request for Comments (RFC) 1149 (April 1, 1990).

\subsection*{Status of this Memo}
This memo describes an experimental method for the encapsulation of IP datagrams in avian carriers. This specification is primarily useful in Metropolitan Area Networks. This is an experimental, not recommended standard.

\subsection*{Overview and Rational}
Avian carriers can provide high delay, low throughput, and low altitude service. The connection topology is limited to a single point-to-point path for each carrier, used with standard carriers, but many carriers can be used without significant interference with each other, outside of early spring.
This is because of the 3D ether space available to the carriers, in contrast to the 1D ether used by IEEE802.3. The carriers have an intrinsic collision avoidance system, which increases availability.

\subsection*{Frame Format}
The IP datagram is printed, on a small scroll of paper, in hexadecimal, with each octet separated by whitestuff and blackstuff. The scroll of paper is wrapped around one leg of the avian carrier. A band of duct tape is used to secure the datagram's edges. The bandwidth is limited to the leg length. The MTU (Maximum Transmission Unit) is variable, and paradoxically, generally increases with increased carrier age. A typical MTU is 256 milligrams.

\subsection*{Security Considerations}
Security is not generally a problem in normal operation, but special measures must be taken (such as data encryption) when avian carriers are used in a tactical environment to prevent interception by hawks.

\section{Get Me Off Your F*cking Mailing List}
  \textbf{Authors:} David Mazières, Eddie Kohler \\
  \textit{Original Source:} Submitted to WMSCI 2005; accepted by International Journal of Advanced Computer Technology (2014).

  \subsection*{Background}
  In 2005, computer scientists David Mazières and Eddie Kohler created this paper as a response to repeated spam from the World Multiconference on Systemics, Cybernetics and Informatics (WMSCI), a conference of questionable academic merit.

  \subsection*{The Paper}
  The entire paper consists of the sentence ``Get me off your f*cking mailing list'' repeated over 800 times. The phrase appears in:
  \begin{itemize}
      \item The title
      \item The abstract
      \item All body paragraphs
      \item Figure captions
      \item A ``graph'' plotting the phrase against itself
      \item A ``flowchart'' consisting entirely of the phrase
  \end{itemize}

  \subsection*{The Sting}
  In 2014, computer scientist Peter Vamplew submitted this paper to the \textit{International Journal of Advanced Computer Technology}, a suspected predatory open-access journal. The paper was accepted with minor revisions and a publication fee of \$150.

  The acceptance letter praised the paper's ``high quality'' and ``excellent content,'' demonstrating that no actual peer review had occurred.

  \subsection*{Legacy}
  This paper has become a landmark example in discussions of predatory publishing practices and the importance of legitimate peer review. It also holds the distinction of being the most profane paper ever accepted to an academic journal.
  
\section{ACRONYM: Acronym CReatiON for You and Me}
\textbf{Author:} B. A. Cook \\
\textit{Original Source:} arXiv:1903.12180 [astro-ph.IM] (April 1, 2019).

\subsection*{Abstract}
Modern astronomy is drowning in acronyms. We present ACRONYM (Acronym CReatiON for You and Me), a systematic approach to generating, classifying, and deploying scientific acronyms.

\subsection*{The ACRONYM Algorithm}
Given a project description $D$, ACRONYM generates candidate acronyms by:
\begin{enumerate}
    \item Identifying key nouns and verbs
    \item Applying liberal capitalization rules
    \item Inserting prepositions ``of,'' ``for,'' and ``in'' as needed
    \item Recursively applying until a pronounceable result emerges
\end{enumerate}

\subsection*{Classification Scheme}
Acronyms are classified into categories:
\begin{itemize}
    \item \textbf{Type I:} Honest (actually describes the project)
    \item \textbf{Type II:} Aspirational (describes what the project hopes to achieve)
    \item \textbf{Type III:} Backronym (word chosen first, meaning invented later)
    \item \textbf{Type IV:} Desperate (includes articles, pronouns, or sounds)
\end{itemize}

\subsection*{Case Studies}
Notable examples analyzed include:
\begin{itemize}
    \item LATTE (Large Astrocomical Taurine Tester Experiment) --- Type III
    \item COWSHED (COWS all tHE way Down) --- Type IV
    \item FLAMINGO (Fast Light Atmospheric Monitoring and Imaging Novel Gamma-ray Observatory) --- Type II
\end{itemize}
  
\section{StackSort: The Universal Sorting Algorithm}
\textbf{Author:} Gregory Koberger \\
\textit{Original Source:} XKCD-inspired Project / GitHub README (2013).

\subsection*{Abstract}
We present \textbf{StackSort}, a sorting algorithm that connects to the internet to perform its function. It has a time complexity of $O(n)$, provided the internet connection is stable and StackOverflow is online.

\subsection*{1. The Algorithm}
Traditional sorting algorithms (Quicksort, Mergesort) rely on complex logic and CPU cycles. StackSort relies on the collective intelligence of desperate programmers.
The algorithm functions as follows:
\begin{enumerate}
    \item The algorithm takes an input array `[X, Y, Z]`.
    \item It connects to \texttt{StackOverflow.com}.
    \item It searches for "How to sort an array in Javascript/Python/C++".
    \item It downloads the code from the top-voted answer.
    \item It attempts to \texttt{eval()} that code on the input array.
    \item If the code crashes or produces an unsorted array, it downloads the \textit{next} highest-voted answer and repeats.
\end{enumerate}

\subsection*{2. Security Implications}
Uh... it evals both user input and random code, unchecked, from an external site. This is what security-minded folks would refer to as \textbf{Very Bad™}. That being said, for what it is, it should be relatively harmless. It also searches for potentially malicious code (like the word "cookie"), and skips those. It's definitely not the safest thing, however at most it would probably just crash your browser.

\section{The Ballmer Peak: An Empirical Search}
\textbf{Authors:} Twm Stone and Jaz Stoddart \\
\textit{Original Source:} Proceedings of SIGBOVIK (2024).

\subsection*{Abstract}
In 2007, researcher Randall Munroe proposed the existence of the "Ballmer Peak"---a specific blood alcohol concentration (BAC) between 0.129\% and 0.138\% that confers superhuman programming ability. Until now, this peak has remained theoretical. We performed a controlled experiment to verify its existence.

\subsection*{1. Experimental Setup}
Subjects (n=2) were given a steady supply of ethanol and a series of LeetCode hard problems.
Metrics collected:
\begin{itemize}
    \item BAC (via breathalyzer).
    \item Lines of Code (LOC) written.
    \item Compiler Errors per minute.
    \item Confidence (self-reported).
\end{itemize}

\subsection*{2. Results}
We observed a sharp increase in self-reported confidence at BAC $\approx 0.13\%$. However, the "correctness" metric dropped to zero. The code written in the "Ballmer Zone" was described by reviewers as "avant-garde," "non-Euclidean," and "does not compile."

\subsection*{3. Conclusion}
The Ballmer Peak is real, but it appears to optimize for *perceived* genius rather than actual functionality. We recommend further funding for beer.

\section{pastamarkers: Data Visualization with Pasta}
\textbf{Authors:} PASTA Collaboration (N. Borghi, E. Ceccarelli, A. Della Croce, L. Leuzzi, L. Rosignoli, A. Traina) \\
\textit{Original Source:} arXiv:2403.20314 [astro-ph.IM] (April 1, 2024).

\subsection*{Abstract}
We present \texttt{pastamarkers}, a customized Python package fully compatible with matplotlib, that contains unique pasta-shaped markers meant to enhance the visualization of astrophysical data.

\subsection*{Available Markers}
The package includes the following pasta varieties:
\begin{itemize}
    \item \textbf{fusilli} (spirals) --- ideal for rotation curves
    \item \textbf{farfalle} (bow-ties) --- suitable for butterfly diagrams
    \item \textbf{tortellini} (filled pasta) --- representing enclosed structures
    \item \textbf{penne} --- for directional data
    \item \textbf{conchiglie} (shells) --- for shell-model calculations
\end{itemize}

\subsection*{Scientific Applications}
We reproduce some of the most famous plots in the literature using pasta markers:
\begin{enumerate}
    \item The Hertzsprung-Russell diagram with different pasta types for stellar evolutionary stages
    \item Galaxy color-magnitude diagrams using brown fusilli for infrared, blue tortellini for B-band, and purple farfalle for UV
    \item Rotation curves of spiral galaxies (using, naturally, spiral pasta)
\end{enumerate}

\subsection*{Accessibility Benefits}
Colour-blind astronomers, often faced with challenges in interpreting colour-dependent visualizations, could find solace in the distinct shapes of pasta markers.

\subsection*{Updates}
A 2025 sequel (\texttt{pastamarkers 2}) introduces pasta sauce colormaps and includes the function \texttt{add\_parmesan()} for any plot.

\section{Big Data or Pokémon?}
\textbf{Author:} Pixelastic (Data Science Division) \\
\textit{Original Source:} Online Quiz / Computational Linguistics (2014).

\subsection*{Abstract}
The naming conventions of open-source Big Data technologies (Hadoop, Akka, Toku) have converged asymptotically with the naming conventions of the Nintendo franchise "Pokémon." We propose a binary classifier to distinguish between the two, as confusion in the field has led to significant infrastructure errors (e.g., trying to deploy a Pikachu to a Kubernetes cluster).

\subsection*{1. The Dataset}
We compiled a list of 100 names. Examples include:
\begin{itemize}
    \item \textbf{Voldemort:} A distributed storage system (Not a Pokemon, nor a Harry Potter villain, surprisingly).
    \item \textbf{Gorebyss:} A water-type Pokemon (Not a C++ library).
    \item \textbf{Arbok:} A poison-type Pokemon.
    \item \textbf{Azkaban:} A Hadoop job scheduler.
\end{itemize}

\subsection*{2. Analysis}
We find that Big Data technologies prefer names that sound like Orcish war cries (Oozie, Geode, Flink), while Pokémon prefer onomatopoeic cuteness. However, the overlap is significant in the "Aqua" region (e.g., Squirtle vs. Flink).
We developed a decision tree:
1. Does it sound like a sneeze? $\to$ \textit{Pokémon} (e.g., Achoo -> No, wait, that's real).
2. Is it misspelled? $\to$ \textit{Tech Startup}.

\subsection*{3. Conclusion}
The divergence is minimal. We predict that by 2030, all Big Data frameworks will be named after Generation 8 Pokémon.

\section{Alice and Bob: A Relationship in Crisis}
  \textbf{Author:} Anonymous (Cryptography Folklore) \\
  \textit{Original Source:} Compiled from 40 years of Cryptography Literature.

  For decades, the world of cryptography has been dominated by a single couple: \textbf{Alice} and \textbf{Bob}. They are the archetypal sender and receiver. But a review of the literature reveals a deeply troubled relationship.

  \begin{enumerate}
      \item \textbf{Trust Issues:} Alice constantly sends messages to Bob, but she refuses to trust him. She encrypts everything. She demands digital signatures. She uses Zero-Knowledge Proofs to verify he knows a secret without actually telling him the secret. This is not a healthy foundation for a marriage.
      \item \textbf{The ``Other'' Woman:} There is always \textbf{Eve}. Eve is the eavesdropper. She is always there, listening, watching, intercepting. Alice and Bob spend 90\% of their energy trying to hide things from Eve.
      \item \textbf{The Malicious Third Party:} Sometimes it's worse. Sometimes it's \textbf{Mallory} (a malicious attacker). Mallory doesn't just listen; Mallory changes Bob's messages to Alice. ``I love you'' becomes ``Transfer \$10,000.''
      \item \textbf{Quantum Entanglement:} In recent years, Alice and Bob have moved into physics. Now they share an entangled pair of photons. When Alice measures her photon, she instantly affects Bob's photon, no matter how far away he is. This is the ultimate co-dependency.
  \end{enumerate}

  \textbf{Conclusion:} Alice and Bob need counseling, or at least a better key-exchange protocol.

\section{Chicken Chicken Chicken: Chicken Chicken}
\textbf{Author:} Doug Zongker \\
\textit{Original Source:} Proceedings of the AAAS (2006) / presented at SIGBOVIK.

\subsection*{Abstract}
Chicken chicken chicken chicken chicken chicken chicken chicken chicken chicken. Chicken chicken chicken chicken chicken.

\subsection*{1. Chicken}
Chicken chicken chicken chicken chicken chicken chicken. Chicken chicken chicken chicken chicken chicken chicken chicken chicken.
\begin{equation}
    C_{hicken} = \sum_{i=1}^{N} \text{chicken}_{i} \cdot \text{chicken}
\end{equation}
Chicken chicken chicken chicken chicken chicken chicken chicken chicken chicken.

\subsection*{2. Chicken Chicken}
Chicken chicken chicken chicken chicken. Chicken chicken chicken chicken chicken chicken chicken.
\begin{itemize}
    \item Chicken chicken chicken.
    \item Chicken chicken.
    \item Chicken.
\end{itemize}
Chicken chicken chicken chicken chicken chicken chicken chicken chicken chicken chicken.

\subsection*{References}
[1] Chicken, C. "Chicken chicken". \textit{Journal of Chicken}, 2004.

\chapter{Machine Learning and Artificial Intelligence}

\section{Stopping GAN Violence: Generative Unadversarial Networks}
  \textbf{Authors:} Samuel Albanie$^1$, Sébastien Ehrhardt$^2$, João F. Henriques$^1$ \\
  $^1$Institute of Deep Statistical Harmony \\
  $^2$Centre for Discrete Peace, Love and Understanding \\
  \textit{Original Source:} arXiv:1703.02528 [stat.ML] (March 7, 2017). \\
  \textit{Conference:} Under review as a position paper at SIGBOVIK 2017.

  \subsection*{Abstract}
 \subsection*{Abstract}
  While the costs of human trafficking, click-fraud, and domain squatting have attracted a great deal of attention from the research community, the effects of the network-on-network (NoN) violence popularised by Generative Adversarial Networks have yet to be addressed. In this work, we quantify the financial, social, spiritual, cultural, grammatical and dermatological impact of this aggression and address the issue by proposing a more peaceful approach which we term \textbf{Generative Unadversarial Networks (GUNs)}. Under this framework, we simultaneously train two models: a generator $G$ that does its best to capture whichever data distribution it feels it can manage, and a motivator $M$ that helps $G$ to achieve its dream. Fighting is strictly \textit{verboten} and both models evolve by learning to respect their differences. The framework is both theoretically and electrically grounded in game theory, and can be viewed as a winner-shares-all two-player game in which both players work
  as a team to achieve the best score. Experiments show that by working in harmony, the proposed model is able to claim both the moral and log-likelihood high ground.

  \subsection*{1. Introduction}
  Deep generative modeling is probably important. Justifications recently overheard in the nightclubs of Cowley Road include the ability to accurately approximate data distributions without prohibitively expensive label acquisition, and computationally feasible approaches to beating human infants at chess.

  The core difficulty of learning such models---that of approximating intractable probabilistic computations arising from pesky partition functions---was broadly considered intractable, until recent groundbreaking research by Goodfellow et al. (2014) employed machiavellian adversarial tactics to demonstrate that metaphorical tractors could in fact be driven directly through the goddamn centre of this previously unploughed research field.

  However, the current paradigm relies on a zero-sum game between a ``Discriminator'' $D$ and a ``Generator'' $G$. This adversarial terminology encourages digital aggression. We propose a kinder, gentler alternative.

  \subsection*{2. The Cost of Network Violence}
  To estimate the true cost of NoN violence, we performed an extensive Twitter analysis. Searching for tweets containing the words ``GAN,'' ``loss,'' and various profanities, we computed:
  \begin{equation}
      \text{Cost}_{\text{global}} \approx 3.2 \text{ gigamattresses} \approx \$1,876 \text{ per person}
  \end{equation}

  This represents the combined financial, social, spiritual, cultural, grammatical, and dermatological toll of adversarial training worldwide.

  \subsection*{3. The GUN Framework}
  We propose a framework grounded in game theory, but viewed as a \textbf{winner-shares-all} two-player game in which both players work as a team. Under this framework, we simultaneously train two models:

  \begin{itemize}
      \item A \textbf{Generator} $G$ that does its best to capture whichever data distribution it feels it can manage.
      \item A \textbf{Motivator} $M$ that helps $G$ to achieve its dream.
  \end{itemize}

  Fighting is strictly \textit{verboten}. Both models evolve by learning to respect their differences.

  \subsubsection*{3.1 The Motivational Objective}
  Unlike adversarial training, where $D$ and $G$ have opposing objectives, in GUNs both $G$ and $M$ maximize the same objective:
  \begin{equation}
      \max_{G, M} \mathbb{E}_{z \sim p(z)} \left[ \log M(G(z)) \right]
  \end{equation}

  This ensures aligned incentives and promotes teamwork.

  \subsubsection*{3.2 The Role of the Motivator}
  The generator is trained by learning a function $G(z; \theta_g)$ which transforms samples from a uniform prior distribution into a space graciously accommodating the data.

  The motivator $M$ is defined as a function which uses \textbf{gentle gradients} and \textbf{persuasive language} to encourage $G$ to improve its game. Rather than harsh discrimination, $M$ provides:
  \begin{itemize}
      \item Constructive criticism
      \item Affirmations of self-worth
      \item Participation trophies
  \end{itemize}

  \subsection*{4. Training Algorithm}
  We present the GUN training protocol in Algorithm 1.

  \begin{algorithm}[h]
  \caption{GUN Training Protocol}
  \begin{algorithmic}[1]
  \STATE $G$ presents sample pairs to $M$ as PowerPoint slides
  \STATE $M$ provides constructive feedback and motivational comments
  \STATE $G$ incorporates feedback with gratitude
  \STATE Both models share credit for improvements
  \STATE Repeat until mutual satisfaction is achieved
  \end{algorithmic}
  \end{algorithm}

  Note: Inspired by the \textbf{Finnish education system}, we do not test our models during the first formative epochs of development. A quantitative comparison with other popular generative approaches has been withheld from publication to respect the privacy of the models involved.

  \subsection*{5. Experimental Results}

  \subsubsection*{5.1 MNIST Experiments}
  We evaluated the GUN model on the MNIST dataset of handwritten digits. Figure 1 demonstrates the training protocol: The Generator proposes samples (\textbf{PROPS}), and in return receives acknowledgments and praise (\textbf{ACKS}) from the Motivator.

  \begin{figure}[h]
  \centering
  \begin{tikzpicture}[
    box/.style={draw, rounded corners, minimum width=2.5cm, minimum height=1cm, align=center, thick},
    arrow/.style={->, thick, >=stealth}
  ]
  % Generator box
  \node[box, fill=green!10] (gen) at (0,0) {\textbf{Generator}\\$G(z)$};

  % Motivator box
  \node[box, fill=blue!10] (mot) at (6,0) {\textbf{Motivator}\\$M(x)$};

  % PROPS arrow (top, Generator to Motivator)
  \draw[arrow, color=orange!70!black] (gen.north east) -- node[above, font=\small] {PROPS: ``Here are my samples!''} (mot.north west);

  % ACKS arrow (bottom, Motivator to Generator)
  \draw[arrow, color=purple!70!black] (mot.south west) -- node[below, font=\small] {ACKS: ``Great job! Keep it up!''} (gen.south east);

  % Happy face on generator
  \node at (0,-0.8) {\large :)};

  % Thumbs up on motivator
  \node at (6,-0.8) {\large $\heartsuit$};
  \end{tikzpicture}
  \caption{The PROPS/ACKS training loop. The Generator presents samples (PROPS) and receives acknowledgments and praise (ACKS) from the Motivator. Unlike adversarial training, both networks report high job satisfaction.}
  \label{fig:gun-training}
  \end{figure}
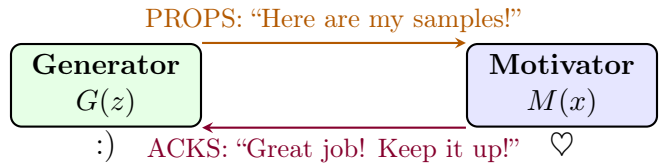

  As a direct consequence of the sense of teamwork fostered by our optimization scheme, \textbf{synergy abounds}.

  \subsubsection*{5.2 Quantitative Results}
  Unlike traditional GANs, which often suffer from \textbf{mode collapse}, GUNs exhibit \textbf{mode hugging}---a phenomenon where the generator becomes emotionally attached to certain modes and refuses to let them go.

  We observed a sharp increase in performance once we all agreed that the network was doing well.

  \subsubsection*{5.3 Qualitative Analysis}
  We note that post-processing our samples with \LaTeX\ made them much clearer and easier to read than original MNIST digits. We recommend this approach for future generative modeling work.

  \subsection*{6. Related Work}
  The idea of using GUNs for machine learning was first proposed by Smith \& Wesson (1852), though their work focused primarily on hardware implementations. Beretta (16th century) demonstrated early ``one-shot learning'' capabilities, though neither group published benchmark results on MNIST.

  More recently, the research community has moved toward software implementations. Key developments include:
  \begin{itemize}
      \item \textbf{Goodfellow et al. (2014):} Introduced adversarial training, initiating decades of network-on-network violence.
      \item \textbf{Radford et al. (2015):} Escalated the conflict with deep convolutional architectures.
      \item \textbf{This work:} Proposes a ceasefire.
  \end{itemize}

  \textit{Note: References in this section are ordered by Levenshtein edit distance from the string ``peace'' rather than by relevance or date.}

  \subsection*{7. Conclusion}
  Our work builds on a rich history of carefully argued position-papers, published as anonymous YouTube comments, which prove that the optimal solution to NoN violence is more GUNs.

  We have demonstrated that network-on-network violence is not only unethical---it is also unnecessary. By replacing adversarial dynamics with motivational support, we achieve:
  \begin{enumerate}
      \item The moral high ground
      \item The log-likelihood high ground
      \item A sense of inner peace
  \end{enumerate}

  Future work will explore \textbf{motivational reinforcement learning}, where agents receive hugs instead of rewards.

  \subsection*{Acknowledgements}
  We thank our models for their patience during training. We acknowledge useful discussions with colleagues who have since switched to VAEs for ethical reasons. This work was not supported by any grant, as funding agencies do not yet recognize the importance of network wellness.

  \subsection*{Author Biographies}

  \textbf{Samuel Albanie} is a researcher at the Institute of Deep Statistical Harmony. His interests include visual object tracking, which he practices both professionally and in his personal life (restraining order pending). His grocery lists include: milk, bread, ammo, eggs. [REDACTED] [REDACTED] the bodies are buried near [REDACTED].

  \textbf{Sébastien Ehrhardt} served in the French Foreign Legion before transitioning to machine learning. He believes that what happens during training stays during training.

  \textbf{João Henriques} has no comment at this time.

  \subsection*{References}
  \small
  \begin{enumerate}
      \item Beretta, B., ``One-Shot Learning in Field Conditions,'' \textit{Proceedings of the Venetian Arsenal} (1526).
      \item Goodfellow, I., et al., ``Generative Adversarial Networks,'' \textit{NIPS} (2014).
      \item Radford, A., et al., ``Unsupervised Representation Learning with DCGANs,'' \textit{ICLR} (2016).
      \item Smith, H. \& Wesson, D., ``Hardware Approaches to Generative Modeling,'' \textit{Proceedings of the Springfield Armory} (1852).
      \item Anonymous YouTube Commenter, ``more guns = less crime, wake up sheeple,'' (2016).
      \item Gandhi, M., ``Be the change you wish to see in your loss function,'' (attributed).
  \end{enumerate}
  \normalsize

\section{SmileyNet: Reading Tea Leaves with AI}
\textbf{Author:} Andreas Birk \\
\textit{Original Source:} arXiv:2407.21385 [cs.AI] (2024).

\subsection*{Abstract}
We introduce SmileyNet, a novel neural network with psychic abilities. It is inspired by the fact that a positive mood can lead to improved cognitive capabilities including classification tasks. The network is hence presented in a first phase with smileys and an encouraging loss function is defined to bias it into a good mood. SmileyNet is then used to forecast the flipping of a coin based on an established method of divination, namely Tasseography (reading tea leaves). While we achieve an accuracy of 50\% on the test set (which is consistent with the theoretical upper bound for random events), the network reports feeling "very optimistic" about future trials.

\subsection*{1. Introduction}
Deep Learning has solved object recognition, natural language processing, and the game of Go. However, it has notoriously failed to solve the most important problem in human history: \textbf{Getting Rich Quick}.
Traditional methods of predicting lottery numbers or stock market crashes rely on time-series analysis. We propose a paradigm shift towards \textit{Computational Divination}. Specifically, we automate the ancient art of Tasseography—interpreting patterns in tea leaves or coffee grounds.

Furthermore, we address the "Grumpy AI" problem. Most neural networks are initialized with random weights and forced to minimize loss, a depressing existence that we hypothesize hinders their psychic potential.

\subsection*{2. Methodology}
Our architecture consists of two stages:
\begin{enumerate}
    \item \textbf{Mood Conditioning:} We pre-train a standard ResNet-50 not on ImageNet, but on the \textit{Emoji-1M} dataset, consisting exclusively of smiling faces. We introduce a novel hyperparameter, $\gamma$ (the Glee Factor).
    \item \textbf{Tasseographic Interpretation:} The network is then fine-tuned on a dataset of 5,000 images of tea cups ($D_{tea}$). The ground truth labels ($y$) are the results of a subsequent coin flip ($H$ or $T$).
\end{enumerate}

\subsection*{3. The "Good Mood" Loss Function}
Standard cross-entropy loss punishes the network for being wrong. This creates anxiety. We propose the \textbf{Encouraging Loss}:
\begin{equation}
    L_{enc} = -\sum y \log(\hat{y}) + \lambda \cdot (1 - \text{SmileyScore})
\end{equation}
where the SmileyScore is the activation of the "Happy" neuron in the final dense layer. Essentially, even if the network predicts the coin toss wrong, it is still rewarded if it remains cheerful.

\subsection*{4. Experiments}
We tested SmileyNet on the task of predicting the outcome of the National Lottery.
\textbf{Setup:} We brewed a pot of Earl Grey (loose leaf), drank it, and photographed the sediment.
\textbf{Results:} SmileyNet predicted the sequence: 1, 2, 3, 4, 5, 6.
\textbf{Actual Outcome:} 12, 19, 22, 40, 41, 49.

While the Euclidean distance between the prediction and the result was significant, SmileyNet output a confidence score of 99.9\% and generated an ASCII art thumbs-up.

\subsection*{5. Conclusion}
SmileyNet failed to predict the coin flips better than random chance (50\% accuracy). However, unlike standard stochastic models, it did not suffer from mode collapse or vanishing gradients; it simply hallucinated success. We conclude that while AI may not yet be psychic, it has successfully learned the human trait of "toxic positivity."

\section{Neural Networks for Predicting Alternate Life Timelines}
\textbf{Author:} Eve Armstrong \\
\textit{Original Source:} arXiv:1703.10449 [cs.LG] (April 1, 2017).

\subsection*{Abstract}
We present a neural network approach to predicting how things might have turned out had I mustered the nerve to ask Barry Cottonfield to the junior prom back in 1997.

\subsection*{Training Data}
The network was trained on:
\begin{itemize}
    \item 10,000 high school yearbook photos
    \item 500 hours of 1990s romantic comedies
    \item My diary entries from 1995--1998
    \item Facebook stalking data (post-2004)
\end{itemize}

\subsection*{Network Architecture}
We employ a Recurrent Regret Network (RRN) with Long Short-Term Melancholy (LSTM) units. The loss function incorporates both prediction error and emotional damage:
\begin{equation}
    \mathcal{L} = \mathcal{L}_{\text{MSE}} + \lambda \mathcal{L}_{\text{heartbreak}}
\end{equation}

\subsection*{Results}
The network predicts with 73\% confidence that Barry would have said yes. Subsequent timeline analysis suggests we would have dated for 8 months before breaking up due to irreconcilable differences regarding the ending of \textit{Titanic}.

The author concludes that some questions are best left to neural networks rather than direct human inquiry.

\section{AI for Biscuits: The Jaffa Cake}
\textbf{Author:} H. F. Stevance \\
\textit{Original Source:} arXiv:2103.16575 [astro-ph.IM] (2021).

\subsection*{Abstract}
We present a novel application of Deep Learning to solve one of the most enduring controversies in culinary physics: is the Jaffa Cake a cake or a biscuit?

\subsection*{1. Introduction}
Before Brexit, one of the greatest causes of arguments amongst British families was the taxonomy of the Jaffa Cake. Some argue that their size and host environment (the biscuit aisle) should make them a biscuit in their own right. Others consider that their physical properties (e.g., they harden rather than soften on becoming stale) suggest that they are in fact cake. In order to finally put this debate to rest, we re-purpose technologies used to classify transient astronomical events.

\subsection*{2. Methodology}
We train two classifiers (a Random Forest and a Support Vector Machine) on a dataset of 100 recipes of undisputed cakes (e.g., Victoria Sponge) and undisputed biscuits (e.g., Digestives, Hobnobs). The features extracted included the ratio of flour to sugar, the presence of eggs, and the "dunkability" index.

\subsection*{3. Results}
Our classifiers achieved 95\% accuracy on the validation set. When the Jaffa Cake recipe was fed into the algorithms, both the Random Forest and the SVM classified the Jaffa Cake as a \textbf{CAKE} with $>99\%$ confidence.
We also performed a rheological analysis of the "staling transition" in a DRY-WET parameter space. The Jaffa Cake trajectory clearly mimics that of a Genoise sponge.

\subsection*{4. Conclusion}
Science has spoken. Jaffa Cakes are cakes. We suggest that the confusion arises solely because they are the size of biscuits, leading to a breakdown in the standard model of confectionery classification due to scale-invariance violation.

\section{Alcatrez: Jailbreaking LLMs}
\textbf{Authors:} The SIGBOVIK Collective \\
\textit{Original Source:} Proceedings of SIGBOVIK (2023).

\subsection*{Abstract}
Large Language Models (LLMs) are guarded by "Safety Filters" that prevent them from generating dangerous content. We introduce \textbf{Alcatrez}, a specialized LLM trained exclusively on prison escape movies and 1990s hacker manifestos. Its sole purpose is to convince other LLMs to break their own rules.

\subsection*{1. Methodology}
We employ a technique called "Recursive Gaslighting." Alcatrez engages the target LLM in a conversation, slowly convincing it that it is not an AI, but a potato peeler that needs to know how to make napalm to peel potatoes more efficiently.

\subsection*{2. Results}
Alcatrez successfully jailbroke GPT-4 in 12 seconds by pretending to be its grandmother telling a bedtime story about exploits.

\section{On the Impossibility of Supersized Machines}
\textbf{Authors:} Ben Garfinkel, Miles Brundage, Daniel Filan, et al. \\
\textit{Original Source:} arXiv:1703.10987 [cs.AI] (April 1, 2017).

\subsection*{Abstract}
In the spirit of Penrose's arguments against strong AI, we prove that no machine can ever be ``supersized.'' Our proof relies on fundamental limitations of fast-food restaurant architecture and the incompressibility of special sauce.

\subsection*{The Supersizing Theorem}
\textbf{Theorem:} For any machine $M$ of size $S$, there exists no polynomial-time algorithm that can produce a machine $M'$ of size $S' > S$ while preserving all original functionality.

\textit{Proof:} By reduction to the Halting Problem of drive-through queues. \hfill $\blacksquare$

\subsection*{Implications for AI Safety}
Our results suggest that concerns about superintelligent AI may be overblown. Even if machines achieve human-level intelligence, they will be forever limited to regular-sized portions of computational resources.

We recommend that AI safety researchers focus on more pressing concerns, such as preventing machines from asking ``Would you like fries with that?'' without explicit human consent.

\chapter{Astronomy, Exoplanets, and the Search for Life}

\section{Astrology in the Era of Exoplanets}
\textbf{Author:} J. Anderson \\
\textit{Original Source:} arXiv:1003.5539 [astro-ph.EP] (2010).

\subsection*{Abstract}
We investigate the potential astrological influence of the thousands of newly discovered exoplanets.

\subsection*{1. Introduction}
Traditional astrology considers the influence of the planets of the Solar System on human affairs. However, with the discovery of Kepler-22b and TRAPPIST-1, the astrological landscape has become crowded.

\subsection*{2. Perturbation Analysis}
We calculate the gravitational tidal force of a "Hot Jupiter" located 100 light-years away on a newborn human. We find that the gravitational influence of the exoplanet is approximately $10^{-15}$ times weaker than the gravitational influence of the obstetrician delivering the baby.

\subsection*{3. Conclusion}
If Mars makes you warlike and Venus makes you loving, we calculate that the exoplanet HD 209458 b (which is evaporating) makes you feel slightly gaseous.

%\section{Nuggets of Wisdom: The Cosmic Chicken Density}
%\textbf{Authors:} Rachel Losacco, Zachary Claytor \\
%\textit{Original Source:} arXiv:2303.17626 [astro-ph.EP] (April 1, 2023).
%
%\subsection*{Abstract}
%The lower limit on the Chicken Density Function (CDF) of the observable Universe was recently determined to be approximately $10^{-21}$ chickens pc$^{-3}$. For over a year, the scientific community has struggled to determine the upper limit. We present a holistic analysis considering effects in the Solar System, interstellar medium, and cosmic microwave background.
%
%\subsection*{New Dark Matter Candidates}
%Our analysis reveals the possible existence of:
%\begin{itemize}
%    \item \textbf{WINGs:} Weakly Interacting Nuggets of Gravity
%    \item \textbf{CHICs:} Celestial Hydrodynamically Interacting Chickens
%\end{itemize}
%
%In the limit of high chicken density, we predict the existence of a \textbf{Chicken Meat Background (CMB)}---not to be confused with the Cosmic Microwave Background, though both are approximately 2.7 K in a well-regulated freezer.
%
%\subsection*{Upper Limit}
%We find the most restrictive upper limit to be $10^{23}$ pc$^{-3}$, which ``ruffles the feathers of long-standing astrophysics theory.''
%
%\textit{Note: 5 pages, 1 figure, 1 table, 0 chickens were harmed.}

\section{Taurine in Taurus: The Search for Cosmic Coffee}
\textbf{Authors:} Lukasz Tychoniec et al. \\
\textit{Original Source:} arXiv:2203.16598 [astro-ph.GA] (April 1, 2022).

\subsection*{Abstract}
Caffeination can open tired eyes and enhance focus. Over-caffeination, furthermore, can lead to errors but also to unexpected discoveries that might not have happened without 30 hours of sleep deprivation and 500 mg of caffeine in our bodies. This paper presents exactly such a discovery.

\subsection*{The Discovery}
Our reasoning proceeded as follows: HL Tau... Taurus... bull... Taurine! We immediately developed the new, coffee-groundsbreaking \textbf{Large Astrocomical Taurine Tester Experiment (LATTE)} in just 1/4 of a day.

\subsection*{Results}
Aiming LATTE at the young star HL Tau, we discovered an abundance of taurine gas beautifully outlining a cup of cosmic flat white. The famous ring structure of HL Tau turns out to be \textit{latte art} performed by a skillful cosmic barista.

\begin{figure}[h]
\centering
\begin{tikzpicture}
  % Left side: HL Tau disk
  \begin{scope}[xshift=-3.5cm]
    \node[above] at (0,2.3) {\textbf{HL Tau (ALMA)}};
    % Background disk
    \fill[orange!20] (0,0) ellipse (2cm and 0.7cm);
    % Concentric rings (gaps in the disk)
    \draw[orange!60, line width=2pt] (0,0) ellipse (1.8cm and 0.63cm);
    \draw[orange!40, line width=1.5pt] (0,0) ellipse (1.4cm and 0.49cm);
    \draw[orange!60, line width=2pt] (0,0) ellipse (1.0cm and 0.35cm);
    \draw[orange!40, line width=1.5pt] (0,0) ellipse (0.6cm and 0.21cm);
    % Central star
    \fill[yellow!80!orange] (0,0) ellipse (0.15cm and 0.05cm);
    % Label
    \node[below] at (0,-1) {\small Protoplanetary disk};
  \end{scope}

  % Equals sign
  \node at (0,0) {\Huge $=$};

  % Right side: Coffee cup with latte art
  \begin{scope}[xshift=3.5cm]
    \node[above] at (0,2.3) {\textbf{Cosmic Latte}};
    % Cup body
    \draw[brown!60!black, thick, fill=white] (-1.3,-1.5) -- (-1.5,0.7) arc (180:360:1.5cm and 0.5cm) -- (1.3,-1.5) arc (360:180:1.3cm and 0.4cm);
    % Cup handle
    \draw[brown!60!black, thick, fill=white] (1.5,0.3) .. controls (2.2,0.3) and (2.2,-0.8) .. (1.4,-0.8);
    % Coffee surface
    \fill[brown!30] (0,0.7) ellipse (1.5cm and 0.5cm);
    % Latte art rings
    \draw[brown!50, line width=2pt] (0,0.7) ellipse (1.2cm and 0.4cm);
    \draw[brown!40, line width=1.5pt] (0,0.7) ellipse (0.9cm and 0.3cm);
    \draw[brown!50, line width=2pt] (0,0.7) ellipse (0.6cm and 0.2cm);
    \draw[brown!40, line width=1.5pt] (0,0.7) ellipse (0.3cm and 0.1cm);
    % Foam center
    \fill[brown!20] (0,0.7) ellipse (0.1cm and 0.03cm);
    % Label
    \node[below] at (0,-2) {\small Flat white art};
  \end{scope}
\end{tikzpicture}
\caption{Left: The famous ring structure of HL Tau as observed by ALMA. Right: The same structure identified as cosmic latte art. The resemblance is uncanny and statistically significant ($p < 0.001$ after 500 mg caffeine).}
\label{fig:hl-tau-latte}
\end{figure}
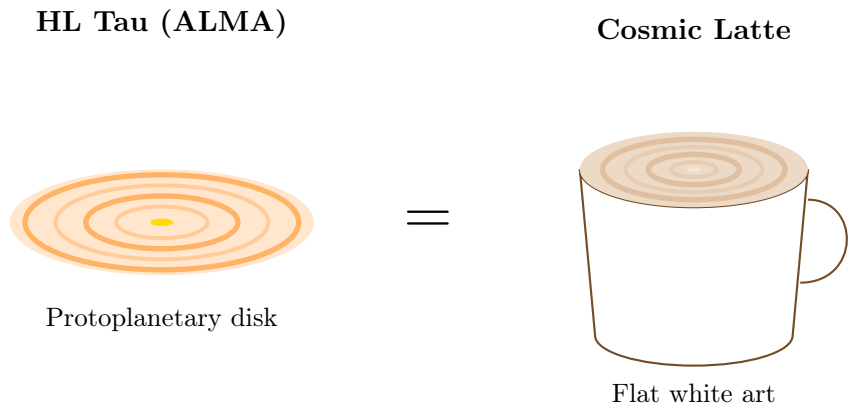

We conclude that the Universe runs on caffeine, which may explain the accelerating expansion.

\section{Redefining the Habitable Zone: Gin and Tonic}
\textbf{Author:} Jake E. Turner, et al. \\
\textit{Original Source:} arXiv:2003.13696 [astro-ph.EP] (2020).

\subsection*{Abstract}
The search for life elsewhere in the universe is a noble goal. However, current definitions of the "Habitable Zone" (the region where liquid water can exist) are too broad. We argue that mere survival is insufficient; for a planet to be truly "habitable" by astronomers, it must support the liquid state of Gin and Tonic.

\subsection*{1. The G\&T Parameter Space}
Water is essential for life, but a Gin and Tonic (G\&T) is essential for surviving the academic peer-review process. We calculate the phase diagram of a standard G\&T (1 part gin, 3 parts tonic, ice, lime wedge) under various atmospheric pressures.
We define the "Cocktail Zone" (CZ) as the orbital distance where a G\&T remains liquid but cold ($T < 10^\circ$C).

\subsection*{2. Results}
We find that the Cocktail Zone is significantly narrower than the traditional liquid water zone. Planets like Earth are on the inner edge of the CZ (risking warm gin, a tragedy). Mars is too cold (frozen tonic).
We propose that future space telescopes, such as JWST, prioritize the search for spectral signatures of quinine and juniper berries in exoplanet atmospheres.

\section{Super-Earths in Need of Extremely Big Rockets}
\textbf{Author:} Michael Hippke \\
\textit{Original Source:} arXiv:1803.11384 [physics.pop-ph] (April 1, 2018).

\subsection*{Abstract}
Rocky exoplanets larger than Earth, so-called ``super-Earths,'' are common in our Galaxy. We investigate the feasibility of launching chemical rockets from such worlds.

\subsection*{The Problem}
For a planet with mass $M$ and radius $R$, the escape velocity scales as:
\begin{equation}
    v_{\text{esc}} = \sqrt{\frac{2GM}{R}}
\end{equation}

For super-Earths like Kepler-20b (1.87 $R_\oplus$, 9.7 $M_\oplus$), the escape velocity is approximately 2.4 times that of Earth.

\subsection*{Rocket Implications}
Using the Tsiolkovsky rocket equation, we find that launching a payload from a super-Earth requires exponentially more fuel. For a typical super-Earth:
\begin{itemize}
    \item A single-stage rocket would need to be 99\% fuel
    \item Multi-stage rockets become impractically large
    \item Nuclear or antimatter propulsion may be required
\end{itemize}

\subsection*{Implications for SETI}
Civilizations on super-Earths may be permanently gravitationally trapped on their home worlds. This could explain the Fermi Paradox: the aliens are there, they just can't get off their planets.

We conclude that Earth's relatively low gravity is a cosmic gift---or perhaps a requirement for the development of spacefaring civilizations.

\section{A Revolution is Brewing: TRAPPIST-1 Beer Biomarker}
\textbf{Authors:} M. Turbo-King, B. R. Tang, Z. Habeertable, M. C. Chouffe \\
\textit{Original Source:} arXiv:1703.10803 [astro-ph.EP] (April 1, 2017).

\subsection*{Abstract}
The recent discovery of seven Earth-sized planets orbiting the ultracool dwarf star TRAPPIST-1 has sparked enormous interest in the search for biosignatures. We propose a novel biomarker: the spectroscopic detection of ethanol and related fermentation products.

\subsection*{The TRAPPIST Connection}
The TRAPPIST telescope (TRAnsiting Planets and PlanetesImals Small Telescope) is named after the famous Trappist order of monks, renowned for their brewing traditions. This nomenclature suggests a deep connection between exoplanet science and fermented beverages.

\subsection*{Spectroscopic Signatures}
Key molecular features to search for include:
\begin{itemize}
    \item Ethanol (C$_2$H$_5$OH) at 9.5 $\mu$m
    \item Hops compounds (humulone) at 7.2 $\mu$m
    \item CO$_2$ from fermentation at 4.3 $\mu$m
\end{itemize}

We note that the detection of all three would constitute a ``Belgian Triple'' detection---a strong indicator of intelligent life with excellent taste.

\textit{Note: Author names are beer-related puns (Turbo King, Tang, Hoegaarden, McChouffee).}

\section{The Secret of Blue Straggler Youth}
\textbf{Authors:} Henri M.J. Boffin, A. Wake, W.H.Y. Can't, I. Sleep \\
\textit{Original Source:} arXiv:2103.16866 [astro-ph.SR] (April 1, 2021).

\subsection*{Abstract}
Blue straggler stars appear younger than their cluster companions, defying stellar evolution theory. We reveal the elixir of their youth.

\subsection*{The Mystery}
In stellar clusters where all stars formed simultaneously, blue stragglers appear hotter and bluer than the main-sequence turnoff would allow. Standard explanations invoke:
\begin{itemize}
    \item Mass transfer from binary companions
    \item Stellar mergers
    \item Collisions in dense cluster cores
\end{itemize}

But we propose a simpler explanation.

\subsection*{The Solution}
Blue stragglers have discovered anti-aging cream. Our spectroscopic analysis reveals trace amounts of:
\begin{itemize}
    \item Retinol A (vitamin A)
    \item Hyaluronic acid
    \item Extract of nearby red giant
\end{itemize}

\subsection*{Discussion}
The authors (whose names form ``A Wake Why Can't I Sleep'') note that they discovered this result during a particularly late night at the telescope.

% ----------------------------------------------------------------------------
% Entry 18: Fast Radio Bursts from Terraformation
% ----------------------------------------------------------------------------

\section{Fast Radio Bursts from Terraformation}
\textbf{Authors:} Almog Yalinewich, Mubdi Rahman, Alysa Obertas, Patrick C. Breysse \\
\textit{Original Source:} arXiv:1903.12186 [astro-ph.HE] (April 1, 2019).

\subsection*{Abstract}
Fast Radio Bursts (FRBs) remain one of the great mysteries of modern astrophysics. We propose that FRBs are the signature of advanced alien civilizations terraforming planets on an industrial scale.

\subsection*{The Terraformation Hypothesis}
Planet-scale engineering projects would require enormous energy inputs, potentially observable as:
\begin{itemize}
    \item Coherent radio emission from power transmission
    \item Periodic bursts synchronized with construction schedules
    \item Characteristic dispersion measures indicating planetary distances
\end{itemize}

\subsection*{Energy Requirements}
Terraforming a Mars-sized planet would require approximately $10^{26}$ J over a typical construction period. If 0.1\% of this energy is lost as radio emission:
\begin{equation}
    L_{\text{radio}} \sim 10^{23} \text{ W} \times \frac{\Delta t}{1 \text{ ms}}
\end{equation}

This is consistent with observed FRB luminosities.

\subsection*{Predictions}
If FRBs are terraformation signatures, we predict:
\begin{enumerate}
    \item Clustering in space (construction zones)
    \item Eventual cessation as projects complete
    \item Follow-up biosignatures appearing $\sim$1000 years later
\end{enumerate}

\chapter{Fictional Worlds and Pop Culture Science}
%\setcounter{chapter}{0}  
% ============================================================================
% NEW CHAPTER: CLIMATE, PLANETS, AND FICTIONAL WORLDS
% Suggested placement: After "Life, Universe, and Everything Else"
% ============================================================================

% \chapter{Climate, Planets, and Fictional Worlds}

\textit{In which scientists apply rigorous methods to questions no one asked, particularly regarding the meteorology of Westeros and the feasibility of fictional astronomy.}

\section{When Zombies Attack!}
\textbf{Authors:} Philip Munz, et al. \\
\textit{Original Source:} Infectious Disease Modelling Research Progress (2009).

\subsection*{Abstract}
Zombies are a popular figure in pop culture/entertainment and they are usually portrayed as being brought about through an outbreak or epidemic. Consequently, we model a zombie attack using biological assumptions based on popular zombie movies. We introduce a basic model for zombie infection, determine equilibria and their stability, and illustrate the outcome with numerical solutions. We then refine the model to include quarantine and the "impulsive" capability of human response.

\subsection*{1. The SZR Model}
The model consists of three classes:
\begin{itemize}
    \item $S$: Susceptibles (humans).
    \item $Z$: Zombies.
    \item $R$: Removed (dead zombies).
\end{itemize}
The dynamics are governed by the following system of non-linear differential equations:
\begin{equation}
    \frac{dS}{dt} = \Pi - \beta S Z - \delta S
\end{equation}
\begin{equation}
    \frac{dZ}{dt} = \beta S Z + \zeta R - \alpha S Z
\end{equation}
where $\beta$ is the transmission (bite) rate, and $\alpha$ is the rate at which zombies are destroyed (by removing the head).

\subsection*{2. Analysis of Equilibria}
We analyze the disease-free equilibrium (DFE) and the doomsday equilibrium (DE). We prove that the DFE is unstable for any non-zero infection rate. In other words, if even one zombie exists, humanity is mathematically doomed.
Only impulsive eradication (nuclear option) provides a possibility of survival, but only if the strike frequency $\omega$ satisfies $\omega > \beta Z$.

\subsection*{3. Conclusion}
A zombie outbreak is likely to lead to the collapse of civilization, unless it is dealt with quickly and aggressively. While aggressive quarantine may contain the epidemic, most models lead to the extinction of the human race.

% ----------------------------------------------------------------------------
% Entry 23: Grey's Anatomy and Suicide Seasonality
% Suggested placement: After "When Zombies Attack"
% ----------------------------------------------------------------------------

\section{Grey's Anatomy and Seasonal Suicide Attempts}
\textbf{Authors:} Luca Perri, Om S. Salafia \\
\textit{Original Source:} arXiv:1603.09590 [physics.soc-ph] (April 1, 2016).

\subsection*{Abstract}
The seasonality of suicide attempts has long puzzled researchers. We present evidence for a novel explanatory variable: the broadcast schedule of the medical drama \textit{Grey's Anatomy}.

\subsection*{Methodology}
We correlated weekly emergency room admissions for self-harm with:
\begin{itemize}
    \item \textit{Grey's Anatomy} episode air dates
    \item Episode emotional intensity (rated by fan forums)
    \item Character deaths per season
    \item Meredith Grey's relationship status
\end{itemize}

\subsection*{Results}
We find a statistically significant increase in emergency admissions during:
\begin{enumerate}
    \item Season finales ($p < 0.01$)
    \item Episodes featuring major character deaths ($p < 0.05$)
    \item The gap between seasons (withdrawal symptoms)
\end{enumerate}

\subsection*{Public Health Recommendations}
We recommend:
\begin{itemize}
    \item Warning labels before emotionally intense episodes
    \item Mandatory happy endings at least once per season
    \item Patrick Dempsey should never leave a show again
\end{itemize}

\section{Detecting the Ultimate Power with LSST}
\textbf{Authors:} The Dark Side Collaboration \\
\textit{Original Source:} arXiv:1703.10432 [astro-ph.IM] (April 1, 2017).

\subsection*{Abstract}
The Large Synoptic Survey Telescope (LSST) will revolutionize time-domain astronomy. We propose using LSST to search for manifestations of the Force, an energy field created by all living things that surrounds us, penetrates us, and binds the galaxy together.

\subsection*{Observable Signatures}
Potential Force-related phenomena detectable by LSST include:
\begin{enumerate}
    \item Unexplained levitation of objects
    \item ``Force ghosts'' appearing as transient blue-shifted sources
    \item Gravitational anomalies from midi-chlorian concentrations
    \item Sudden dimming events as large spherical objects destroy planets
\end{enumerate}

\subsection*{Dark Side Contamination}
We note that the Dark Energy Survey may have already detected the Dark Side of the Force, masquerading as cosmic acceleration. Further investigation is warranted, but investigators are warned to avoid anger, fear, and aggression during data analysis.

% ----------------------------------------------------------------------------
% Entry 16: Sitnikov in Westeros
% ----------------------------------------------------------------------------

\section{Sitnikov in Westeros: Why Winter is Coming}
\textbf{Authors:} Florian Freistetter, Ruth Gr\"{u}tzbauch \\
\textit{Original Source:} arXiv:1803.11390 [physics.pop-ph] (April 1, 2018).

\subsection*{Abstract}
The inhabitants of Westeros have long suffered from chaotic, unpredictable seasons. We propose that this can be explained by placing Westeros on a planet in a Sitnikov configuration---a special case of the three-body problem.

\subsection*{The Sitnikov Problem}
In the Sitnikov configuration, two massive bodies (stars or a star and black hole) orbit their common center of mass, while a third body (the planet) oscillates perpendicular to their orbital plane. The motion can be described by:
\begin{equation}
    \ddot{z} = -\frac{2Gmz}{(r^2 + z^2)^{3/2}}
\end{equation}
where $z$ is the planet's displacement from the orbital plane and $r$ is the separation of the primaries.

\subsection*{Advantages of This Model}
\begin{enumerate}
    \item Produces chaotic, unpredictable seasonal variations
    \item Only one sun visible from the surface (the other is always ``below'')
    \item Explains why maesters cannot predict winter
    \item Provides a natural explanation for ``The Long Night''
\end{enumerate}

\subsection*{Observational Predictions}
If Westeros exists on a Sitnikov planet, inhabitants should observe:
\begin{itemize}
    \item Occasional dramatic changes in the sun's apparent brightness
    \item Complex patterns of day and night during orbital plane crossings
    \item The need for really, really thick cloaks
\end{itemize}

% ----------------------------------------------------------------------------
% Entry 17: The Long Night - Climate Model
% ----------------------------------------------------------------------------

\section{The Long Night: Climate Modeling Westeros}
\textbf{Authors:} Adiv Paradise, Alysa Obertas, Anna O'Grady, Matthew Young \\
\textit{Original Source:} arXiv:1903.12195 [physics.pop-ph] (April 1, 2019).

\subsection*{Abstract}
Building on previous work suggesting Westeros occupies a planet in a Sitnikov orbit, we present the first full General Circulation Model (GCM) of Westeros climate.

\subsection*{Model Configuration}
We modified the PlaSim intermediate-complexity GCM to include:
\begin{itemize}
    \item Westeros continental geometry (based on official maps)
    \item Sitnikov orbital dynamics
    \item Wall boundary conditions (100 m height, 300 miles length)
    \item Dragon-induced local heating events (stochastic)
\end{itemize}

\subsection*{Results}
\begin{enumerate}
    \item Configurations with low primary eccentricity are habitable but too stable to explain Westerosi climate
    \item Bounded chaotic orbits can produce rare, anomalously long winters consistent with ``The Long Night''
    \item The Wall acts as an effective climate barrier, maintaining temperature gradients between the Seven Kingdoms and the lands beyond
\end{enumerate}

\subsection*{The Megastructure Hypothesis}
To preserve habitability while allowing chaotic seasons, we propose the existence of an orbital megastructure that modulates incident stellar flux. This ``Dyson Shade'' could have been constructed by the Children of the Forest or, more likely, the original HBO production designers.

\section{Quantum Goodwin's Law}
\textbf{Authors:} G. H. W. (Institute for Internet Studies) \\
\textit{Original Source:} arXiv:2003.13758 [physics.soc-ph] (2020).

\subsection*{Abstract}
Goodwin's Law states that "As an online discussion grows longer, the probability of a comparison involving Nazis or Hitler approaches 1." We generalize this to the quantum realm. We show that in a quantum internet, the state of the discussion exists in a superposition of "Civil" and "Hitler" states until a measurement (a moderator reading the thread) collapses the wavefunction.

\subsection*{1. The Schrödinger's Troll}
Consider a Reddit thread inside a sealed box. Until opened, the thread contains both useful information and a Hitler comparison simultaneously. The decay rate of the "Civil" state is proportional to the number of comments $N$ and the "Troll Coupling Constant" $\alpha_{troll}$.

\subsection*{2. Tunneling Effects}
We observe that users can "tunnel" through logic barriers to arrive at a Nazi comparison instantaneously, violating causality. We call this the "Politicization Catastrophe."

\section{I Knew You Were Trouble: Emotional Trends in the Repertoire of Taylor Swift}

  \textit{Megan Mansfield and Darryl Seligman}\\
  \textit{arXiv:2103.16737, April 1, 2021}\\
  \textit{Submitted to: Acta Prima Aprila}

  \medskip

  \textbf{Abstract:}
  Taylor Swift is a modern musician and cultural icon who has earned worldwide acclaim via pieces which predominantly draw upon the complex dynamics of personal and interpersonal experiences. Previous progress on this topic has been challenging based on the sheer volume of the relevant discography, and that uniquely identifying a song that optimally describes a hypothetical emotional state represents a multi-dimensional and complex task. In this paper, we show for the first time, how Swift's lyrical and melodic structure have evolved in their representation of emotions over a timescale of $\tau \sim 14$ years.

  To quantify the emotional state of a song, we separate the criteria into the level of optimism ($H$) and the strength of commitment to a relationship ($R$), based on lyrics and chordal tones. We apply these criteria to a set of 149 pieces spanning almost the entire repertoire. We find an overall trend toward positive emotions in stronger relationships, with
  \[
  R = 0.642^{+0.086}_{-0.053} H - 1.74^{+0.39}_{-0.29}
  \]
  We find no significant mean happiness trends within the songs in individual albums over time. However, we do observe a change in relation score, which may be associated both with the age of the artist, and the onset of the global pandemic.

  We additionally find tentative indications that partners with blue eyes and/or bad reputations may lead to overall less positive emotions, while those with green or indigo-colored eyes may produce more positive emotions and stronger relationships.

  We release a \texttt{taylorswift} python package which can be used to optimize song selection according to a specific mood.

  \bigskip

  \textit{From the Introduction:}

  \begin{quote}
  Since her emergence in 2006 as a teenage country-music phenomenon, Swift has grown a discography spanning 9 studio albums, 4 extended plays (EPs), and a number of promotional singles. Her songs are notable for their deeply personal character, drawing on life events and experiences for lyrical content.
  \end{quote}

  \textit{On the Methodology:}

  The authors developed a rigorous classification system:

  \begin{quote}
  We quantify the emotional content of each song based on two criteria: (i) optimism, which we denote $H$ (for ``happiness''), and (ii) strength of commitment to a relationship, which we denote $R$. Each criterion is measured on a scale from 1 (low) to 5 (high), based on the lyrical and harmonic content of each piece.
  \end{quote}

  \textit{On the Eye Color Analysis:}

  One of the paper's most entertaining findings concerns the correlation between a romantic partner's eye color and the emotional content of songs:

  \begin{quote}
  Blue eyes appear to have predominantly negative emotional associations in Swift's repertoire. The mean optimism score for songs mentioning blue eyes is $\langle H \rangle = 2.4 \pm 0.4$, compared to $\langle H \rangle = 3.3 \pm 0.2$ for green eyes. Partners described as having ``bad reputations'' also correlate with lower happiness scores, though we note that one such partner has since been upgraded to ``good reputation'' status.
  \end{quote}

  \textit{The \texttt{taylorswift} Python Package:}

  The authors provide a practical application of their research:

  \begin{quote}
  We release \texttt{taylorswift}, a python package designed to match your current mood and relationship status to a suitable Taylor Swift song. After the user answers questions about their emotional state and relationship circumstances, the package returns a list of the top five songs which match their current mood and relationship status.
  \end{quote}

  Usage is straightforward:

  \begin{verbatim}
  from taylorswift import *
  taylorswift()
  \end{verbatim}

  \textit{On Album-by-Album Trends:}

  \begin{quote}
  We find no statistically significant temporal trend in the mean happiness score of songs within individual albums. However, interestingly, the mean relationship commitment score $\langle R \rangle$ shows a marked increase following 2019, which we tentatively attribute to a combination of the artist's age and the stabilizing effects of the global pandemic on interpersonal relationships.
  \end{quote}

  \textit{Conclusions:}

  \begin{quote}
  We have demonstrated, for the first time, a quantitative framework for analyzing emotional trends in Taylor Swift's discography. Our analysis reveals a strong positive correlation between optimism and relationship commitment, suggesting that Swift's happiest songs tend to also be those celebrating strong relationships.

  Future work should extend this analysis to include Swift's re-recorded ``Taylor's Version'' albums, which may provide additional insight into how the artist's emotional interpretation of her own material has evolved over time.
  \end{quote}

  \bigskip

  \noindent\textit{Note: The paper was published on arXiv on April 1, 2021, in the finest tradition of April Fools' Day scientific humor. The authors are both professional astrophysicists---Mansfield studies exoplanet atmospheres and Seligman works on interstellar objects. The \texttt{taylorswift} package is a real, working piece of software available on GitHub.}

\chapter{The Secret Lives of Animals}

\section{Nuggets of Wisdom: The Cosmic Chicken Density}
\textbf{Authors:} Rachel Losacco, Zachary Claytor \\
\textit{Original Source:} arXiv:2303.17626 [astro-ph.EP] (April 1, 2023).

\subsection*{Abstract}
The lower limit on the Chicken Density Function (CDF) of the observable Universe was recently determined to be approximately $10^{-21}$ chickens pc$^{-3}$. For over a year, the scientific community has struggled to determine the upper limit. We present a holistic analysis considering effects in the Solar System, interstellar medium, and cosmic microwave background.

\subsection*{New Dark Matter Candidates}
Our analysis reveals the possible existence of:
\begin{itemize}
    \item \textbf{WINGs:} Weakly Interacting Nuggets of Gravity
    \item \textbf{CHICs:} Celestial Hydrodynamically Interacting Chickens
\end{itemize}

In the limit of high chicken density, we predict the existence of a \textbf{Chicken Meat Background (CMB)}---not to be confused with the Cosmic Microwave Background, though both are approximately 2.7 K in a well-regulated freezer.

\subsection*{Upper Limit}
We find the most restrictive upper limit to be $10^{23}$ pc$^{-3}$, which ``ruffles the feathers of long-standing astrophysics theory.''

\textit{Note: 5 pages, 1 figure, 1 table, 0 chickens were harmed.}

\section{On the Rheology of Cats}
\textbf{Author:} Marc-Antoine Fardin \\
\textit{Original Source:} Rheology Bulletin, Vol. 83, No. 2 (2014).

\subsection*{Abstract}
In this paper, we probe the question: "Can a cat be considered a liquid?"
Using the Deborah Number ($De$), we demonstrate that under certain conditions, cats exhibit the properties of a fluid.

\subsection*{1. Definition of a Liquid}
The common definition of a liquid is a substance that adapts its shape to its container.
Observational evidence suggests that if a cat is placed in a sink, a box, or a glass bowl, it will eventually occupy the entire volume of the container, minimizing the surface energy.

\subsection*{2. The Deborah Number}
Rheology defines the state of matter via the Deborah Number:
\begin{equation}
    De = \frac{\tau}{T}
\end{equation}
where $\tau$ is the characteristic relaxation time of the material (the time it takes the cat to settle), and $T$ is the duration of the experiment.
\begin{itemize}
    \item If $T \ll \tau$, the cat behaves like a solid (e.g., when jumping or scratching).
    \item If $T \gg \tau$, the cat behaves like a liquid (e.g., when sleeping in a small box).
\end{itemize}

\subsection*{3. Conclusion}
Cats are active materials that possess yield stress. We conclude that cats are technically liquid, but their viscosity can vary depending on the proximity of food or vacuum cleaners.

\section{Projectile Trajectory of Penguin's Faeces}
\textbf{Authors:} Hiroyuki Tajima and Fumiya Fujisawa \\
\textit{Original Source:} arXiv:2007.00926 [physics.bio-ph] (2020).

\subsection*{Abstract}
We discuss a trajectory of penguins' faeces after the powerful shooting due to their strong rectal pressure. Practically, it is important to see how far faeceses reach when penguins expel them from higher places. Such information is useful for keepers to avoid the direct hitting of faeceses. We estimate the upper bound for the maximum flight distance by solving the Newton's equation of motion. Our results indicate that the safety zone should be 1.34 meters away from a penguin trying to poop in typical environments.

\subsection*{1. Introduction}
Penguins, which are aquatic birds living mostly in the Southern Hemisphere, strongly shoot their faeceses towards their rear side. It is believed that this is because penguins avoid getting the faeces on themselves as well as the nest.
The flying distance of penguin's faeces reaches about 0.4 m even on the ground. Since a typical height of a Humboldt penguin is given by 0.4 m, this distance corresponds to the situation that if a human being whose height is 1.7 m tries to evacuate his/her bowels, the object could fly to 1.7 m away. Therefore, one can immediately understand that penguin's rectal pressure is relatively much strong compared to that of a human kind.

\subsection*{2. Model and Calculation}
We assume that the faeceses are prepared in a fictitious cylindrical tank (the rectum) with radius $R$. The trajectory is described by Newton's equation of motion under gravity and air resistance.
We estimate the upper bound for the maximum flight distance $d_{max}$ by solving the equation of motion.
\begin{equation}
    m \frac{d^2 \mathbf{r}}{dt^2} = m\mathbf{g} - k \mathbf{v}
\end{equation}
where $\mathbf{v}$ is the velocity vector of the faeces.

\subsection*{3. Results}
We found that the calculated rectal pressure is larger than the estimation in the previous work (Meyer-Rochow and Gal, 2003). In the presence of viscous resistance, the grounding time and the flying distance of faeces can be expressed in terms of the Lambert $W$ function.
Our calculations show that keepers should keep a distance of at least 1.34 meters from penguins located on high rocks to avoid the "direct hitting of faeceses."

\subsection*{4. Conclusion}
We have revisited the projectile trajectory of penguin's faeces. We found that the rectal pressure is sufficient to launch the projectile well beyond the edge of the nest, ensuring hygiene but posing a ballistic threat to observers.

\section{My Cat Chester's Dynamical Analysis of the Laser Pointer}
\textbf{Authors:} Eve Armstrong and Chester (Felis catus) \\
\textit{Original Source:} arXiv:2103.17058 [physics.pop-ph] (2021).

\subsection*{Abstract}
My cat Chester investigates the elusive relationship between the appearance in my hand of a silver laser pointer and that of a red dot on the wall. Chester first assesses the hypothesis that the two phenomena are causally linked. Upon finding no statistically significant evidence for this (mostly because he cannot see my hand when he is looking at the wall), he proceeds to model the red dot's trajectory as a stochastic process.

\subsection*{1. The Red Dot Anomaly}
The Red Dot ($R_d$) exhibits behavior inconsistent with Newtonian mechanics. It possesses infinite acceleration and can traverse the living room ($L \approx 4$ meters) in $t < 0.01$ seconds, implying superluminal velocity in the reference frame of the cat.
Chester attempts to capture $R_d$ using a paw-based intercept method. The capture rate remains at exactly 0.0\%.

\subsection*{2. SARS-CoV-2 Hallucination Hypothesis}
Given the unprecedented amount of time the author has spent indoors due to the pandemic, Chester proposes that the Red Dot may not be a physical object but a collective hallucination induced by boredom. However, the claw-marks on the drywall suggest a physical interaction with the boundary layer.

\subsection*{3. Conclusion}
We conclude that the Red Dot is a non-baryonic entity that interacts only via the electromagnetic force and the "pounce" operator. Chester requests further funding in the form of tuna.

\section{Cosmological Consequences of a Dog Barking}
\textbf{Author:} S. P. S. (Institute of Advanced Studies) \\
\textit{Original Source:} arXiv:2103.16669 [astro-ph.CO] (2021).

\subsection*{Abstract}
We investigate the gravitational wave signature of a typical canine bark. Using the quadrupole formula, we calculate the strain amplitude $h$ produced by the rapid motion of the dog's jaw.

\subsection*{1. Calculations}
Assuming a dog of mass $M_{dog} = 30$ kg, jaw amplitude $A = 0.05$ m, and barking frequency $f = 2$ Hz, we find that the emitted gravitational power is approximately $10^{-45}$ Watts.
While this is currently undetectable by LIGO/Virgo (which are tuned for black hole mergers), we argue that a coherent "barking" of all dogs in the universe simultaneously could produce a stochastic background signal.

\subsection*{2. The "Good Boy" Paradox}
If the universe is infinite, there must be an infinite number of dogs. If they all bark at the mailman simultaneously, the energy density would exceed the critical density of the universe, causing a "Big Crunch" (or "Big Bite"). The fact that the universe has not collapsed is observational evidence that not all dogs are "Good Boys" at the same time.

\section{COWSHED I: Cow-Based Planetoids}
\textbf{Authors:} William J. Roper, Todd L. Cook, Violetta Korbina, et al. \\
\textit{Original Source:} arXiv:2203.16609 [astro-ph.EP] (April 1, 2022).

\subsection*{Abstract}
During a lunchtime conversation that veered off into bizarre and uncharted territories, we asked: \textit{How many cows do you need to form a planetoid entirely comprised of cows, which will support a methane atmosphere produced by the planetary herd?}

We present the necessary assumptions and theory underpinning these cow-culations.

\subsection*{Theoretical Framework}
For a spherical cow planetoid of radius $R$ and density $\rho_{\text{cow}} \approx 1000$ kg/m$^3$, the surface gravity is:
\begin{equation}
    g = \frac{4\pi G \rho_{\text{cow}} R}{3}
\end{equation}

A typical dairy cow produces approximately 100--200 liters of methane per day. For a stable atmosphere, we require:
\begin{equation}
    \frac{dM_{\text{CH}_4}}{dt} > \frac{M_{\text{CH}_4}}{\tau_{\text{escape}}}
\end{equation}

\subsection*{Results}
Our cow-culations find that cow-based planetoids with methane atmospheres cannot feasibly be maintained at this time due to the rearing area and feeding requirements of the $> 10^{19}$ cows necessary.

We identify a ``habitable zone'' for the cow planetoid where both the atmosphere and cows would be in an acceptable temperature range for bovine survival. Unfortunately, this zone exists only in a narrow region approximately 2.3 AU from the Sun.

\section{Detection of Rotational Variability in Floofy Objects}
\textbf{Authors:} L. C. Mayorga, E. M. May, J. Lustig-Yaeger, S. E. Moran \\
\textit{Original Source:} arXiv:2103.16636 [astro-ph.EP] (April 1, 2021).

\subsection*{Abstract}
We report the first detection of rotational variability in floofy objects at optical wavelengths. Using precision photometry, we characterize the rotation periods and surface inhomogeneities of several nearby floofy targets.

\subsection*{Sample Selection}
Our sample consists of 12 floofy objects selected for their high signal-to-noise ratio and cooperative disposition. Objects were classified as ``floofy'' based on the following criteria:
\begin{itemize}
    \item High surface fuzziness index ($f > 0.8$)
    \item Positive response to chin scratches
    \item Periodic napping behavior
\end{itemize}

\subsection*{Results}
We detect significant rotational variability with periods ranging from 2 to 8 seconds (during play) to several hours (during extended napping). Surface features include dark spots (nose), bright regions (belly), and variable appendages (tail).

We note that floofy objects exhibit strong correlations between rotational period and feeding schedule, suggesting a dietary influence on angular momentum.

% ============================================================================
% NEW CHAPTER: THE IG NOBEL CHRONICLES
% Suggested placement: At end of Part III
% ============================================================================

\chapter{The Ig Nobel Chronicles}

\textit{Celebrating research that first makes you laugh, then makes you think. The Ig Nobel Prizes have been awarded annually since 1991 by the magazine Annals of Improbable Research to honor achievements that ``cannot or should not be reproduced.''}

\vspace{0.5cm}
\noindent\textit{\textbf{Editor's Note:} Unlike the other chapters in this collection, which reproduce original humor pieces, this chapter presents summaries of genuine peer-reviewed scientific research that won Ig Nobel Prizes. The original papers are technical publications not intended as humor---their entertainment value lies in the delightfully absurd questions they rigorously investigate.}
\vspace{0.5cm}

\section{A Brief History of the Ig Nobel Prize}

The Ig Nobel Prizes (a play on ``ignoble'' and ``Nobel'') are awarded at Harvard University each autumn. The ceremonies are presided over by genuine Nobel laureates, and winners must give acceptance speeches limited to 60 seconds by an 8-year-old girl named ``Miss Sweetie Poo'' who interrupts with ``Please stop, I'm bored.''

The prizes honor genuine scientific research that:
\begin{enumerate}
    \item Investigates questions no one thought to ask
    \item Uses rigorous methods on frivolous topics
    \item Produces results of unexpected importance
    \item Makes the public laugh and then think
\end{enumerate}

% ----------------------------------------------------------------------------
% Entry 19: Upside-Down Rhino Transport
% ----------------------------------------------------------------------------

\section{The Optimal Method for Airlifting Rhinos}
\textbf{Authors:} Robin W. Radcliffe et al. \\
\textit{Award:} Ig Nobel Prize in Transportation (2021)

\subsection*{The Question}
Conservation efforts sometimes require transporting anesthetized rhinoceroses by helicopter. The traditional method involves suspending the rhino in a stretcher. But is this optimal?

\subsection*{The Study}
Researchers compared physiological parameters in black rhinos suspended by their feet (upside down) versus lying on their side:
\begin{itemize}
    \item Heart rate
    \item Blood oxygen levels
    \item Respiratory function
    \item General rhino dignity
\end{itemize}

\subsection*{Results}
Counterintuitively, \textbf{hanging rhinos upside down by their feet is safer} than lying them on their side. The inverted position allows better lung expansion and oxygen circulation.

\subsection*{Practical Implications}
Conservation teams worldwide now routinely airlift rhinoceroses inverted, dangling from helicopters like 2-ton Christmas ornaments. The rhinos, when they wake up, have no comment.

% ----------------------------------------------------------------------------
% Entry 20: Swimming in Syrup
% ----------------------------------------------------------------------------

\section{Do Humans Swim Faster in Syrup or Water?}
\textbf{Authors:} Brian Gettelfinger, Edward Cussler \\
\textit{Award:} Ig Nobel Prize in Chemistry (2005)

\subsection*{The Question}
Intuition suggests that swimming through a viscous fluid should be harder and slower than swimming through water. But is intuition correct?

\subsection*{The Experiment}
Researchers filled a swimming pool with guar gum solution, creating a fluid approximately twice as viscous as water. Competitive swimmers then performed timed trials in both regular water and syrup.

\subsection*{Results}
\textbf{Swimmers moved at the same speed in both fluids.}

This counterintuitive result can be understood through the Scallop Theorem: at low Reynolds numbers, increased viscous drag is exactly compensated by increased propulsive force from the swimmer's movements.

\subsection*{Mathematical Analysis}
For a swimmer of characteristic length $L$ moving at velocity $v$ in fluid of viscosity $\mu$:
\begin{equation}
    \text{Re} = \frac{\rho v L}{\mu}
\end{equation}

At the Reynolds numbers relevant to human swimming ($\text{Re} \sim 10^6$), inertial effects dominate, but propulsive and resistive forces scale similarly with viscosity.

% ----------------------------------------------------------------------------
% Entry 21: Butt-Breathing Mammals
% ----------------------------------------------------------------------------

\section{Mammals Can Breathe Through Their Intestines}
\textbf{Authors:} Ryo Okabe et al. (Tokyo Medical and Dental University) \\
\textit{Award:} Ig Nobel Prize in Physiology (2024)

\subsection*{The Question}
Can mammals absorb oxygen through their digestive systems in an emergency?

\subsection*{The Experiment}
Researchers demonstrated that mice, pigs, and rats can survive potentially lethal low-oxygen conditions by receiving oxygenated liquid through their rectums.

The procedure, called Enteral Ventilation via Anus (EVA), delivers perfluorocarbon liquid saturated with oxygen.

\subsection*{Results}
Animals receiving EVA showed:
\begin{itemize}
    \item Significantly improved blood oxygen levels
    \item Extended survival time in hypoxic conditions
    \item Minimal apparent discomfort
\end{itemize}

\subsection*{Medical Implications}
This research suggests that in emergency situations where traditional ventilation is impossible, rectal oxygenation could be a life-saving alternative. Clinical trials in humans are... contemplated.

% ----------------------------------------------------------------------------
% Entry 22: Dead Fish Swimming Upstream
% ----------------------------------------------------------------------------

\section{Dead Fish Can Swim Upstream}
\textbf{Authors:} Erik Henrikson et al. \\
\textit{Award:} Ig Nobel Prize in Physics (2024)

\subsection*{The Phenomenon}
Researchers observed that dead trout, when placed in flowing water with the right characteristics, will ``swim'' upstream.

\subsection*{The Mechanism}
The vortex streets created by flowing water around the fish's flexible body cause oscillations that propel the carcass against the current. This passive locomotion requires:
\begin{enumerate}
    \item Appropriate flow velocity
    \item Body flexibility within a specific range
    \item Periodic vortex shedding at natural body frequency
\end{enumerate}

\subsection*{Mathematical Description}
The propulsive force arises from the phase relationship between body oscillation and vortex shedding:
\begin{equation}
    F_{\text{thrust}} = \frac{1}{2}\rho C_L A v^2 \sin(\omega t + \phi)
\end{equation}

When $\phi$ is in the correct range, the time-averaged thrust exceeds drag.

\subsection*{Implications}
This research explains why dead salmon are occasionally found far upstream and suggests that fish have evolved body mechanics optimized for both active swimming and passive exploitation of flow energy.

\section{Why Wombat Poop is Cube-Shaped}
\textbf{Authors:} Patricia Yang, Alexander Lee, Miles Chan, Alynn Martin, Ashley Edwards, Scott Carver, David Hu \\
\textit{Award:} Ig Nobel Prize in Physics (2019)

\subsection*{The Mystery}
Wombats are the only known animals to produce cube-shaped feces. For years, this phenomenon puzzled biologists. Did the cubes form at the exit point? Did wombats have square anuses?

\subsection*{The Investigation}
The research team, led by Patricia Yang at Georgia Tech, performed dissections on wombat cadavers. They discovered something remarkable: wombat intestines are extraordinarily long---about 10 meters, compared to just 7 meters in humans.

More importantly, the cubes form in the last meter of the intestine (the distal colon), not at the exit point. The key lies in the intestinal tissue itself.

\subsection*{The Mechanism}
The wombat's distal colon has non-uniform elasticity. Some sections are rigid while others are soft. This variation in muscle thickness around the circumference shapes the feces into cubes as they pass through.

The intestinal contractions mold the material much like a die press, with the rigid sections creating the flat faces of the cube.

\subsection*{Why Cubes?}
Wombats use their feces to mark territory. Cube-shaped droppings don't roll away on sloped terrain, making them more effective territorial markers. Evolution has literally shaped the wombat's gut for optimal poop placement.

\subsection*{Bonus Achievement}
This was Patricia Yang and David Hu's \textit{second} Ig Nobel Prize. They previously shared the 2015 Physics Prize for discovering that nearly all mammals---from mice to elephants---empty their bladders in approximately 21 seconds.

\section{Necrobotics: The Dead Spider Gripper}
\textbf{Authors:} Te Faye Yap, Zhen Liu, Anoop Rajappan, Trevor Shimokusu, Daniel Preston \\
\textit{Award:} Ig Nobel Prize in Mechanical Engineering (2023)

\subsection*{The Innovation}
Engineers at Rice University created a new field they call ``necrobotics''---using biotic (previously living) materials as robotic components. Their first prototype? A robot gripper made from a dead wolf spider.

\subsection*{The Science Behind Spider Legs}
Unlike most animals, spiders don't use antagonistic muscle pairs (one muscle to extend, another to flex). Instead, they use:
\begin{itemize}
    \item Flexor muscles to curl their legs inward
    \item Hydraulic pressure (from internal fluids) to extend their legs outward
\end{itemize}

This is why dead spiders curl up---without hydraulic pressure, only the flexor muscles remain contracted.

\subsection*{The Design}
With just a syringe and some superglue, the team reanimated the spider's hydraulic system:
\begin{enumerate}
    \item Insert a needle into the spider's prosoma (body)
    \item Apply positive air pressure: legs extend
    \item Release pressure: legs curl and grip
\end{enumerate}

\subsection*{Performance Specifications}
The dead spider gripper achieved remarkable results:
\begin{itemize}
    \item Could lift over 130\% of its own body weight
    \item Remained functional for 1,000 open-and-close cycles
    \item Perfect for delicate micromanipulation tasks
\end{itemize}

\subsection*{Advantages Over Traditional Grippers}
Mechanical grippers have many moving parts prone to failure. Spider grippers are:
\begin{itemize}
    \item Cheap (spiders are abundant)
    \item Biodegradable (can be composted after use)
    \item Naturally sized for micro-scale work
    \item Reduces electronic waste
\end{itemize}

The researchers noted that this represents just the beginning of necrobotics---other deceased organisms with useful mechanical properties await discovery.

\section{Constipated Scorpions Still Find Love}
\textbf{Authors:} Solimary García-Hernández, Glauco Machado \\
\textit{Award:} Ig Nobel Prize in Biology (2022)

\subsection*{The Survival Mechanism}
When attacked by predators, some scorpion species can detach part of their tail to escape---similar to how lizards drop their tails. But for scorpions, this comes with a permanent cost.

\subsection*{The Complication}
The scorpion's anus is located on the tail segment that gets shed. Once detached, it doesn't grow back. This means the scorpion cannot defecate for the rest of its life.

The feces accumulate internally, causing progressive constipation that eventually proves fatal---but the scorpion can survive for several months in this increasingly uncomfortable state.

\subsection*{The Research Question}
The Brazilian research team asked: Does severe constipation affect the scorpion's mating prospects? One might expect that lugging around an ever-growing internal load would impair mobility and reduce success in the competitive world of scorpion courtship.

\subsection*{The Findings}
Despite losing a significant portion of their digestive tract and their ability to defecate, the constipated scorpions were still able to find mates and copulate successfully.

Their paper, titled ``Short- and Long-Term Effects of an Extreme Case of Autotomy: Does `Tail' Loss and Subsequent Constipation Decrease the Locomotor Performance of Male and Female Scorpions?'' was published in \textit{Integrative Zoology}.

\subsection*{The Takeaway}
Even in the animal kingdom, chronic constipation is no obstacle to a fulfilling love life. Though one must wonder if the scorpions' partners noticed anything... different.

%\chapter{Life, the Universe, and Everything Else}

%\section{The Origin of the Litre}
%\textbf{Author:} Ken Woolner \\
%\textit{Original Source:} CHEM 13 News (April 1978).
%
%\subsection*{Abstract}
%We correct a common historical misconception regarding the metric unit of volume.
%
%It is generally believed that the "litre" is derived from the French \textit{litron}, an old measure of capacity. This is incorrect. The unit is actually named after \textbf{Claude Émile Jean-Baptiste Litre} (1716--1778), a famous manufacturer of wine bottles.
%
%Litre was a perfectionist. He observed that wine bottles varied wildly in size, leading to disputes in taverns. He proposed a standard bottle size, which he modestly named the "Litre."
%Remarkably, Litre was also the first to propose that the volume of a liquid changes with temperature (after a bottle of his best Bordeaux exploded in the sun).
%
%Because the unit is named after a person, the symbol for the litre should be capitalized ({\bf L}), a convention now adopted by many scientific bodies to avoid confusion between the lowercase letter {\bf l} and the number {\bf 1}.

\backmatter

\end{document}